\documentclass[
 aps,prb,reprint,superscriptaddress,
 amsmath,amssymb,amsfonts,
 floatfix,longbibliography
]{revtex4-2}
\usepackage{graphicx}
\usepackage{bm}
\usepackage{appendix}
\usepackage[dvipsnames]{xcolor}
\usepackage{hyperref}
\hypersetup{
    colorlinks=true,
    citecolor=Green,
    linkcolor=Red,
    urlcolor=NavyBlue
}
\usepackage[capitalise]{cleveref}
\usepackage[italicdiff]{physics}
\usepackage{relsize}
\DeclareUnicodeCharacter{2212}{-}

\begin{document}

\title{Probing the pseudogap and beyond: Examining single-particle properties of the hole- and electron-doped Hubbard model}

\author{Wen O. Wang}
\email{wenwang.physics@gmail.com}
\affiliation{Department of Applied Physics, Stanford University, Stanford, CA 94305, USA}
\affiliation{Stanford Institute for Materials and Energy Sciences,
SLAC National Accelerator Laboratory, 2575 Sand Hill Road, Menlo Park, CA 94025, USA}
\affiliation{Kavli Institute for Theoretical Physics, University of California, Santa Barbara, CA 93106, USA}

\author{Edwin W. Huang}
\affiliation{Department of Physics and Astronomy, University of Notre Dame, Notre Dame, IN 46556, USA}
\affiliation{Stavropoulos Center for Complex Quantum Matter, University of Notre Dame, Notre Dame, IN 46556, USA}

\author{Brian Moritz}
\affiliation{Stanford Institute for Materials and Energy Sciences,
SLAC National Accelerator Laboratory, 2575 Sand Hill Road, Menlo Park, CA 94025, USA}

\author{Thomas P. Devereaux}
\email{tpd@stanford.edu}
\affiliation{Stanford Institute for Materials and Energy Sciences,
SLAC National Accelerator Laboratory, 2575 Sand Hill Road, Menlo Park, CA 94025, USA}
\affiliation{
Department of Materials Science and Engineering, Stanford University, Stanford, CA 94305, USA}
\affiliation{
Geballe Laboratory for Advanced Materials, Stanford University, Stanford, CA 94305, USA}

\begin{abstract}
We compute high-resolution angle-resolved photoemission spectroscopy of the Hubbard model using the unbiased determinant quantum Monte Carlo algorithm, revealing an asymmetry between electron and hole doping.
Electron doping exhibits more coherent quasiparticles and stronger antiferromagnetic correlations compared to hole doping.  
At low doping, a nodal-antinodal dichotomy on the Fermi surface is observed, similar to cuprate experiments. 
The dichotomy reflects the momentum dependence of the Mott gap, as manifested in both the spectral function and the self-energy.
For hole doping, we observe a transition towards the pseudogap, without signature of pocket formation.
The simulated nuclear magnetic resonance pseudogap temperatures do not necessarily agree with the temperature determined by spectroscopy. These findings collectively suggest the pseudogap is a smooth crossover driven by strong correlations.
\end{abstract}

\maketitle

\section{Introduction}

Angle-resolved photoemission spectroscopy (ARPES) \cite{RevModPhys.93.025006,Vishik_2018} and scanning-tunneling microscopy (STM)~\cite{RevModPhys.79.353}, which directly probe single-particle properties, have been powerful tools for investigating exotic phenomena in complex quantum materials, such as the high-$T_c$ cuprates.
The rich phase diagram of the cuprates exhibits an asymmetry between hole and electron doping, including antiferromagnetism near half-filling, superconductivity at larger doping, a pseudogap regime, a broad range of strange-metal physics in the normal state, as well as various intertwined orders~\cite{Keimer2015,RevModPhys.87.457}. 
In the normal state, extensive regions of the phase diagram are characterized by low-energy gaps that open at specific portions of the Fermi surface. 
For electron-doped cuprates such as Nd$_{2−x}$Ce$_x$CuO$_4$, gaps appear at ``hot spots'', where the Fermi surface crosses the antiferromagnetic (AFM) zone boundary~\cite{RevModPhys.93.025006,PhysRevLett.87.147003,PhysRevLett.88.257001,RevModPhys.75.473,PhysRevLett.94.047005,PhysRevB.75.224514,RevModPhys.82.2421,doi:10.1073/pnas.1816121116,annurev:/content/journals/10.1146/annurev-conmatphys-031218-013210,Xu2023}, and are clearly associated with the AFM order.
In hole-doped cuprates, 
a gap-like feature persists above the superconducting transition temperature, observed as a suppression of the local density of states near the Fermi level in STM experiments~\cite{PhysRevLett.80.149,ye2023visualizing}, and is referred to as the pseudogap.
From ARPES experiments, the pseudogap is characterized by the coexistence of coherent spectral weight in the nodal region (around $\mathbf{k}=(\pi/2,\pi/2)$) with a gap in the antinodal region (around $\mathbf{k}=(\pi,0)$), producing a ``Fermi arc,'' or a ``nodal-antinodal dichotomy''.
The pseudogap emerges below a characteristic temperature $T^*$, which decreases with increased doping.
Unlike the electron-doped case, it is unclear whether the hole-doped pseudogap is directly tied to AFM correlations, rendering its origin more elusive and motivating extensive research efforts~\cite{Keimer2015,RevModPhys.87.457,RevModPhys.93.025006}.

A number of experimental techniques beyond ARPES and STM have also identified features associated with the pseudogap.
Historically, nuclear magnetic resonance (NMR) experiments provided the earliest evidence for the pseudogap, observed in the temperature dependence of both the Knight shift and the spin-relaxation rate~\cite{PhysRevLett.63.1700,PhysRevB.58.R5960,Tom_Timusk_1999}.
Various transport properties, including both charge and thermal transport, exhibit pseudogap features~\cite{annurev:/content/journals/10.1146/annurev-conmatphys-031218-013210}.
For example, in YBa$_2$Cu$_3$O$_{7−x}$, the pseudogap manifests as deviations from the linear-$T$ dependence of the dc resistivity~\cite{PhysRevLett.70.3995} and suppression of the low-frequency interplane conductivity~\cite{PhysRevLett.71.1645}.
The ab-plane optical conductivity also shows pseudogap behavior in underdoped cuprates such as Bi$_2$Sr$_2$CaCu$_2$O$_8$~\cite{PhysRevLett.77.3212}.
As thermodynamic indicators, a sharp peak in the normal-state electronic specific heat $C_{el}$ near the pseudogap termination doping, and a logarithmic divergence of $C_{el}/T$ as $T$ approaches zero are observed in La$_{2−x}$Sr$_x$CuO$_4$, suggesting a possible link between the pseudogap and a quantum critical point~\cite{Michon2019}.
Neutron scattering experiments reveal a distinct hourglass-shaped spin excitation spectrum within the pseudogap regime~\cite{Tranquada2004,Hayden2004,
Vignolle2007,Xu2009,PhysRevLett.102.167002}. 
Raman spectroscopy identifies carrier-related properties in the B$_{2g}$ channel but not the B$_{1g}$ channel as they reflect different regions of the Brillouin zone, which is another hallmark of the pseudogap~\cite{PhysRevLett.78.4837}.

We note that these experimental observations reveal that the pseudogap features do not correspond to a single characteristic temperature, but instead span multiple, probe-dependent temperature scales. Some features are linked to a ``high-temperature'' pseudogap, while others are associated with a lower-temperature pseudogap. These different pseudogaps are likely characterized by distinct gap energy scales. In particular, NMR Knight shift measurements have identified two distinct crossover temperatures~\cite{PhysRevB.58.R5960}, each of which aligns more closely with different features observed in other experiments.
Indeed, recent NMR studies suggest that the lower-temperature crossover may be less directly tied to the primary pseudogap onset and instead track the emergence of short-range charge-density-wave correlations~\cite{Zhou2025,zhou2025robust}.

From the theoretical side, numerous proposals exist for the origin of pseudogap phenomena,
including various types of orders or stripes~\cite{PhysRevB.56.6120,PhysRevX.4.031017,RevModPhys.87.457,malcolms2024rise,PhysRevB.109.045116,Keimer2015}, 
preformed pairs~\cite{Emery1995,Tom_Timusk_1999,Keimer2015}, time-reversal symmetry breaking~\cite{PhysRevB.55.14554,PhysRevLett.83.3538,PhysRevB.73.155113,PhysRevB.81.064515}, spin-liquid-based models~\cite{PhysRevB.65.165113,RevModPhys.78.17,PhysRevB.93.165139,PhysRevB.94.205117,doi:10.1073/pnas.2302701120,Christos2024},
and fermiology, such as Lifshitz transitions~\cite{PhysRevX.8.021048,PhysRevLett.120.067002,PhysRevLett.114.147001,PhysRevB.96.094525,Doiron-Leyraud2017,PhysRevResearch.2.033067}.
Meanwhile, other perspectives emphasize the importance of strong correlation effects that cannot be adequately captured within effective weak-coupling or mean-field theories, 
for example proximity to the Mott insulator or some underlying quantum critical point, without explicitly invoking symmetry breaking or competing orders~\cite{Keimer2015,PhysRevLett.102.056404,PhysRevB.82.134505,PhysRevB.98.195109,horio2018common}.
In these scenarios, the evolution of quasiparticle coherence may be key to understanding the pseudogap.

Despite extensive experimental and theoretical effort, the origin of the pseudogap remains unclear. A foundational debate, which remains ongoing, involves whether the pseudogap onset temperature $T^*$ is (1) a sharply defined phase transition, (2) a sharp transition in a disorder-free system that becomes rounded into a crossover in the presence of disorder, or (3) truly a crossover, even in the absence of disorder~\cite{RevModPhys.87.457,RevModPhys.93.025006}.
Comparing pseudogap temperatures across different experiments and simulations helps address this debate. 
If the pseudogap is a phase of matter characterized by explicit symmetry breaking,
one expects different experimental techniques to detect consistent pseudogap transition temperatures $T^*$.
If the pseudogap is instead a smooth crossover driven by strong correlations, then the definition of $T^*$ is naturally ambiguous and different experiments don't have to agree.

To investigate the pseudogap temperature scales, provide unbiased numerical evidence for testing different theoretical scenarios, and ultimately shed light on the pseudogap's nature,
we conduct large-scale, systematic numerical simulations of the Hubbard model, which has been widely studied numerically~\cite{annurev:/content/journals/10.1146/annurev-conmatphys-031620-102024,annurev:/content/journals/10.1146/annurev-conmatphys-090921-033948,PhysRevLett.72.705,PhysRevLett.79.1122,Maier2000,PhysRevLett.86.139,M.Jarrell_2001,PhysRevLett.91.017002,PhysRevLett.92.126401,PhysRevB.72.155105,PhysRevB.74.125110,10.1063/1.2199446,PhysRevB.73.165114,PhysRevLett.97.036401,Gull_2008,PhysRevB.80.165126,PhysRevLett.102.206407,PhysRevB.80.064501,PhysRevLett.102.056404,Ferrero_2009,PhysRevB.82.134505,PhysRevLett.108.076401,PhysRevLett.108.216401,Sordi2012,PhysRevLett.110.216405,PhysRevB.87.041101,PhysRevB.90.035111,PhysRevB.90.224507,PhysRevLett.114.236402,PhysRevB.93.081107,Chen2017,PhysRevB.96.041105,PhysRevB.98.195109,PhysRevX.8.021048,PhysRevLett.120.067002,PhysRevResearch.2.033067,PhysRevB.101.115141,huang2020strong,Krien2022,PhysRevB.106.245132,PhysRevB.107.085126,doi:10.1126/science.ade9194,PhysRevLett.133.166501,10.21468/SciPostPhys.16.2.059} to understand interacting electrons, particularly in cuprate superconductors; since different numerical approaches rely on distinct approximations, they can yield different or even seemingly conflicting conclusions, making it essential to establish a controlled reference standard for comparing and reconciling results across methods.
We use the numerically exact determinant quantum Monte Carlo (DQMC)~\cite{DQMC1,DQMC2} algorithm, combined with maximum entropy (MaxEnt) analytic continuation~\cite{maxent1,maxent2}, to compute the single-particle spectral function and other properties of the Hubbard model with hole and electron doping away from the half-filled Mott insulating phase. 
In particular, the momentum resolution of the single-particle spectral function $A(\mathbf{k},\omega)$ is significantly enhanced through the application of twisted boundary conditions~\cite{PhysRevB.53.6865,PhysRevE.64.016702,doi:10.1143/JPSJ.76.074719,PhysRevB.94.085103,PhysRevB.96.205145,PhysRevB.98.075156,huang2020strong,Mai2025}.

Our state-of-the-art DQMC study provides a systematic and unbiased characterization of electron-hole asymmetry and pseudogap physics in the doped Hubbard model.
We assemble an internally consistent dataset that combines extensive twisted-boundary-condition calculations with single-particle spectra, spin correlations and NMR response, and thermodynamic and transport observables across a broad parameter regime.
We focus on investigating and benchmarking the high-temperature pseudogap observed in experiments, which is associated with temperatures of at least several hundred kelvin.
When quasiparticles are coherent, as on the electron-doped side, fermiology together with $(\pi,\pi)$ spin fluctuations is sufficient to produce ``hot spots'' on the Fermi surface, but no appreciable pseudogap. In contrast, on the hole-doped side, proximity to the Mott gap renders quasiparticles incoherent near the antinodes, as manifested by a $\Re \Sigma \sim 1/\omega$ self-energy, giving rise to ``Fermi arcs'' and a pseudogap.
In either case, the magnetic correlation length remains short and does not control the pseudogap. 

The results are organized into four sections.
In \cref{magcohere}, we compare the magnetism and coherence between hole and electron doping.
Specifically, electron doping produces sharper Fermi surfaces and ``hot spots'', indicating the presence of more coherent quasiparticles and stronger AFM correlations relative to the hole-doped case.
In \cref{secdichotomy}, we focus on the nodal-antinodal dichotomy at low doping, which is asymmetric between electron and hole doping.
In particular, for hole doping, we observe the emergence of ``Fermi arcs'', and the temperature dependence of the spectral weight indicates a transition toward the pseudogap.
Leveraging the unbiased nature of our DQMC simulations, we find that the observed nodal-antinodal dichotomy arises from the momentum dependence of the Mott gap, which is manifested in the behavior of both the spectral function and self-energy near the Mott gap.
This finding contrasts with the notion that ``Fermi arcs'' arise from pockets whose back side is damped~\cite{PhysRevB.73.174501,PhysRevB.81.115129,Christos2024}, as we observe no explicit signature of pocket formation in the real part of the self-energy.
In \cref{sec:NMR}, the pseudogap termination temperature determined by NMR measurements does not necessarily coincide with that determined from the single-particle properties, 
emphasizing the ambiguity in defining a pseudogap temperature $T^*$ using different measurements. This suggests that there is not a well-defined energy scale tied to broken symmetry.
As our simulations of the Hubbard model include no disorder, our results indicate that the pseudogap manifests as a true smooth crossover and highlight the strong coupling nature of this phenomenon.
In \cref{transportandthermodyanmicssec}, we analyze transport and thermodynamic signatures associated with the pseudogap and find no evidence of nonanalyticity, consistent with a crossover interpretation.

Throughout the main text, we express energies in units of the nearest-neighbor hopping energy $t$.
We fix the on-site interaction at $U=6t$, the next-nearest-neighbor hopping at $t'=-0.25t$, and perform our simulations on $8\times 8$ lattices, unless specified otherwise.
For low temperatures $T/t \leq 1/3$, we combine data from simulations with $15$ different twisted boundary conditions (including the untwisted one) in order to achieve a momentum resolution equal to that of a $64 \times 64$ simulation.
Additionally, we consider a third-nearest-neighbor hopping $t''$, set to either $0$ or $0.15t$, and discuss any resulting differences where relevant.
The formalism and other details are provided in \cref{supp1,supp2}, and supplementary data and the MaxEnt systematic error analysis are available in \cref{suppdata}.

\section{Magnetism and Coherence} \label{magcohere}

\begin{figure}[tbp]
    \centering
    \includegraphics[width=\linewidth]{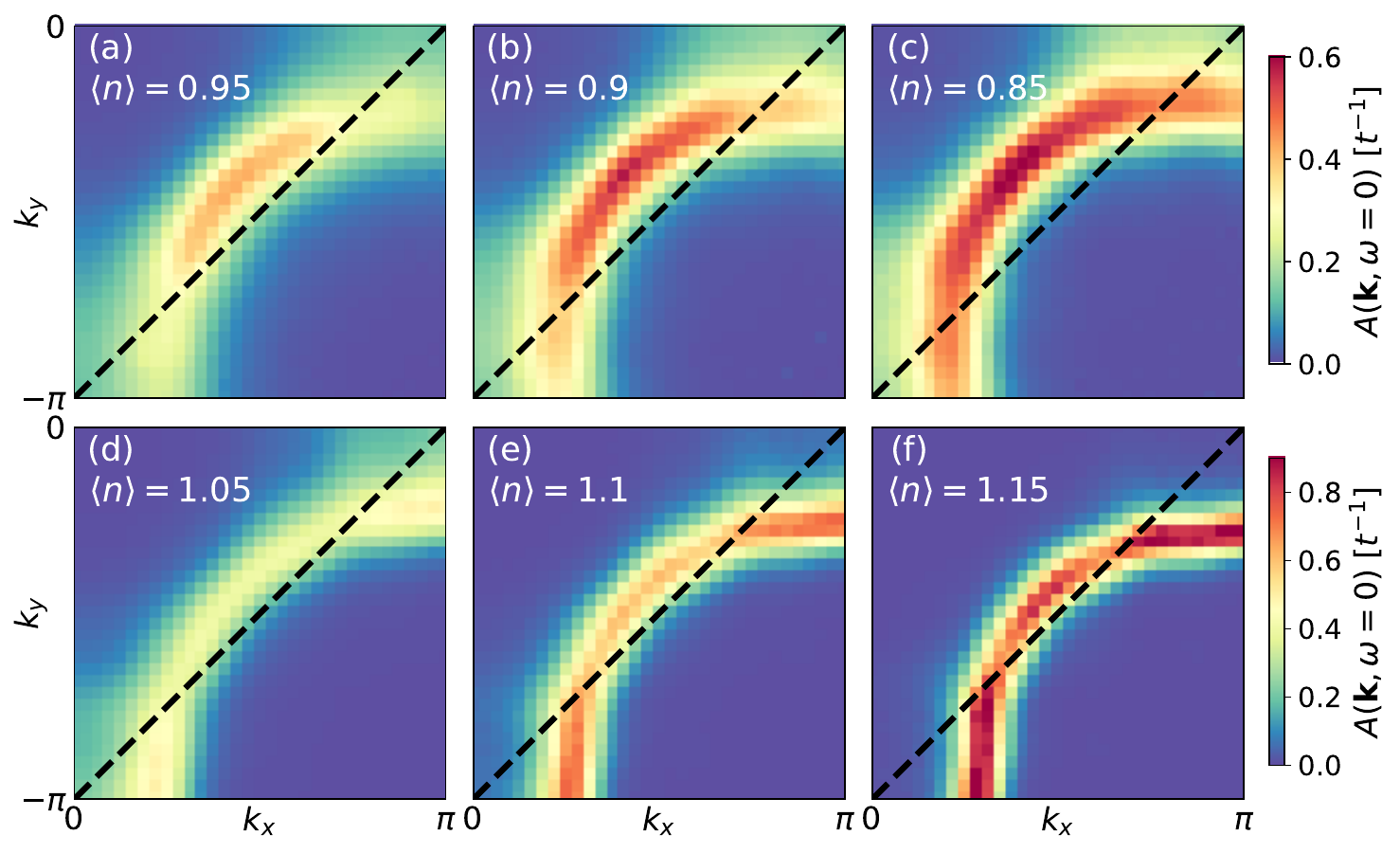}
    \caption{Interacting Fermi surfaces illustrated by the spectral function $A(\mathbf{k},\omega=0)$. Dashed lines mark the AFM zone boundary. Third-nearest-neighbor hopping $t''=0.15t$ and temperature $T=t/4$.
}
    \label{fig:FS}
\end{figure}

In this section we explore connections between antiferromagnetic correlations and electron coherence observed in the single-particle spectral function, contrasting hole-doped with electron-doped scenarios.
Figure~\ref{fig:FS} shows the spectral function at the Fermi level, $A(\mathbf{k}, \omega=0)$.
We fix the next- and third-nearest-neighbor hoppings ($t'$, $t''$) and vary the electron density (filling) $\langle n \rangle=1-x$ to distinguish between hole doping ($x>0$) and electron doping ($x<0$).
Here, the interacting Fermi surface is defined as the momentum points of highest intensity. As doping increases, this Fermi surface shows higher intensity and becomes narrower in momentum space, suggesting enhanced coherence. 
Comparing electron and hole doping at the same doping density $|x|$, we observe two differences: first, the Fermi surface is sharper with larger spectral weight on the electron-doped side, characterizing more coherent quasiparticles; second, the angular dependence of the spectral weight differs. 

This angular dependence we observe is consistent with the nodal-antinodal dichotomy in experiments:
the incoherent regions of the Fermi surface (i.e., with suppressed spectral weight) lie in the antinodal region for hole doping, but in the nodal region for electron doping.
The suppressed intensity near $\mathbf{k} = (\pi,0)$ in Figs.~\ref{fig:FS}~(a) to (c) matches a ``Fermi arc'', typically observed in hole-doped cuprates~\cite{doi:10.1126/science.1103627,Lee2007,Hashimoto2010,RevModPhys.93.025006}. 
Under low ($x=-0.05$) electron doping in Fig.~\ref{fig:FS}~(d), 
the suppression instead appears near $\mathbf{k} = (\pi/2,\pi/2)$.
With further electron doping, reduced intensity moves away from the nodal point and forms ``hot spots'' where the AFM zone boundary intersects the Fermi surface [see Fig.~\ref{fig:FS}~(f)], qualitatively similar to observations in electron-doped cuprates~\cite{RevModPhys.93.025006,PhysRevLett.87.147003,PhysRevLett.88.257001,RevModPhys.75.473,PhysRevLett.94.047005,PhysRevB.75.224514,RevModPhys.82.2421,doi:10.1073/pnas.1816121116,Xu2023}.
In contrast,
the AFM zone boundary holds no similar special significance for hole doping. 
As shown in Appendix \ref{hubbardandspectral}, we observe similar behavior in \cref{proxybetaover2} for $-\beta G(\mathbf{k}, \tau=\beta/2)$, the imaginary-time Green’s function multiplied by the inverse temperature, which serves as a proxy for $A(\mathbf{k},\omega=0)$ and does not require analytic continuation.
This asymmetry, in which hot spots appear only for electron doping rather than hole doping, suggests a more dominant effect of $(\pi,\pi)$ scattering with electron doping than hole doping.

\begin{figure}[htbp]
    \centering
    \includegraphics[width=\linewidth]{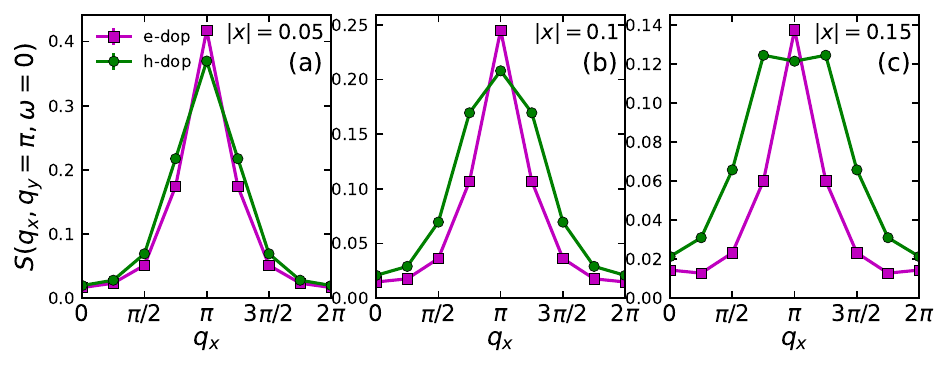}
    \caption{Spin structure factor $S(\mathbf{q},\omega=0)$ [in arbitrary units] along the momentum cut $q_y=\pi$, 
    comparing hole doping (green lines with circles) and electron doping (purple lines with squares) at doping (a) $|x|=0.05$, (b) $0.1$, and (c) $0.15$. The parameters are the same as in \cref{fig:FS}. 
}
    \label{fig:Sqw}
\end{figure}

To directly compare magnetic correlations, the zero-energy spin structure factor $S(\mathbf{q},\omega=0)$, defined in \cref{nmrmeasurements}, 
is plotted in \cref{fig:Sqw}. 
Increasing either hole or electron doping leads to a reduction in antiferromagnetism, evident from the decreased $S(\mathbf{q}=(\pi,\pi),\omega=0)$.
At $x=0.15$ hole doping in \cref{fig:Sqw}~(c), the emergence of a double-peak profile indicates incommensurate spin correlations, while
for electron doping a single peak at $\mathbf{q} = (\pi,\pi)$ persists across all considered doping levels.
Moreover, the $(\pi,\pi)$ peak is sharper with higher magnitude for electron doping than for hole doping at equivalent doping levels. These observations suggest stronger AFM spin correlations and more pronounced $(\pi,\pi)$ scattering under electron doping, aligning with observations of ``hot spots'' in \cref{fig:FS}. 
The findings agree with both prior DQMC studies~\cite{doi:10.1126/science.aak9546,Huang2018} and neutron scattering experiments in cuprates~\cite{Tranquada2004,Hayden2004,PhysRevB.71.024522,
Vignolle2007,Xu2009,PhysRevLett.102.167002,doi:10.1146/annurev-conmatphys-031115-011401,doi:10.1143/JPSJ.81.011007,Motoyama2007,PhysRevLett.90.137004,RevModPhys.82.2421}.

\begin{figure*}[htbp]
    \centering
    \includegraphics[width=\linewidth]{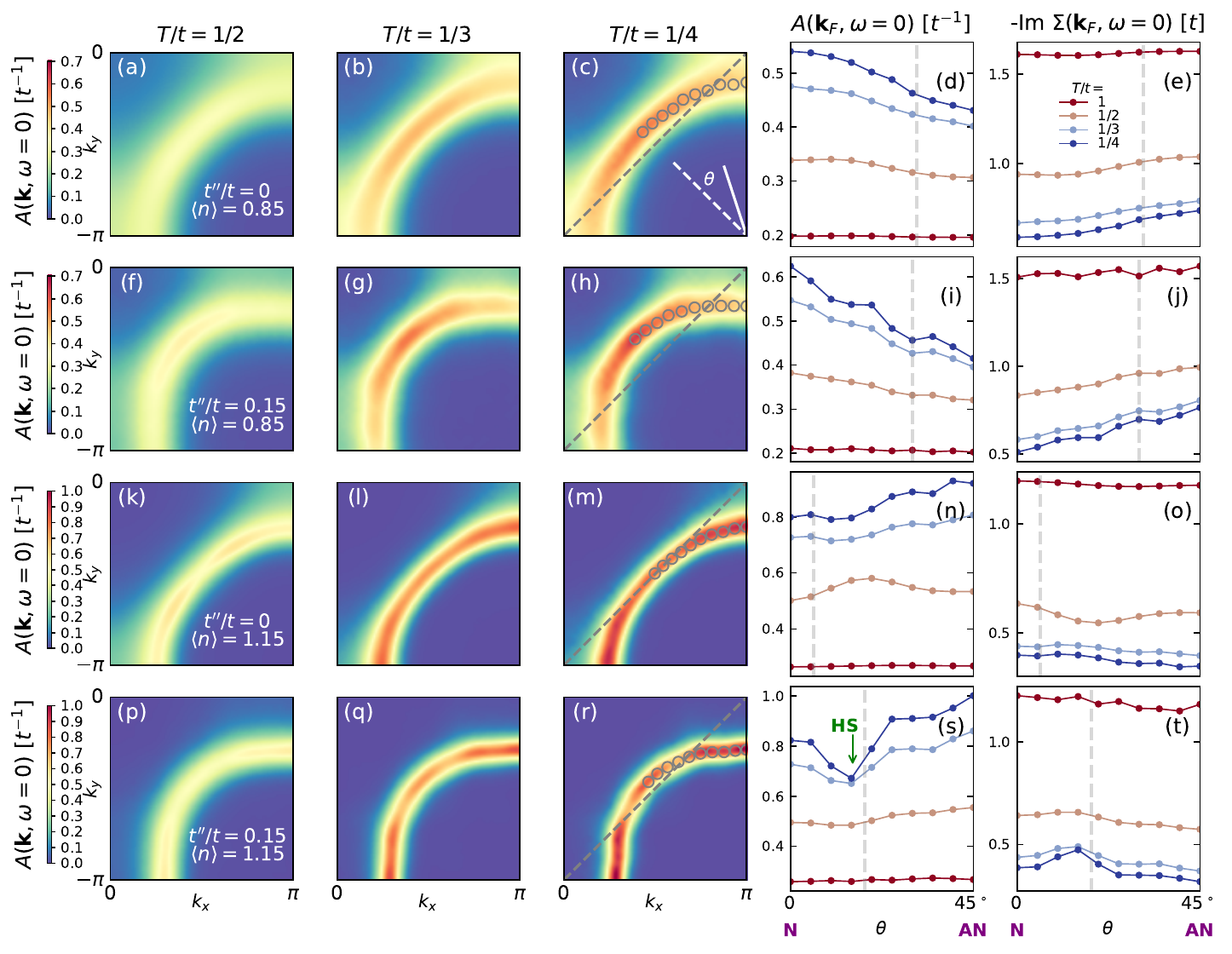}
    \caption{
    Temperature evolution of the Fermi surfaces, as well as spectral weight and the imaginary part of the self-energy around the Fermi surface for $|x|=0.15$ with both hole and electron doping.
    Each row corresponds to a fixed third-nearest neighbor hopping $t''$.
    The three left columns show the temperature dependence of the Fermi surface, illustrated by the spectral function $A(\mathbf{k},\omega=0)$ interpolated in momentum space.
    In the third column ($T/t=1/4$), gray dashed lines indicate the AFM zone boundary, and gray empty circles highlight the Fermi surfaces, taken from peak $A(\mathbf{k},\omega=0)$ intensities for each momentum cut, where $10$ $\mathbf{k}_F$ points are selected at uniformly spaced angular orientations $\theta$ from the node (N) to the antinode (AN).
    The two right-hand columns display the angular variation of $A(\mathbf{k},\omega=0)$ and $-\Im \Sigma(\mathbf{k}_F,\omega=0)$ along the Fermi surface at different temperatures, with the Fermi surface locations determined independently at each temperature.
    Grey vertical dashed lines mark the intersections of the $T/t=1/4$ Fermi surface with the AFM zone boundary. 
    The dip in $A(\mathbf{k},\omega=0)$ is highlighted and identified with the ``hot spot'' (HS).
}
    \label{fig:Tdep_highdop}
\end{figure*}

\begin{figure}[htbp]
    \centering
    \includegraphics[width=0.8\linewidth]{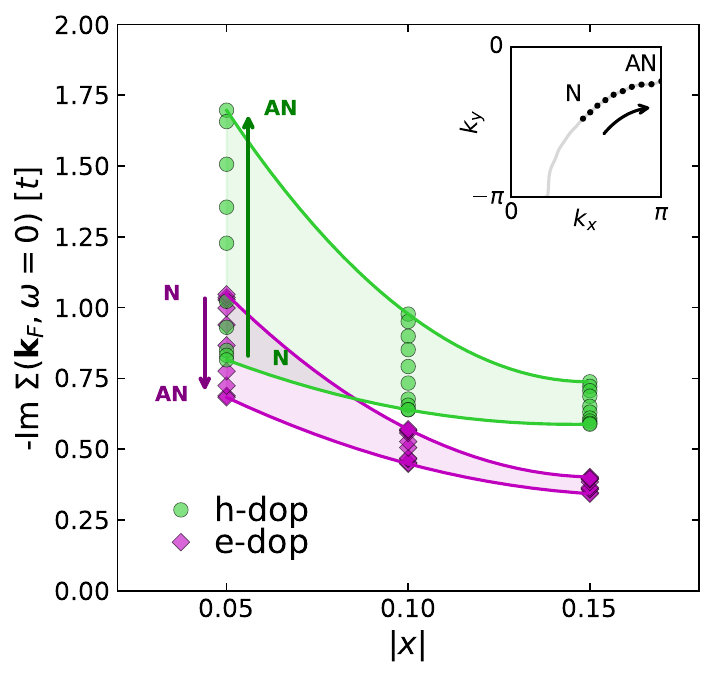}
\caption{Negative imaginary part of the zero-frequency self-energy along the Fermi surface $-\Im \Sigma(\mathbf{k}_F,\omega=0)$ at various hole and electron dopings at a temperature $T/t=1/4$ for $t''/t=0$. 
Colored curves are a visual guide to the maximum and minimum values for hole (green) and electron (purple) doping.
Arrows indicate the general trend of the distribution of the $\mathbf{k}_F$, though not strictly monotonic for all doping values, as exemplified by Fig.~\ref{fig:Tdep_highdop}~(o).  The inset shows a schematic of the $\mathbf{k}_F$ points selected on the Fermi surface. As in Figs.~\ref{fig:Tdep_highdop} and \ref{fig:Tdep_lowdop}, $10$ points are selected from the nodal (N) to antinodal (AN) region at uniformly spaced angles. 
}
    \label{fig:imaginaryselfenergy}
\end{figure}

Having observed a connection between spin correlations and ``hot spots'' on the Fermi surface, we now focus on $|x|=0.15$ for different temperatures and parameters in \cref{fig:Tdep_highdop}, in order to identify the conditions under which these hot spots emerge. The three left-most columns track the evolution of the spectral function at the Fermi energy ($\omega=0$) for the selected parameters, while the two right-most columns track the angular dependence of the spectral function and of the imaginary part of the self-energy around the Fermi surface, revealing the distribution of single-particle states and the scattering intensity or loss of coherence, respectively, as indicated for the lowest temperature by the light gray circles in \cref{fig:Tdep_highdop}~(c), (h), (m), and (r).
At high temperatures $T/t\gtrsim 1/2$,
the Fermi surfaces, as shown in Fig.~\ref{fig:Tdep_highdop}~(a), (f), (k) and (p), possess a ``fuzzy'' appearance due to thermal excitations, and both $A$ and $-\Im \Sigma$ are nearly uniform as a function of angle $\theta$, especially for $T/t = 1$, as shown by the evolution around the Fermi surface in the right-hand columns.
As the temperature decreases, both $A$ and $-\Im \Sigma$ develop an angular dependence, as the Fermi surfaces become sharper and the spectra approach those seen in Fig.~\ref{fig:FS}.

Addressing hole-doping behavior (shown in the upper two rows), Fig.~\ref{fig:Tdep_highdop}~(a) to (j) reveal ``Fermi arcs'' that emerge at lower $T$, where $A$ is suppressed and $-\Im \Sigma$ is largest near the antinode. 
For electron doping without $t''$, as shown in Fig.~\ref{fig:Tdep_highdop}~(k) to (o), the Fermi surface and the AFM zone boundary are nearly tangent at the nodal point, such that distinct ``hot spots'' fail to emerge. 
Adding $t''/t=0.15$ produces sufficient curvature, such that the AFM zone boundary intersects the Fermi surface, and the ``hot spots'' become readily apparent as illustrated in Fig.~\ref{fig:Tdep_highdop}~(p) to (t). 
As discussed earlier, 
the Fermi surface in the electron-doped case is generally more distinct, with a larger spectral weight and smaller negative imaginary self-energy at $\omega=0$.
This is further shown in the direct comparison of the imaginary self-energy along the Fermi surface in Fig.~\ref{fig:imaginaryselfenergy} for various doping values, suggesting overall a higher degree of coherence for electron doping, which, combined with more pronounced $(\pi,\pi)$ scattering, leads to the formation of ``hot spots''.

\begin{figure}[htbp]
    \centering
    \includegraphics[width=\linewidth]{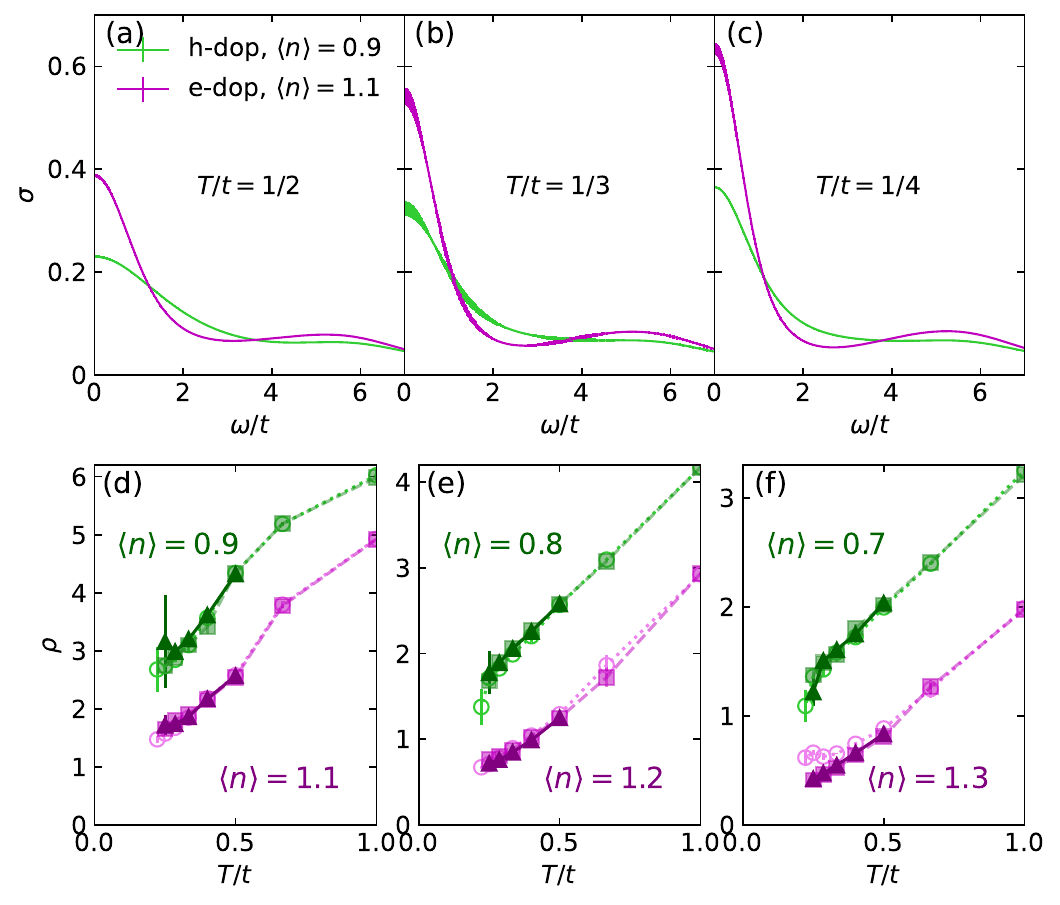}
\caption{Comparison of optical conductivity and resistivity for hole and electron doping with $t''/t=0$.
(a) to (c) display the optical conductivity $\sigma(\omega)$ for $10\%$ hole and electron doping at different temperatures, simulated on an $8 \times 8$ lattice.
(d) to (f) show the temperature dependence of the DC resistivity $\rho(T)$ for three different doping values. Results are shown for three lattice sizes: $8\times 8$ (circles on dotted lines), $10\times 10$ (squares on dashed lines), and $12\times 12$ (triangles on solid lines).
}
    \label{fig:resistivity}
\end{figure}

The finding of stronger coherence for electron doping compared to hole doping is not isolated to spectral data; it is also evident in transport and thermodynamic properties. 
For these analyses, we set $t''=0$, as it is not expected to significantly alter qualitative trends.
Figure~\ref{fig:resistivity} compares the electronic transport,  
showing a sharper low-frequency Drude peak for electron-doping, indicating a longer effective relaxation time, and  
a DC resistivity [$\rho=\sigma^{-1}(\omega=0)$] that is consistently lower by a factor $\sim 2 - 3$ for electron doping at the lowest temperatures.
Additionally, hole doping reproduces a previously reported~\cite{doi:10.1126/science.aau7063} linear-in-$T$ $\rho$, characteristic of strange metallic behavior.
In contrast, electron doping hints at a curvature in $\rho$ at the lowest temperatures, suggesting a potential shift toward Fermi-liquid-like $T^2$ behavior.
This electron-hole contrast is consistent with both experiments and calculations of related models~\cite{Shastry_2018,PhysRevB.72.075108,PhysRevLett.117.197001,PhysRevB.69.024504,PhysRevLett.60.2194,PhysRevB.40.2254,PhysRevLett.69.2975,PhysRevLett.70.3995,PhysRevLett.93.267001,Daou2009,doi:10.1126/science.1165015,doi:10.1073/pnas.1301989110,PhysRevLett.113.177005,PhysRevB.95.224517,Barisic_2019}. In particular, the work of Shastry and Mai~\cite{Shastry_2018} reported a similarly lower and more strongly curved resistivity on the electron-doped side in the $t$-$J$ model.
Finite-size analysis also suggests a higher degree of delocalization for electron doping, as expected for more coherent excitations: while $8\times 8$ is generally sufficient to show no discernible difference from larger lattices, at $30\%$ electron doping in Fig.~\ref{fig:resistivity}~(f) the finite-size effect becomes negligible only for $10\times 10$ and larger.
Inverse diffusivity, detailed in \cref{suppdata:Dinv}, \cref{fig:invdiffusivity}, 
shows similar behaviors in support of these observations.

\begin{figure}[htbp]
    \centering
    \includegraphics[width=\linewidth]{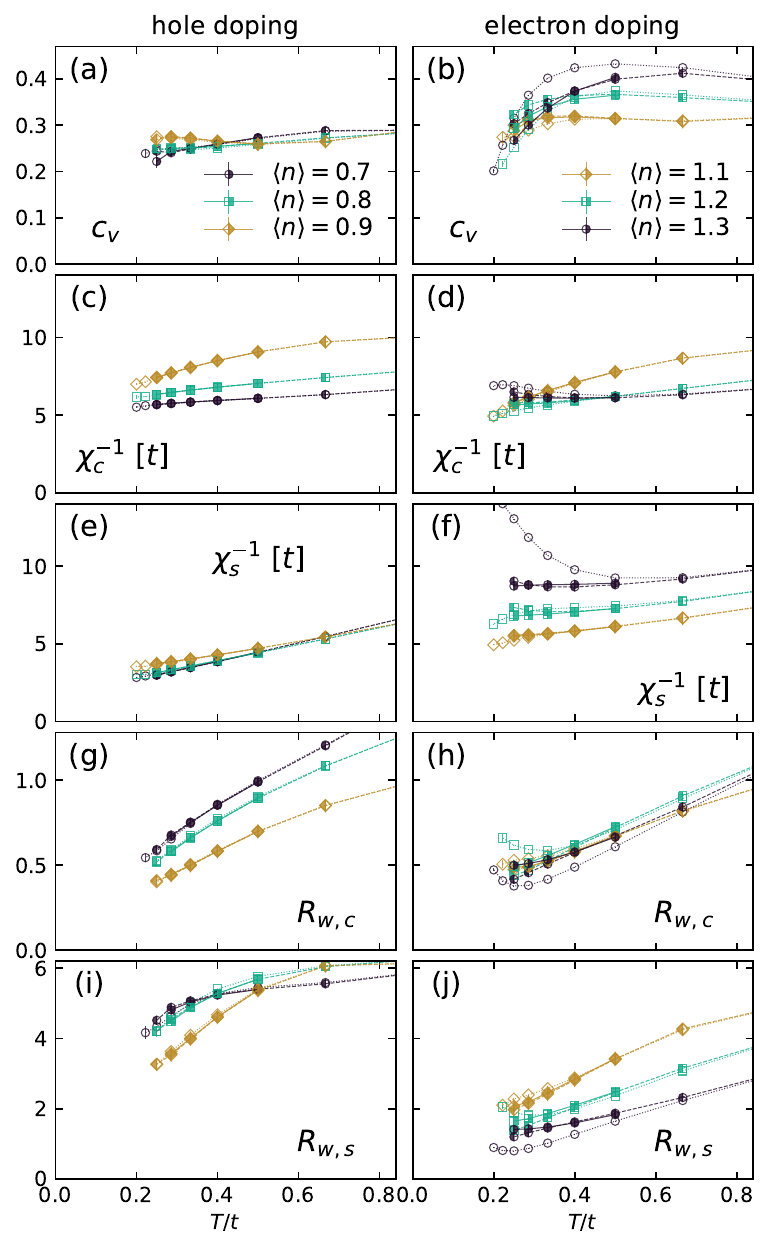}
    \caption{Comparison of hole and electron doping on various thermodynamic properties for $t''/t=0$.
    The properties include specific heat $c_v$ [(a) and (b)], inverse charge compressibility $\chi^{-1}_c$ [(c) and (d)], inverse spin susceptibility $\chi^{-1}_s$ [(e) and (f)], ``charge'' Wilson ratio $R_{w,c}\equiv\pi^2 T\chi_c /(3c_v)$ [(g) and (h)], and ``spin'' Wilson ratio $R_{w,s}\equiv 4\pi^2 T\chi_s /(3c_v)$ [(i) and (j)]. 
    Lattice sizes:  $8\times 8$ (dotted lines with open symbols),  $10\times 10$ (dashed lines with left-filled symbols), and $12\times 12$ (solid lines with right-filled symbols).
    }
    \label{fig:thermodynamics}
\end{figure}

Figure~\ref{fig:thermodynamics} compares several thermodynamic properties, which should exhibit characteristic scaling at temperatures much lower than the Fermi energy~\cite{Coleman_2015} for a Fermi liquid:  the specific heat $c_v$ should exhibit a linear temperature dependence at low temperatures; and charge compressibility $\chi_c$ and spin susceptibility $\chi_s$ are predicted to approach constants.
Derived from these behaviors, both the ``charge'' and ``spin'' Wilson ratios, $R_{w,c}\equiv\pi^2 T\chi_c /(3c_v)$ and $R_{w,s}\equiv 4\pi^2 T\chi_s /(3c_v)$, 
are expected to approach constants.
Figure~\ref{fig:thermodynamics}~(a) and (b) show $c_v$, which for hole doping has a weak temperature dependence but for electron doping shows a stronger low-$T$ trend that is consistent with an approximately linear-in-$T$ behavior upon cooling.
In Fig.~\ref{fig:thermodynamics}~(c) to (f), 
$\chi^{-1}_c$ and $\chi^{-1}_s$ show no significant contrast between hole and electron doping; both are weakly temperature dependent and are consistent with extrapolating to sizable finite intercepts as $T\rightarrow 0$, although electron doping exhibits slightly weaker temperature dependence at sufficiently large sizes especially for $\chi^{-1}_s$.
Consequently, Fig.~\ref{fig:thermodynamics}~(g) to (j) indicate that, for electron doping at sufficiently large sizes, the ``charge'' and ``spin'' Wilson ratios show a clearer tendency to level off as $T\rightarrow 0$.
In particular, the spin Wilson ratio $R_{w,s}$ is much closer to $1$ for electron doping, indicating weaker Landau parameter corrections from interactions.
The thermodynamic properties also display stronger finite-size effects,
and the local moment lies closer to the non-interacting value (see \cref{fig:mz2} in \cref{suppdata:mz2}), both reflecting a higher degree of delocalization, and coherence, with electron doping.

These results, showing that the electron-doped regime is more coherent, or effectively more ``weakly'' interacting, suggest that while weak-coupling approaches -- combining Fermiology and nesting at the AFM wavevector $\mathbf{q}=(\pi,\pi)$ -- may work well for understanding electron-doped ``hot spots,'' one should not naively apply the same methods to understand hole-doped cuprates, especially with regard to the pseudogap phenomenon.
One can heuristically understand why the AFM is more robust and extends to higher doping, and why there is greater coherence for electron doping compared to hole doping by considering the simple motion of a hole in a locally AFM background~\cite{PhysRevLett.63.680,PhysRevB.44.317,PhysRevB.75.035106,Jia2014}. 
A particle-hole transformation converts the Hubbard Hamiltonian with negative $t'$ and filling $\langle n \rangle=1+|x|>1$ to one with positive $t'$ and filling $\langle n \rangle=1-|x|<1$. 
The effective positive $t'$ for electron doping favors hopping within the same sublattice, which preserves the local AFM pattern, such that a hole can move more easily. 
In contrast, nearest-neighbor hopping between different magnetic sublattices disrupts the local AFM pattern, such that  
one would expect hole doping (negative $t'$, so same-sublattice diagonal hopping is less energetically favorable and the motion relies more on nearest-neighbor hops that scramble the local AFM pattern) to exhibit stronger scattering and more incoherence.

\section{Nodal-antinodal Dichotomy and Pseudogap} \label{secdichotomy}

\begin{figure*}[htbp]
    \centering
    \includegraphics[width=\linewidth]{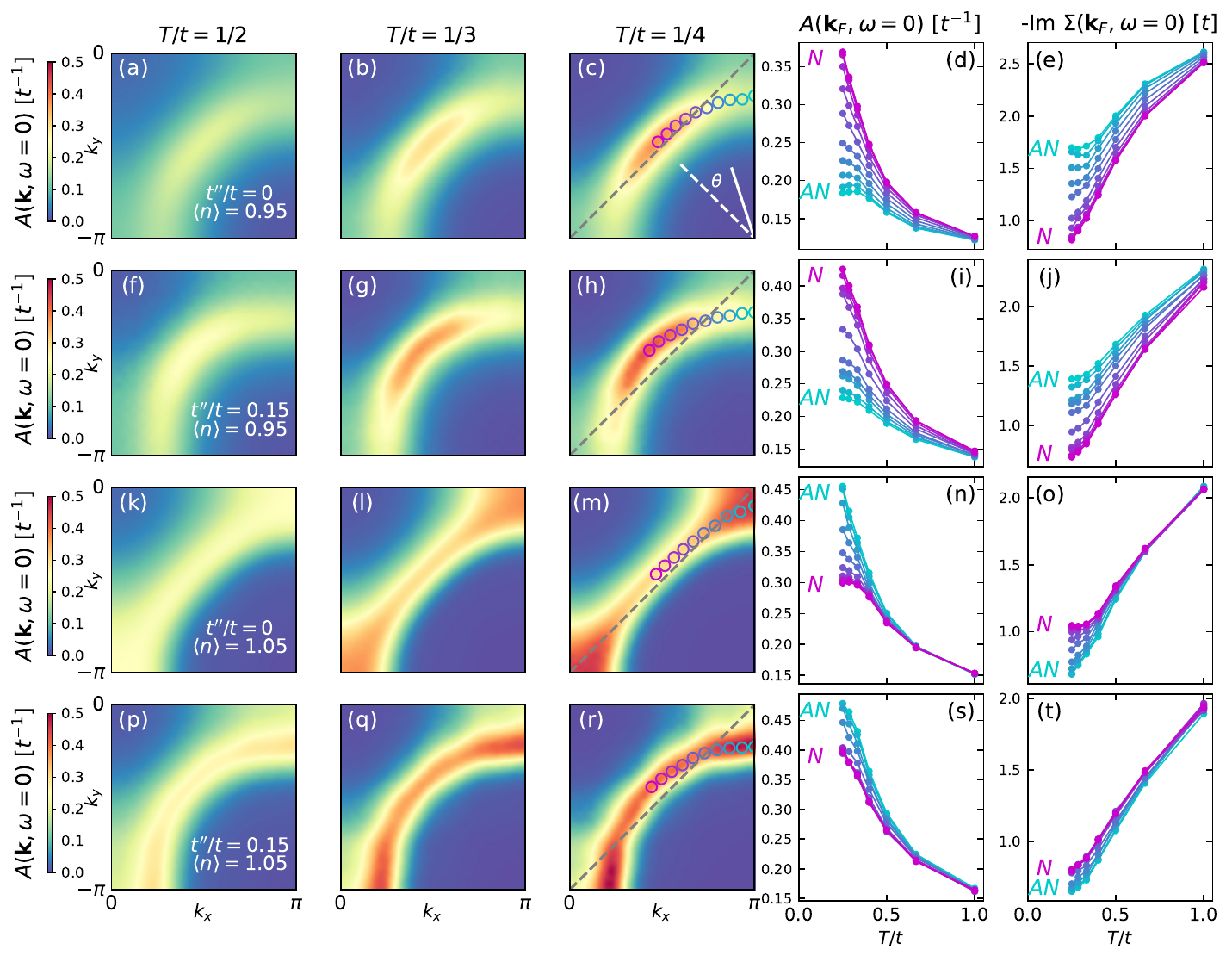}
    \caption{
Temperature evolution of the particle-hole asymmetric Fermi surface and negative imaginary self-energies at low doping ($|x|= 0.05$).
The three columns on the left have the same layout as those in Fig.~\ref{fig:Tdep_highdop}, but the data corresponds to $|x|= 0.05$. 
In the third column, color-coded empty circles mark positions on the Fermi surface at specific angles $\theta$, from the nodal (N) to the antinodal (AN) region. 
Each circle matches curves (color-coded) in the two right-hand columns to illustrate the temperature dependence of $A(\mathbf{k},\omega=0)$ (fourth column) and $-\Im \Sigma(\mathbf{k}_F,\omega=0)$ (fifth column) around the Fermi surface.    
Other than $t''$ and $\langle n\rangle$ specified in the first column, the other parameters are the same as in Fig.~\ref{fig:Tdep_highdop}.
}
    \label{fig:Tdep_lowdop}
\end{figure*}

We now focus on the nodal-antinodal dichotomy. As discussed previously, the dichotomy is more pronounced at lower doping, and in Fig.~\ref{fig:Tdep_lowdop} it is highlighted by using $|x|=0.05$ as a representative example.
The impact of temperature on the Fermi surface 
is similar to that observed at higher doping ($|x|=0.15$), as in Fig.~\ref{fig:Tdep_highdop}:
the Fermi surface becomes sharper and better-defined with decreasing temperature, thereby enhancing the dichotomy, with opposite trends observed between hole and electron doping. At $5\%$ hole doping ($\langle n\rangle=0.95$) without $t''$ [\cref{fig:Tdep_lowdop}~(d)], nodal
$A$ increases with decreasing temperatures while antinodal $A$ increases more slowly and actually peaks near $T/t \sim 0.33$, while 
$-\Im \Sigma$ shows the opposite trend with an antinodal minimum at the same temperature [\cref{fig:Tdep_lowdop}~(e)]. 
The antinodal trend suggests progression towards a gap near the antinode, \textit{i.e.} the pseudogap, with
the peak in $A$ defining the pseudogap onset temperature $T^*_{A}=t/3$.
It is worth noting that $t''$ affects this behavior.
Neither the peak or low-temperature reduction in the antinodal spectral weight nor the increase in $-\Im \Sigma$ is observed, and in general, the spectral weight is higher and $-\Im \Sigma$ is lower for $t''/t=0.15$ compared to that for $t''/t=0$.
One may speculate that a nonzero $t''$ slightly enhances coherence, potentially shifting the onset of the pseudogap to lower temperatures. Similar observations can be made for electron doping, but with a reversal in the nodal-antinodal dichotomy and larger $A$ and smaller $-\Im \Sigma$ compared to hole doping.

\begin{figure*}[htbp]
    \centering
    \includegraphics[width=\linewidth]{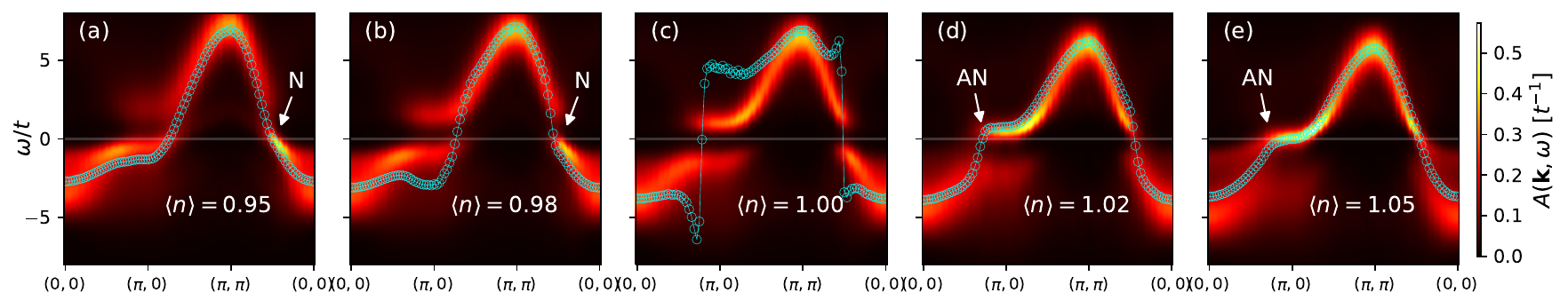}
\caption{$A(\mathbf{k},\omega)$ (color density plots) and $\Re\Sigma_{\mathbf{k}}(\omega=0) + (\epsilon_{\mathbf{k}}-\mu)$ (open circles) along the momentum cut $(0,0)$-$(\pi,0)$-$(\pi,\pi)$-$(0,0)$ for
$t''/t=0$ at temperature $T/t=1/4$. Here, $\epsilon_\mathbf{k}$ is the non-interacting band energy and $\mu$ is the chemical potential.
Arrows highlight the region that gains spectral weight first (nodal for hole doping, antinodal for electron doping) upon doping away from half-filling in either direction.
}
    \label{fig:realselfenergy_kU6}
\end{figure*}

The reversed nodal-antinodal dichotomy between hole and electron doping can be understood by examining $A(\mathbf{k},\omega)$ along the high-symmetry momentum cut, as shown in \cref{fig:realselfenergy_kU6}.
At half-filling the Fermi level lies within the Mott gap, which separates the upper and lower Hubbard bands.
While the gap size is relatively uniform, the gap center is momentum-dependent. In the nodal region, the Fermi level lies closer to the lower Hubbard band, whereas in the antinodal region it approaches the upper Hubbard band more closely.
Consequently, for hole doping, as in Fig.~\ref{fig:realselfenergy_kU6}~(a) and (b), one dopes into the lower Hubbard band and the nodal region acquires spectral weight more quickly. Conversely, upon electron doping, as shown in Fig.~\ref{fig:realselfenergy_kU6}~(d) and (e), the chemical potential moves into the upper Hubbard band, causing the antinodal region to gain spectral weight earlier than the nodal region. Thus, the general behavior of the nodal-antinodal dichotomy at the Fermi level reflects the underlying momentum dependence of the Mott gap.

The connection between the dichotomy and the Mott gap is also reflected in the behavior of the real part of the self-energy,
which is overlaid on the spectral function in Fig.~\ref{fig:realselfenergy_kU6}. Dyson's equation
\begin{equation}
G^R(\mathbf{k},\omega) = \frac{1}{\omega-(\epsilon_\mathbf{k}-\mu)+i\delta-\Sigma_\mathbf{k}(\omega)}
\end{equation}
tells us that when scattering is weak (i.e., when $-\Im \Sigma_\mathbf{k}(\omega)$ is small), the dispersion identified by high-intensity regions in $A(\mathbf{k},\omega) = -\Im G^R(\mathbf{k},\omega)/\pi$ should follow $\omega - (\Re\Sigma_\mathbf{k}(\omega)+(\epsilon_\mathbf{k}-\mu))=0$. At the larger doping $|x|=0.05$ in Fig.~\ref{fig:realselfenergy_kU6}~(a) and (e), we observe that the spectral intensity roughly tracks $\omega = \Re\Sigma_\mathbf{k}(\omega=0)+(\epsilon_\mathbf{k}-\mu)$, which agrees with a weaker frequency dependence of $\Re\Sigma_\mathbf{k}(\omega)$.
In momentum space, $\Re\Sigma_\mathbf{k}(\omega=0)+(\epsilon_\mathbf{k}-\mu)$ crosses zero smoothly, and it changes sign at the momentum points that mark a well-defined Fermi surface.
As doping decreases toward half-filling, this behavior changes, as shown in Fig.~\ref{fig:realselfenergy_kU6}~(c), where a gap opens at the Fermi level: $\Re\Sigma_\mathbf{k}(\omega=0)+(\epsilon_\mathbf{k}-\mu)$ develops a momentum dependence that deviates significantly from the dispersion and changes sign through a singularity associated with the Mott gap rather than through zero.
This behavior is reinforced by results for larger $U$ ($U/t=8$ presented in \cref{fig:realselfenergy_kU8}), which lead to a more pronounced Mott gap and a sharper singularity. 
Under light hole doping, shown in \cref{fig:realselfenergy_kU6}~(b), the nodal singularity disappears, 
whereas near the antinode, $\Re\Sigma_\mathbf{k}(\omega=0)+(\epsilon_\mathbf{k}-\mu)$ no longer shows a sharp singularity but instead a continuous, non-monotonic variation associated with a gap edge close to the Fermi level. 
The behavior corresponds to the nodal-antinodal dichotomy observed for hole doping, and is reversed for electron doping, shown in \cref{fig:realselfenergy_kU6}~(d).

\begin{figure*}[htbp]
    \centering
    \includegraphics[width=\linewidth]{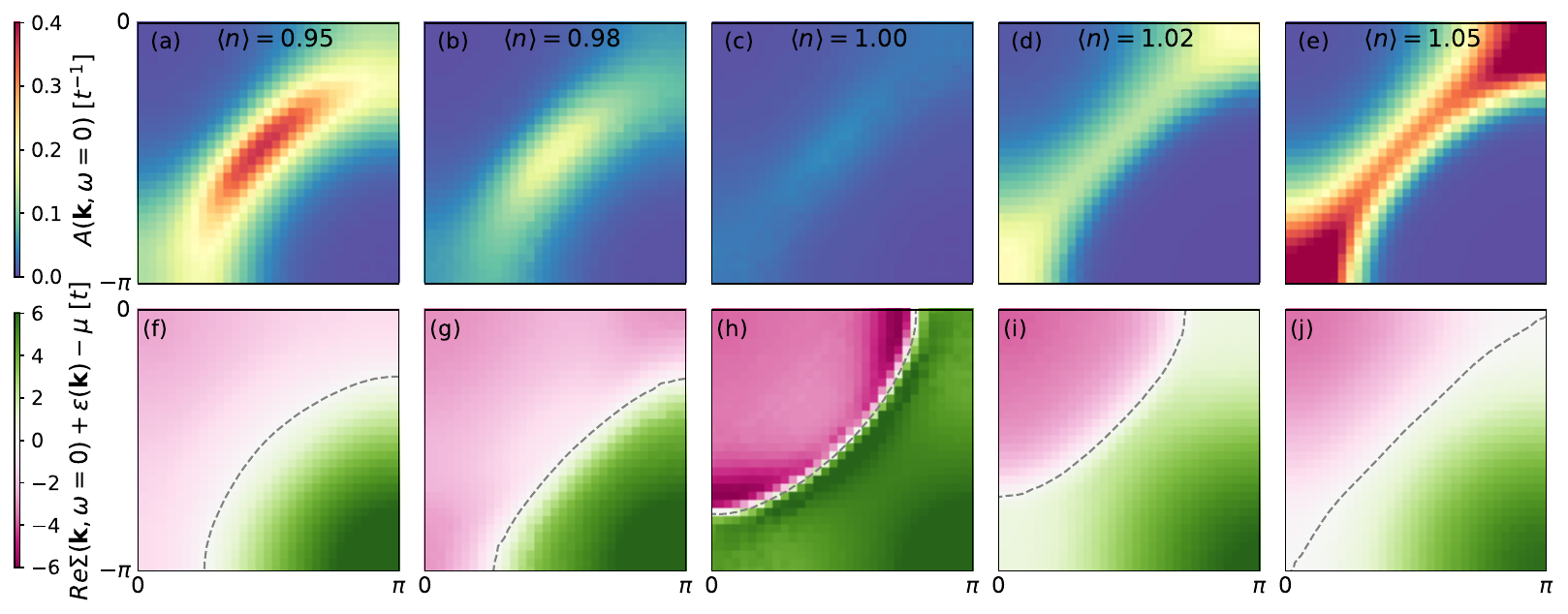}
\caption{
$A(\mathbf{k},\omega=0)$ [(a) to (e)] and $\Re\Sigma_{\mathbf{k}}(\omega=0) + (\epsilon_{\mathbf{k}}-\mu)$ [(f) to (j)] throughout the Brillouin zone for
$t''/t=0$ at temperature $T/t=0.25$.
Dashed lines highlight the change in sign of $\Re\Sigma_{\mathbf{k}}(\omega=0) + (\epsilon_{\mathbf{k}}-\mu)$.
}
    \label{fig:FS_realselfenergy_kU6}
\end{figure*}

We extend the connection between spectral weight suppression and the self-energy singularity to the entire Brillouin zone in \cref{fig:FS_realselfenergy_kU6}.
At $|x|=0.05$, the Fermi surface defined by the highest intensity of $A(\mathbf{k},\omega)$ [Fig.~\ref{fig:FS_realselfenergy_kU6}, (a) and (e)] roughly corresponds to the momentum points at which $\Re\Sigma_\mathbf{k}(\omega=0)+(\epsilon_\mathbf{k}-\mu)$ changes sign [Fig.~\ref{fig:FS_realselfenergy_kU6}, (f) and (j)].
At low hole doping, the ``Fermi arc'' is prominent in Fig.~\ref{fig:FS_realselfenergy_kU6} (b), with a weakly non-monotonic momentum dependence of $\Re\Sigma_\mathbf{k}(\omega=0)+(\epsilon_\mathbf{k}-\mu)$ around the antinodal regions, shown in Fig.~\ref{fig:FS_realselfenergy_kU6} (g).
At half-filling, the entire Fermi surface is gapped out, and a prominent singularity appears in $\Re\Sigma_\mathbf{k}(\omega=0)+(\epsilon_\mathbf{k}-\mu)$ around the momentum points marking what would have been the Fermi surface.

No enclosed Fermi-surface pockets appear within the doping and temperature range of this study. 
This observation is important for the ongoing debate about the origin of ``Fermi arcs''~\cite{Keimer2015,PhysRevB.105.035117,PhysRevB.79.245116,PhysRevB.80.220513,PhysRevB.108.235156}.
Some proposals suggest that these arcs arise as parts of underlying pockets~\cite{PhysRevB.73.174501,PhysRevB.81.115129,Christos2024}, whose back sides are suppressed, \textit{e.g.} by AFM correlations.
Fermi pockets have been observed in multilayer cuprates~\cite{doi:10.1126/science.aay7311}, where interlayer coupling stabilizes the AFM near half-filling; in those cases, the back and front sides of the pockets lie symmetrically about the AFM zone boundary, rather than placing the back side directly on it. The observations imply that when no pockets are seen in lower-layer systems, they likely do not exist.
At an onset temperature of the pseudogap $T^*$, which we defined above,
our self-energy analysis also reveals no pockets, contradicting the notion that arcs arise from whole pockets whose back sides are damped~\cite{PhysRevB.73.174501,PhysRevB.81.115129,Christos2024}.
Instead, it is consistent with recent advanced numerics, including cluster approaches (e.g. DCA), diagrammatic extensions of DMFT, and dynamical variational Monte Carlo~\cite{PhysRevB.74.125110,PhysRevLett.102.056404,Krien2022,PhysRevLett.133.166501,PhysRevB.106.245132,PhysRevLett.97.036401,PhysRevLett.110.216405}.

\begin{figure*}[htbp]
    \centering
    \includegraphics[width=\linewidth]{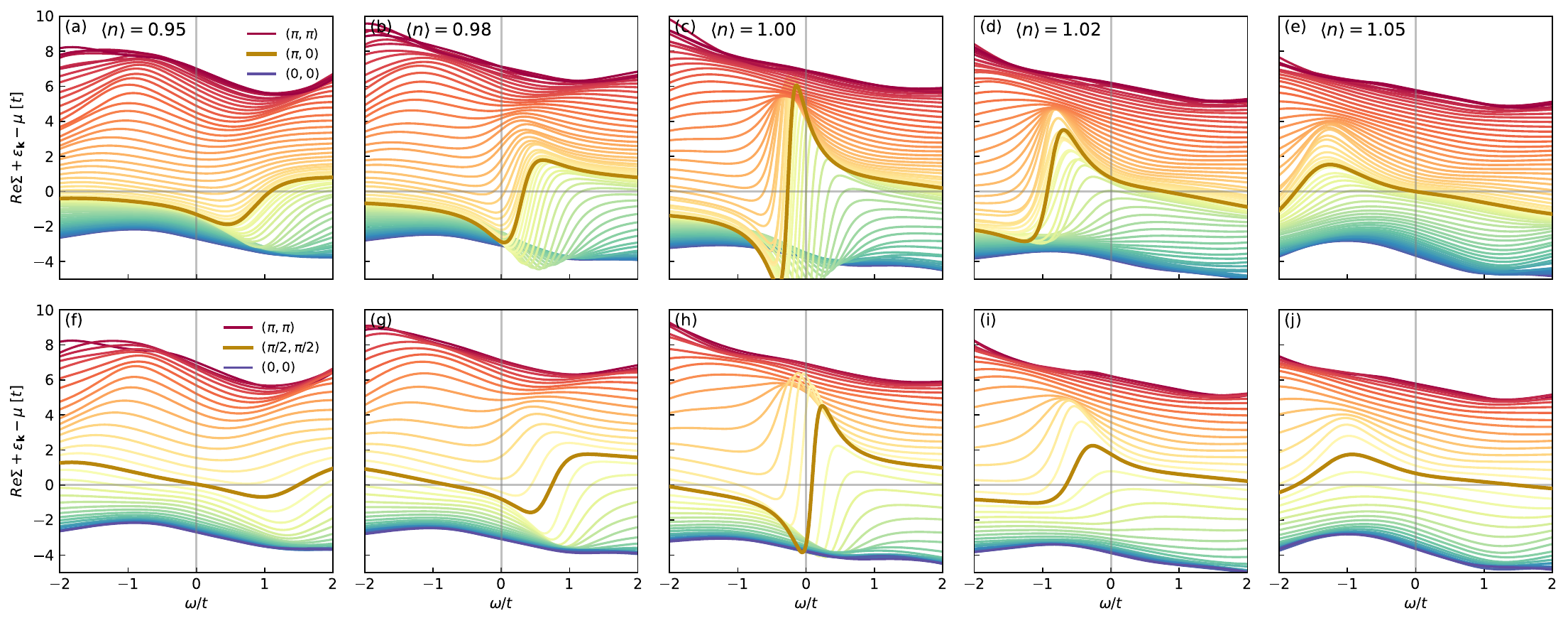}
\caption{Frequency dependence of $\Re\Sigma_{\mathbf{k}}(\omega) + (\epsilon_{\mathbf{k}}-\mu)$ along the antinodal [$(0,0)$-$(\pi,0)$-$(\pi,\pi)$, plots (a) to (e)] and nodal [$(0,0)$-$(\pi/2,\pi/2)$-$(\pi,\pi)$, plots (f) to (j)] cuts for $t''/t=0$ at temperature $T/t=0.25$.
}
    \label{fig:freq_realself_U6}
\end{figure*}

We next investigate the frequency dependence of the self-energy based on the following arguments: if the real part of the self-energy has a pole-like dependence [$\Re\Sigma_\mathbf{k}(\omega)\sim 1/( \omega-\omega_0 )$], the corresponding spectral function develops a gap; and if $\Re\Sigma_\mathbf{k}$ varies linearly with frequency [$\Re\Sigma_\mathbf{k}(\omega)\sim \omega$] and has a negative slope, it leads to a gapless dispersion.
\cref{sec:selfsupp} discusses the mathematical reasoning behind these arguments.
In Fig.~\ref{fig:freq_realself_U6}, 
the Mott gap shows a divergence reminiscent of $1/( \omega-\omega_0 )$, but there exists a region of positive slope to maintain continuity.
At half-filling, as shown in Fig.~\ref{fig:freq_realself_U6}~(c) and (h), the gap is located near the Fermi level ($\omega = 0$).
As doping increases, the region with a positive-slope shifts away from the Fermi level and becomes less steep, as the remnant Mott gap moves above or below the Fermi energy with hole or electron doping, respectively. 
For hole doping, the center of the region with positive-slope is closer to the Fermi level near the antinode, while it is closer to the Fermi level near the node with electron doping, resulting in the observed dichotomy in the spectral function.

\begin{figure*}[htbp]
    \centering
    \includegraphics[width=0.9\linewidth]{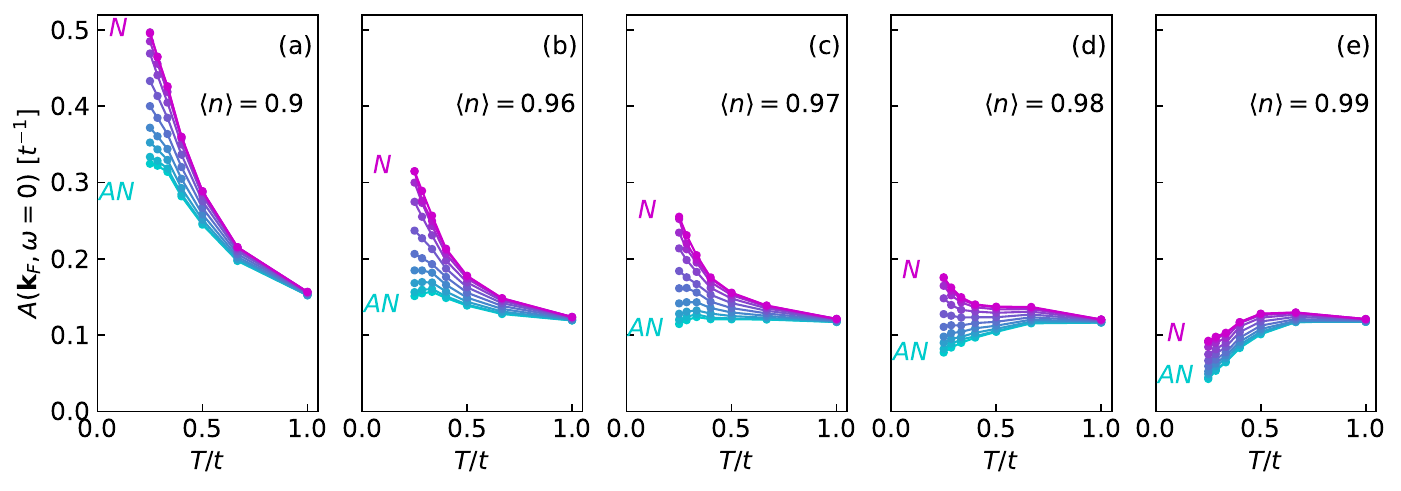}
    \caption{The temperature dependence of the spectral function $A(\mathbf{k},\omega=0)$ along the Fermi surface, presented in the same way as in \cref{fig:Tdep_lowdop} (d) for various doping values with $t''/t=0$. 
}
    \label{TAdop}
\end{figure*}

\section{The pseudogap from NMR} \label{sec:NMR}

The connection to the Mott gap suggests that the pseudogap onsets as a crossover.
We examine this issue from the perspective of $T^*$ taken from different measurements: do pseudogaps identified by other techniques align with that extracted from the spectral function and how do they align in parameter space?
If the pseudogap is indeed a crossover, it is reasonable to expect that different measurement techniques will give different answers to the question.
In addition to $T^*_A$, we examine the temperature dependence of the Knight shift $K_s$ and the spin-lattice relaxation time $T_1$, from which we determine $T^*_{K_s}$ and $T^*_{T_1}$, and compare them with $T^*_{A}$.

We first analyze $T^*_A$, determined from the spectral function at various doping values in \cref{TAdop}, which shows a relatively sharp crossover from Mott to pseudogap behavior. 
For $\langle n\rangle\leq 0.9$ [\cref{TAdop} (a)], the doping is too high for pseudogap features to appear within the accessible temperature window: the spectral weight increases monotonically along the entire Fermi surface upon cooling.
For $0.95\leq\langle n\rangle\leq 0.97$ [\cref{fig:Tdep_lowdop} (d), \cref{TAdop} (b) and (c)],  the spectral weight increases at the node while decreasing at the antinode, at least at lower temperatures around $T_A^*\sim t/3$, indicating onset of the pseudogap. Unfortunately for even lower doping $\langle n \rangle \geq 0.98$ [\cref{TAdop} (d) and (e)], the spectral weight along the entire Fermi surface is suppressed as $T$ decreases, at least for these temperatures, as the Fermi level is close to the edge of the Mott gap.
It is the selective suppression of the spectral weight at the antinode, without affecting the node, that characterizes the pseudogap; therefore, we distinguish the pseudogap onset and identify $T_A^*$ for $0.95\leq\langle n\rangle\leq 0.97$.
In \cref{lowelectron}, Fig.~\ref{TAdop_e}, we present the same analysis for electron doping. The behavior is qualitatively similar, except that the node-antinode contrast is reversed and weaker.

\begin{figure*}[htbp]
    \centering
    \includegraphics[width=0.7\linewidth]{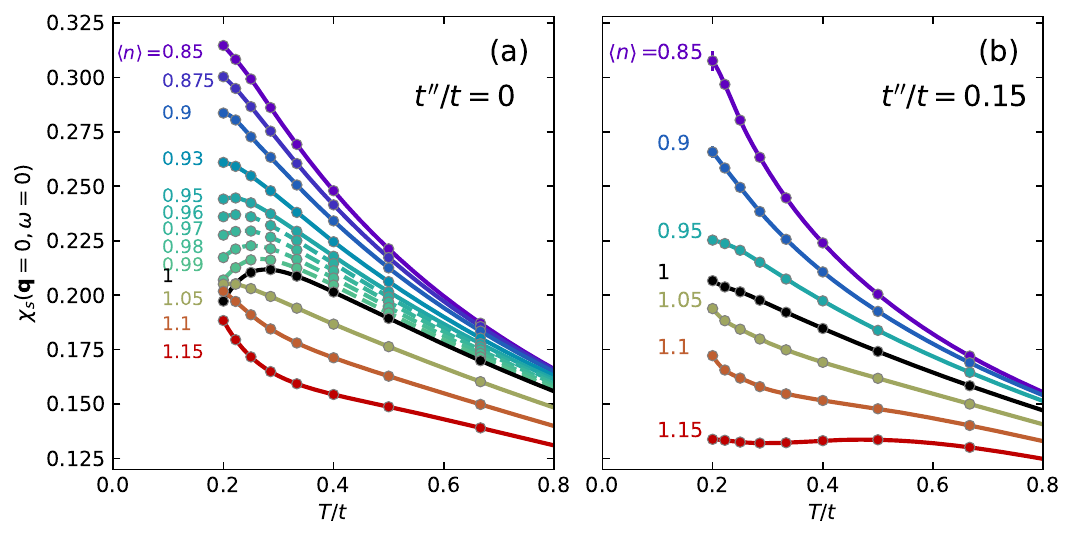}
    \caption{Knight shift $K_s$, defined as the spin (Pauli) susceptibility $\chi_s(\mathbf{q} = \mathbf{0},\omega = 0)$ [in arbitrary units], for both (a) $t''/t=0$ and (b) $t''/t=0.15$. 
    }\label{fig:Knightshift}
\end{figure*}

\begin{figure*}[htbp]
    \centering
    \includegraphics[width=0.9\linewidth]{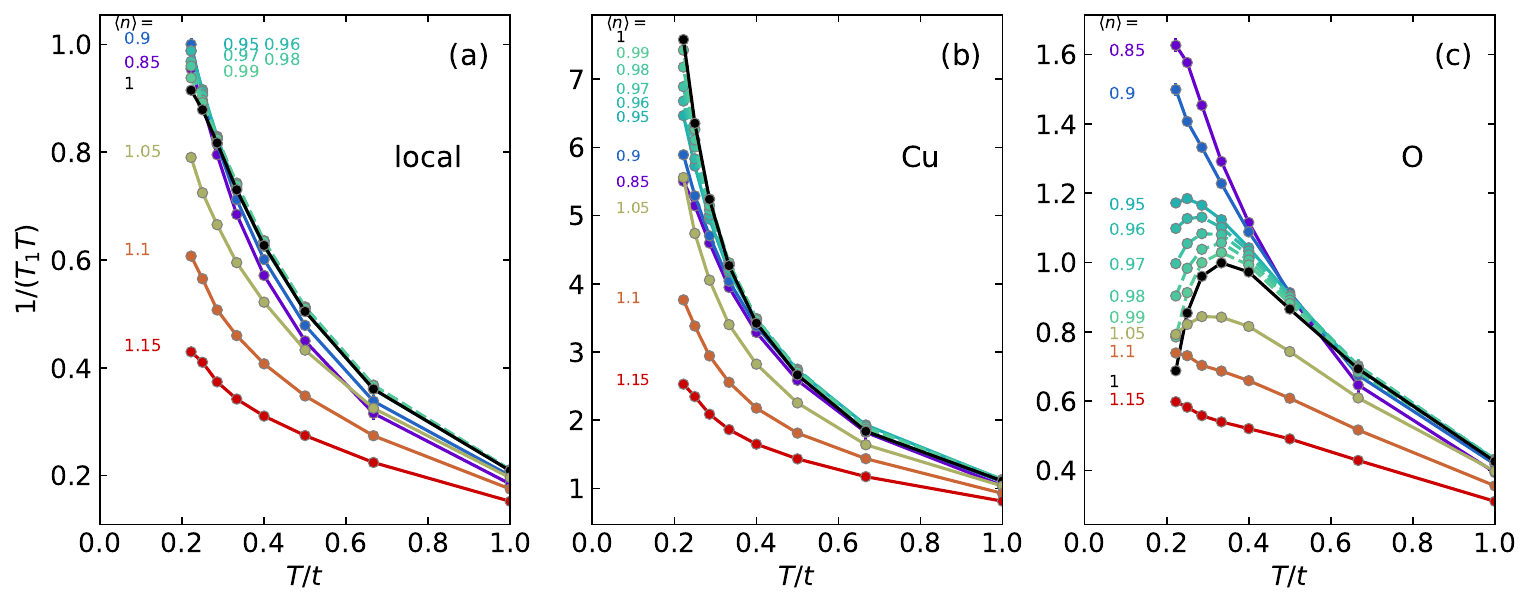}
    \caption{Spin-lattice relaxation rate divided by temperature, $(T_1 T)^{-1}$ [in arbitrary units], for
    $t''/t=0$, calculated using different form factors appropriate for Cu and O nuclei, as detailed in \cref{nmrmeasurements}. 
}
    \label{fig:T1T}
\end{figure*}

From NMR measurements, the pseudogap onset temperature $T^*$ can be identified either by a drop in $K_s$ or in $(T_1 T)^{-1}$ on oxygen sites as temperature decreases. 
The definitions and details about obtaining these quantities in DQMC calculations are provided in \cref{nmrmeasurements}.
Results for the temperature-dependent $K_s$ and $(T_1 T)^{-1}$ are shown in \cref{fig:Knightshift} and \cref{fig:T1T}, respectively.
At low hole doping and $t''/t=0$, $K_s$ has a clearly identifiable peak; however, consistent with the previously discussed $t''$ dependence of the pseudogap temperature from $A(\mathbf{k},\omega=0)$, this peak occurs at much lower temperatures, even outside the range of our study, as seen by comparing the results in \cref{fig:Knightshift}~(a) and (b).
In Fig.~\ref{fig:T1T}, $(T_1 T)^{-1}$ is analyzed for $t''/t=0$ using different form factors that are considered to be appropriate for different nuclear sites~\cite{PhysRevB.41.1797,PhysRevB.42.167,PhysRevB.43.258,PhysRevB.52.13585,Chen2017}. 
No downturn is observed in Fig.~\ref{fig:T1T}~(a) or (b) for a ``local" or Cu form factor, as neither filters out the substantial contribution from $\mathbf{q}=(\pi,\pi)$~\cite{PhysRevB.41.1797}.
However, the form factor for O does suppress the AFM spin fluctuations and reveals a downturn in Fig.~\ref{fig:T1T}~(c), where the peak defines $T^*_{T_1}$. 
The simulated $T^*_{K_s}$ and $T^*_{T_1}$ temperature scales are reasonably comparable to the experimentally measured range of a few hundred Kelvin~\cite{PhysRevLett.63.1700,PhysRevB.58.R5960,Tom_Timusk_1999}.
A unique aspect of the NMR Knight shift in probing the pseudogap is that two crossover temperatures have been observed in several compounds; and our results likely correspond to the higher-temperature crossover.

\begin{figure}[htbp]
    \centering
    \includegraphics[width=0.8\linewidth]{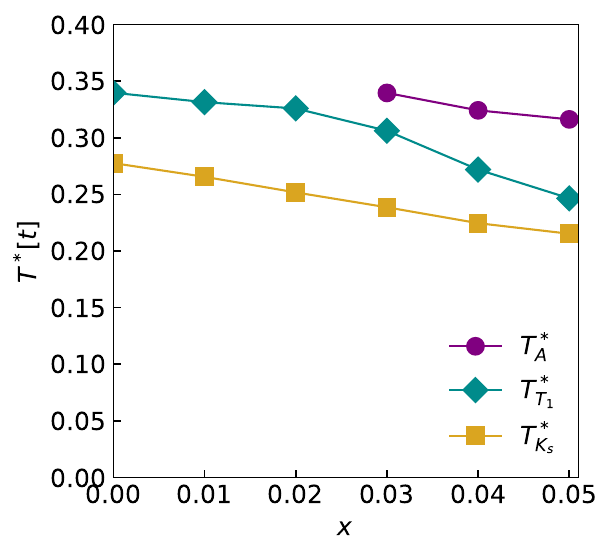}
    \caption{Doping dependence of $T^*_A$, $T^*_{K_s}$, and $T^*_{T_1}$ for $t''/t=0$, as obtained from the peak temperature of the antinodal spectral weight, the Knight shift, and the spin-lattice relaxation rate divided by temperature, respectively. Cubic spline fitting is used on the discrete temperature data to determine the peak temperatures.
}
    \label{fig:Tstar}
\end{figure}

We summarize the doping dependence of $T^*_A$, $T^*_{K_s}$, and $T^*_{T_1}$ (for O) in \cref{fig:Tstar}. 
While the temperatures from these methods are not wildly different, they also do not necessarily coincide. 
The discrepancy in $T^*$ from NMR and the spectral function contrasts with prior numerical studies using different algorithms (such as DCA and cellular DMFT)~\cite{PhysRevLett.102.206407,PhysRevB.87.041101,Chen2017}, and emphasizes the somewhat arbitrary nature of defining a pseudogap temperature from different measurements~\cite{RevModPhys.87.457}. This certainly does not indicate a well-defined temperature scale tied to some form of broken symmetry.
The pseudogap defined from the spectral function is not directly associated with the temperature scale associated with the formation of short-range fluctuating spin correlations as suggested by NMR.
Moreover, the nodal-antinodal dichotomy gradually disappears at higher temperatures and higher doping without a sharp termination.

The differing $T^*$ values and gradual change in the dichotomy lend support to the interpretation that the pseudogap is a crossover phenomenon, and this is fully consistent with numerous Hubbard model studies that have found no clear connection between $T^*$ and a well-defined phase transition. 
For example, it is difficult to stabilize a time reversal symmetry-breaking loop-current state in the Hubbard model~\cite{PhysRevB.90.224507}.
Additionally, calculations have shown that fluctuating intertwined stripes~\cite{PhysRevB.107.085126,annurev:/content/journals/10.1146/annurev-conmatphys-031620-102024,annurev:/content/journals/10.1146/annurev-conmatphys-090921-033948} exist without forming static long-range order. 
Some recent studies using different methods, such as cellular DMFT~\cite{10.21468/SciPostPhys.16.2.059} and diagrammatic Monte Carlo~\cite{doi:10.1126/science.ade9194}, link the pseudogap to magnetic correlations; we do not see such a direct link in our DQMC window. 
Differences between methods can reflect different parameter regimes and method-specific analysis choices and modeling assumptions.
Nonetheless, they also describe the pseudogap as a finite-$T$ crossover, consistent with our interpretation.

Efforts to elucidate the impacts of Fermi surface topology have proposed a connection between the pseudogap termination and a Lifshitz transition~\cite{PhysRevX.8.021048,PhysRevLett.120.067002,PhysRevLett.114.147001,PhysRevB.96.094525,Doiron-Leyraud2017,PhysRevResearch.2.033067}; however, this connection is in doubt even based on experimental evidence~\cite{RevModPhys.93.025006,PhysRevB.69.094515,PhysRevB.73.174511,doi:10.1126/science.aar3394,Drozdov2018,PhysRevLett.121.077004,Michon2019}.
We observe no direct association between the pseudogap and a Lifshitz transition down to the lowest temperatures in our study, and for typical parameters, the Lifshitz doping~\cite{thermopower} is much higher than the doping levels where we observe pseudogap onsets $T^*_A$, $T^*_{K_s}$, and $T^*_{T_1}$.
Taken as a whole, our results imply that the pseudogap cannot be understood solely based on fermiology, and instead they emphasize the dominant role played by the effects of strong interactions.

\section{Transport and other thermodynamics within the pseudogap region} \label{transportandthermodyanmicssec}

\begin{figure}[htbp]
    \centering
    \includegraphics[width=\linewidth]{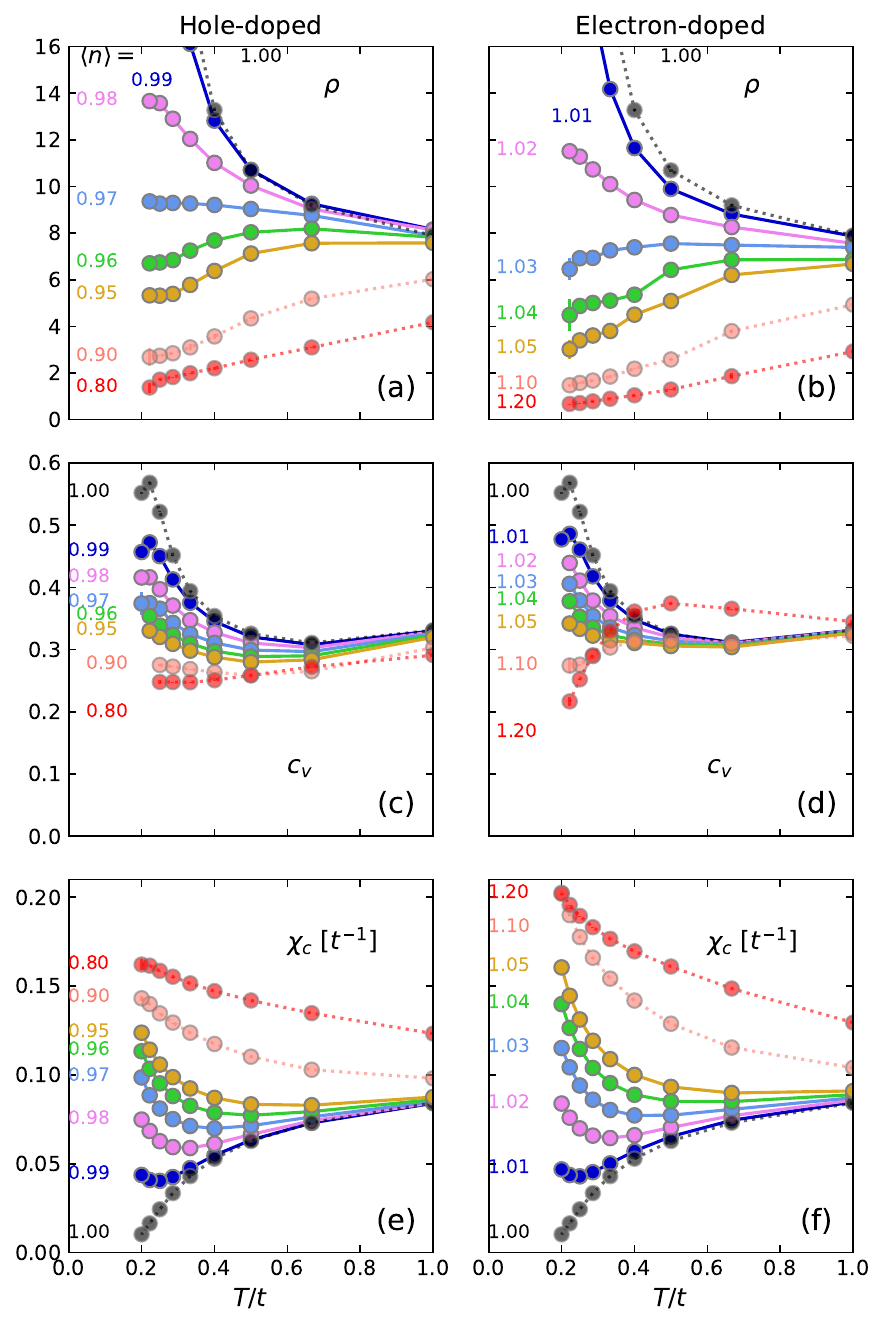}
    \caption{Resistivity $\rho$ (a,b), specific heat $c_v$ (c,d), and charge compressibility $\chi_c$ (e,f), for $t''/t=0$, highlighting the low-doping regime relevant to the pseudogap. Panels (a,c,e) show hole doping and (b,d,f) show electron doping.
}
    \label{fig:lowdopingtransportandthermo}
\end{figure}

We now discuss how the crossover nature of the observed pseudogap manifests in transport and thermodynamic responses near the pseudogap temperature scale.
In Fig.~\ref{fig:lowdopingtransportandthermo}, results are shown for different electron fillings, $\langle n \rangle = 1 - x$, including both hole doping ($x>0$) and electron doping ($x<0$).
The resistivity $\rho$, specific heat $c_v$, and charge susceptibility $\chi_c$ vary smoothly through $T^*$, showing no indication of nonanalytic behavior.
The same holds for the spin susceptibility $\chi_s$, which is equivalent to the Knight shift $K_s$ (see Fig.~\ref{fig:Knightshift} and \cref{sec:thermodynamics}).
Although no sharp transition feature is present, within the low hole-doping regime $0<x<5\%$ the detailed trends show suggestive correspondence with pseudogap behavior.

In Fig.~\ref{fig:lowdopingtransportandthermo}~(a), $\rho$ for $x<5\%$ exhibits an intermediate temperature dependence between the divergent behavior at half filling and the nearly linear-in-$T$ resistivity at higher doping.
If we extrapolate our accessible temperatures toward $T\to 0$, $\rho$ appears to approach a sizable intercept.
In a clean, translationally invariant system without disorder or phonons, however, the dc resistivity must vanish as $T\to 0$, since electron-electron Umklapp scattering has vanishing phase space at zero temperature.
This indicates a deviation from linear-in-$T$ resistivity for $x<5\%$ relative to the higher-doping regime, echoing experimental observations in cuprates where $\rho(T)$ departs from linearity in the pseudogap regime~\cite{PhysRevLett.69.2975,PhysRevLett.70.3995,PhysRevLett.93.267001,Daou2009,doi:10.1126/science.1165015,doi:10.1073/pnas.1301989110,PhysRevLett.113.177005,PhysRevB.95.224517,Barisic_2019}.
While phonons and disorder, which are absent in our model, clearly influence experimental transport, our results show that interactions alone can already produce a pronounced deviation from linear-$T$ resistivity in this low-doping regime.
For electron doping in Fig.~\ref{fig:lowdopingtransportandthermo}~(b), the deviation is weaker at comparable $|x|$, consistent with a reduced dichotomy and a more coherent single-particle spectrum (cf. \cref{lowelectron} and \cref{TAdop_e}).
Transverse transport properties such as the Hall and thermal Hall responses, while challenging to compute, are natural directions for future work and would provide valuable complements to our longitudinal transport results. 
Experimentally, Hall-related observables (e.g., the Hall coefficient or Hall conductivity) show pronounced evolution across much of the underdoped/pseudogap regime, whereas the cotangent of the Hall angle $\cot\theta_H$ is often found to follow an approximately $T^2$ dependence over wide regions of the phase diagram~\cite{Barisic_2019,PhysRevLett.67.2088,PhysRevLett.69.2855,PhysRevLett.72.2636,KUBO1992378,Badoux2016,PhysRevB.102.075114}.
Future numerical studies would be important for clarifying this behavior and its connection to pseudogap physics.

How do thermodynamic properties (besides $\chi_s$ analyzed in Sec.~\ref{sec:NMR}) behave in the same low-doping regime?
In Figs.~\ref{fig:lowdopingtransportandthermo} (c,d), specific heat $c_v$ at small doping remains close to the half-filled peak associated with the antiferromagnetic correlations~\cite{PhysRevB.105.L161103} (consistent with a diverging low-energy structure factor at $(\pi,\pi)$ in Fig.~\ref{fig:pipi}).
We notice that a clear asymmetry between hole and electron doping in $c_v$ only becomes apparent for $|x|>5\%$.
In contrast, the compressibility $\chi_c$  in Figs.~\ref{fig:lowdopingtransportandthermo} (e,f) deviates immediately from the half-filling insulating behavior upon doping.
For $|x|\leq 5\%$, the temperature dependence of $\chi_c$ qualitatively tracks the evolution of low-energy spectral weight on the portions of the Fermi surface that remain most coherent (cf. Figs.~\ref{TAdop} and \ref{TAdop_e}).
At comparable $|x|$, $\chi_c$ is larger for electron doping, consistent with stronger coherence.

It is useful to interpret the small-doping regime in terms of a downfolded $t$-$t'$-$J$ description for $|x|<5\%$, where $c_v$ is largely controlled by the exchange scale $J$, while single-particle coherence is governed by a distinct scale captured more directly by the spectral function and $\chi_c$.
The near insensitivity of $c_v$ to changing the sign of doping suggests that $t'$ plays a relatively minor role in the dominant contribution to $c_v$, even though it affects the spectral function and $\chi_c$.
The exchange-driven spin contribution to $c_v$ can remain prominent in the pseudogap regime, again implying that different observables need not identify the same crossover temperature as in \cref{sec:NMR}.
If lower temperatures and higher doping become accessible, where the exchange-driven peak in $c_v$ is no longer the dominant feature, it will be interesting to test whether $c_v$ shows a suppression correlated with pseudogap spectral-weight, as reported in cuprate experiments~\cite{PhysRevLett.71.1740,LORAMA19982091,LORAM200159,doi:10.1143/JPSJ.73.2232,PhysRevB.103.214506}, which merits future study.

\section{Summary}

Using the numerically exact DQMC algorithm, we compute the high-resolution single-particle spectral function of the Hubbard model, which is asymmetric between hole and electron doping and, for the temperatures accessible using our methods, is qualitatively similar to the normal-state Fermi surface observed in ARPES on cuprates.
For high doping, we observe stronger AFM correlations, as manifested in the spin structure factor, and more coherent quasiparticles, reflected in differences in transport and thermodynamic properties, for electron doping compared to hole doping; 
this is attributed to electron doping (which can be mapped to hole doping with a sign change of $t'$) effectively favoring hopping within the same sublattice, thereby preserving AFM order, while hole doping favors inter-sublattice hopping that disrupts it. 
For electron doping, coherent quasiparticles and $(\pi,\pi)$ spin fluctuations generate Fermi-surface hot spots but no pseudogap.

For low doping, the momentum and doping dependence of the remnant Mott gap and the $\Re \Sigma \sim 1/\omega$ self-energy generate a nodal-antinodal dichotomy.
On the hole-doped side, proximity to the Mott gap suppresses coherence at the antinodes while the nodal regions remain coherent, producing ``Fermi arcs'' and the onset of a pseudogap within our simulation temperature range. 
The antiferromagnetic correlation length is short and does not control the pseudogap, and our self-energy analysis shows no signature of small Fermi pockets.
The pseudogap we observe likely corresponds to the high-temperature pseudogap, visible only for dopings near half-filling. 

For this high-temperature pseudogap, our results indicate that pseudogap temperatures defined from NMR measurements do not necessarily coincide with a pseudogap temperature taken from the single-particle spectral function, and suggest that the pseudogap onset $T^*$ is a crossover, rather than a true phase transition, even in the absence of disorder. This crossover interpretation is further supported by transport and thermodynamic signatures in the low-doping regime.

Our results do not preclude the possible emergence of a ``lower-temperature'' pseudogap at dopings away from half-filling, which would be distinct from the Mott gap and characterized by a much smaller gap energy scale, as suggested by prior simulations using dynamical mean-field theory and cluster perturbation theory~\cite{PhysRevLett.91.017002,PhysRevB.101.115141}.
This scale could potentially be manifest as an anomaly in transport and thermodynamic properties at the corresponding temperature.

Broadly speaking, our results provide an unbiased and internally consistent DQMC dataset for the doped Hubbard model. 
By systematically combining momentum-resolved single-particle spectra with spin correlations, NMR response, and thermodynamic and transport properties, we establish a common reference for interpreting electron-hole asymmetry and pseudogap crossovers. 
We expect this dataset and the associated diagnostics to be useful both for direct comparison with experiments and for calibrating and constraining complementary theoretical and numerical approaches.

\section{Data and Code availability}

The data and analysis routines (Jupyter/Python) needed to reproduce the figures,  as well as the DQMC source code, can be found in Ref.~\cite{PhotoemissionDataZenodo}.
Raw simulation data that support the findings of this study are available from the corresponding authors upon reasonable request.

\begin{acknowledgments}
We acknowledge helpful and stimulating discussions with J. K. Ding, M.-H. Julien, S. A. Kivelson, D.-H. Lee, Z. X. Shen, and J.-X. Zhang.
\emph{Funding:}
Work at Stanford University and SLAC was supported by the U.S. Department of Energy (DOE), Office of Basic Energy Sciences,
Division of Materials Sciences and Engineering. 
E.W.H. was supported by the Gordon and Betty Moore Foundation EPiQS Initiative through the grants GBMF 4305 and GBMF 8691.
Computational work was performed on the Sherlock cluster at Stanford University and on resources of the National Energy Research Scientific Computing Center, a U.S. Department of Energy Office of Science User Facility, using NERSC Award No.~BES-ERCAP0031425.
W.O.W. also acknowledges support from the Gordon and Betty Moore Foundation through Grant GBMF8690 to the University of California, Santa Barbara, to the Kavli Institute for Theoretical Physics (KITP), and from the Simons Collaboration on Ultra-Quantum Matter, which is a grant from the Simons Foundation (Grant No. 651440).
This research was supported in part by grant NSF PHY-2309135 to the Kavli Institute for Theoretical Physics (KITP).
\end{acknowledgments}

\clearpage
\appendix

\setcounter{figure}{0}
\renewcommand{\thefigure}{A\arabic{figure}}

\makeatletter
\renewcommand{\theHfigure}{A\arabic{figure}}
\makeatother

\section{Model, Formalism, and Methods} \label{supp1}
\subsection{Hubbard Model and Spectral Function} \label{hubbardandspectral}

We investigate the $2$-dimensional single-band Hubbard model with spin $S=1/2$ on a square lattice with linear size $L$.
The Hamiltonian is
\begin{align}
&\hat{H} = - t \mathlarger{\sum}\limits_{\langle \mathbf{i},\mathbf{j} \rangle, \sigma} \hat{c}^\dagger_{\mathbf{i},\sigma}\hat{c}_{\mathbf{j},\sigma}
  - t' \mathlarger{\sum}\limits_{\langle \langle \mathbf{i},\mathbf{j} \rangle \rangle, \sigma} \hat{c}^\dagger_{\mathbf{i},\sigma}\hat{c}_{\mathbf{j},\sigma}
- t'' \mathlarger{\sum}\limits_{\langle \langle \langle \mathbf{i},\mathbf{j} \rangle \rangle \rangle, \sigma} \hat{c}^\dagger_{\mathbf{i},\sigma}\hat{c}_{\mathbf{j},\sigma} \nonumber \\
&+ \mathrm{h.c.} + U\mathlarger{\sum}\limits_{\mathbf{i}} (\hat{n}_{\mathbf{i},\uparrow}-\frac{1}{2})(\hat{n}_{\mathbf{i},\downarrow}-\frac{1}{2}), \label{hubbard}
\end{align}
where $t$ ($t'$, $t''$) is the nearest-neighbor (next-nearest-neighbor, third-nearest-neighbor) hopping, $U$ is the on-site Coulomb interaction, $\mathit{\hat{c}}_{\mathbf{i},\mathit{\sigma}}^{\dagger}$ $(\mathit{\hat{c}}_{\mathbf{i},\mathit{\sigma}})$ is the creation (annihilation) operator for an electron at lattice site $\mathbf{i}=i_x\mathbf{e}_x + i_y \mathbf{e}_y$ $(i_x, i_y = 0, ..., L-1)$, with $\mathbf{e}_x$ and $\mathbf{e}_y$ the unit vectors along the $x$ and $y$ directions, and
$\mathit{\hat{n}}_{\mathbf{i},\mathit{\sigma}} \equiv \mathit{\hat{c}}_{\mathbf{i},\mathit{\sigma}}^{\dagger} \mathit{\hat{c}}_{\mathbf{i},\mathit{\sigma}}$ is the number operator at site $\mathbf{i}$ with spin $\sigma$. 

In this paper, we adopt a unit system where the reduced Planck constant $\hbar$, the Boltzmann constant $k_B$, the elementary charge $e$, and the lattice constant are set to $1$. 
All quantities are expressed in units derived from this convention. Unless explicitly stated otherwise for arbitrary units, any quantity without a specified unit should be understood as either dimensionless or expressed in units of $1$ under this system. 

The simulation is conducted within the grand canonical ensemble, where the expectation value of an arbitrary operator $\hat{O}$ is given by $\langle \hat{O} \rangle = \Tr(e^{-\beta(\hat{H}-\mu \hat{N})}\hat{O})/\Tr(e^{-\beta(\hat{H}-\mu \hat{N})})$ with $\mu$ representing the chemical potential, $\beta=1/T$ the inverse temperature, and $\hat{N}=\sum_{\mathbf{i},\sigma}\hat{n}_{\mathbf{i},\mathit{\sigma}}$ the total particle number operator. 
With a negative $t'$, we investigate both hole doping with filling $\langle n \rangle \equiv \langle \hat{N} \rangle/L^2=1-x<1$ and electron doping with filling $\langle n \rangle =1-x>1$, where $x$ is the doping value. 

The retarded Green's function is defined as
\begin{equation}
    G^R(\mathbf{k},\sigma,t) \equiv -i \Theta(t) \langle \{ \hat{c}_{\mathbf{k},\sigma}(t),\hat{c}_{\mathbf{k},\sigma}^\dagger \} \rangle. \label{gkr}
\end{equation}
Here, the time evolution of an operator $\hat{O}$ at real time $t$ is given by
\begin{equation}
\hat{O}(t) = e^{i(\hat{H}-\mu \hat{N})t} \hat{O}  e^{-i(\hat{H}-\mu \hat{N})t}.
\end{equation}
Applying a Fourier transformation yields the retarded Green's function in the frequency domain:
\begin{equation}
    G^R(\mathbf{k},\sigma,\omega) = \int_{-\infty}^\infty dt e^{i(\omega+i\delta^+) t}G^R(\mathbf{k},\sigma,t), 
\end{equation}
where $i\delta^+$ is added to the frequency to ensure convergence.
The imaginary-time Green's function is given by
\begin{equation}
    G(\mathbf{k},\sigma,\tau) = -\langle T_\tau \hat{c}_{\mathbf{k},\sigma}(\tau) \hat{c}_{\mathbf{k},\sigma}^\dagger \rangle. \label{Gk}
\end{equation}
An arbitrary operator $\hat{O}$ at imaginary time $\tau$ is 
\begin{equation}
\hat{O}(\tau) = e^{(\hat{H}-\mu \hat{N})\tau} \hat{O}  e^{-(\hat{H}-\mu \hat{N})\tau}. \label{otau}
\end{equation}
Given that the spin-up and spin-down components yield identical values, we hereafter omit $\sigma$ in the Green's functions for simplicity.
From Eqs.~\eqref{gkr}-\eqref{otau}, it is derived that the imaginary-time Green's function is related to the retarded Green's function by~\cite{mahan2013many}:
\begin{equation}
G(\mathbf{k},\tau) = -\int_{-\infty}^{\infty}d\omega\frac{e^{-\tau\omega}}{1+e^{-\beta\omega}} A(\mathbf{k},\omega), \label{howA}
\end{equation}
where $A(\mathbf{k},\omega)$ is the spectral function, 
\begin{equation}
A(\mathbf{k},\omega) = -\frac{1}{\pi} \Im G^R(\mathbf{k},\omega). \label{defA}
\end{equation}
In our simulations, we use \cref{howA} to calculate $A(\mathbf{k},\omega)$ from $G(\mathbf{k},\tau)$ using maximum entropy (MaxEnt) analytic continuation~\cite{maxent1,maxent2}.
However, analytic continuation is generally challenging. The integral in \cref{howA} is ill-conditioned: even small numerical errors in $G(\mathbf{k}, \tau)$ can be greatly amplified, introducing systematic errors in $A(\mathbf{k}, \omega)$. 
We discuss these potential systematic errors of the MaxEnt procedure in \cref{MAXENT}.

\begin{figure}[htbp]
    \centering
    \includegraphics[width=\linewidth]{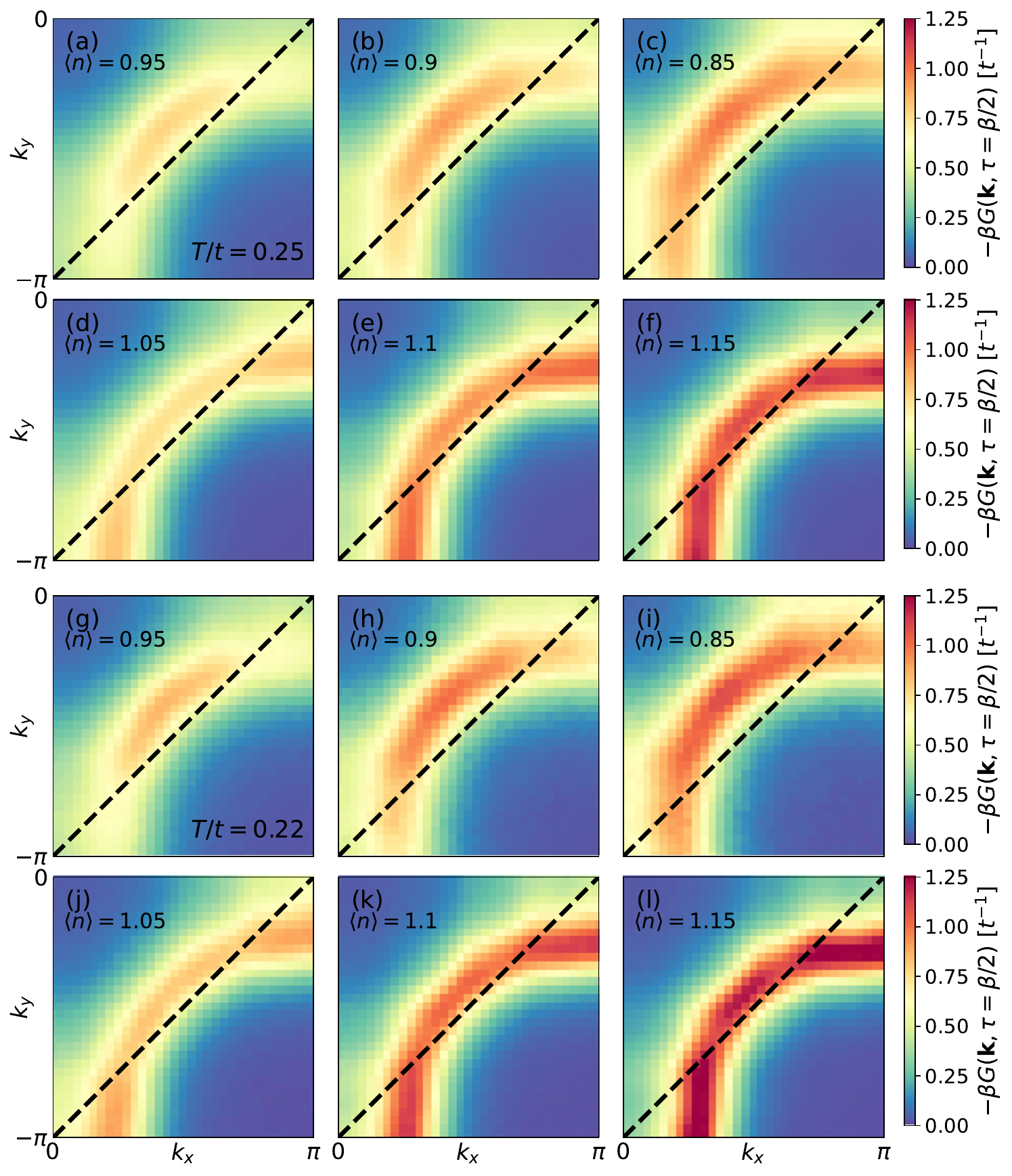}
    \caption{The Fermi surface probed by $-\beta G(\mathbf{k}, \tau=\beta/2)$, which serves as a proxy for $A(\mathbf{k},\omega=0)$.
    All parameters are the same as in \cref{fig:FS}, except that the temperature is $T/t=0.25$ in (a) to (f) and $T/t=0.22$ in (g) to (l).
}
    \label{proxybetaover2}
\end{figure}

To support our MaxEnt-based conclusions, we compute $-\beta G(\mathbf{k}, \tau=\beta/2)$ as a proxy for $A(\mathbf{k}, \omega=0)$, which avoids the need for analytic continuation.
This proxy is based on the fact that $G(\mathbf{k}, \tau=\beta/2)$ incorporates the largest contributions from $A(\mathbf{k}, \omega)$ near $\omega=0$. 
Specifically, from \cref{howA}, 
\begin{equation}
-\beta G(\mathbf{k}, \tau=\frac{\beta}{2}) = \beta \int^{\infty}_{-\infty}d\omega\frac{1}{2\cosh(\omega\beta/2)}A(\mathbf{k},\omega). 
\end{equation}
As shown in \cref{proxybetaover2}, the resulting Fermi surfaces are qualitatively similar to those in \cref{fig:FS}, showing key features such as Fermi arcs and hot spots at corresponding dopings. 
Moreover, the proxy method is more tolerant of numerical errors in $G(\mathbf{k}, \tau)$ than MaxEnt, so we can access lower temperatures. In \cref{proxybetaover2}, in addition to presenting results at $T/t=0.25$, we also show those at $T/t=0.22$. At the lower temperature, the Fermi surface is sharper and exhibits more pronounced angular dependence.

\begin{figure}[htbp]
    \centering
    \includegraphics[width=\linewidth]{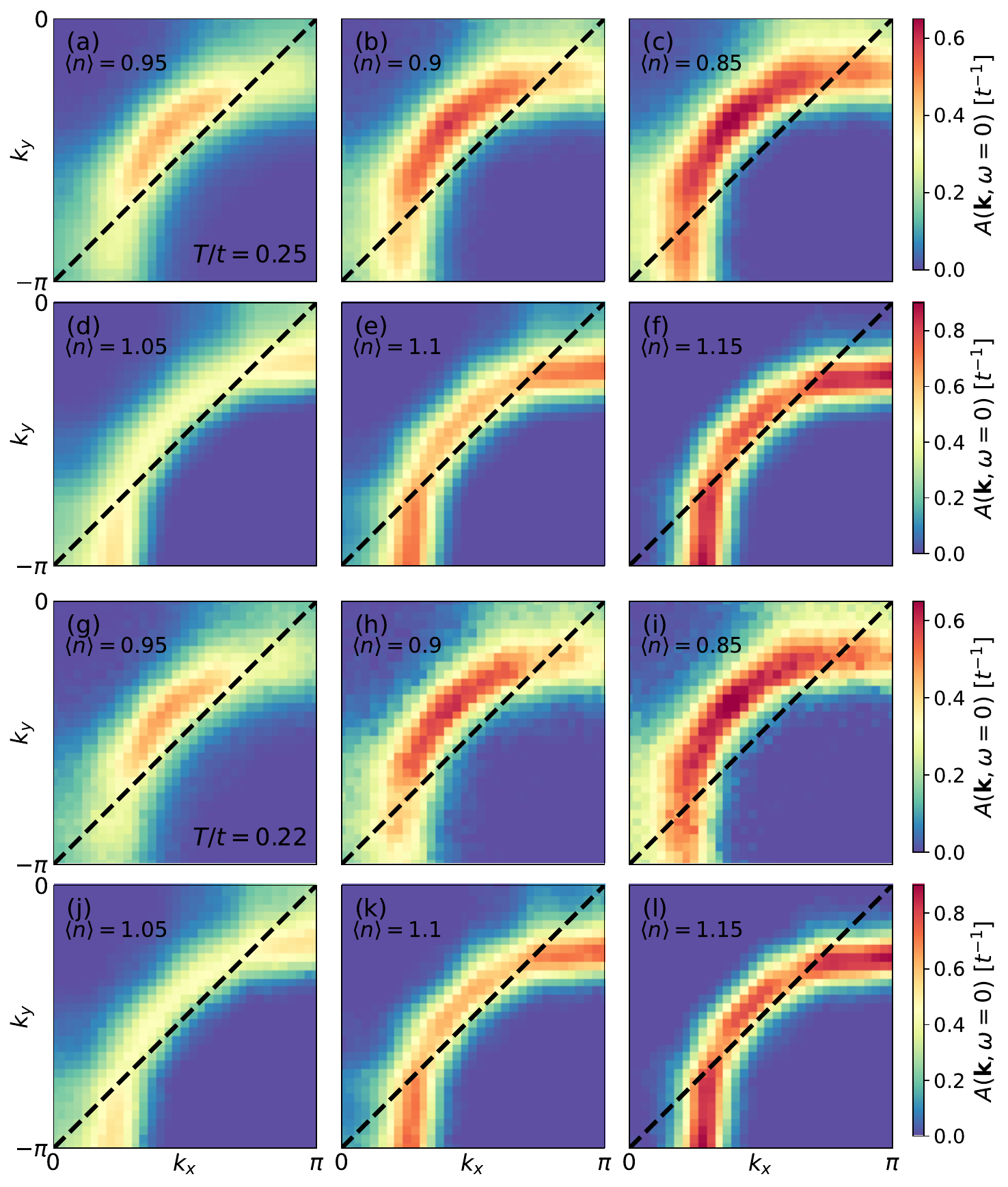}
\caption{Zero-frequency spectral function
$A(\mathbf{k},\omega=0)$ obtained from Pad\'e analytic continuation of
$G(\mathbf{k},i\omega_n)$.
All parameters match those in \cref{proxybetaover2}; only the color scale is adjusted.
}
    \label{anapade}
\end{figure}

As an additional cross-check of our MaxEnt continuation, we also applied Pad\'e analytic continuation to the single-particle Green's function on the fermionic Matsubara axis.
We first compute
\begin{align}
G(\mathbf{k},i\omega_n) = \int_0^\beta d\tau\, e^{i\omega_n\tau}\, G(\mathbf{k},\tau), 
\qquad \omega_n=(2n+1)\pi/\beta .
\label{eqmatsubara}
\end{align}
Together with Eqs.~\eqref{gkr}-\eqref{otau}, Eq.~\eqref{eqmatsubara} implies that \(G(\mathbf{k},i\omega_n)\) is the restriction to the imaginary axis of a function \(G(\mathbf{k},z)\) that is analytic in the upper half of the complex-frequency plane, and that its boundary value on the real axis yields the retarded Green's function,
\begin{equation}
G^R(\mathbf{k},\omega)=G(\mathbf{k},\omega+i\delta^+).
\end{equation}
In practice, we approximate \(G(\mathbf{k},z)\) by a Pad\'e rational function fitted to the Matsubara data \(G(\mathbf{k},i\omega_n)\), and evaluate it at \(z=\omega+i\delta\).
We then obtain \(A(\mathbf{k},\omega)\) from Eq.~\eqref{defA}; in particular, we extract \(A(\mathbf{k},0)\).
To suppress spurious sharp features that can arise in Pad\'e continuation of noisy data, we use a finite broadening \(\delta=0.2\).
The resulting spectral function is shown in Fig.~\ref{anapade}. The main features, including Fermi arcs and hot spots, are qualitatively consistent with Figs.~\ref{fig:FS} and \ref{proxybetaover2}.

\subsection{Twisted Boundary Conditions} \label{twistedsec}

When simulating the Hamiltonian in Eq.~\eqref{hubbard} with periodic boundary conditions applied along both the $x$ and $y$ directions, such that
\begin{align}
\hat{c}^\dagger_{\mathbf{i}+L\mathbf{e}_x,\sigma} = \hat{c}^\dagger_{\mathbf{i},\sigma}, \nonumber \\
\hat{c}^\dagger_{\mathbf{i}+L\mathbf{e}_y,\sigma} = \hat{c}^\dagger_{\mathbf{i},\sigma},
\end{align}
we are able to compute the Green's function in Eq.~\eqref{Gk} at discrete momentum points defined as 
\begin{align}
\mathbf{k}_0 = \frac{2\pi}{L} (m_x \mathbf{e}_x + m_y \mathbf{e}_y), \label{k0}
\end{align}
for $m_x, m_y = 0, ..., L-1$.
To access additional momentum points,
\begin{align}
\mathbf{k} = \mathbf{k}_0 +\bm{\theta} \label{condition}
\end{align}
for non-zero arbitrary $\bm{\theta}$, without resorting to more costly larger-lattice simulations, we employ ``twisted'' boundary conditions:
\begin{align}
\hat{c}^\dagger_{\mathbf{i}+L \mathbf{e}_x,\sigma} = e^{-iL\bm{\theta}\cdot \mathbf{e}_x} \hat{c}^\dagger_{\mathbf{i},\sigma}, \label{twist1}\\
\hat{c}^\dagger_{\mathbf{i}+L \mathbf{e}_y,\sigma} = e^{-iL\bm{\theta}\cdot \mathbf{e}_y} \hat{c}^\dagger_{\mathbf{i},\sigma}. \label{twist2}
\end{align}
Applying \cref{twist1}, we have 
\begin{align}
&\hat{c}_{\mathbf{k},\sigma}^\dagger  = \frac{1}{L}\sum_{\mathbf{i}}e^{i\mathbf{k}\cdot\mathbf{i}}\hat{c}^\dagger_{\mathbf{i},\sigma} \nonumber \\
& = \frac{1}{L}\sum_{\mathbf{i}}e^{i\mathbf{k}\cdot(\mathbf{i}+L\mathbf{e}_x)}\hat{c}^\dagger_{\mathbf{i}+L\mathbf{e}_x,\sigma} = \frac{1}{L}\sum_{\mathbf{i}}e^{i\mathbf{k}\cdot\mathbf{i}}e^{iL(\mathbf{k}-\bm{\theta})\cdot\mathbf{e}_x} \hat{c}^\dagger_{\mathbf{i},\sigma} \nonumber \\
& = e^{iL(\mathbf{k}-\bm{\theta})\cdot\mathbf{e}_x} \hat{c}_{\mathbf{k},\sigma}^\dagger.
\end{align}
Similarly, for the $y$ direction, $\hat{c}_{\mathbf{k},\sigma}^\dagger = e^{iL(\mathbf{k}-\bm{\theta})\cdot\mathbf{e}_y} \hat{c}_{\mathbf{k},\sigma}^\dagger$ using \cref{twist2}.
For $\hat{c}_{\mathbf{k},\sigma}^\dagger$ to be well-defined under the twisted boundary conditions, the following must hold:
\begin{align}
e^{iL(\mathbf{k}-\bm{\theta})\cdot\mathbf{e}_x} = 1, \\
e^{iL(\mathbf{k}-\bm{\theta})\cdot\mathbf{e}_y} = 1,
\end{align}
which are satisfied by the momentum points $\mathbf{k}$ defined in Eq.~\eqref{condition}.

The ``twisted'' conditions in Eqs.~\eqref{twist1} and \eqref{twist2} modify translational symmetry as the hopping energy between the lattice's ending sites along either direction now includes an additional phase.
For example, the term $t\hat{c}^\dagger_{L\mathbf{e}_x,\sigma}\hat{c}_{(L-1)\mathbf{e}_x,\sigma}=te^{-iL\theta_x}\hat{c}^\dagger_{0,\sigma}\hat{c}_{(L-1)\mathbf{e}_x,\sigma}$ under the twisted boundary. To make translational symmetry manifest in the Hamiltonian, we define alternative operators as
\begin{align}
\hat{\tilde{c}}_{\mathbf{i},\sigma}^\dagger \equiv e^{i\bm{\theta}\cdot\mathbf{i}}\hat{c}_{\mathbf{i},\sigma}^\dagger,
\end{align}
which satisfy:
\begin{align}
\hat{\tilde{c}}_{\mathbf{i}+L\mathbf{e}_x,\sigma}^\dagger = e^{i\bm{\theta}\cdot(\mathbf{i}+L\mathbf{e}_x)} \hat{c}_{\mathbf{i}+L\mathbf{e}_x,\sigma}^\dagger = e^{i\bm{\theta}\cdot\mathbf{i}}\hat{c}_{\mathbf{i},\sigma}^\dagger = \hat{\tilde{c}}_{\mathbf{i},\sigma}^\dagger.
\end{align}
Accordingly,
\begin{align}
\hat{c}^\dagger_{\mathbf{i},\sigma}\hat{c}_{\mathbf{j},\sigma} = e^{-i\bm{\theta}\cdot(\mathbf{i}-\mathbf{j})} \hat{\tilde{c}}^\dagger_{\mathbf{i},\sigma}\hat{\tilde{c}}_{\mathbf{j},\sigma}.
\end{align}
Therefore, the Hamiltonian in Eq.~\eqref{hubbard}, when expressed in terms of $\hat{\tilde{c}}^\dagger$ and $\hat{\tilde{c}}$, preserves translational symmetry and is used in our simulations. 
Complex number calculations are required when $\theta \neq 0$, and these are more computationally expensive compared to the $\theta = 0$ case.

We compute $A(\mathbf{k}_0+\bm{\theta},\omega)$ by using \cref{howA} and calculating $G(\mathbf{k}_0+\bm{\theta},\tau)$, defined in \cref{Gk}, through 
\begin{align}
G(\mathbf{k}_0+\bm{\theta},\sigma,\tau) = - \langle T_\tau \hat{\tilde{c}}_{\mathbf{k}_0,\sigma}(\tau) \hat{\tilde{c}}_{\mathbf{k}_0,\sigma}^\dagger \rangle, 
\end{align}
because
\begin{align}
\hat{c}^\dagger_{\mathbf{k}_0+\bm{\theta},\sigma} = \frac{1}{L} \sum_\mathbf{i} e^{i(\mathbf{k}_0+\bm{\theta})\cdot\mathbf{i}} \hat{c}^\dagger_{\mathbf{i},\sigma} = \frac{1}{L} \sum_\mathbf{i} e^{i \mathbf{k}_0 \cdot \mathbf{i} } \hat{\tilde{c}}_{\mathbf{i},\sigma}^\dagger = \hat{\tilde{c}}^\dagger_{\mathbf{k}_0,\sigma}.
\end{align}

In the paper, we primarily use an $8\times 8$ lattice size for our simulations. 
Given the spectral function's more significant variation across momentum space at lower temperatures, we apply twisted boundary conditions for $T/t \leq 1/3$ to enhance momentum resolution. 
We employ $15$ different twist angles (including $\theta=0$) to achieve a momentum resolution equal to that of a $64\times 64$ simulation.
We further apply the interpolation procedure, as described in the following section, based on these momentum points, which additionally enhances the resolution. The interpolated Fermi surfaces are shown in \cref{fig:Tdep_highdop} and \cref{fig:Tdep_lowdop} in the main text.
\footnote{At high temperatures ($T/t > 1/3$), we apply $15$ twisted boundary conditions for $t''/t=0.15$, but only the standard periodic boundary condition ($\theta=0$) for $t''/t=0$. This choice does not affect the interpolated results because the spectral weight varies smoothly at these high temperatures.}

\subsection{Self-Energy and Interpolation}

For the non-interacting limit, the Green's function is given by
\begin{equation}
G^R(\mathbf{k},\omega) = \frac{1}{\omega-(\epsilon_\mathbf{k}-\mu)+i\delta},
\end{equation}
where $\epsilon_\mathbf{k}$ represents the band energy at momentum $\mathbf{k}$ and is defined by $\hat{H}(U=0)=\sum\limits_{\mathbf{k},\sigma}\epsilon_{\mathbf{k}}\hat{c}_{\mathbf{k},\sigma}^\dagger \hat{c}_{\mathbf{k},\sigma}$. The expression for $\epsilon_\mathbf{k}$ is
\begin{align}
\epsilon_\mathbf{k} = &-2t (\cos(k_x)+\cos(k_y)) -4t'\cos(k_x)\cos(k_y) \nonumber \\ &- 2t''(\cos(2k_x)+\cos(2k_y)).
\end{align}
When interactions are non-zero, as per Dyson's equation~\cite{mahan2013many,negele2018quantum,abrikosov2012methods}, the Green's function is modified to:
\begin{equation}
G^R(\mathbf{k},\omega) = \frac{1}{\omega-(\epsilon_\mathbf{k}-\mu)+i\delta-\Sigma_\mathbf{k}(\omega)}, \label{Dyson}
\end{equation}
where $\Sigma_\mathbf{k}(\omega)$ denotes the self-energy. 
After computing the spectral function $A(\mathbf{k},\omega)$ via MaxEnt, we obtain $\Im G^R(\mathbf{\mathbf{k}},\omega)$ through Eq.~\eqref{defA}. 
The real part of $ G^R(\mathbf{\mathbf{k}},\omega)$ is then calculated using the Kramers-Kronig relation. After computing the full $ G^R(\mathbf{\mathbf{k}},\omega)$, the self-energy $\Sigma_\mathbf{k}(\omega)$ is determined using Eq.~\eqref{Dyson}.

While the spectral function $A(\mathbf{k},\omega)$ can vary rapidly between momentum points, it is generally assumed that $\Sigma_\mathbf{k}(\omega)$ changes more gradually. Thus, to interpolate $A(\mathbf{k},\omega)$ within the Brillouin zone, 
we interpolate both the real and imaginary parts of $\Sigma_\mathbf{k}(\omega)$ to a finer grid of momentum points using two-dimensional spline interpolation, therefore achieving interpolated $A(\mathbf{k},\omega)$. 

\subsection{NMR Measurements and Spin Structure Factor} \label{nmrmeasurements}

The Knight shift, defined as the spin susceptibility at zero momentum ($\mathbf{q}=\mathbf{0}$) and zero frequency ($\omega=0$), is expressed as
\begin{align}
&\chi_s(\mathbf{q} = \mathbf{0},\omega = 0) = \int_0^{\beta} d\tau \chi_s(\mathbf{q}=\mathbf{0},\tau), \label{knightshifteq}
\end{align}
where $\chi_s(\mathbf{q},\tau)$ is related to its real-space counterpart $\chi_s(\mathbf{r},\tau)$ through a discrete Fourier transform,
\begin{align}
\chi_s(\mathbf{q},\tau) = \sum_\mathbf{r} \chi_s(\mathbf{r},\tau) e^{-i\mathbf{q}\cdot\mathbf{r}}. \label{fourierq}
\end{align}
The spin-spin correlation function in real space and imaginary time is defined as
\begin{align}
\chi_s(\mathbf{r},\tau)=\frac{1}{4}\ev{(\hat{n}_{\mathbf{r},\uparrow}(\tau)-\hat{n}_{\mathbf{r},\downarrow}(\tau))(\hat{n}_{\mathbf{0},\uparrow}(0)-\hat{n}_{\mathbf{0},\downarrow}(0))}. \label{eq:chis}
\end{align}
To examine the spin-lattice relaxation time ($T_1$) for a nucleus $\alpha$ responding to a magnetic field parallel to the $c$ axis (indicated by the ``$\|$'' label), we use
~\cite{PhysRevB.41.1797,PhysRevB.42.167,PhysRevB.43.258,PhysRevB.52.13585,Chen2017}
\begin{align}
{(^\alpha T_{1 \|}T)}^{-1} \propto {\sum_{\mathbf{q}}} \frac{{^\alpha F}_{\|}(\mathbf{q})}{L^2}\lim_{\omega \rightarrow 0}\frac{\Im \chi_s(\mathbf{q},\omega)}{\omega}. \label{formfactor}
\end{align}
The symbol ``$\propto$'' refers to proportionality up to a constant factor. 
We choose to disregard constant prefactors to focus on the temperature dependence of both the Knight shift, as evaluated from \eqref{knightshifteq}, and ${(^\alpha T_{1 \|}T)}^{-1}$, as calculated from the right-hand side of \eqref{formfactor}, allowing us to present both results in arbitrary units. 
${^\alpha F}_{\|}(\mathbf{q})$ denotes the form factor.
The imaginary part of the dynamical spin susceptibility $\Im \chi_s(\mathbf{q},\omega)$ is related to the correlation function in imaginary time $\tau$ by~\cite{Huang2018}
\begin{align}
\chi_s(\mathbf{q},\tau) = \int_0^\infty \frac{d\omega}{\pi} \frac{e^{-\tau\omega}+e^{-(\beta-\tau)\omega}}{1-e^{-\beta\omega}}\Im \chi_s(\mathbf{q},\omega). \label{maxent}
\end{align}
We apply MaxEnt analytic continuation to convert $\sum_\mathbf{q} {^\alpha F}_{\|}(\mathbf{q}) \chi_s(\mathbf{q},\tau)$ to $\sum_\mathbf{q} {^\alpha F}_{\|}(\mathbf{q}) \Im \chi_s(\mathbf{q},\omega)/\omega$ following the relationship in \eqref{maxent}, thereby determining the right-hand side of \eqref{formfactor}.

The dynamical spin structure factor $S(\mathbf{q},\omega)$, shown in Fig.~\ref{fig:Sqw} for $q_y=\pi$ in the $\omega\rightarrow0$ limit, is calculated by 
\begin{equation}
S(\mathbf{q},\omega) = \frac{1}{\pi}\frac{\Im \chi_s(\mathbf{q},\omega)}{1-e^{-\beta\omega}},
\end{equation}
with $\Im \chi_s(\mathbf{q},\omega)/\omega$ for individual $\mathbf{q}$ obtained using MaxEnt and \eqref{maxent}.

For Cu nuclei, by considering isotropic on-site hyperfine coupling and disregarding the overall constant, which is not of interest, we adopt a constant form factor ${^{\text{Cu}} F}_{\|}(\mathbf{q})=1$~\cite{PhysRevB.41.1797} for Fig.~\ref{fig:T1T}~(a). In this case, using \eqref{fourierq}, the right-hand side of \eqref{formfactor} corresponds to the zero-frequency slope of the imaginary part of the local dynamical spin susceptibility~\cite{PhysRevB.91.155111,PhysRevB.103.075136},
\begin{align}
{\sum_{\mathbf{q}}} \frac{1}{L^2}\lim_{\omega \rightarrow 0}\frac{\Im \chi_s(\mathbf{q},\omega)}{\omega} = \lim_{\omega \rightarrow 0}\frac{\Im \chi_s(\mathbf{r}=\mathbf{0},\omega)}{\omega}. \label{local}
\end{align}

Following the Shastry-Mila-Rice Hamiltonian~\cite{MILA1989561,PhysRevLett.63.1288}, we also employ the following form factor in Fig.~\ref{fig:T1T}~(b) for Cu sites~\cite{PhysRevB.41.1797,PhysRevB.42.167,PhysRevB.43.258,PhysRevB.52.13585,Chen2017}:
\begin{align}
{^{\text{Cu}} F}_{\|} = \left\{ A_{\perp} + 2B \left[ \cos(q_x) + \cos(q_y) \right] \right\}^2,
\end{align}
where we choose the ratio of the hyperfine constants as $A_{\perp}=0.84B$, as determined by experimental measurements
according to Refs.~\cite{PhysRevB.42.167,PhysRevB.43.258,PhysRevB.52.13585}.
The choice is considered suitably appropriate for both the YBa$_{2}$Cu$_{3}$O$_{6+x}$ and La$_{2-x}$Sr$_{x}$CuO$_{4}$ families~\cite{PhysRevB.52.13585}.
Ignoring units and overall constants, we set $B=1$ for simplicity.
For oxygen sites in Fig.~\ref{fig:T1T}~(c), we use~\cite{PhysRevB.41.1797,PhysRevB.52.13585}
\begin{align}
{^{\text{O}} F}_{\|} =  2 \left\{ 1 + \frac{1}{2} \left[ \cos(q_x) + \cos(q_y)  \right] \right\}, \label{oxygenfactor}
\end{align}
where once again, we omit the overall constant~\cite{PhysRevB.41.1797,PhysRevB.42.167,PhysRevB.43.258,PhysRevB.52.13585,Chen2017} in our calculations.
Note that all considered form factors ${^\alpha F}_{\|}$ are non-negative across all $\mathbf{q}$ values. 
One can derive that $\Im \chi_s(\mathbf{q},\omega)/\omega \geq 0$;
therefore, the summation $\sum_\mathbf{q} {^\alpha F}_{\|}(\mathbf{q}) \Im \chi_s(\mathbf{q},\omega)/\omega$ is non-negative (positive definite), 
which is a necessary condition for the applicability of the MaxEnt method to \eqref{maxent}.
For simplicity, symbols $\alpha$ and $\|$ are omitted in the main text.

\begin{figure}[htbp]
    \centering
    \includegraphics[width=\linewidth]{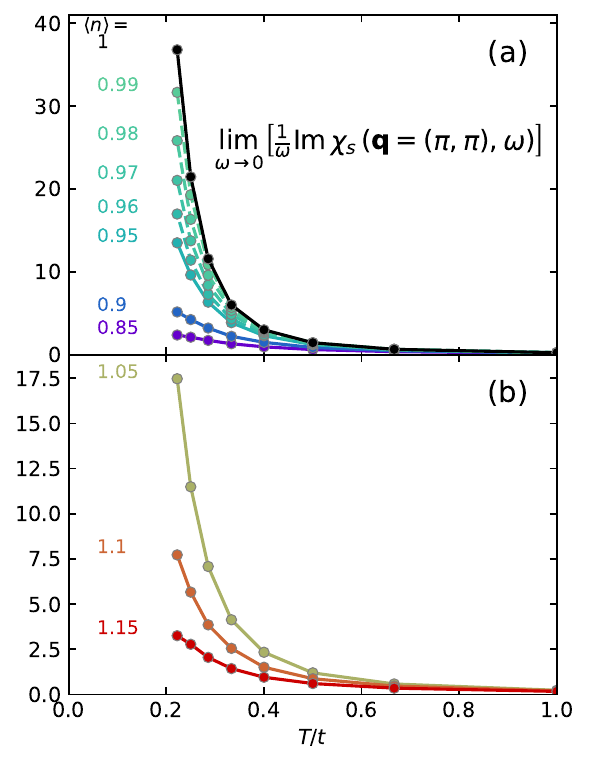}
    \caption{Temperature dependence of $\lim_{\omega \rightarrow 0}(\Im \chi_s(\mathbf{q} = (\pi,\pi),\omega )/\omega)$ [in arbitrary units] for filling range (a) $\langle n \rangle=0.85$ to $1$ and (b) $1.05$ to $1.15$.  
    Parameters are $U/t=6$, $t'/t=-0.25$, and $t''/t=0$, for a lattice size $8\times 8$.
}
    \label{fig:pipi}
\end{figure}

The form factor for O sites in \eqref{oxygenfactor} filters out the contribution from the AFM spin fluctuations at $\mathbf{q}=(\pi,\pi)$, while this contribution does not vanish for Cu sites. As a result, the large-momentum spin fluctuations, rapidly increasing with decreasing temperature as shown in Fig.~\ref{fig:pipi}, play a substantial role in determining the temperature dependence of ${(^{\text{Cu}} T_{1 \|}T)}^{-1}$, shown in Fig.~\ref{fig:T1T}~(a) and (b). 

\subsection{Transport and Thermodynamic Properties}
\label{sec:thermodynamics}

Detailed formalism and methodology for transport measurements can be found in Refs.~\cite{doi:10.1126/science.aau7063,PhysRevB.105.L161103,thermopower,doi:10.1126/science.ade3232}.
Systematic uncertainties associated with MaxEnt analytic continuation are discussed in \cref{MAXENT} and in the supplementary materials of some of these works.
Thermodynamic properties are calculated through equal-time correlation functions. 
We define $\Lambda_{\hat{O}_1 \hat{O}_2} = \beta(\langle{\hat{O}_1 \hat{O}_2}\rangle-\langle{\hat{O}_1}\rangle\langle{\hat{O}_2}\rangle)$ for arbitrary operators $\hat{O}_1$ and $\hat{O}_2$.
Specifically, the specific heat $c_v$, charge compressibility $\chi_c$ and spin susceptibility $\chi_s$ in Fig.~\ref{fig:thermodynamics} are calculated using the following formalism:
\begin{align}
&c_v = \frac{\beta}{L^2} \left(\Lambda_{\hat{H} \hat{H}} - \frac{\Lambda_{\hat{H} \hat{N}}^2}{\Lambda_{\hat{N}\hat{N}}}\right)  \nonumber \\
&\chi_c=\frac{1}{L^2} \Lambda_{\hat{N}\hat{N}} \nonumber \\
&\chi_s=\frac{1}{L^2} \Lambda_{\hat{S}_z\hat{S}_z},
\end{align}
where $\hat{S}_z\equiv\sum_{\mathbf{r}}(\hat{n}_{\mathbf{r},\uparrow}-\hat{n}_{\mathbf{r},\downarrow})/2$ is the $z$-component of the total spin operator.
Details about formalism for some of these measurements can be found in Refs.~\cite{PhysRevB.105.L161103,thermopower,doi:10.1126/science.ade3232}.
To clarify the relationship between the thermodynamic property $\chi_s$ and previous notations that use the same symbol, the spin susceptibility $\chi_s$ as discussed here is defined as the uniform and static limit of the correlation function, 
$\chi_s(\mathbf{q}=\mathbf{0},\tau=0)$ (definition given in \eqref{fourierq} and \eqref{eq:chis}), multiplied by $\beta$. 
Since the total $\hat{S}_z$ is a conserved quantity, $\chi_s(\mathbf{q}=\mathbf{0},\tau)$ is independent of $\tau$, and therefore the equal-time spin susceptibility $\chi_s =\Lambda_{\hat{S}_z\hat{S}_z}/L^2$ is equal to the Knight shift $K_s \equiv\chi_s(\mathbf{q} = \mathbf{0},\omega = 0)$.

\setcounter{figure}{0}
\renewcommand{\thefigure}{B\arabic{figure}}

\makeatletter
\renewcommand{\theHfigure}{B\arabic{figure}}
\makeatother

\section{Other Simulation Details} \label{supp2}

\begin{figure*}[htbp]
    \centering
    \includegraphics[width=\linewidth]{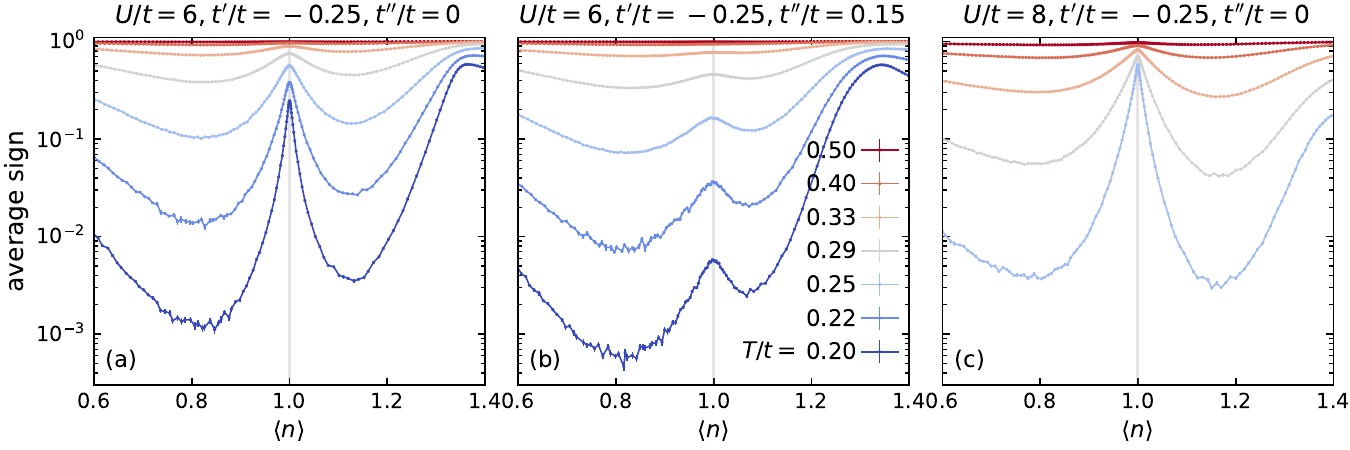}
    \caption{Average fermion sign of the DQMC simulations on an $8\times 8$ lattice.
    Trotter discretization is $d\tau = 0.02/t$.
    Error bars are included for both density $\langle n\rangle$ and average sign, although they are mostly invisible for density.
}
    \label{fig:fermion_sign}
\end{figure*}

Most technical details of the DQMC algorithm and MaxEnt procedure are similar to those in Refs.~\cite{doi:10.1126/science.aau7063,PhysRevB.105.L161103,thermopower,doi:10.1126/science.ade3232}.
Simulations are performed on $8\times 8$ square lattice clusters unless specified otherwise. 
Twisted boundary conditions are applied in certain single-particle measurements, as described in \cref{twistedsec}, to enhance momentum resolution, while original periodic boundary conditions are used for other measurements.

The imaginary time Trotter discretization, denoted as $d\tau$, is selected to be sufficiently small to effectively control the Trotter errors. 
Unless specified, all quantities are measured using $d\tau=0.05/t$. 
At high temperatures, the minimum number of imaginary time slices used in the Trotter decomposition was $\tilde{L}=\beta/d\tau = 20$.
The chemical potential tuning process follows the methodology described in the Supplementary Materials of Ref.~\cite{doi:10.1126/science.ade3232} to determine the chemical potential $\mu$ for a designated target filling $\langle n\rangle$.
The maximum imaginary time discretization $d\tau$ used in this tuning process is set at $0.02/t$. 
The chemical potential is tuned independently for different parameters.
For various twisted boundary conditions under the same parameters, we use the value of $\mu$ tuned with the original periodic boundary conditions.

Using DQMC for the Hubbard model without particle-hole symmetry, we encounter the fermion sign problem.
Figure~\ref{fig:fermion_sign} displays the average fermion sign for both hole- and electron-doping.
The sign problem becomes worse with larger $U$, lower temperature, and intermediate doping. 
For the majority of data discussed in this paper with $U/t=6$, 
the lowest temperature is typically between $T/t = 1/4$ and $1/4.5$ to ensure small statistical error at a reasonable computational cost, particularly for measurements that involve the MaxEnt analytic continuation procedure, where sufficiently good statistics are critical.

To achieve acceptable statistical errors, extensive measurement data are collected. 
Each Markov chain in the Monte Carlo process includes $5\times 10^4$ warm-up sweeps followed by $10^6$ measurement sweeps through space-time. 
Unequal-time measurements are typically taken every $4$ sweeps.
At low temperatures, for each parameter set and specific boundary condition, we typically make on the order of $10^8$ to $10^9$ unequal-time measurements.
A greater number of equal-time measurements are obtained as they are collected multiple times during a single space-time sweep.

For the MaxEnt procedure, to determine the adjustable parameter which assigns statistical weights and entropy in the maximized function in MaxEnt, we use the method of Ref.~\cite{PhysRevE.94.023303}.
A ``model'' function is required in MaxEnt to regularize the real-frequency correlation function. 
For the single-particle spectral functions, the temperature annealing procedure is used at $T/t \leq 1/3$, where the outcome from $T/t = 1/3$ serves as the initial model function for lower temperatures, and a flat model is used for higher temperatures ($T/t \geq 1/3$).
For transport measurements, the model function is determined in the same way as in Refs.~\cite{thermopower,doi:10.1126/science.ade3232}. 
As in these works, for lattice sizes larger than $8\times 8$, the highest temperature model functions are chosen to be the spectral functions obtained on an $8\times 8$ lattice at one higher temperature.
For calculations of $S(\mathbf{q},\omega)$ and $(T_1 T)^{-1}$, we use MaxEnt to obtain $\Im \chi_s(\mathbf{q},\omega)/\omega$ or its summation with the appropriate form factor, where a flat model is used for all parameters.

Uncertainties in our measurements include sampling errors from the Monte Carlo data and systematic errors due to Trotter error, finite-size effects, and the limitations of analytic continuation.
Beyond those discussed here, detailed discussions of these systematic errors can be found in prior publications~\cite{doi:10.1126/science.aau7063,PhysRevB.105.L161103,thermopower,doi:10.1126/science.ade3232}. 
In this work, error bars representing statistical sampling errors are displayed for measurements, except for those of single-particle spectral functions and self-energies, which involve multiple twisted boundary conditions or interpolation processes that complicate their determination.
In many cases, the error bars are smaller than the symbols, making them not visible.
For measurements involving MaxEnt, including those related to $\Im \chi_s(\mathbf{q},\omega)$ (such as $S(\mathbf{q},\omega)$ and $1/(T_1T)$) and transport measurements (such as $\sigma(\omega)$, $\rho$, and $D^{-1}$), error bars are determined through bootstrap resampling. 
We calculated $100$ bootstraps and used the standard deviation of the distribution as the standard error of the mean. 
The mean values represent the averages obtained from these bootstrap samples.
For measurements not requiring MaxEnt, error bars are calculated using the jackknife resampling method.

\setcounter{figure}{0}
\renewcommand{\thefigure}{C\arabic{figure}}

\makeatletter
\renewcommand{\theHfigure}{C\arabic{figure}}
\makeatother

\section{Supplementary data} \label{suppdata}

\subsection{Inverse Charge Diffusivity} \label{suppdata:Dinv}

\begin{figure*}[htbp]
    \centering
    \includegraphics[width=0.9\linewidth]{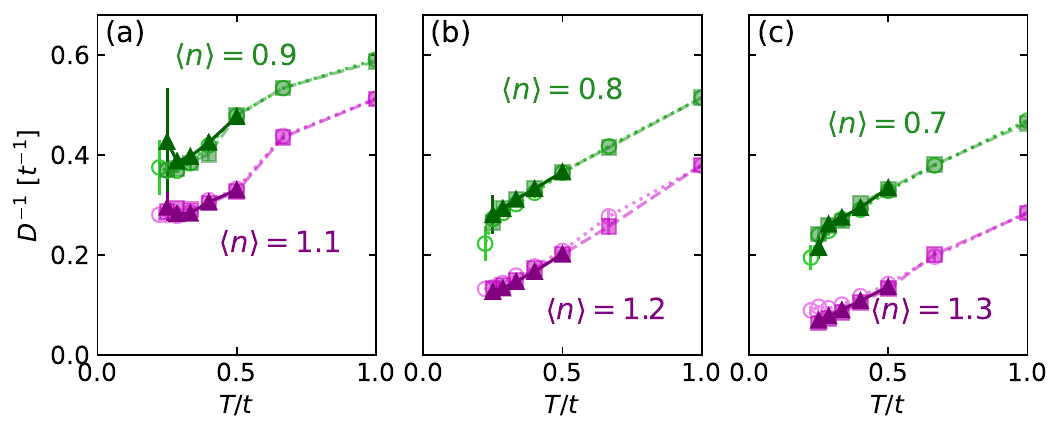}
    \caption{The inverse diffusivity $D^{-1}=\chi_c/\sigma(\omega=0)$ plotted in a similar way as in Fig.~\ref{fig:resistivity}~(d) to (f) in the main text, but with a unified $y$-axis scale in this plot. Parameters are the same as in Fig.~\ref{fig:resistivity}.
}
    \label{fig:invdiffusivity}
\end{figure*}

According to the Einstein relation, charge diffusivity is $D=\sigma/\chi_c$.
The inverse diffusivity $D^{-1}=\chi_c/\sigma(\omega=0)$ is shown in Fig.~\ref{fig:invdiffusivity}. The contrasting features observed between electron and hole doping are similar to those discussed for the DC resistivity $\rho$ in Fig.~\ref{fig:resistivity}~(d) to (f).

\subsection{Local Moment} \label{suppdata:mz2}

The local moment is
\begin{align}
&\langle \hat{m}_z^2 \rangle \equiv 4\chi_s(\mathbf{r}=\mathbf{0},\tau=0) \nonumber \\
&= \ev{(\hat{n}_{\uparrow}-\hat{n}_{\downarrow})^2} =\ev{\hat{n}_{\uparrow}+\hat{n}_{\downarrow}-2 \hat{n}_{\uparrow} \hat{n}_{\downarrow}}, \label{mz2}
\end{align}
where $\chi_s(\mathbf{r}=\mathbf{0},\tau=0)$ is defined in \eqref{eq:chis}.
For simplicity, we omit the specification $\mathbf{r}=\mathbf{0},\tau=0$ for the density operators. Both the non-interacting ($U/t=0$) limit and the strong-coupling ($U/t \rightarrow \infty$) limit can be solved analytically.

\begin{figure}[htbp]
    \centering
    \includegraphics[width=\linewidth]{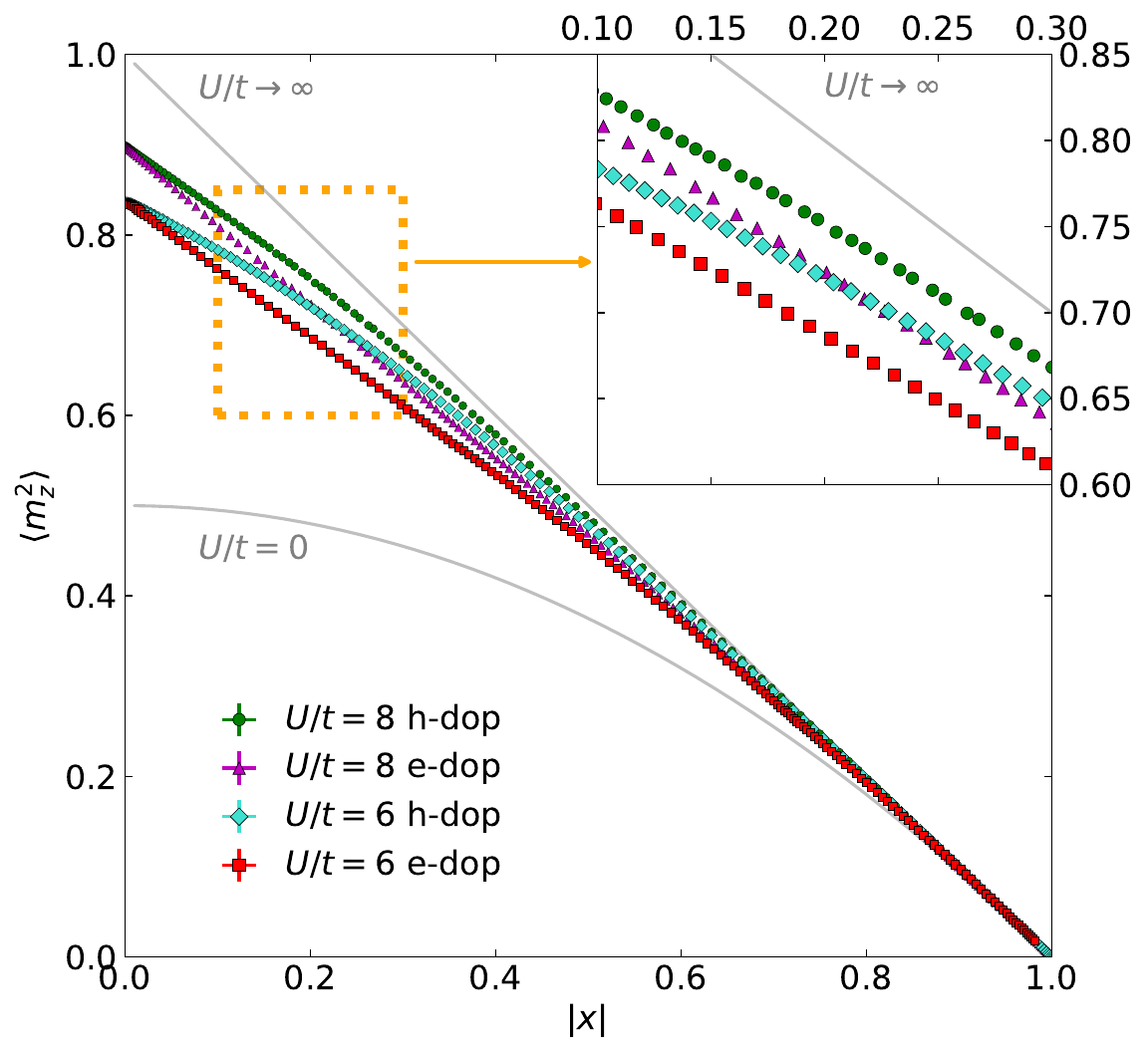}
    \caption{Local moment $\langle \hat{m}_z^2 \rangle$ for both hole and electron doping at temperature $T/t=1/4$ and with two interaction strengths $U/t=8$ and $6$. 
    Grey lines represent the $U/t=0$ limit ($\langle \hat{m}_z^2 \rangle = (1-x^2)/2$) and $U/t \rightarrow \infty$ limit ($\langle \hat{m}_z^2 \rangle = 1-|x|$) for reference.
    The inset provides a zoomed-in view of the highlighted area. 
    Parameters are $t'/t=-0.25$ and $t''/t=0$, for a lattice size $8\times 8$ with imaginary time discretization $d\tau=0.02/t$.
}
    \label{fig:mz2}
\end{figure}

In the $U/t=0$ limit, 
$\ev{\hat{n}_{\uparrow} \hat{n}_{\downarrow}} =\ev{\hat{n}_{\uparrow}} \ev{\hat{n}_{\downarrow}}$ and $\ev{\hat{n}_{\uparrow}}=\ev{\hat{n}_{\downarrow}}=\langle n\rangle/2=(1-x)/2$.
Eq.~\eqref{mz2} becomes
\begin{align}
 \ev{\hat{n}_{\uparrow}}+\ev{\hat{n}_{\downarrow}} - 2 \ev{\hat{n}_{\uparrow}} \ev{\hat{n}_{\downarrow}}
 =\langle  n\rangle-\frac{\langle n\rangle^2}{2} = \frac{1}{2}(1-x^2).
\end{align}

In the $U/t \rightarrow \infty$ limit, one can solve the grand canonical atomic-limit Hamiltonian $\hat{H}(t=0)-\mu \hat{N}$ for a single site to obtain the result $\langle \hat{m}_z^2 \rangle=1-|x|$.
The derivation of the result based on the minimum double occupancy imposed by energy penalties is discussed below.
For hole doping, double occupancy is zero ($\ev{\hat{n}_{\uparrow} \hat{n}_{\downarrow}} =0$).
Eq.~\eqref{mz2} simplifies to
\begin{align}
 & \ev{\hat{n}_{\uparrow}}+\ev{\hat{n}_{\downarrow}} = \ev{n} =1-x.
\end{align}
For electron doping, double occupancy is minimized, $\ev{\hat{n}_{\uparrow} \hat{n}_{\downarrow}} = \langle n\rangle-1 = -x$.
Eq.~\eqref{mz2} simplifies to
\begin{align}
 & \ev{\hat{n}_{\uparrow}}+\ev{\hat{n}_{\downarrow}} -2 \ev{\hat{n}_{\uparrow} \hat{n}_{\downarrow}} = 1-x+2x=1+x.
\end{align}
Summarizing both doping cases, we obtain $\langle \hat{m}_z^2 \rangle=1-|x|$.

Figure~\ref{fig:mz2} shows results for $\langle \hat{m}_z^2 \rangle$ under both hole and electron doping.
At $U/t=6$ and $8$, $\langle \hat{m}_z^2 \rangle$ deviates from the non-interacting limit and is closer to the strong-coupling (atomic) limit.
The hole-doped results are closer to the atomic limit compared to the electron-doped case, suggesting electrons are more localized in the hole-doped regime.

\begin{figure*}[htbp]
    \centering
    \includegraphics[width=\linewidth]{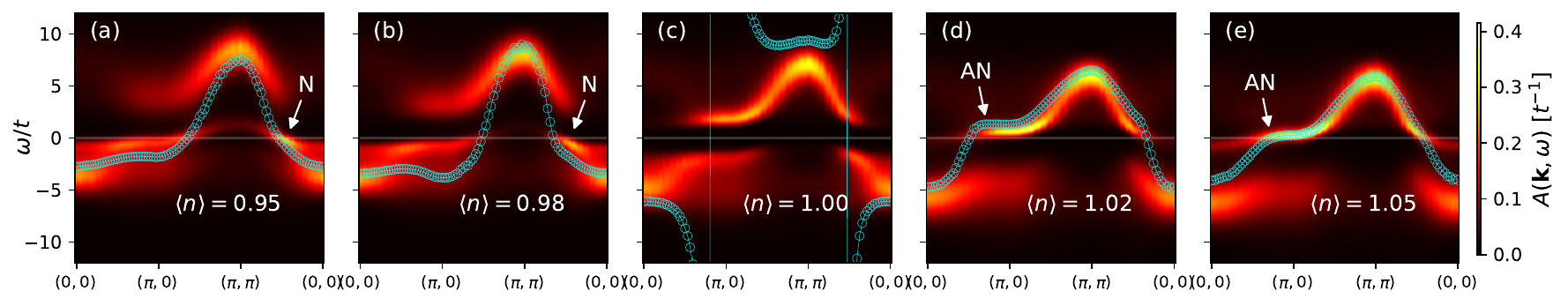}
\caption{Similar analysis to Fig.~\ref{fig:realselfenergy_kU6} in the main text but for $U/t=8$, $t'/t=-0.25$, and $t''/t=0$ at temperature $T/t=1/3.5$.
}
    \label{fig:realselfenergy_kU8}
\end{figure*}

\begin{figure*}[htbp]
    \centering
    \includegraphics[width=\linewidth]{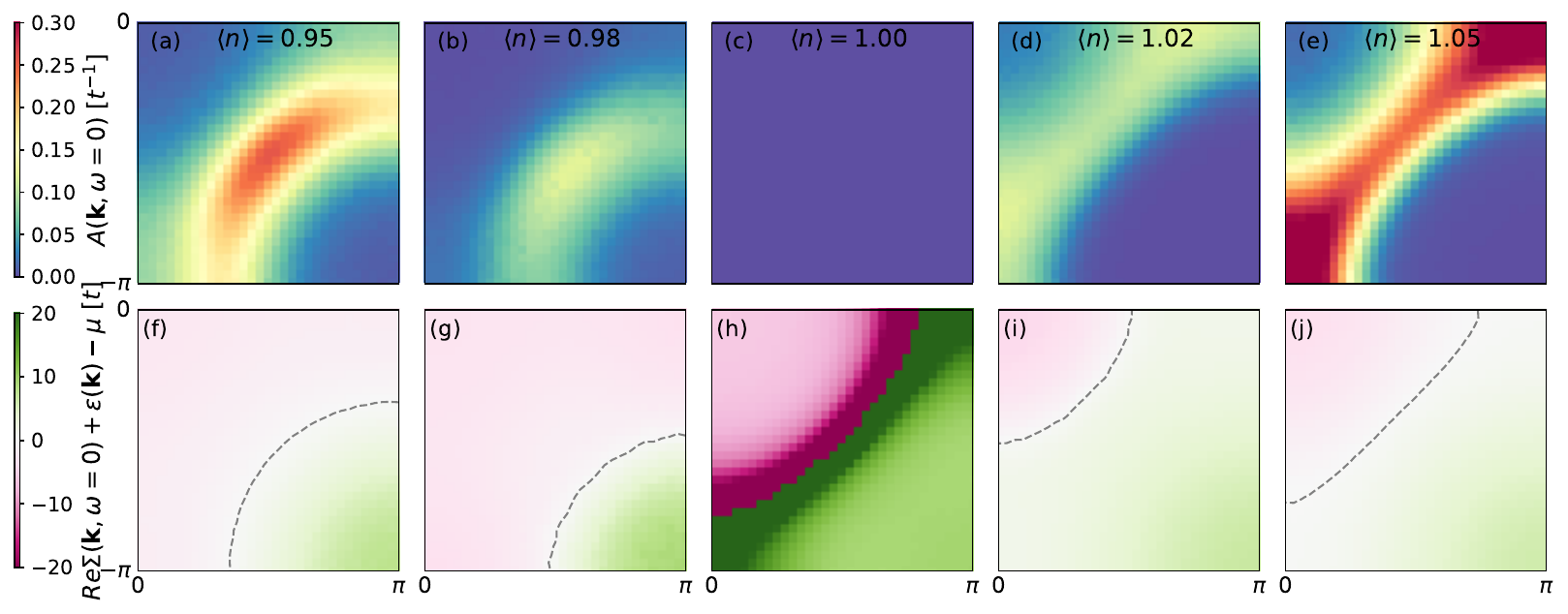}
\caption{Similar analysis to Fig.~\ref{fig:FS_realselfenergy_kU6} in the main text but for $U/t=8$, $t'/t=-0.25$, and $t''/t=0$ at temperature $T/t=1/3.5$.
}
    \label{fig:FS_realselfenergy_kU8}
\end{figure*}

\subsection{Supplementary Self-energy Analysis} \label{sec:selfsupp}

\cref{fig:realselfenergy_kU8} and \cref{fig:FS_realselfenergy_kU8} present a similar analysis to \cref{fig:realselfenergy_kU6} and \cref{fig:FS_realselfenergy_kU6} in the main text, but for $U/t=8$ and temperature $T/t=1/3.5$.
As $U$ is greater, the gap is more pronounced, and accordingly, the singularity in $\Re\Sigma_{\mathbf{k}}(\omega=0) + (\epsilon_{\mathbf{k}}-\mu)$ is sharper.

\begin{figure*}[htbp]
    \centering
    \includegraphics[width=\linewidth]{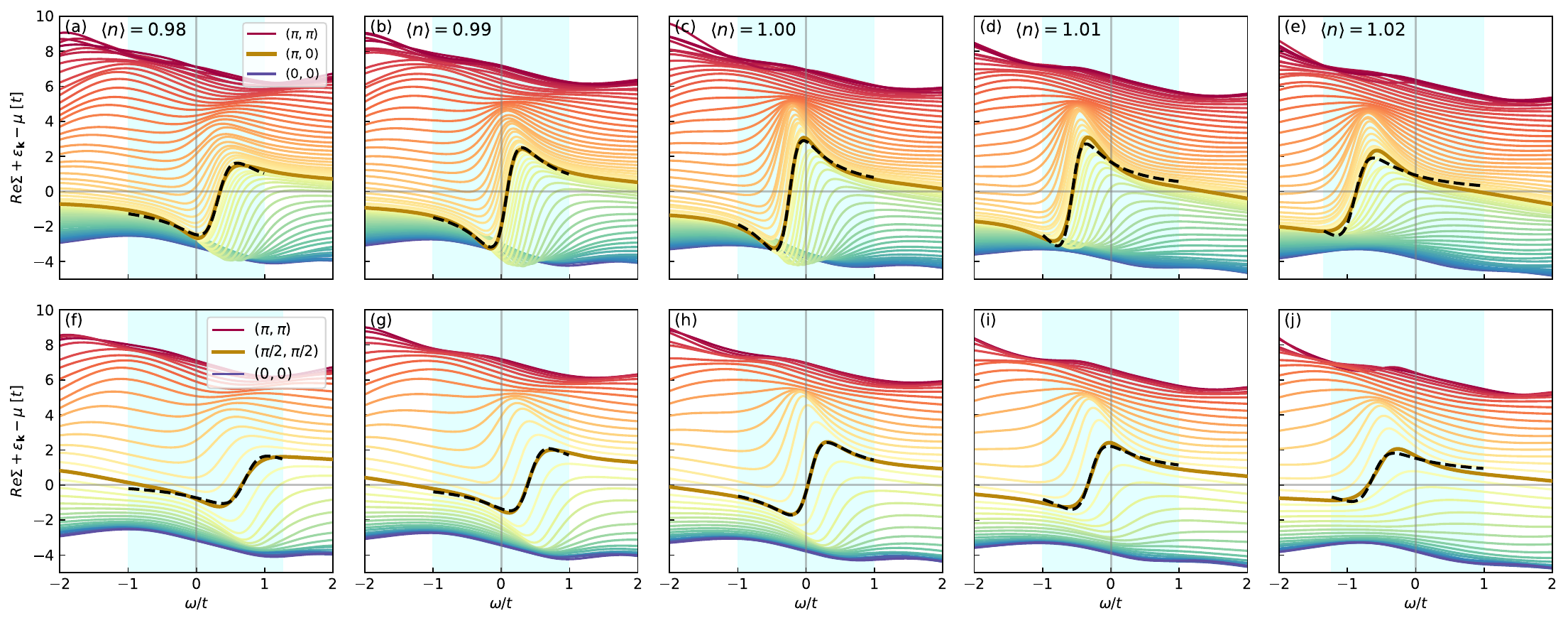}
\caption{Frequency dependence of $\Re\Sigma_{\mathbf{k}}(\omega) + (\epsilon_{\mathbf{k}}-\mu)$ along the antinodal and nodal cuts, plotted in a similar way to Fig.~\ref{fig:freq_realself_U6} for $U/t=6$, $t'/t=-0.25$, $t''/t=0$ but at temperature $T/t=1/3.5$ and up to $|x|= 0.02$. The shaded region marks where the data is fitted to \cref{eq:fitting_eq} and the black dashed lines are the resulting best-fit curves.
The selected region includes at least the interval $\omega \in [-t,t]$ and the frequencies at which the maximum and minimum occur.
Simulation lattice size is $8\times 8$.
}
    \label{fig:fittingU6}
\end{figure*}

\begin{figure*}[htbp]
    \centering
    \includegraphics[width=\linewidth]{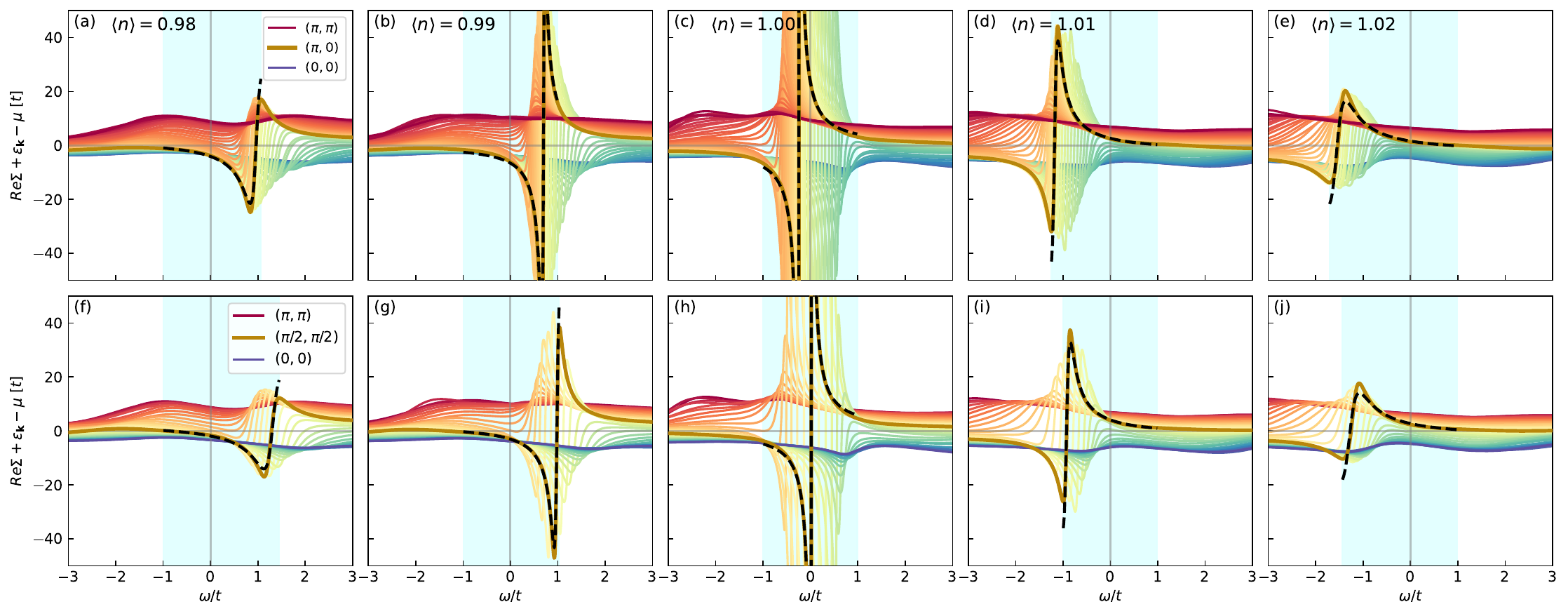}
\caption{Similar analysis to Fig.~\ref{fig:fittingU6} but for $U/t=8$.
}
    \label{fig:fittingU8}
\end{figure*}

\begin{figure}[htbp]
    \centering
    \includegraphics[width=\linewidth]{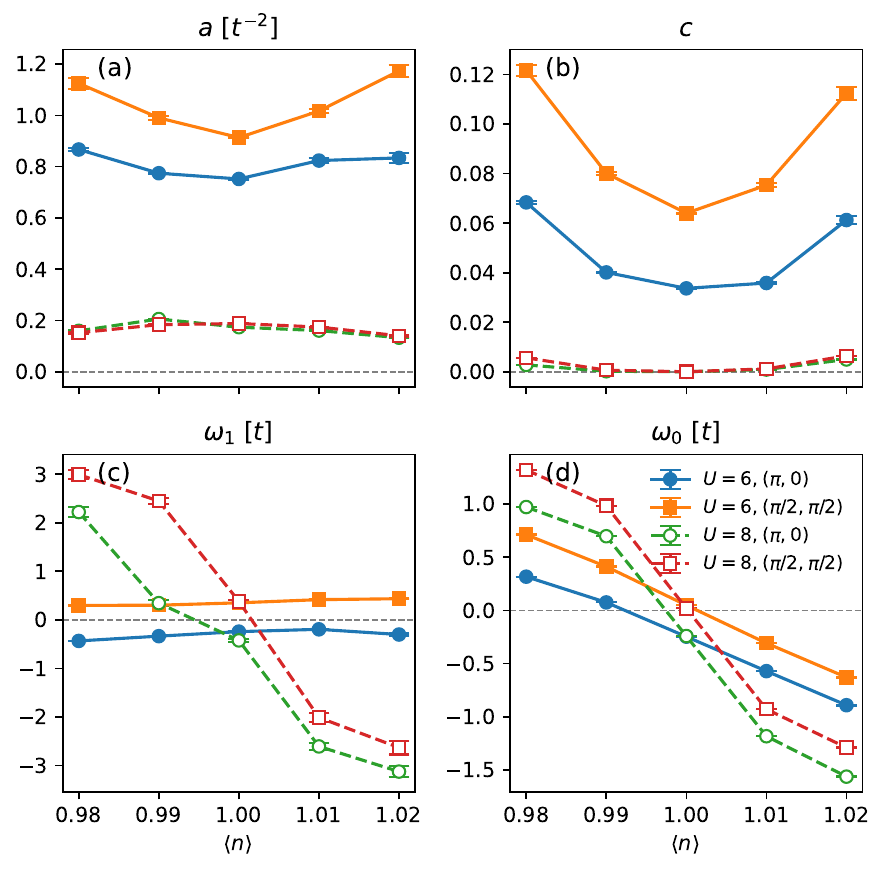}
\caption{The resulting fitting parameters. Best-fit curves are shown in \cref{fig:fittingU6} and \cref{fig:fittingU8} for $U/t=6$ and $U/t=8$, respectively.
}
    \label{fig:fittingresults}
\end{figure}

We now demonstrate the arguments about the frequency dependence of the real part of the self-energy, as introduced in the main text.
First, consider the case where $\Re \Sigma_\mathbf{k}(\omega)$ has a pole-like dependence, i.e., $\Re \Sigma_\mathbf{k}(\omega)= [a( \omega-\omega_0 )]^{-1}$.
Upon substitution into \cref{Dyson} (neglecting $\Im\Sigma$), the Green's function has two poles at
\begin{align}
\omega = \frac{\omega_0+\epsilon_\mathbf{k}-\mu}{2} \pm \frac{1}{2}\sqrt{[\omega_0-(\epsilon_\mathbf{k}-\mu)]^2+\frac{4}{a}},
\end{align}
implying a gap centered at $(\omega_0+\epsilon_\mathbf{k}-\mu)/2$. For the case in which $\omega_0 \approx 0$ and $\mathbf{k}$ satisfies $\epsilon_\mathbf{k}-\mu \approx 0$, the gap size is described by $2a^{-1/2}$.
On the other hand, if the self-energy varies linearly with frequency and has a negative slope, i.e.,
$\Sigma_\mathbf{k}(\omega)= -b\omega$, then \cref{Dyson} (neglecting $\Im\Sigma_\mathbf{k}$) becomes
\begin{align}
G^R(\mathbf{k},\omega) = \frac{(1+b)^{-1}}{\omega-(1+b)^{-1}(\epsilon_\mathbf{k}-\mu)}.
\end{align}
In this case, the effect is a renormalization of the band without the formation of a gap.
In Fig.~\ref{fig:freq_realself_U6} in the main text,
we see this negative slope happen as part of the tail of the pole-like $(\omega - \omega_0)^{-1}$ profile when doping is sufficiently high and the gap is far from the Fermi level.

For a more quantitative analysis,
we fit the frequency-dependent $\Re\Sigma_{\mathbf{k}}(\omega) + (\epsilon_{\mathbf{k}}-\mu)$ at $\mathbf{k}=(\pi,0)$ and $(\pi/2,\pi/2)$ to the following function: 
\begin{align}
\frac{\omega - \omega_0}{a (\omega- \omega_0)^ 2 + c}+\omega_1.\label{eq:fitting_eq}
\end{align}
Parameter $a$ is related to the gap size, and when $a$ is small, the gap is large.
Large magnitudes of $c$ indicate a deviation from the pole-like behavior associated with the gap.
The parameters $\omega_0$ and $\omega_1$ are related to the gap center and energy dispersion.
We apply this fitting to both $U/t=6$ and $U/t=8$, shown in \cref{fig:fittingU6} and \cref{fig:fittingU8}, respectively, using identical $t'$, $t''$, and temperature, and extending up to $|x|= 0.02$. 
The fitting performs reasonably well, and as the $U/t=8$ case has a more pronounced gap, the pole-like behavior is much sharper compared to that for $U/t=6$.

Fitting results that focus on $a$ and $c$ are shown in \cref{fig:fittingresults}~(a) and (b), respectively.
As expected, both $a$ and $c$ are smaller for $U/t=8$, where the gap is more pronounced, compared to $U/t=6$, and they increase (or show little variation) with doping, consistent with the diminishing influence of the gap as doping increases.

\subsection{Finite-Size Effects}

\begin{figure}[htbp]
    \centering
    \includegraphics[width=0.92\linewidth]{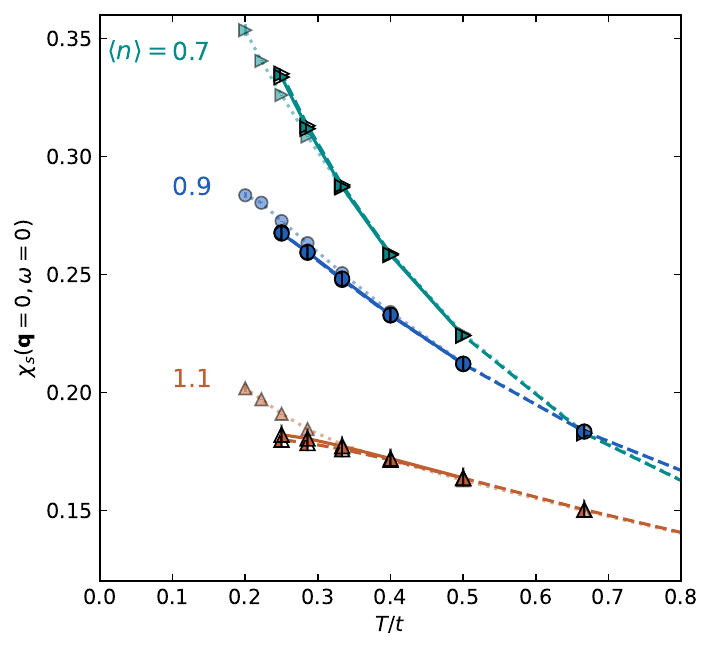}
    \caption{Finite size analysis of the knight shift $\chi_s(\mathbf{q} = \mathbf{0},\omega = 0)$ [in arbitrary units] at several representative dopings. Parameters are $U/t=6$, $t'/t=-0.25$, and $t''/t=0$. Lattice sizes are $8\times 8$ (dotted lines with semi-transparent symbols),  $10\times 10$ (dashed lines with left-filled symbols), and $12\times 12$ (solid lines with right-filled symbols).
}
    \label{fig:knightshift_finitesize}
\end{figure}

In the main text, we analyze finite-size effects on transport and thermodynamic properties, observing them to be typically more pronounced for electron doping. 
Similarly, the finite-size effects on the NMR knight shift are analyzed in Fig.~\ref{fig:knightshift_finitesize}, where finite-size effects do not qualitatively alter the observed behaviors or influence our conclusions.

As indicated by these measurements, finite-size effects persist for electron doping on the $8\times 8$ lattice at our lowest accessible temperatures. 
These finite-size effects are likely to extend to other properties not explicitly analyzed here, such as the single-particle spectral function, self-energy, and $(T_1 T)^{-1}$. However, while these effects may modify quantitative assessments, they are not expected to qualitatively impact our conclusions.

\subsection{Spectral function at low electron doping}\label{lowelectron}

\begin{figure*}[htbp]
    \centering
    \includegraphics[width=\linewidth]{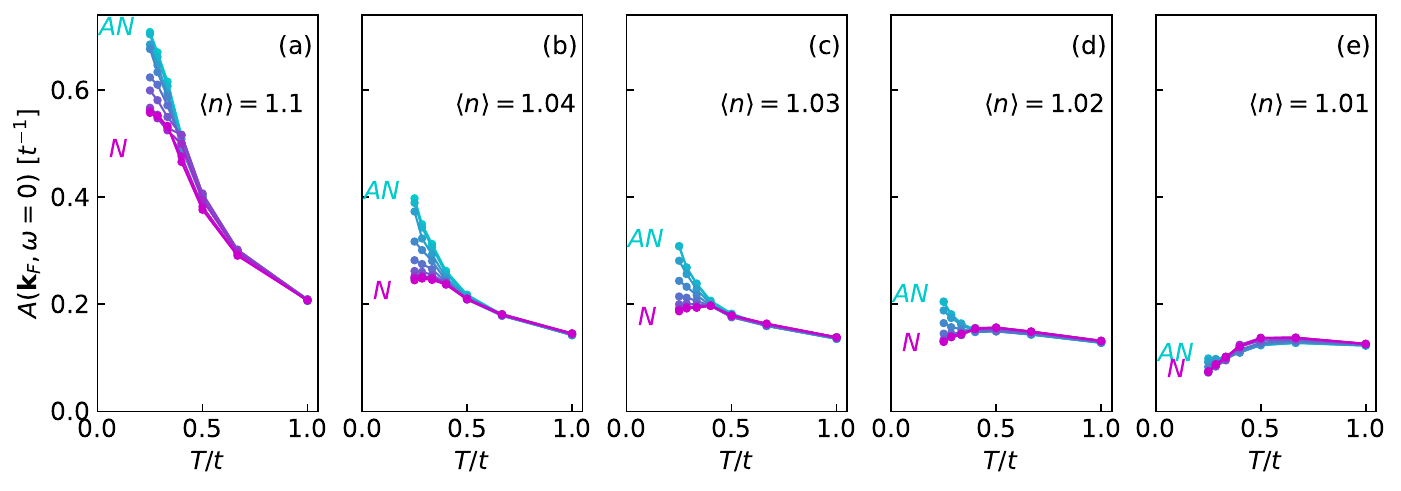}
   \caption{Temperature dependence of the spectral function $A(\mathbf{k},\omega=0)$ along the Fermi surface for electron doping, analogous to \cref{TAdop}. All parameters are the same as in \cref{TAdop}.}
    \label{TAdop_e}
\end{figure*}

\Cref{TAdop_e} shows the temperature evolution of $A(\mathbf{k},\omega=0)$ along the Fermi surface for electron doping, in direct analogy to \cref{TAdop}. At the lowest electron dopings, the spectra remain Mott-dominated: the spectral weight is suppressed along the entire Fermi surface upon cooling, similar to the very low hole-doping case ($x\sim 0.01$). The node-antinode dichotomy is reversed compared to hole doping. The characteristic temperature scale is comparable to that on the hole-doped side, but the dichotomy is weaker (the difference between the nodal and antinodal spectra is smaller).

\subsection{Systematic Error from MaxEnt}
\label{MAXENT}

\begin{figure}[htbp]
    \centering
    \includegraphics[width=\linewidth]{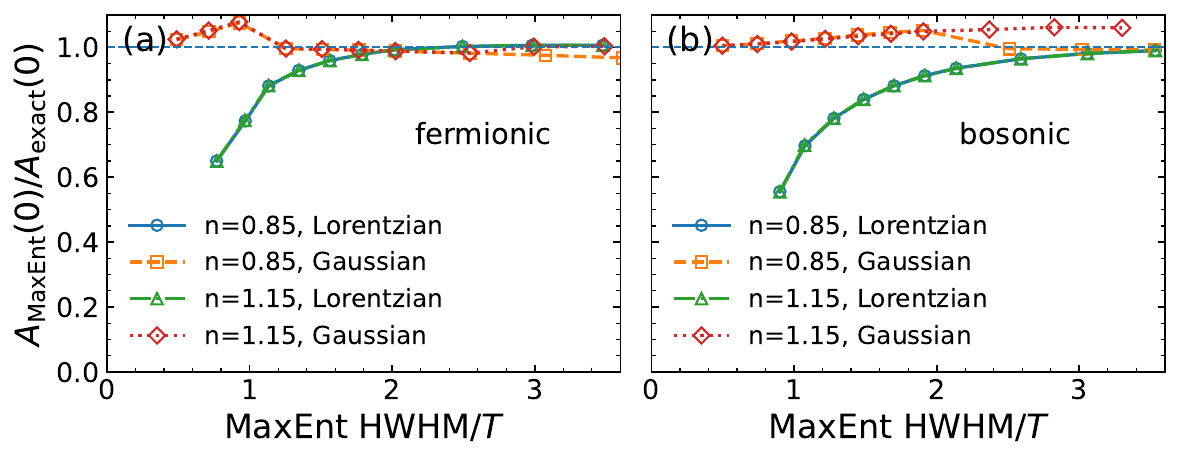}
\caption{
MaxEnt stability test using constructed spectra with DQMC-level noise. Parameters used to extract the DQMC noise are $U/t=6$, $t'/t=-0.25$, $t''/t=0$, and $T/t=1/3$, on an $8\times 8$ lattice.
We use a flat model in MaxEnt. For the fermionic (single-particle) test, we use the noise statistics of the local Green's function $G(\mathbf{r}=\mathbf{0},\tau)$, which corresponds to the local density of states after analytic continuation. 
For the bosonic (transport) test, we use the noise statistics of the $\mathbf{q}=\mathbf{0}$ current-current correlator, which yields the optical conductivity after analytic continuation.
}
    \label{fig:supp_fig_maxentcheck}
\end{figure}

Analytic continuation is intrinsically ill-conditioned, so it is important to assess the size of potential systematic errors introduced by MaxEnt.
For hole-doped transport, a detailed stability analysis was performed in the supplementary materials of Ref.~\cite{doi:10.1126/science.aau7063} for the bosonic kernel relevant to conductivity. The analysis found that MaxEnt reconstructions are quantitatively reliable when the half width at half maximum (HWHM) of the relevant spectral feature satisfies $\mathrm{HWHM}\gtrsim 2T$.
Here we follow the same strategy and extend it in two directions: (1) to the fermionic kernel in \cref{howA} relevant for single-particle spectra, and (2) to electron doping, where we observe sharper features that could, in principle, be harder to resolve.

Our procedure is as follows.
We construct a normalized trial spectrum centered at $\omega_0=0$ with a controlled width.
Specifically, we consider either a Lorentzian
\begin{align}
A_{\mathrm{L}}(\omega;\gamma)
= \frac{1}{\pi}\frac{\gamma}{(\omega-\omega_0)^2+\gamma^2},
\end{align}
whose half width at half maximum (HWHM) is $\gamma$, or a Gaussian
\begin{align}
A_{\mathrm{G}}(\omega;\sigma)
= \frac{1}{\sqrt{2\pi}\sigma}\exp\!\left[-\frac{(\omega-\omega_0)^2}{2\sigma^2}\right],
\end{align}
with $\mathrm{HWHM}=\sqrt{2\ln 2}\,\sigma$.
We vary $\gamma$ (or $\sigma$) to tune the feature width, generate the corresponding imaginary-time correlator using the appropriate kernel (fermionic or bosonic), and add noise drawn from DQMC.
We then apply MaxEnt and quantify the reconstruction accuracy via the ratio
$A_{\mathrm{MaxEnt}}(0)/A_{\mathrm{exact}}(0)$.

The results are summarized in \cref{fig:supp_fig_maxentcheck}.
For Gaussian test spectra, the reconstruction is robust over the range of widths we tested, consistent with Ref.~\cite{doi:10.1126/science.aau7063}.
For Lorentzian spectra, we reproduce the bosonic criterion $\mathrm{HWHM}\gtrsim 2T$ for quantitatively controlled reconstructions.
For the fermionic kernel, the requirement is slightly less stringent: the reconstruction becomes reliable already for $\mathrm{HWHM}\gtrsim 1.5T$ in our tests.
Using noise statistics taken from either hole- or electron-doped simulations does not qualitatively change these conclusions.

In the DQMC data analyzed in this work, the Drude peak in the conductivity is generally broad enough to satisfy $\mathrm{HWHM}\gtrsim 2T$, with only marginal cases at the lowest temperatures for the highest electron doping ($\langle n\rangle=1.3$), where the width remains $\gtrsim 1.5T$.
The dominant peaks in the single-particle spectra typically satisfy $\mathrm{HWHM}\gtrsim 1.5T$, with a few marginal cases (e.g., $\langle n\rangle=1.15$ at the lowest temperatures) approaching $\sim T$.
These checks support that the MaxEnt continuations used here are reliable for the trends and temperature scales discussed in the main text.

\bibliography{main}

\begin{thebibliography}{175}%
\makeatletter
\providecommand \@ifxundefined [1]{%
 \@ifx{#1\undefined}
}%
\providecommand \@ifnum [1]{%
 \ifnum #1\expandafter \@firstoftwo
 \else \expandafter \@secondoftwo
 \fi
}%
\providecommand \@ifx [1]{%
 \ifx #1\expandafter \@firstoftwo
 \else \expandafter \@secondoftwo
 \fi
}%
\providecommand \natexlab [1]{#1}%
\providecommand \enquote  [1]{``#1''}%
\providecommand \bibnamefont  [1]{#1}%
\providecommand \bibfnamefont [1]{#1}%
\providecommand \citenamefont [1]{#1}%
\providecommand \href@noop [0]{\@secondoftwo}%
\providecommand \href [0]{\begingroup \@sanitize@url \@href}%
\providecommand \@href[1]{\@@startlink{#1}\@@href}%
\providecommand \@@href[1]{\endgroup#1\@@endlink}%
\providecommand \@sanitize@url [0]{\catcode `\\12\catcode `\$12\catcode
  `\&12\catcode `\#12\catcode `\^12\catcode `\_12\catcode `\%12\relax}%
\providecommand \@@startlink[1]{}%
\providecommand \@@endlink[0]{}%
\providecommand \url  [0]{\begingroup\@sanitize@url \@url }%
\providecommand \@url [1]{\endgroup\@href {#1}{\urlprefix }}%
\providecommand \urlprefix  [0]{URL }%
\providecommand \Eprint [0]{\href }%
\providecommand \doibase [0]{https://doi.org/}%
\providecommand \selectlanguage [0]{\@gobble}%
\providecommand \bibinfo  [0]{\@secondoftwo}%
\providecommand \bibfield  [0]{\@secondoftwo}%
\providecommand \translation [1]{[#1]}%
\providecommand \BibitemOpen [0]{}%
\providecommand \bibitemStop [0]{}%
\providecommand \bibitemNoStop [0]{.\EOS\space}%
\providecommand \EOS [0]{\spacefactor3000\relax}%
\providecommand \BibitemShut  [1]{\csname bibitem#1\endcsname}%
\let\auto@bib@innerbib\@empty
\bibitem [{\citenamefont {Sobota}\ \emph {et~al.}(2021)\citenamefont {Sobota},
  \citenamefont {He},\ and\ \citenamefont {Shen}}]{RevModPhys.93.025006}%
  \BibitemOpen
  \bibfield  {author} {\bibinfo {author} {\bibfnamefont {J.~A.}\ \bibnamefont
  {Sobota}}, \bibinfo {author} {\bibfnamefont {Y.}~\bibnamefont {He}},\ and\
  \bibinfo {author} {\bibfnamefont {Z.-X.}\ \bibnamefont {Shen}},\ }\bibfield
  {title} {\bibinfo {title} {Angle-resolved photoemission studies of quantum
  materials},\ }\href {https://doi.org/10.1103/RevModPhys.93.025006} {\bibfield
   {journal} {\bibinfo  {journal} {Rev. Mod. Phys.}\ }\textbf {\bibinfo
  {volume} {93}},\ \bibinfo {pages} {025006} (\bibinfo {year}
  {2021})}\BibitemShut {NoStop}%
\bibitem [{\citenamefont {Vishik}(2018)}]{Vishik_2018}%
  \BibitemOpen
  \bibfield  {author} {\bibinfo {author} {\bibfnamefont {I.~M.}\ \bibnamefont
  {Vishik}},\ }\bibfield  {title} {\bibinfo {title} {Photoemission perspective
  on pseudogap, superconducting fluctuations, and charge order in cuprates: a
  review of recent progress},\ }\href
  {https://doi.org/10.1088/1361-6633/aaba96} {\bibfield  {journal} {\bibinfo
  {journal} {Reports on Progress in Physics}\ }\textbf {\bibinfo {volume}
  {81}},\ \bibinfo {pages} {062501} (\bibinfo {year} {2018})}\BibitemShut
  {NoStop}%
\bibitem [{\citenamefont {Fischer}\ \emph {et~al.}(2007)\citenamefont
  {Fischer}, \citenamefont {Kugler}, \citenamefont {Maggio-Aprile},
  \citenamefont {Berthod},\ and\ \citenamefont {Renner}}]{RevModPhys.79.353}%
  \BibitemOpen
  \bibfield  {author} {\bibinfo {author} {\bibfnamefont {O.}~\bibnamefont
  {Fischer}}, \bibinfo {author} {\bibfnamefont {M.}~\bibnamefont {Kugler}},
  \bibinfo {author} {\bibfnamefont {I.}~\bibnamefont {Maggio-Aprile}}, \bibinfo
  {author} {\bibfnamefont {C.}~\bibnamefont {Berthod}},\ and\ \bibinfo {author}
  {\bibfnamefont {C.}~\bibnamefont {Renner}},\ }\bibfield  {title} {\bibinfo
  {title} {Scanning tunneling spectroscopy of high-temperature
  superconductors},\ }\href {https://doi.org/10.1103/RevModPhys.79.353}
  {\bibfield  {journal} {\bibinfo  {journal} {Rev. Mod. Phys.}\ }\textbf
  {\bibinfo {volume} {79}},\ \bibinfo {pages} {353} (\bibinfo {year}
  {2007})}\BibitemShut {NoStop}%
\bibitem [{\citenamefont {Keimer}\ \emph {et~al.}(2015)\citenamefont {Keimer},
  \citenamefont {Kivelson}, \citenamefont {Norman}, \citenamefont {Uchida},\
  and\ \citenamefont {Zaanen}}]{Keimer2015}%
  \BibitemOpen
  \bibfield  {author} {\bibinfo {author} {\bibfnamefont {B.}~\bibnamefont
  {Keimer}}, \bibinfo {author} {\bibfnamefont {S.~A.}\ \bibnamefont
  {Kivelson}}, \bibinfo {author} {\bibfnamefont {M.~R.}\ \bibnamefont
  {Norman}}, \bibinfo {author} {\bibfnamefont {S.}~\bibnamefont {Uchida}},\
  and\ \bibinfo {author} {\bibfnamefont {J.}~\bibnamefont {Zaanen}},\
  }\bibfield  {title} {\bibinfo {title} {From quantum matter to
  high-temperature superconductivity in copper oxides},\ }\href
  {https://doi.org/10.1038/nature14165} {\bibfield  {journal} {\bibinfo
  {journal} {Nature}\ }\textbf {\bibinfo {volume} {518}},\ \bibinfo {pages}
  {179} (\bibinfo {year} {2015})}\BibitemShut {NoStop}%
\bibitem [{\citenamefont {Fradkin}\ \emph {et~al.}(2015)\citenamefont
  {Fradkin}, \citenamefont {Kivelson},\ and\ \citenamefont
  {Tranquada}}]{RevModPhys.87.457}%
  \BibitemOpen
  \bibfield  {author} {\bibinfo {author} {\bibfnamefont {E.}~\bibnamefont
  {Fradkin}}, \bibinfo {author} {\bibfnamefont {S.~A.}\ \bibnamefont
  {Kivelson}},\ and\ \bibinfo {author} {\bibfnamefont {J.~M.}\ \bibnamefont
  {Tranquada}},\ }\bibfield  {title} {\bibinfo {title} {{Colloquium: Theory of
  intertwined orders in high temperature superconductors}},\ }\href
  {https://doi.org/10.1103/RevModPhys.87.457} {\bibfield  {journal} {\bibinfo
  {journal} {Rev. Mod. Phys.}\ }\textbf {\bibinfo {volume} {87}},\ \bibinfo
  {pages} {457} (\bibinfo {year} {2015})}\BibitemShut {NoStop}%
\bibitem [{\citenamefont {Armitage}\ \emph {et~al.}(2001)\citenamefont
  {Armitage}, \citenamefont {Lu}, \citenamefont {Kim}, \citenamefont
  {Damascelli}, \citenamefont {Shen}, \citenamefont {Ronning}, \citenamefont
  {Feng}, \citenamefont {Bogdanov}, \citenamefont {Shen}, \citenamefont
  {Onose}, \citenamefont {Taguchi}, \citenamefont {Tokura}, \citenamefont
  {Mang}, \citenamefont {Kaneko},\ and\ \citenamefont
  {Greven}}]{PhysRevLett.87.147003}%
  \BibitemOpen
  \bibfield  {author} {\bibinfo {author} {\bibfnamefont {N.~P.}\ \bibnamefont
  {Armitage}}, \bibinfo {author} {\bibfnamefont {D.~H.}\ \bibnamefont {Lu}},
  \bibinfo {author} {\bibfnamefont {C.}~\bibnamefont {Kim}}, \bibinfo {author}
  {\bibfnamefont {A.}~\bibnamefont {Damascelli}}, \bibinfo {author}
  {\bibfnamefont {K.~M.}\ \bibnamefont {Shen}}, \bibinfo {author}
  {\bibfnamefont {F.}~\bibnamefont {Ronning}}, \bibinfo {author} {\bibfnamefont
  {D.~L.}\ \bibnamefont {Feng}}, \bibinfo {author} {\bibfnamefont
  {P.}~\bibnamefont {Bogdanov}}, \bibinfo {author} {\bibfnamefont {Z.-X.}\
  \bibnamefont {Shen}}, \bibinfo {author} {\bibfnamefont {Y.}~\bibnamefont
  {Onose}}, \bibinfo {author} {\bibfnamefont {Y.}~\bibnamefont {Taguchi}},
  \bibinfo {author} {\bibfnamefont {Y.}~\bibnamefont {Tokura}}, \bibinfo
  {author} {\bibfnamefont {P.~K.}\ \bibnamefont {Mang}}, \bibinfo {author}
  {\bibfnamefont {N.}~\bibnamefont {Kaneko}},\ and\ \bibinfo {author}
  {\bibfnamefont {M.}~\bibnamefont {Greven}},\ }\bibfield  {title} {\bibinfo
  {title} {Anomalous electronic structure and pseudogap effects in
  {${\mathrm{Nd}}_{1.85}{\mathrm{Ce}}_{0.15}{\mathrm{CuO}}_{4}$}},\ }\href
  {https://doi.org/10.1103/PhysRevLett.87.147003} {\bibfield  {journal}
  {\bibinfo  {journal} {Phys. Rev. Lett.}\ }\textbf {\bibinfo {volume} {87}},\
  \bibinfo {pages} {147003} (\bibinfo {year} {2001})}\BibitemShut {NoStop}%
\bibitem [{\citenamefont {Armitage}\ \emph {et~al.}(2002)\citenamefont
  {Armitage}, \citenamefont {Ronning}, \citenamefont {Lu}, \citenamefont {Kim},
  \citenamefont {Damascelli}, \citenamefont {Shen}, \citenamefont {Feng},
  \citenamefont {Eisaki}, \citenamefont {Shen}, \citenamefont {Mang},
  \citenamefont {Kaneko}, \citenamefont {Greven}, \citenamefont {Onose},
  \citenamefont {Taguchi},\ and\ \citenamefont
  {Tokura}}]{PhysRevLett.88.257001}%
  \BibitemOpen
  \bibfield  {author} {\bibinfo {author} {\bibfnamefont {N.~P.}\ \bibnamefont
  {Armitage}}, \bibinfo {author} {\bibfnamefont {F.}~\bibnamefont {Ronning}},
  \bibinfo {author} {\bibfnamefont {D.~H.}\ \bibnamefont {Lu}}, \bibinfo
  {author} {\bibfnamefont {C.}~\bibnamefont {Kim}}, \bibinfo {author}
  {\bibfnamefont {A.}~\bibnamefont {Damascelli}}, \bibinfo {author}
  {\bibfnamefont {K.~M.}\ \bibnamefont {Shen}}, \bibinfo {author}
  {\bibfnamefont {D.~L.}\ \bibnamefont {Feng}}, \bibinfo {author}
  {\bibfnamefont {H.}~\bibnamefont {Eisaki}}, \bibinfo {author} {\bibfnamefont
  {Z.-X.}\ \bibnamefont {Shen}}, \bibinfo {author} {\bibfnamefont {P.~K.}\
  \bibnamefont {Mang}}, \bibinfo {author} {\bibfnamefont {N.}~\bibnamefont
  {Kaneko}}, \bibinfo {author} {\bibfnamefont {M.}~\bibnamefont {Greven}},
  \bibinfo {author} {\bibfnamefont {Y.}~\bibnamefont {Onose}}, \bibinfo
  {author} {\bibfnamefont {Y.}~\bibnamefont {Taguchi}},\ and\ \bibinfo {author}
  {\bibfnamefont {Y.}~\bibnamefont {Tokura}},\ }\bibfield  {title} {\bibinfo
  {title} {Doping dependence of an $\mathit{n}$-type cuprate superconductor
  investigated by angle-resolved photoemission spectroscopy},\ }\href
  {https://doi.org/10.1103/PhysRevLett.88.257001} {\bibfield  {journal}
  {\bibinfo  {journal} {Phys. Rev. Lett.}\ }\textbf {\bibinfo {volume} {88}},\
  \bibinfo {pages} {257001} (\bibinfo {year} {2002})}\BibitemShut {NoStop}%
\bibitem [{\citenamefont {Damascelli}\ \emph {et~al.}(2003)\citenamefont
  {Damascelli}, \citenamefont {Hussain},\ and\ \citenamefont
  {Shen}}]{RevModPhys.75.473}%
  \BibitemOpen
  \bibfield  {author} {\bibinfo {author} {\bibfnamefont {A.}~\bibnamefont
  {Damascelli}}, \bibinfo {author} {\bibfnamefont {Z.}~\bibnamefont
  {Hussain}},\ and\ \bibinfo {author} {\bibfnamefont {Z.-X.}\ \bibnamefont
  {Shen}},\ }\bibfield  {title} {\bibinfo {title} {Angle-resolved photoemission
  studies of the cuprate superconductors},\ }\href
  {https://doi.org/10.1103/RevModPhys.75.473} {\bibfield  {journal} {\bibinfo
  {journal} {Rev. Mod. Phys.}\ }\textbf {\bibinfo {volume} {75}},\ \bibinfo
  {pages} {473} (\bibinfo {year} {2003})}\BibitemShut {NoStop}%
\bibitem [{\citenamefont {Matsui}\ \emph {et~al.}(2005)\citenamefont {Matsui},
  \citenamefont {Terashima}, \citenamefont {Sato}, \citenamefont {Takahashi},
  \citenamefont {Wang}, \citenamefont {Yang}, \citenamefont {Ding},
  \citenamefont {Uefuji},\ and\ \citenamefont
  {Yamada}}]{PhysRevLett.94.047005}%
  \BibitemOpen
  \bibfield  {author} {\bibinfo {author} {\bibfnamefont {H.}~\bibnamefont
  {Matsui}}, \bibinfo {author} {\bibfnamefont {K.}~\bibnamefont {Terashima}},
  \bibinfo {author} {\bibfnamefont {T.}~\bibnamefont {Sato}}, \bibinfo {author}
  {\bibfnamefont {T.}~\bibnamefont {Takahashi}}, \bibinfo {author}
  {\bibfnamefont {S.-C.}\ \bibnamefont {Wang}}, \bibinfo {author}
  {\bibfnamefont {H.-B.}\ \bibnamefont {Yang}}, \bibinfo {author}
  {\bibfnamefont {H.}~\bibnamefont {Ding}}, \bibinfo {author} {\bibfnamefont
  {T.}~\bibnamefont {Uefuji}},\ and\ \bibinfo {author} {\bibfnamefont
  {K.}~\bibnamefont {Yamada}},\ }\bibfield  {title} {\bibinfo {title}
  {Angle-resolved photoemission spectroscopy of the antiferromagnetic
  superconductor
  {${\mathrm{N}\mathrm{d}}_{1.87}{\mathrm{C}\mathrm{e}}_{0.13}{\mathrm{C}\mathrm{u}\mathrm{O}}_{4}$}:
  Anisotropic spin-correlation gap, pseudogap, and the induced quasiparticle
  mass enhancement},\ }\href {https://doi.org/10.1103/PhysRevLett.94.047005}
  {\bibfield  {journal} {\bibinfo  {journal} {Phys. Rev. Lett.}\ }\textbf
  {\bibinfo {volume} {94}},\ \bibinfo {pages} {047005} (\bibinfo {year}
  {2005})}\BibitemShut {NoStop}%
\bibitem [{\citenamefont {Matsui}\ \emph {et~al.}(2007)\citenamefont {Matsui},
  \citenamefont {Takahashi}, \citenamefont {Sato}, \citenamefont {Terashima},
  \citenamefont {Ding}, \citenamefont {Uefuji},\ and\ \citenamefont
  {Yamada}}]{PhysRevB.75.224514}%
  \BibitemOpen
  \bibfield  {author} {\bibinfo {author} {\bibfnamefont {H.}~\bibnamefont
  {Matsui}}, \bibinfo {author} {\bibfnamefont {T.}~\bibnamefont {Takahashi}},
  \bibinfo {author} {\bibfnamefont {T.}~\bibnamefont {Sato}}, \bibinfo {author}
  {\bibfnamefont {K.}~\bibnamefont {Terashima}}, \bibinfo {author}
  {\bibfnamefont {H.}~\bibnamefont {Ding}}, \bibinfo {author} {\bibfnamefont
  {T.}~\bibnamefont {Uefuji}},\ and\ \bibinfo {author} {\bibfnamefont
  {K.}~\bibnamefont {Yamada}},\ }\bibfield  {title} {\bibinfo {title}
  {{Evolution of the pseudogap across the magnet-superconductor phase boundary
  of ${\mathrm{Nd}}_{2-x}{\mathrm{Ce}}_{x}\mathrm{Cu}{\mathrm{O}}_{4}$}},\
  }\href {https://doi.org/10.1103/PhysRevB.75.224514} {\bibfield  {journal}
  {\bibinfo  {journal} {Phys. Rev. B}\ }\textbf {\bibinfo {volume} {75}},\
  \bibinfo {pages} {224514} (\bibinfo {year} {2007})}\BibitemShut {NoStop}%
\bibitem [{\citenamefont {Armitage}\ \emph {et~al.}(2010)\citenamefont
  {Armitage}, \citenamefont {Fournier},\ and\ \citenamefont
  {Greene}}]{RevModPhys.82.2421}%
  \BibitemOpen
  \bibfield  {author} {\bibinfo {author} {\bibfnamefont {N.~P.}\ \bibnamefont
  {Armitage}}, \bibinfo {author} {\bibfnamefont {P.}~\bibnamefont {Fournier}},\
  and\ \bibinfo {author} {\bibfnamefont {R.~L.}\ \bibnamefont {Greene}},\
  }\bibfield  {title} {\bibinfo {title} {Progress and perspectives on
  electron-doped cuprates},\ }\href
  {https://doi.org/10.1103/RevModPhys.82.2421} {\bibfield  {journal} {\bibinfo
  {journal} {Rev. Mod. Phys.}\ }\textbf {\bibinfo {volume} {82}},\ \bibinfo
  {pages} {2421} (\bibinfo {year} {2010})}\BibitemShut {NoStop}%
\bibitem [{\citenamefont {He}\ \emph {et~al.}(2019)\citenamefont {He},
  \citenamefont {Rotundu}, \citenamefont {Scheurer}, \citenamefont {He},
  \citenamefont {Hashimoto}, \citenamefont {Xu}, \citenamefont {Wang},
  \citenamefont {Huang}, \citenamefont {Jia}, \citenamefont {Chen},
  \citenamefont {Moritz}, \citenamefont {Lu}, \citenamefont {Lee},
  \citenamefont {Devereaux},\ and\ \citenamefont
  {Shen}}]{doi:10.1073/pnas.1816121116}%
  \BibitemOpen
  \bibfield  {author} {\bibinfo {author} {\bibfnamefont {J.}~\bibnamefont
  {He}}, \bibinfo {author} {\bibfnamefont {C.~R.}\ \bibnamefont {Rotundu}},
  \bibinfo {author} {\bibfnamefont {M.~S.}\ \bibnamefont {Scheurer}}, \bibinfo
  {author} {\bibfnamefont {Y.}~\bibnamefont {He}}, \bibinfo {author}
  {\bibfnamefont {M.}~\bibnamefont {Hashimoto}}, \bibinfo {author}
  {\bibfnamefont {K.-J.}\ \bibnamefont {Xu}}, \bibinfo {author} {\bibfnamefont
  {Y.}~\bibnamefont {Wang}}, \bibinfo {author} {\bibfnamefont {E.~W.}\
  \bibnamefont {Huang}}, \bibinfo {author} {\bibfnamefont {T.}~\bibnamefont
  {Jia}}, \bibinfo {author} {\bibfnamefont {S.}~\bibnamefont {Chen}}, \bibinfo
  {author} {\bibfnamefont {B.}~\bibnamefont {Moritz}}, \bibinfo {author}
  {\bibfnamefont {D.}~\bibnamefont {Lu}}, \bibinfo {author} {\bibfnamefont
  {Y.~S.}\ \bibnamefont {Lee}}, \bibinfo {author} {\bibfnamefont {T.~P.}\
  \bibnamefont {Devereaux}},\ and\ \bibinfo {author} {\bibfnamefont {Z.-X.}\
  \bibnamefont {Shen}},\ }\bibfield  {title} {\bibinfo {title} {Fermi surface
  reconstruction in electron-doped cuprates without antiferromagnetic
  long-range order},\ }\href {https://doi.org/10.1073/pnas.1816121116}
  {\bibfield  {journal} {\bibinfo  {journal} {Proceedings of the National
  Academy of Sciences}\ }\textbf {\bibinfo {volume} {116}},\ \bibinfo {pages}
  {3449} (\bibinfo {year} {2019})}\BibitemShut {NoStop}%
\bibitem [{\citenamefont {Proust}\ and\ \citenamefont
  {Taillefer}(2019)}]{annurev:/content/journals/10.1146/annurev-conmatphys-031218-013210}%
  \BibitemOpen
  \bibfield  {author} {\bibinfo {author} {\bibfnamefont {C.}~\bibnamefont
  {Proust}}\ and\ \bibinfo {author} {\bibfnamefont {L.}~\bibnamefont
  {Taillefer}},\ }\bibfield  {title} {\bibinfo {title} {The remarkable
  underlying ground states of cuprate superconductors},\ }\href
  {https://doi.org/10.1146/annurev-conmatphys-031218-013210} {\bibfield
  {journal} {\bibinfo  {journal} {Annual Review of Condensed Matter Physics}\
  }\textbf {\bibinfo {volume} {10}},\ \bibinfo {pages} {409} (\bibinfo {year}
  {2019})}\BibitemShut {NoStop}%
\bibitem [{\citenamefont {Xu}\ \emph {et~al.}(2023)\citenamefont {Xu},
  \citenamefont {Guo}, \citenamefont {Hashimoto}, \citenamefont {Li},
  \citenamefont {Chen}, \citenamefont {He}, \citenamefont {He}, \citenamefont
  {Li}, \citenamefont {Berntsen}, \citenamefont {Rotundu}, \citenamefont {Lee},
  \citenamefont {Devereaux}, \citenamefont {Rydh}, \citenamefont {Lu},
  \citenamefont {Lee}, \citenamefont {Tjernberg},\ and\ \citenamefont
  {Shen}}]{Xu2023}%
  \BibitemOpen
  \bibfield  {author} {\bibinfo {author} {\bibfnamefont {K.-J.}\ \bibnamefont
  {Xu}}, \bibinfo {author} {\bibfnamefont {Q.}~\bibnamefont {Guo}}, \bibinfo
  {author} {\bibfnamefont {M.}~\bibnamefont {Hashimoto}}, \bibinfo {author}
  {\bibfnamefont {Z.-X.}\ \bibnamefont {Li}}, \bibinfo {author} {\bibfnamefont
  {S.-D.}\ \bibnamefont {Chen}}, \bibinfo {author} {\bibfnamefont
  {J.}~\bibnamefont {He}}, \bibinfo {author} {\bibfnamefont {Y.}~\bibnamefont
  {He}}, \bibinfo {author} {\bibfnamefont {C.}~\bibnamefont {Li}}, \bibinfo
  {author} {\bibfnamefont {M.~H.}\ \bibnamefont {Berntsen}}, \bibinfo {author}
  {\bibfnamefont {C.~R.}\ \bibnamefont {Rotundu}}, \bibinfo {author}
  {\bibfnamefont {Y.~S.}\ \bibnamefont {Lee}}, \bibinfo {author} {\bibfnamefont
  {T.~P.}\ \bibnamefont {Devereaux}}, \bibinfo {author} {\bibfnamefont
  {A.}~\bibnamefont {Rydh}}, \bibinfo {author} {\bibfnamefont {D.-H.}\
  \bibnamefont {Lu}}, \bibinfo {author} {\bibfnamefont {D.-H.}\ \bibnamefont
  {Lee}}, \bibinfo {author} {\bibfnamefont {O.}~\bibnamefont {Tjernberg}},\
  and\ \bibinfo {author} {\bibfnamefont {Z.-X.}\ \bibnamefont {Shen}},\
  }\bibfield  {title} {\bibinfo {title} {{Bogoliubov quasiparticle on the
  gossamer \uppercase{F}ermi surface in electron-doped cuprates}},\ }\href
  {https://doi.org/10.1038/s41567-023-02209-x} {\bibfield  {journal} {\bibinfo
  {journal} {Nature Physics}\ }\textbf {\bibinfo {volume} {19}},\ \bibinfo
  {pages} {1834} (\bibinfo {year} {2023})}\BibitemShut {NoStop}%
\bibitem [{\citenamefont {Renner}\ \emph {et~al.}(1998)\citenamefont {Renner},
  \citenamefont {Revaz}, \citenamefont {Genoud}, \citenamefont {Kadowaki},\
  and\ \citenamefont {Fischer}}]{PhysRevLett.80.149}%
  \BibitemOpen
  \bibfield  {author} {\bibinfo {author} {\bibfnamefont {C.}~\bibnamefont
  {Renner}}, \bibinfo {author} {\bibfnamefont {B.}~\bibnamefont {Revaz}},
  \bibinfo {author} {\bibfnamefont {J.-Y.}\ \bibnamefont {Genoud}}, \bibinfo
  {author} {\bibfnamefont {K.}~\bibnamefont {Kadowaki}},\ and\ \bibinfo
  {author} {\bibfnamefont {O.}~\bibnamefont {Fischer}},\ }\bibfield  {title}
  {\bibinfo {title} {Pseudogap precursor of the superconducting gap in under-
  and overdoped
  ${\mathrm{\uppercase{b}i}}_{2}{\mathrm{\uppercase{s}r}}_{2}{\mathrm{\uppercase{c}a\uppercase{c}u}}_{2}{\uppercase{o}}_{8+\ensuremath{\delta}}$},\
  }\href {https://doi.org/10.1103/PhysRevLett.80.149} {\bibfield  {journal}
  {\bibinfo  {journal} {Phys. Rev. Lett.}\ }\textbf {\bibinfo {volume} {80}},\
  \bibinfo {pages} {149} (\bibinfo {year} {1998})}\BibitemShut {NoStop}%
\bibitem [{\citenamefont {Ye}\ \emph {et~al.}(2026)\citenamefont {Ye},
  \citenamefont {Zhao}, \citenamefont {Yao}, \citenamefont {Chen},
  \citenamefont {Yu}, \citenamefont {Dong}, \citenamefont {Hao}, \citenamefont
  {Li}, \citenamefont {Shi}, \citenamefont {Liu}, \citenamefont {Jin},\ and\
  \citenamefont {Wang}}]{ye2023visualizing}%
  \BibitemOpen
  \bibfield  {author} {\bibinfo {author} {\bibfnamefont {S.}~\bibnamefont
  {Ye}}, \bibinfo {author} {\bibfnamefont {J.}~\bibnamefont {Zhao}}, \bibinfo
  {author} {\bibfnamefont {Z.}~\bibnamefont {Yao}}, \bibinfo {author}
  {\bibfnamefont {S.}~\bibnamefont {Chen}}, \bibinfo {author} {\bibfnamefont
  {R.}~\bibnamefont {Yu}}, \bibinfo {author} {\bibfnamefont {Z.}~\bibnamefont
  {Dong}}, \bibinfo {author} {\bibfnamefont {Z.}~\bibnamefont {Hao}}, \bibinfo
  {author} {\bibfnamefont {X.}~\bibnamefont {Li}}, \bibinfo {author}
  {\bibfnamefont {L.}~\bibnamefont {Shi}}, \bibinfo {author} {\bibfnamefont
  {Q.}~\bibnamefont {Liu}}, \bibinfo {author} {\bibfnamefont {C.}~\bibnamefont
  {Jin}},\ and\ \bibinfo {author} {\bibfnamefont {Y.}~\bibnamefont {Wang}},\
  }\bibfield  {title} {\bibinfo {title} {{Visualization of the Zhang--Rice
  singlet, electronic molecules and Cooper pair formation in a cuprate
  superconductor}},\ }\bibfield  {journal} {\bibinfo  {journal} {Nat. Phys.}\
  }\href {https://doi.org/10.1038/s41567-026-03375-4}
  {10.1038/s41567-026-03375-4} (\bibinfo {year} {2026})\BibitemShut {NoStop}%
\bibitem [{\citenamefont {Alloul}\ \emph {et~al.}(1989)\citenamefont {Alloul},
  \citenamefont {Ohno},\ and\ \citenamefont {Mendels}}]{PhysRevLett.63.1700}%
  \BibitemOpen
  \bibfield  {author} {\bibinfo {author} {\bibfnamefont {H.}~\bibnamefont
  {Alloul}}, \bibinfo {author} {\bibfnamefont {T.}~\bibnamefont {Ohno}},\ and\
  \bibinfo {author} {\bibfnamefont {P.}~\bibnamefont {Mendels}},\ }\bibfield
  {title} {\bibinfo {title} {$^{89}\mathrm{Y}$ {NMR} evidence for a
  {Fermi}-liquid behavior in
  {${\mathrm{YBa}}_{2}$${\mathrm{Cu}}_{3}$${\mathrm{O}}_{6+\mathrm{x}}$}},\
  }\href {https://doi.org/10.1103/PhysRevLett.63.1700} {\bibfield  {journal}
  {\bibinfo  {journal} {Phys. Rev. Lett.}\ }\textbf {\bibinfo {volume} {63}},\
  \bibinfo {pages} {1700} (\bibinfo {year} {1989})}\BibitemShut {NoStop}%
\bibitem [{\citenamefont {Ishida}\ \emph {et~al.}(1998)\citenamefont {Ishida},
  \citenamefont {Yoshida}, \citenamefont {Mito}, \citenamefont {Tokunaga},
  \citenamefont {Kitaoka}, \citenamefont {Asayama}, \citenamefont {Nakayama},
  \citenamefont {Shimoyama},\ and\ \citenamefont {Kishio}}]{PhysRevB.58.R5960}%
  \BibitemOpen
  \bibfield  {author} {\bibinfo {author} {\bibfnamefont {K.}~\bibnamefont
  {Ishida}}, \bibinfo {author} {\bibfnamefont {K.}~\bibnamefont {Yoshida}},
  \bibinfo {author} {\bibfnamefont {T.}~\bibnamefont {Mito}}, \bibinfo {author}
  {\bibfnamefont {Y.}~\bibnamefont {Tokunaga}}, \bibinfo {author}
  {\bibfnamefont {Y.}~\bibnamefont {Kitaoka}}, \bibinfo {author} {\bibfnamefont
  {K.}~\bibnamefont {Asayama}}, \bibinfo {author} {\bibfnamefont
  {Y.}~\bibnamefont {Nakayama}}, \bibinfo {author} {\bibfnamefont
  {J.}~\bibnamefont {Shimoyama}},\ and\ \bibinfo {author} {\bibfnamefont
  {K.}~\bibnamefont {Kishio}},\ }\bibfield  {title} {\bibinfo {title}
  {{Pseudogap behavior in single-crystal
  ${\mathrm{Bi}}_{2}{\mathrm{Sr}}_{2}{\mathrm{CaCu}}_{2}{\mathrm{O}}_{8+\ensuremath{\delta}}$
  probed by Cu NMR}},\ }\href {https://doi.org/10.1103/PhysRevB.58.R5960}
  {\bibfield  {journal} {\bibinfo  {journal} {Phys. Rev. B}\ }\textbf {\bibinfo
  {volume} {58}},\ \bibinfo {pages} {R5960} (\bibinfo {year}
  {1998})}\BibitemShut {NoStop}%
\bibitem [{\citenamefont {Timusk}\ and\ \citenamefont
  {Statt}(1999)}]{Tom_Timusk_1999}%
  \BibitemOpen
  \bibfield  {author} {\bibinfo {author} {\bibfnamefont {T.}~\bibnamefont
  {Timusk}}\ and\ \bibinfo {author} {\bibfnamefont {B.}~\bibnamefont {Statt}},\
  }\bibfield  {title} {\bibinfo {title} {The pseudogap in high-temperature
  superconductors: an experimental survey},\ }\href
  {https://doi.org/10.1088/0034-4885/62/1/002} {\bibfield  {journal} {\bibinfo
  {journal} {Reports on Progress in Physics}\ }\textbf {\bibinfo {volume}
  {62}},\ \bibinfo {pages} {61} (\bibinfo {year} {1999})}\BibitemShut {NoStop}%
\bibitem [{\citenamefont {Ito}\ \emph {et~al.}(1993)\citenamefont {Ito},
  \citenamefont {Takenaka},\ and\ \citenamefont
  {Uchida}}]{PhysRevLett.70.3995}%
  \BibitemOpen
  \bibfield  {author} {\bibinfo {author} {\bibfnamefont {T.}~\bibnamefont
  {Ito}}, \bibinfo {author} {\bibfnamefont {K.}~\bibnamefont {Takenaka}},\ and\
  \bibinfo {author} {\bibfnamefont {S.}~\bibnamefont {Uchida}},\ }\bibfield
  {title} {\bibinfo {title} {Systematic deviation from \uppercase{T}-linear
  behavior in the in-plane resistivity of
  ${\mathrm{\uppercase{yb}a}}_{2}$${\mathrm{\uppercase{c}u}}_{3}$${\mathrm{\uppercase{o}}}_{7\mathrm{\ensuremath{-}}\mathit{y}}$:
  \uppercase{E}vidence for dominant spin scattering},\ }\href
  {https://doi.org/10.1103/PhysRevLett.70.3995} {\bibfield  {journal} {\bibinfo
   {journal} {Phys. Rev. Lett.}\ }\textbf {\bibinfo {volume} {70}},\ \bibinfo
  {pages} {3995} (\bibinfo {year} {1993})}\BibitemShut {NoStop}%
\bibitem [{\citenamefont {Homes}\ \emph {et~al.}(1993)\citenamefont {Homes},
  \citenamefont {Timusk}, \citenamefont {Liang}, \citenamefont {Bonn},\ and\
  \citenamefont {Hardy}}]{PhysRevLett.71.1645}%
  \BibitemOpen
  \bibfield  {author} {\bibinfo {author} {\bibfnamefont {C.~C.}\ \bibnamefont
  {Homes}}, \bibinfo {author} {\bibfnamefont {T.}~\bibnamefont {Timusk}},
  \bibinfo {author} {\bibfnamefont {R.}~\bibnamefont {Liang}}, \bibinfo
  {author} {\bibfnamefont {D.~A.}\ \bibnamefont {Bonn}},\ and\ \bibinfo
  {author} {\bibfnamefont {W.~N.}\ \bibnamefont {Hardy}},\ }\bibfield  {title}
  {\bibinfo {title} {Optical conductivity of c axis oriented
  ${\mathrm{\uppercase{yb}a}}_{2}$${\mathrm{\uppercase{c}u}}_{3}$${\mathrm{\uppercase{o}}}_{6.70}$:
  Evidence for a pseudogap},\ }\href
  {https://doi.org/10.1103/PhysRevLett.71.1645} {\bibfield  {journal} {\bibinfo
   {journal} {Phys. Rev. Lett.}\ }\textbf {\bibinfo {volume} {71}},\ \bibinfo
  {pages} {1645} (\bibinfo {year} {1993})}\BibitemShut {NoStop}%
\bibitem [{\citenamefont {Puchkov}\ \emph {et~al.}(1996)\citenamefont
  {Puchkov}, \citenamefont {Fournier}, \citenamefont {Basov}, \citenamefont
  {Timusk}, \citenamefont {Kapitulnik},\ and\ \citenamefont
  {Kolesnikov}}]{PhysRevLett.77.3212}%
  \BibitemOpen
  \bibfield  {author} {\bibinfo {author} {\bibfnamefont {A.~V.}\ \bibnamefont
  {Puchkov}}, \bibinfo {author} {\bibfnamefont {P.}~\bibnamefont {Fournier}},
  \bibinfo {author} {\bibfnamefont {D.~N.}\ \bibnamefont {Basov}}, \bibinfo
  {author} {\bibfnamefont {T.}~\bibnamefont {Timusk}}, \bibinfo {author}
  {\bibfnamefont {A.}~\bibnamefont {Kapitulnik}},\ and\ \bibinfo {author}
  {\bibfnamefont {N.~N.}\ \bibnamefont {Kolesnikov}},\ }\bibfield  {title}
  {\bibinfo {title} {Evolution of the pseudogap state of high- ${T}_{c}$
  superconductors with doping},\ }\href
  {https://doi.org/10.1103/PhysRevLett.77.3212} {\bibfield  {journal} {\bibinfo
   {journal} {Phys. Rev. Lett.}\ }\textbf {\bibinfo {volume} {77}},\ \bibinfo
  {pages} {3212} (\bibinfo {year} {1996})}\BibitemShut {NoStop}%
\bibitem [{\citenamefont {Michon}\ \emph {et~al.}(2019)\citenamefont {Michon},
  \citenamefont {Girod}, \citenamefont {Badoux}, \citenamefont
  {Ka{\v{c}}mar{\v{c}}{\'i}k}, \citenamefont {Ma}, \citenamefont {Dragomir},
  \citenamefont {Dabkowska}, \citenamefont {Gaulin}, \citenamefont {Zhou},
  \citenamefont {Pyon}, \citenamefont {Takayama}, \citenamefont {Takagi},
  \citenamefont {Verret}, \citenamefont {Doiron-Leyraud}, \citenamefont
  {Marcenat}, \citenamefont {Taillefer},\ and\ \citenamefont
  {Klein}}]{Michon2019}%
  \BibitemOpen
  \bibfield  {author} {\bibinfo {author} {\bibfnamefont {B.}~\bibnamefont
  {Michon}}, \bibinfo {author} {\bibfnamefont {C.}~\bibnamefont {Girod}},
  \bibinfo {author} {\bibfnamefont {S.}~\bibnamefont {Badoux}}, \bibinfo
  {author} {\bibfnamefont {J.}~\bibnamefont {Ka{\v{c}}mar{\v{c}}{\'i}k}},
  \bibinfo {author} {\bibfnamefont {Q.}~\bibnamefont {Ma}}, \bibinfo {author}
  {\bibfnamefont {M.}~\bibnamefont {Dragomir}}, \bibinfo {author}
  {\bibfnamefont {H.~A.}\ \bibnamefont {Dabkowska}}, \bibinfo {author}
  {\bibfnamefont {B.~D.}\ \bibnamefont {Gaulin}}, \bibinfo {author}
  {\bibfnamefont {J.-S.}\ \bibnamefont {Zhou}}, \bibinfo {author}
  {\bibfnamefont {S.}~\bibnamefont {Pyon}}, \bibinfo {author} {\bibfnamefont
  {T.}~\bibnamefont {Takayama}}, \bibinfo {author} {\bibfnamefont
  {H.}~\bibnamefont {Takagi}}, \bibinfo {author} {\bibfnamefont
  {S.}~\bibnamefont {Verret}}, \bibinfo {author} {\bibfnamefont
  {N.}~\bibnamefont {Doiron-Leyraud}}, \bibinfo {author} {\bibfnamefont
  {C.}~\bibnamefont {Marcenat}}, \bibinfo {author} {\bibfnamefont
  {L.}~\bibnamefont {Taillefer}},\ and\ \bibinfo {author} {\bibfnamefont
  {T.}~\bibnamefont {Klein}},\ }\bibfield  {title} {\bibinfo {title}
  {Thermodynamic signatures of quantum criticality in cuprate
  superconductors},\ }\href {https://doi.org/10.1038/s41586-019-0932-x}
  {\bibfield  {journal} {\bibinfo  {journal} {Nature}\ }\textbf {\bibinfo
  {volume} {567}},\ \bibinfo {pages} {218} (\bibinfo {year}
  {2019})}\BibitemShut {NoStop}%
\bibitem [{\citenamefont {Tranquada}\ \emph {et~al.}(2004)\citenamefont
  {Tranquada}, \citenamefont {Woo}, \citenamefont {Perring}, \citenamefont
  {Goka}, \citenamefont {Gu}, \citenamefont {Xu}, \citenamefont {Fujita},\ and\
  \citenamefont {Yamada}}]{Tranquada2004}%
  \BibitemOpen
  \bibfield  {author} {\bibinfo {author} {\bibfnamefont {J.~M.}\ \bibnamefont
  {Tranquada}}, \bibinfo {author} {\bibfnamefont {H.}~\bibnamefont {Woo}},
  \bibinfo {author} {\bibfnamefont {T.~G.}\ \bibnamefont {Perring}}, \bibinfo
  {author} {\bibfnamefont {H.}~\bibnamefont {Goka}}, \bibinfo {author}
  {\bibfnamefont {G.~D.}\ \bibnamefont {Gu}}, \bibinfo {author} {\bibfnamefont
  {G.}~\bibnamefont {Xu}}, \bibinfo {author} {\bibfnamefont {M.}~\bibnamefont
  {Fujita}},\ and\ \bibinfo {author} {\bibfnamefont {K.}~\bibnamefont
  {Yamada}},\ }\bibfield  {title} {\bibinfo {title} {Quantum magnetic
  excitations from stripes in copper oxide superconductors},\ }\href
  {https://doi.org/10.1038/nature02574} {\bibfield  {journal} {\bibinfo
  {journal} {Nature}\ }\textbf {\bibinfo {volume} {429}},\ \bibinfo {pages}
  {534} (\bibinfo {year} {2004})}\BibitemShut {NoStop}%
\bibitem [{\citenamefont {Hayden}\ \emph {et~al.}(2004)\citenamefont {Hayden},
  \citenamefont {Mook}, \citenamefont {Dai}, \citenamefont {Perring},\ and\
  \citenamefont {Do{\u{g}}an}}]{Hayden2004}%
  \BibitemOpen
  \bibfield  {author} {\bibinfo {author} {\bibfnamefont {S.~M.}\ \bibnamefont
  {Hayden}}, \bibinfo {author} {\bibfnamefont {H.~A.}\ \bibnamefont {Mook}},
  \bibinfo {author} {\bibfnamefont {P.}~\bibnamefont {Dai}}, \bibinfo {author}
  {\bibfnamefont {T.~G.}\ \bibnamefont {Perring}},\ and\ \bibinfo {author}
  {\bibfnamefont {F.}~\bibnamefont {Do{\u{g}}an}},\ }\bibfield  {title}
  {\bibinfo {title} {The structure of the high-energy spin excitations in a
  high-transition-temperature superconductor},\ }\href
  {https://doi.org/10.1038/nature02576} {\bibfield  {journal} {\bibinfo
  {journal} {Nature}\ }\textbf {\bibinfo {volume} {429}},\ \bibinfo {pages}
  {531} (\bibinfo {year} {2004})}\BibitemShut {NoStop}%
\bibitem [{\citenamefont {Vignolle}\ \emph {et~al.}(2007)\citenamefont
  {Vignolle}, \citenamefont {Hayden}, \citenamefont {McMorrow}, \citenamefont
  {R{\o}nnow}, \citenamefont {Lake}, \citenamefont {Frost},\ and\ \citenamefont
  {Perring}}]{Vignolle2007}%
  \BibitemOpen
  \bibfield  {author} {\bibinfo {author} {\bibfnamefont {B.}~\bibnamefont
  {Vignolle}}, \bibinfo {author} {\bibfnamefont {S.~M.}\ \bibnamefont
  {Hayden}}, \bibinfo {author} {\bibfnamefont {D.~F.}\ \bibnamefont
  {McMorrow}}, \bibinfo {author} {\bibfnamefont {H.~M.}\ \bibnamefont
  {R{\o}nnow}}, \bibinfo {author} {\bibfnamefont {B.}~\bibnamefont {Lake}},
  \bibinfo {author} {\bibfnamefont {C.~D.}\ \bibnamefont {Frost}},\ and\
  \bibinfo {author} {\bibfnamefont {T.~G.}\ \bibnamefont {Perring}},\
  }\bibfield  {title} {\bibinfo {title} {{Two energy scales in the spin
  excitations of the high-temperature superconductor
  La$_{2−x}$Sr$_x$CuO$_4$}},\ }\href {https://doi.org/10.1038/nphys546}
  {\bibfield  {journal} {\bibinfo  {journal} {Nature Physics}\ }\textbf
  {\bibinfo {volume} {3}},\ \bibinfo {pages} {163} (\bibinfo {year}
  {2007})}\BibitemShut {NoStop}%
\bibitem [{\citenamefont {Xu}\ \emph {et~al.}(2009)\citenamefont {Xu},
  \citenamefont {Gu}, \citenamefont {H{\"u}cker}, \citenamefont {Fauqu{\'e}},
  \citenamefont {Perring}, \citenamefont {Regnault},\ and\ \citenamefont
  {Tranquada}}]{Xu2009}%
  \BibitemOpen
  \bibfield  {author} {\bibinfo {author} {\bibfnamefont {G.}~\bibnamefont
  {Xu}}, \bibinfo {author} {\bibfnamefont {G.~D.}\ \bibnamefont {Gu}}, \bibinfo
  {author} {\bibfnamefont {M.}~\bibnamefont {H{\"u}cker}}, \bibinfo {author}
  {\bibfnamefont {B.}~\bibnamefont {Fauqu{\'e}}}, \bibinfo {author}
  {\bibfnamefont {T.~G.}\ \bibnamefont {Perring}}, \bibinfo {author}
  {\bibfnamefont {L.~P.}\ \bibnamefont {Regnault}},\ and\ \bibinfo {author}
  {\bibfnamefont {J.~M.}\ \bibnamefont {Tranquada}},\ }\bibfield  {title}
  {\bibinfo {title} {{Testing the itinerancy of spin dynamics in
  superconducting Bi$_2$Sr$_2$CaCu$_2$O$_{8+\delta}$}},\ }\href
  {https://doi.org/10.1038/nphys1360} {\bibfield  {journal} {\bibinfo
  {journal} {Nature Physics}\ }\textbf {\bibinfo {volume} {5}},\ \bibinfo
  {pages} {642} (\bibinfo {year} {2009})}\BibitemShut {NoStop}%
\bibitem [{\citenamefont {Lipscombe}\ \emph {et~al.}(2009)\citenamefont
  {Lipscombe}, \citenamefont {Vignolle}, \citenamefont {Perring}, \citenamefont
  {Frost},\ and\ \citenamefont {Hayden}}]{PhysRevLett.102.167002}%
  \BibitemOpen
  \bibfield  {author} {\bibinfo {author} {\bibfnamefont {O.~J.}\ \bibnamefont
  {Lipscombe}}, \bibinfo {author} {\bibfnamefont {B.}~\bibnamefont {Vignolle}},
  \bibinfo {author} {\bibfnamefont {T.~G.}\ \bibnamefont {Perring}}, \bibinfo
  {author} {\bibfnamefont {C.~D.}\ \bibnamefont {Frost}},\ and\ \bibinfo
  {author} {\bibfnamefont {S.~M.}\ \bibnamefont {Hayden}},\ }\bibfield  {title}
  {\bibinfo {title} {Emergence of coherent magnetic excitations in the high
  temperature underdoped
  {${\mathrm{La}}_{2-x}{\mathrm{Sr}}_{x}{\mathrm{CuO}}_{4}$} superconductor at
  low temperatures},\ }\href {https://doi.org/10.1103/PhysRevLett.102.167002}
  {\bibfield  {journal} {\bibinfo  {journal} {Phys. Rev. Lett.}\ }\textbf
  {\bibinfo {volume} {102}},\ \bibinfo {pages} {167002} (\bibinfo {year}
  {2009})}\BibitemShut {NoStop}%
\bibitem [{\citenamefont {Nemetschek}\ \emph {et~al.}(1997)\citenamefont
  {Nemetschek}, \citenamefont {Opel}, \citenamefont {Hoffmann}, \citenamefont
  {M\"uller}, \citenamefont {Hackl}, \citenamefont {Berger}, \citenamefont
  {Forr\'o}, \citenamefont {Erb},\ and\ \citenamefont
  {Walker}}]{PhysRevLett.78.4837}%
  \BibitemOpen
  \bibfield  {author} {\bibinfo {author} {\bibfnamefont {R.}~\bibnamefont
  {Nemetschek}}, \bibinfo {author} {\bibfnamefont {M.}~\bibnamefont {Opel}},
  \bibinfo {author} {\bibfnamefont {C.}~\bibnamefont {Hoffmann}}, \bibinfo
  {author} {\bibfnamefont {P.~F.}\ \bibnamefont {M\"uller}}, \bibinfo {author}
  {\bibfnamefont {R.}~\bibnamefont {Hackl}}, \bibinfo {author} {\bibfnamefont
  {H.}~\bibnamefont {Berger}}, \bibinfo {author} {\bibfnamefont
  {L.}~\bibnamefont {Forr\'o}}, \bibinfo {author} {\bibfnamefont
  {A.}~\bibnamefont {Erb}},\ and\ \bibinfo {author} {\bibfnamefont
  {E.}~\bibnamefont {Walker}},\ }\bibfield  {title} {\bibinfo {title}
  {Pseudogap and superconducting gap in the electronic \uppercase{R}aman
  spectra of underdoped cuprates},\ }\href
  {https://doi.org/10.1103/PhysRevLett.78.4837} {\bibfield  {journal} {\bibinfo
   {journal} {Phys. Rev. Lett.}\ }\textbf {\bibinfo {volume} {78}},\ \bibinfo
  {pages} {4837} (\bibinfo {year} {1997})}\BibitemShut {NoStop}%
\bibitem [{\citenamefont {Zhou}\ \emph
  {et~al.}(2025{\natexlab{a}})\citenamefont {Zhou}, \citenamefont {Vinograd},
  \citenamefont {Hirata}, \citenamefont {Wu}, \citenamefont {Mayaffre},
  \citenamefont {Kr{\"a}mer}, \citenamefont {Hardy}, \citenamefont {Liang},
  \citenamefont {Bonn}, \citenamefont {Loew}, \citenamefont {Porras},
  \citenamefont {Keimer},\ and\ \citenamefont {Julien}}]{Zhou2025}%
  \BibitemOpen
  \bibfield  {author} {\bibinfo {author} {\bibfnamefont {R.}~\bibnamefont
  {Zhou}}, \bibinfo {author} {\bibfnamefont {I.}~\bibnamefont {Vinograd}},
  \bibinfo {author} {\bibfnamefont {M.}~\bibnamefont {Hirata}}, \bibinfo
  {author} {\bibfnamefont {T.}~\bibnamefont {Wu}}, \bibinfo {author}
  {\bibfnamefont {H.}~\bibnamefont {Mayaffre}}, \bibinfo {author}
  {\bibfnamefont {S.}~\bibnamefont {Kr{\"a}mer}}, \bibinfo {author}
  {\bibfnamefont {W.~N.}\ \bibnamefont {Hardy}}, \bibinfo {author}
  {\bibfnamefont {R.}~\bibnamefont {Liang}}, \bibinfo {author} {\bibfnamefont
  {D.~A.}\ \bibnamefont {Bonn}}, \bibinfo {author} {\bibfnamefont
  {T.}~\bibnamefont {Loew}}, \bibinfo {author} {\bibfnamefont {J.}~\bibnamefont
  {Porras}}, \bibinfo {author} {\bibfnamefont {B.}~\bibnamefont {Keimer}},\
  and\ \bibinfo {author} {\bibfnamefont {M.-H.}\ \bibnamefont {Julien}},\
  }\bibfield  {title} {\bibinfo {title} {Signatures of two gaps in the spin
  susceptibility of a cuprate superconductor},\ }\href
  {https://doi.org/10.1038/s41567-024-02692-w} {\bibfield  {journal} {\bibinfo
  {journal} {Nature Physics}\ }\textbf {\bibinfo {volume} {21}},\ \bibinfo
  {pages} {97} (\bibinfo {year} {2025}{\natexlab{a}})}\BibitemShut {NoStop}%
\bibitem [{\citenamefont {Zhou}\ \emph
  {et~al.}(2025{\natexlab{b}})\citenamefont {Zhou}, \citenamefont {Vinograd},
  \citenamefont {Mayaffre}, \citenamefont {Porras}, \citenamefont {Kim},
  \citenamefont {Loew}, \citenamefont {Liu}, \citenamefont {Le~Tacon},
  \citenamefont {Keimer},\ and\ \citenamefont {Julien}}]{zhou2025robust}%
  \BibitemOpen
  \bibfield  {author} {\bibinfo {author} {\bibfnamefont {R.}~\bibnamefont
  {Zhou}}, \bibinfo {author} {\bibfnamefont {I.}~\bibnamefont {Vinograd}},
  \bibinfo {author} {\bibfnamefont {H.}~\bibnamefont {Mayaffre}}, \bibinfo
  {author} {\bibfnamefont {J.}~\bibnamefont {Porras}}, \bibinfo {author}
  {\bibfnamefont {H.-H.}\ \bibnamefont {Kim}}, \bibinfo {author} {\bibfnamefont
  {T.}~\bibnamefont {Loew}}, \bibinfo {author} {\bibfnamefont {Y.}~\bibnamefont
  {Liu}}, \bibinfo {author} {\bibfnamefont {M.}~\bibnamefont {Le~Tacon}},
  \bibinfo {author} {\bibfnamefont {B.}~\bibnamefont {Keimer}},\ and\ \bibinfo
  {author} {\bibfnamefont {M.-H.}\ \bibnamefont {Julien}},\ }\bibfield  {title}
  {\bibinfo {title} {Robust charge-density wave correlations in optimally-doped
  $\mathrm{\uppercase{yb}a}_{2}\mathrm{\uppercase{c}u}_{3}\mathrm{\uppercase{o}}_y$},\
  }\href {https://doi.org/10.1103/8klr-4wc5} {\bibfield  {journal} {\bibinfo
  {journal} {Phys. Rev. Lett.}\ }\textbf {\bibinfo {volume} {135}},\ \bibinfo
  {pages} {106503} (\bibinfo {year} {2025}{\natexlab{b}})}\BibitemShut
  {NoStop}%
\bibitem [{\citenamefont {Emery}\ \emph {et~al.}(1997)\citenamefont {Emery},
  \citenamefont {Kivelson},\ and\ \citenamefont {Zachar}}]{PhysRevB.56.6120}%
  \BibitemOpen
  \bibfield  {author} {\bibinfo {author} {\bibfnamefont {V.~J.}\ \bibnamefont
  {Emery}}, \bibinfo {author} {\bibfnamefont {S.~A.}\ \bibnamefont
  {Kivelson}},\ and\ \bibinfo {author} {\bibfnamefont {O.}~\bibnamefont
  {Zachar}},\ }\bibfield  {title} {\bibinfo {title} {Spin-gap proximity effect
  mechanism of high-temperature superconductivity},\ }\href
  {https://doi.org/10.1103/PhysRevB.56.6120} {\bibfield  {journal} {\bibinfo
  {journal} {Phys. Rev. B}\ }\textbf {\bibinfo {volume} {56}},\ \bibinfo
  {pages} {6120} (\bibinfo {year} {1997})}\BibitemShut {NoStop}%
\bibitem [{\citenamefont {Lee}(2014)}]{PhysRevX.4.031017}%
  \BibitemOpen
  \bibfield  {author} {\bibinfo {author} {\bibfnamefont {P.~A.}\ \bibnamefont
  {Lee}},\ }\bibfield  {title} {\bibinfo {title} {Amperean pairing and the
  pseudogap phase of cuprate superconductors},\ }\href
  {https://doi.org/10.1103/PhysRevX.4.031017} {\bibfield  {journal} {\bibinfo
  {journal} {Phys. Rev. X}\ }\textbf {\bibinfo {volume} {4}},\ \bibinfo {pages}
  {031017} (\bibinfo {year} {2014})}\BibitemShut {NoStop}%
\bibitem [{\citenamefont {Malcolms}\ \emph {et~al.}(2026)\citenamefont
  {Malcolms}, \citenamefont {Menke}, \citenamefont {Tseng}, \citenamefont
  {Jacob}, \citenamefont {Held}, \citenamefont {Hansmann},\ and\ \citenamefont
  {Sch{\"a}fer}}]{malcolms2024rise}%
  \BibitemOpen
  \bibfield  {author} {\bibinfo {author} {\bibfnamefont {M.~O.}\ \bibnamefont
  {Malcolms}}, \bibinfo {author} {\bibfnamefont {H.}~\bibnamefont {Menke}},
  \bibinfo {author} {\bibfnamefont {Y.-T.}\ \bibnamefont {Tseng}}, \bibinfo
  {author} {\bibfnamefont {E.}~\bibnamefont {Jacob}}, \bibinfo {author}
  {\bibfnamefont {K.}~\bibnamefont {Held}}, \bibinfo {author} {\bibfnamefont
  {P.}~\bibnamefont {Hansmann}},\ and\ \bibinfo {author} {\bibfnamefont
  {T.}~\bibnamefont {Sch{\"a}fer}},\ }\bibfield  {title} {\bibinfo {title}
  {{Rise and fall of the pseudogap in the Emery model, insights for
  cuprates}},\ }\href {https://doi.org/10.1038/s42005-026-02685-6} {\bibfield
  {journal} {\bibinfo  {journal} {Commun. Phys.}\ }\textbf {\bibinfo {volume}
  {9}},\ \bibinfo {pages} {179} (\bibinfo {year} {2026})}\BibitemShut {NoStop}%
\bibitem [{\citenamefont {Markiewicz}\ and\ \citenamefont
  {Bansil}(2024)}]{PhysRevB.109.045116}%
  \BibitemOpen
  \bibfield  {author} {\bibinfo {author} {\bibfnamefont {R.~S.}\ \bibnamefont
  {Markiewicz}}\ and\ \bibinfo {author} {\bibfnamefont {A.}~\bibnamefont
  {Bansil}},\ }\bibfield  {title} {\bibinfo {title} {Theory of cuprate
  pseudogap as antiferromagnetic order with charged domain walls},\ }\href
  {https://doi.org/10.1103/PhysRevB.109.045116} {\bibfield  {journal} {\bibinfo
   {journal} {Phys. Rev. B}\ }\textbf {\bibinfo {volume} {109}},\ \bibinfo
  {pages} {045116} (\bibinfo {year} {2024})}\BibitemShut {NoStop}%
\bibitem [{\citenamefont {Emery}\ and\ \citenamefont
  {Kivelson}(1995)}]{Emery1995}%
  \BibitemOpen
  \bibfield  {author} {\bibinfo {author} {\bibfnamefont {V.~J.}\ \bibnamefont
  {Emery}}\ and\ \bibinfo {author} {\bibfnamefont {S.~A.}\ \bibnamefont
  {Kivelson}},\ }\bibfield  {title} {\bibinfo {title} {Importance of phase
  fluctuations in superconductors with small superfluid density},\ }\href
  {https://doi.org/10.1038/374434a0} {\bibfield  {journal} {\bibinfo  {journal}
  {Nature}\ }\textbf {\bibinfo {volume} {374}},\ \bibinfo {pages} {434}
  (\bibinfo {year} {1995})}\BibitemShut {NoStop}%
\bibitem [{\citenamefont {Varma}(1997)}]{PhysRevB.55.14554}%
  \BibitemOpen
  \bibfield  {author} {\bibinfo {author} {\bibfnamefont {C.~M.}\ \bibnamefont
  {Varma}},\ }\bibfield  {title} {\bibinfo {title} {{Non-Fermi-liquid states
  and pairing instability of a general model of copper oxide metals}},\ }\href
  {https://doi.org/10.1103/PhysRevB.55.14554} {\bibfield  {journal} {\bibinfo
  {journal} {Phys. Rev. B}\ }\textbf {\bibinfo {volume} {55}},\ \bibinfo
  {pages} {14554} (\bibinfo {year} {1997})}\BibitemShut {NoStop}%
\bibitem [{\citenamefont {Varma}(1999)}]{PhysRevLett.83.3538}%
  \BibitemOpen
  \bibfield  {author} {\bibinfo {author} {\bibfnamefont {C.~M.}\ \bibnamefont
  {Varma}},\ }\bibfield  {title} {\bibinfo {title} {Pseudogap phase and the
  quantum-critical point in copper-oxide metals},\ }\href
  {https://doi.org/10.1103/PhysRevLett.83.3538} {\bibfield  {journal} {\bibinfo
   {journal} {Phys. Rev. Lett.}\ }\textbf {\bibinfo {volume} {83}},\ \bibinfo
  {pages} {3538} (\bibinfo {year} {1999})}\BibitemShut {NoStop}%
\bibitem [{\citenamefont {Varma}(2006)}]{PhysRevB.73.155113}%
  \BibitemOpen
  \bibfield  {author} {\bibinfo {author} {\bibfnamefont {C.~M.}\ \bibnamefont
  {Varma}},\ }\bibfield  {title} {\bibinfo {title} {Theory of the pseudogap
  state of the cuprates},\ }\href {https://doi.org/10.1103/PhysRevB.73.155113}
  {\bibfield  {journal} {\bibinfo  {journal} {Phys. Rev. B}\ }\textbf {\bibinfo
  {volume} {73}},\ \bibinfo {pages} {155113} (\bibinfo {year}
  {2006})}\BibitemShut {NoStop}%
\bibitem [{\citenamefont {Aji}\ \emph {et~al.}(2010)\citenamefont {Aji},
  \citenamefont {Shekhter},\ and\ \citenamefont {Varma}}]{PhysRevB.81.064515}%
  \BibitemOpen
  \bibfield  {author} {\bibinfo {author} {\bibfnamefont {V.}~\bibnamefont
  {Aji}}, \bibinfo {author} {\bibfnamefont {A.}~\bibnamefont {Shekhter}},\ and\
  \bibinfo {author} {\bibfnamefont {C.~M.}\ \bibnamefont {Varma}},\ }\bibfield
  {title} {\bibinfo {title} {Theory of the coupling of quantum-critical
  fluctuations to fermions and $d$-wave superconductivity in cuprates},\ }\href
  {https://doi.org/10.1103/PhysRevB.81.064515} {\bibfield  {journal} {\bibinfo
  {journal} {Phys. Rev. B}\ }\textbf {\bibinfo {volume} {81}},\ \bibinfo
  {pages} {064515} (\bibinfo {year} {2010})}\BibitemShut {NoStop}%
\bibitem [{\citenamefont {Wen}(2002)}]{PhysRevB.65.165113}%
  \BibitemOpen
  \bibfield  {author} {\bibinfo {author} {\bibfnamefont {X.-G.}\ \bibnamefont
  {Wen}},\ }\bibfield  {title} {\bibinfo {title} {Quantum orders and symmetric
  spin liquids},\ }\href {https://doi.org/10.1103/PhysRevB.65.165113}
  {\bibfield  {journal} {\bibinfo  {journal} {Phys. Rev. B}\ }\textbf {\bibinfo
  {volume} {65}},\ \bibinfo {pages} {165113} (\bibinfo {year}
  {2002})}\BibitemShut {NoStop}%
\bibitem [{\citenamefont {Lee}\ \emph {et~al.}(2006)\citenamefont {Lee},
  \citenamefont {Nagaosa},\ and\ \citenamefont {Wen}}]{RevModPhys.78.17}%
  \BibitemOpen
  \bibfield  {author} {\bibinfo {author} {\bibfnamefont {P.~A.}\ \bibnamefont
  {Lee}}, \bibinfo {author} {\bibfnamefont {N.}~\bibnamefont {Nagaosa}},\ and\
  \bibinfo {author} {\bibfnamefont {X.-G.}\ \bibnamefont {Wen}},\ }\bibfield
  {title} {\bibinfo {title} {{Doping a \uppercase{M}ott insulator: Physics of
  high-temperature superconductivity}},\ }\href
  {https://doi.org/10.1103/RevModPhys.78.17} {\bibfield  {journal} {\bibinfo
  {journal} {Rev. Mod. Phys.}\ }\textbf {\bibinfo {volume} {78}},\ \bibinfo
  {pages} {17} (\bibinfo {year} {2006})}\BibitemShut {NoStop}%
\bibitem [{\citenamefont {Patel}\ \emph {et~al.}(2016)\citenamefont {Patel},
  \citenamefont {Chowdhury}, \citenamefont {Allais},\ and\ \citenamefont
  {Sachdev}}]{PhysRevB.93.165139}%
  \BibitemOpen
  \bibfield  {author} {\bibinfo {author} {\bibfnamefont {A.~A.}\ \bibnamefont
  {Patel}}, \bibinfo {author} {\bibfnamefont {D.}~\bibnamefont {Chowdhury}},
  \bibinfo {author} {\bibfnamefont {A.}~\bibnamefont {Allais}},\ and\ \bibinfo
  {author} {\bibfnamefont {S.}~\bibnamefont {Sachdev}},\ }\bibfield  {title}
  {\bibinfo {title} {Confinement transition to density wave order in metallic
  doped spin liquids},\ }\href {https://doi.org/10.1103/PhysRevB.93.165139}
  {\bibfield  {journal} {\bibinfo  {journal} {Phys. Rev. B}\ }\textbf {\bibinfo
  {volume} {93}},\ \bibinfo {pages} {165139} (\bibinfo {year}
  {2016})}\BibitemShut {NoStop}%
\bibitem [{\citenamefont {Chatterjee}\ and\ \citenamefont
  {Sachdev}(2016)}]{PhysRevB.94.205117}%
  \BibitemOpen
  \bibfield  {author} {\bibinfo {author} {\bibfnamefont {S.}~\bibnamefont
  {Chatterjee}}\ and\ \bibinfo {author} {\bibfnamefont {S.}~\bibnamefont
  {Sachdev}},\ }\bibfield  {title} {\bibinfo {title} {{Fractionalized Fermi
  liquid with bosonic chargons as a candidate for the pseudogap metal}},\
  }\href {https://doi.org/10.1103/PhysRevB.94.205117} {\bibfield  {journal}
  {\bibinfo  {journal} {Phys. Rev. B}\ }\textbf {\bibinfo {volume} {94}},\
  \bibinfo {pages} {205117} (\bibinfo {year} {2016})}\BibitemShut {NoStop}%
\bibitem [{\citenamefont {Christos}\ \emph {et~al.}(2023)\citenamefont
  {Christos}, \citenamefont {Luo}, \citenamefont {Shackleton}, \citenamefont
  {Zhang}, \citenamefont {Scheurer},\ and\ \citenamefont
  {Sachdev}}]{doi:10.1073/pnas.2302701120}%
  \BibitemOpen
  \bibfield  {author} {\bibinfo {author} {\bibfnamefont {M.}~\bibnamefont
  {Christos}}, \bibinfo {author} {\bibfnamefont {Z.-X.}\ \bibnamefont {Luo}},
  \bibinfo {author} {\bibfnamefont {H.}~\bibnamefont {Shackleton}}, \bibinfo
  {author} {\bibfnamefont {Y.-H.}\ \bibnamefont {Zhang}}, \bibinfo {author}
  {\bibfnamefont {M.~S.}\ \bibnamefont {Scheurer}},\ and\ \bibinfo {author}
  {\bibfnamefont {S.}~\bibnamefont {Sachdev}},\ }\bibfield  {title} {\bibinfo
  {title} {A model of $d$-wave superconductivity, antiferromagnetism, and
  charge order on the square lattice},\ }\href
  {https://doi.org/10.1073/pnas.2302701120} {\bibfield  {journal} {\bibinfo
  {journal} {Proceedings of the National Academy of Sciences}\ }\textbf
  {\bibinfo {volume} {120}},\ \bibinfo {pages} {e2302701120} (\bibinfo {year}
  {2023})}\BibitemShut {NoStop}%
\bibitem [{\citenamefont {Christos}\ and\ \citenamefont
  {Sachdev}(2024)}]{Christos2024}%
  \BibitemOpen
  \bibfield  {author} {\bibinfo {author} {\bibfnamefont {M.}~\bibnamefont
  {Christos}}\ and\ \bibinfo {author} {\bibfnamefont {S.}~\bibnamefont
  {Sachdev}},\ }\bibfield  {title} {\bibinfo {title} {{Emergence of nodal
  Bogoliubov quasiparticles across the transition from the pseudogap metal to
  the $d$-wave superconductor}},\ }\href
  {https://doi.org/10.1038/s41535-023-00608-0} {\bibfield  {journal} {\bibinfo
  {journal} {npj Quantum Materials}\ }\textbf {\bibinfo {volume} {9}},\
  \bibinfo {pages} {4} (\bibinfo {year} {2024})}\BibitemShut {NoStop}%
\bibitem [{\citenamefont {Wu}\ \emph {et~al.}(2018)\citenamefont {Wu},
  \citenamefont {Scheurer}, \citenamefont {Chatterjee}, \citenamefont
  {Sachdev}, \citenamefont {Georges},\ and\ \citenamefont
  {Ferrero}}]{PhysRevX.8.021048}%
  \BibitemOpen
  \bibfield  {author} {\bibinfo {author} {\bibfnamefont {W.}~\bibnamefont
  {Wu}}, \bibinfo {author} {\bibfnamefont {M.~S.}\ \bibnamefont {Scheurer}},
  \bibinfo {author} {\bibfnamefont {S.}~\bibnamefont {Chatterjee}}, \bibinfo
  {author} {\bibfnamefont {S.}~\bibnamefont {Sachdev}}, \bibinfo {author}
  {\bibfnamefont {A.}~\bibnamefont {Georges}},\ and\ \bibinfo {author}
  {\bibfnamefont {M.}~\bibnamefont {Ferrero}},\ }\bibfield  {title} {\bibinfo
  {title} {Pseudogap and {Fermi}-surface topology in the two-dimensional
  \uppercase{H}ubbard model},\ }\href
  {https://doi.org/10.1103/PhysRevX.8.021048} {\bibfield  {journal} {\bibinfo
  {journal} {Phys. Rev. X}\ }\textbf {\bibinfo {volume} {8}},\ \bibinfo {pages}
  {021048} (\bibinfo {year} {2018})}\BibitemShut {NoStop}%
\bibitem [{\citenamefont {Bragan\ifmmode~\mbox{\c{c}}\else \c{c}\fi{}a}\ \emph
  {et~al.}(2018)\citenamefont {Bragan\ifmmode~\mbox{\c{c}}\else \c{c}\fi{}a},
  \citenamefont {Sakai}, \citenamefont {Aguiar},\ and\ \citenamefont
  {Civelli}}]{PhysRevLett.120.067002}%
  \BibitemOpen
  \bibfield  {author} {\bibinfo {author} {\bibfnamefont {H.}~\bibnamefont
  {Bragan\ifmmode~\mbox{\c{c}}\else \c{c}\fi{}a}}, \bibinfo {author}
  {\bibfnamefont {S.}~\bibnamefont {Sakai}}, \bibinfo {author} {\bibfnamefont
  {M.~C.~O.}\ \bibnamefont {Aguiar}},\ and\ \bibinfo {author} {\bibfnamefont
  {M.}~\bibnamefont {Civelli}},\ }\bibfield  {title} {\bibinfo {title}
  {Correlation-driven {Lifshitz} transition at the emergence of the pseudogap
  phase in the two-dimensional \uppercase{H}ubbard model},\ }\href
  {https://doi.org/10.1103/PhysRevLett.120.067002} {\bibfield  {journal}
  {\bibinfo  {journal} {Phys. Rev. Lett.}\ }\textbf {\bibinfo {volume} {120}},\
  \bibinfo {pages} {067002} (\bibinfo {year} {2018})}\BibitemShut {NoStop}%
\bibitem [{\citenamefont {Benhabib}\ \emph {et~al.}(2015)\citenamefont
  {Benhabib}, \citenamefont {Sacuto}, \citenamefont {Civelli}, \citenamefont
  {Paul}, \citenamefont {Cazayous}, \citenamefont {Gallais}, \citenamefont
  {M\'easson}, \citenamefont {Zhong}, \citenamefont {Schneeloch}, \citenamefont
  {Gu}, \citenamefont {Colson},\ and\ \citenamefont
  {Forget}}]{PhysRevLett.114.147001}%
  \BibitemOpen
  \bibfield  {author} {\bibinfo {author} {\bibfnamefont {S.}~\bibnamefont
  {Benhabib}}, \bibinfo {author} {\bibfnamefont {A.}~\bibnamefont {Sacuto}},
  \bibinfo {author} {\bibfnamefont {M.}~\bibnamefont {Civelli}}, \bibinfo
  {author} {\bibfnamefont {I.}~\bibnamefont {Paul}}, \bibinfo {author}
  {\bibfnamefont {M.}~\bibnamefont {Cazayous}}, \bibinfo {author}
  {\bibfnamefont {Y.}~\bibnamefont {Gallais}}, \bibinfo {author} {\bibfnamefont
  {M.-A.}\ \bibnamefont {M\'easson}}, \bibinfo {author} {\bibfnamefont {R.~D.}\
  \bibnamefont {Zhong}}, \bibinfo {author} {\bibfnamefont {J.}~\bibnamefont
  {Schneeloch}}, \bibinfo {author} {\bibfnamefont {G.~D.}\ \bibnamefont {Gu}},
  \bibinfo {author} {\bibfnamefont {D.}~\bibnamefont {Colson}},\ and\ \bibinfo
  {author} {\bibfnamefont {A.}~\bibnamefont {Forget}},\ }\bibfield  {title}
  {\bibinfo {title} {Collapse of the normal-state pseudogap at a {Lifshitz}
  transition in the
  {${\mathrm{Bi}}_{2}{\mathrm{Sr}}_{2}{\mathrm{CaCu}}_{2}{\mathrm{O}}_{8+\ensuremath{\delta}}$}
  cuprate superconductor},\ }\href
  {https://doi.org/10.1103/PhysRevLett.114.147001} {\bibfield  {journal}
  {\bibinfo  {journal} {Phys. Rev. Lett.}\ }\textbf {\bibinfo {volume} {114}},\
  \bibinfo {pages} {147001} (\bibinfo {year} {2015})}\BibitemShut {NoStop}%
\bibitem [{\citenamefont {Loret}\ \emph {et~al.}(2017)\citenamefont {Loret},
  \citenamefont {Sakai}, \citenamefont {Benhabib}, \citenamefont {Gallais},
  \citenamefont {Cazayous}, \citenamefont {M\'easson}, \citenamefont {Zhong},
  \citenamefont {Schneeloch}, \citenamefont {Gu}, \citenamefont {Forget},
  \citenamefont {Colson}, \citenamefont {Paul}, \citenamefont {Civelli},\ and\
  \citenamefont {Sacuto}}]{PhysRevB.96.094525}%
  \BibitemOpen
  \bibfield  {author} {\bibinfo {author} {\bibfnamefont {B.}~\bibnamefont
  {Loret}}, \bibinfo {author} {\bibfnamefont {S.}~\bibnamefont {Sakai}},
  \bibinfo {author} {\bibfnamefont {S.}~\bibnamefont {Benhabib}}, \bibinfo
  {author} {\bibfnamefont {Y.}~\bibnamefont {Gallais}}, \bibinfo {author}
  {\bibfnamefont {M.}~\bibnamefont {Cazayous}}, \bibinfo {author}
  {\bibfnamefont {M.~A.}\ \bibnamefont {M\'easson}}, \bibinfo {author}
  {\bibfnamefont {R.~D.}\ \bibnamefont {Zhong}}, \bibinfo {author}
  {\bibfnamefont {J.}~\bibnamefont {Schneeloch}}, \bibinfo {author}
  {\bibfnamefont {G.~D.}\ \bibnamefont {Gu}}, \bibinfo {author} {\bibfnamefont
  {A.}~\bibnamefont {Forget}}, \bibinfo {author} {\bibfnamefont
  {D.}~\bibnamefont {Colson}}, \bibinfo {author} {\bibfnamefont
  {I.}~\bibnamefont {Paul}}, \bibinfo {author} {\bibfnamefont {M.}~\bibnamefont
  {Civelli}},\ and\ \bibinfo {author} {\bibfnamefont {A.}~\bibnamefont
  {Sacuto}},\ }\bibfield  {title} {\bibinfo {title} {{Vertical temperature
  boundary of the pseudogap under the superconducting dome in the phase diagram
  of
  ${\mathrm{Bi}}_{2}{\mathrm{Sr}}_{2}{\mathrm{CaCu}}_{2}{\mathrm{O}}_{8+\ensuremath{\delta}}$}},\
  }\href {https://doi.org/10.1103/PhysRevB.96.094525} {\bibfield  {journal}
  {\bibinfo  {journal} {Phys. Rev. B}\ }\textbf {\bibinfo {volume} {96}},\
  \bibinfo {pages} {094525} (\bibinfo {year} {2017})}\BibitemShut {NoStop}%
\bibitem [{\citenamefont {Doiron-Leyraud}\ \emph {et~al.}(2017)\citenamefont
  {Doiron-Leyraud}, \citenamefont {Cyr-Choini{\`e}re}, \citenamefont {Badoux},
  \citenamefont {Ataei}, \citenamefont {Collignon}, \citenamefont {Gourgout},
  \citenamefont {Dufour-Beaus{\'e}jour}, \citenamefont {Tafti}, \citenamefont
  {Lalibert{\'e}}, \citenamefont {Boulanger}, \citenamefont {Matusiak},
  \citenamefont {Graf}, \citenamefont {Kim}, \citenamefont {Zhou},
  \citenamefont {Momono}, \citenamefont {Kurosawa}, \citenamefont {Takagi},\
  and\ \citenamefont {Taillefer}}]{Doiron-Leyraud2017}%
  \BibitemOpen
  \bibfield  {author} {\bibinfo {author} {\bibfnamefont {N.}~\bibnamefont
  {Doiron-Leyraud}}, \bibinfo {author} {\bibfnamefont {O.}~\bibnamefont
  {Cyr-Choini{\`e}re}}, \bibinfo {author} {\bibfnamefont {S.}~\bibnamefont
  {Badoux}}, \bibinfo {author} {\bibfnamefont {A.}~\bibnamefont {Ataei}},
  \bibinfo {author} {\bibfnamefont {C.}~\bibnamefont {Collignon}}, \bibinfo
  {author} {\bibfnamefont {A.}~\bibnamefont {Gourgout}}, \bibinfo {author}
  {\bibfnamefont {S.}~\bibnamefont {Dufour-Beaus{\'e}jour}}, \bibinfo {author}
  {\bibfnamefont {F.~F.}\ \bibnamefont {Tafti}}, \bibinfo {author}
  {\bibfnamefont {F.}~\bibnamefont {Lalibert{\'e}}}, \bibinfo {author}
  {\bibfnamefont {M.-E.}\ \bibnamefont {Boulanger}}, \bibinfo {author}
  {\bibfnamefont {M.}~\bibnamefont {Matusiak}}, \bibinfo {author}
  {\bibfnamefont {D.}~\bibnamefont {Graf}}, \bibinfo {author} {\bibfnamefont
  {M.}~\bibnamefont {Kim}}, \bibinfo {author} {\bibfnamefont {J.-S.}\
  \bibnamefont {Zhou}}, \bibinfo {author} {\bibfnamefont {N.}~\bibnamefont
  {Momono}}, \bibinfo {author} {\bibfnamefont {T.}~\bibnamefont {Kurosawa}},
  \bibinfo {author} {\bibfnamefont {H.}~\bibnamefont {Takagi}},\ and\ \bibinfo
  {author} {\bibfnamefont {L.}~\bibnamefont {Taillefer}},\ }\bibfield  {title}
  {\bibinfo {title} {{Pseudogap phase of cuprate superconductors confined by
  \uppercase{F}ermi surface topology}},\ }\href
  {https://doi.org/10.1038/s41467-017-02122-x} {\bibfield  {journal} {\bibinfo
  {journal} {Nature Communications}\ }\textbf {\bibinfo {volume} {8}},\
  \bibinfo {pages} {2044} (\bibinfo {year} {2017})}\BibitemShut {NoStop}%
\bibitem [{\citenamefont {Wu}\ \emph {et~al.}(2020)\citenamefont {Wu},
  \citenamefont {Scheurer}, \citenamefont {Ferrero},\ and\ \citenamefont
  {Georges}}]{PhysRevResearch.2.033067}%
  \BibitemOpen
  \bibfield  {author} {\bibinfo {author} {\bibfnamefont {W.}~\bibnamefont
  {Wu}}, \bibinfo {author} {\bibfnamefont {M.~S.}\ \bibnamefont {Scheurer}},
  \bibinfo {author} {\bibfnamefont {M.}~\bibnamefont {Ferrero}},\ and\ \bibinfo
  {author} {\bibfnamefont {A.}~\bibnamefont {Georges}},\ }\bibfield  {title}
  {\bibinfo {title} {{Effect of Van Hove singularities in the onset of
  pseudogap states in \uppercase{M}ott insulators}},\ }\href
  {https://doi.org/10.1103/PhysRevResearch.2.033067} {\bibfield  {journal}
  {\bibinfo  {journal} {Phys. Rev. Res.}\ }\textbf {\bibinfo {volume} {2}},\
  \bibinfo {pages} {033067} (\bibinfo {year} {2020})}\BibitemShut {NoStop}%
\bibitem [{\citenamefont {Sakai}\ \emph {et~al.}(2009)\citenamefont {Sakai},
  \citenamefont {Motome},\ and\ \citenamefont
  {Imada}}]{PhysRevLett.102.056404}%
  \BibitemOpen
  \bibfield  {author} {\bibinfo {author} {\bibfnamefont {S.}~\bibnamefont
  {Sakai}}, \bibinfo {author} {\bibfnamefont {Y.}~\bibnamefont {Motome}},\ and\
  \bibinfo {author} {\bibfnamefont {M.}~\bibnamefont {Imada}},\ }\bibfield
  {title} {\bibinfo {title} {Evolution of electronic structure of doped
  {\uppercase{m}ott} insulators: Reconstruction of poles and zeros of {Green}'s
  function},\ }\href {https://doi.org/10.1103/PhysRevLett.102.056404}
  {\bibfield  {journal} {\bibinfo  {journal} {Phys. Rev. Lett.}\ }\textbf
  {\bibinfo {volume} {102}},\ \bibinfo {pages} {056404} (\bibinfo {year}
  {2009})}\BibitemShut {NoStop}%
\bibitem [{\citenamefont {Sakai}\ \emph {et~al.}(2010)\citenamefont {Sakai},
  \citenamefont {Motome},\ and\ \citenamefont {Imada}}]{PhysRevB.82.134505}%
  \BibitemOpen
  \bibfield  {author} {\bibinfo {author} {\bibfnamefont {S.}~\bibnamefont
  {Sakai}}, \bibinfo {author} {\bibfnamefont {Y.}~\bibnamefont {Motome}},\ and\
  \bibinfo {author} {\bibfnamefont {M.}~\bibnamefont {Imada}},\ }\bibfield
  {title} {\bibinfo {title} {{Doped high-${T}_{c}$ cuprate superconductors
  elucidated in the light of zeros and poles of the electronic Green's
  function}},\ }\href {https://doi.org/10.1103/PhysRevB.82.134505} {\bibfield
  {journal} {\bibinfo  {journal} {Phys. Rev. B}\ }\textbf {\bibinfo {volume}
  {82}},\ \bibinfo {pages} {134505} (\bibinfo {year} {2010})}\BibitemShut
  {NoStop}%
\bibitem [{\citenamefont {Sakai}\ \emph {et~al.}(2018)\citenamefont {Sakai},
  \citenamefont {Civelli},\ and\ \citenamefont {Imada}}]{PhysRevB.98.195109}%
  \BibitemOpen
  \bibfield  {author} {\bibinfo {author} {\bibfnamefont {S.}~\bibnamefont
  {Sakai}}, \bibinfo {author} {\bibfnamefont {M.}~\bibnamefont {Civelli}},\
  and\ \bibinfo {author} {\bibfnamefont {M.}~\bibnamefont {Imada}},\ }\bibfield
   {title} {\bibinfo {title} {Direct connection between {\uppercase{m}ott}
  insulators and $d$-wave high-temperature superconductors revealed by
  continuous evolution of self-energy poles},\ }\href
  {https://doi.org/10.1103/PhysRevB.98.195109} {\bibfield  {journal} {\bibinfo
  {journal} {Phys. Rev. B}\ }\textbf {\bibinfo {volume} {98}},\ \bibinfo
  {pages} {195109} (\bibinfo {year} {2018})}\BibitemShut {NoStop}%
\bibitem [{\citenamefont {Horio}\ \emph {et~al.}(2025)\citenamefont {Horio},
  \citenamefont {Sakai}, \citenamefont {Suzuki}, \citenamefont {Nonaka},
  \citenamefont {Hashimoto}, \citenamefont {Lu}, \citenamefont {Shen},
  \citenamefont {Ohgi}, \citenamefont {Konno}, \citenamefont {Adachi},
  \citenamefont {Koike}, \citenamefont {Imada},\ and\ \citenamefont
  {Fujimori}}]{horio2018common}%
  \BibitemOpen
  \bibfield  {author} {\bibinfo {author} {\bibfnamefont {M.}~\bibnamefont
  {Horio}}, \bibinfo {author} {\bibfnamefont {S.}~\bibnamefont {Sakai}},
  \bibinfo {author} {\bibfnamefont {H.}~\bibnamefont {Suzuki}}, \bibinfo
  {author} {\bibfnamefont {Y.}~\bibnamefont {Nonaka}}, \bibinfo {author}
  {\bibfnamefont {M.}~\bibnamefont {Hashimoto}}, \bibinfo {author}
  {\bibfnamefont {D.}~\bibnamefont {Lu}}, \bibinfo {author} {\bibfnamefont
  {Z.-X.}\ \bibnamefont {Shen}}, \bibinfo {author} {\bibfnamefont
  {T.}~\bibnamefont {Ohgi}}, \bibinfo {author} {\bibfnamefont {T.}~\bibnamefont
  {Konno}}, \bibinfo {author} {\bibfnamefont {T.}~\bibnamefont {Adachi}},
  \bibinfo {author} {\bibfnamefont {Y.}~\bibnamefont {Koike}}, \bibinfo
  {author} {\bibfnamefont {M.}~\bibnamefont {Imada}},\ and\ \bibinfo {author}
  {\bibfnamefont {A.}~\bibnamefont {Fujimori}},\ }\bibfield  {title} {\bibinfo
  {title} {Pseudogap in electron-doped cuprates: Strong correlation leading to
  band splitting},\ }\href {https://doi.org/10.1073/pnas.2406624122} {\bibfield
   {journal} {\bibinfo  {journal} {Proceedings of the National Academy of
  Sciences}\ }\textbf {\bibinfo {volume} {122}},\ \bibinfo {pages}
  {e2406624122} (\bibinfo {year} {2025})}\BibitemShut {NoStop}%
\bibitem [{\citenamefont {Arovas}\ \emph {et~al.}(2022)\citenamefont {Arovas},
  \citenamefont {Berg}, \citenamefont {Kivelson},\ and\ \citenamefont
  {Raghu}}]{annurev:/content/journals/10.1146/annurev-conmatphys-031620-102024}%
  \BibitemOpen
  \bibfield  {author} {\bibinfo {author} {\bibfnamefont {D.~P.}\ \bibnamefont
  {Arovas}}, \bibinfo {author} {\bibfnamefont {E.}~\bibnamefont {Berg}},
  \bibinfo {author} {\bibfnamefont {S.~A.}\ \bibnamefont {Kivelson}},\ and\
  \bibinfo {author} {\bibfnamefont {S.}~\bibnamefont {Raghu}},\ }\bibfield
  {title} {\bibinfo {title} {The \uppercase{H}ubbard model},\ }\href
  {https://doi.org/10.1146/annurev-conmatphys-031620-102024} {\bibfield
  {journal} {\bibinfo  {journal} {Annual Review of Condensed Matter Physics}\
  }\textbf {\bibinfo {volume} {13}},\ \bibinfo {pages} {239} (\bibinfo {year}
  {2022})}\BibitemShut {NoStop}%
\bibitem [{\citenamefont {Qin}\ \emph {et~al.}(2022)\citenamefont {Qin},
  \citenamefont {Schäfer}, \citenamefont {Andergassen}, \citenamefont
  {Corboz},\ and\ \citenamefont
  {Gull}}]{annurev:/content/journals/10.1146/annurev-conmatphys-090921-033948}%
  \BibitemOpen
  \bibfield  {author} {\bibinfo {author} {\bibfnamefont {M.}~\bibnamefont
  {Qin}}, \bibinfo {author} {\bibfnamefont {T.}~\bibnamefont {Schäfer}},
  \bibinfo {author} {\bibfnamefont {S.}~\bibnamefont {Andergassen}}, \bibinfo
  {author} {\bibfnamefont {P.}~\bibnamefont {Corboz}},\ and\ \bibinfo {author}
  {\bibfnamefont {E.}~\bibnamefont {Gull}},\ }\bibfield  {title} {\bibinfo
  {title} {The \uppercase{H}ubbard model: A computational perspective},\ }\href
  {https://doi.org/10.1146/annurev-conmatphys-090921-033948} {\bibfield
  {journal} {\bibinfo  {journal} {Annual Review of Condensed Matter Physics}\
  }\textbf {\bibinfo {volume} {13}},\ \bibinfo {pages} {275} (\bibinfo {year}
  {2022})}\BibitemShut {NoStop}%
\bibitem [{\citenamefont {Bulut}\ \emph {et~al.}(1994)\citenamefont {Bulut},
  \citenamefont {Scalapino},\ and\ \citenamefont {White}}]{PhysRevLett.72.705}%
  \BibitemOpen
  \bibfield  {author} {\bibinfo {author} {\bibfnamefont {N.}~\bibnamefont
  {Bulut}}, \bibinfo {author} {\bibfnamefont {D.~J.}\ \bibnamefont
  {Scalapino}},\ and\ \bibinfo {author} {\bibfnamefont {S.~R.}\ \bibnamefont
  {White}},\ }\bibfield  {title} {\bibinfo {title} {One-electron spectral
  weight of the doped two-dimensional \uppercase{H}ubbard model},\ }\href
  {https://doi.org/10.1103/PhysRevLett.72.705} {\bibfield  {journal} {\bibinfo
  {journal} {Phys. Rev. Lett.}\ }\textbf {\bibinfo {volume} {72}},\ \bibinfo
  {pages} {705} (\bibinfo {year} {1994})}\BibitemShut {NoStop}%
\bibitem [{\citenamefont {Preuss}\ \emph {et~al.}(1997)\citenamefont {Preuss},
  \citenamefont {Hanke}, \citenamefont {Gr\"ober},\ and\ \citenamefont
  {Evertz}}]{PhysRevLett.79.1122}%
  \BibitemOpen
  \bibfield  {author} {\bibinfo {author} {\bibfnamefont {R.}~\bibnamefont
  {Preuss}}, \bibinfo {author} {\bibfnamefont {W.}~\bibnamefont {Hanke}},
  \bibinfo {author} {\bibfnamefont {C.}~\bibnamefont {Gr\"ober}},\ and\
  \bibinfo {author} {\bibfnamefont {H.~G.}\ \bibnamefont {Evertz}},\ }\bibfield
   {title} {\bibinfo {title} {Pseudogaps and their interplay with magnetic
  excitations in the doped 2\uppercase{D} \uppercase{H}ubbard model},\ }\href
  {https://doi.org/10.1103/PhysRevLett.79.1122} {\bibfield  {journal} {\bibinfo
   {journal} {Phys. Rev. Lett.}\ }\textbf {\bibinfo {volume} {79}},\ \bibinfo
  {pages} {1122} (\bibinfo {year} {1997})}\BibitemShut {NoStop}%
\bibitem [{\citenamefont {Maier}\ \emph {et~al.}(2000)\citenamefont {Maier},
  \citenamefont {Jarrell}, \citenamefont {Pruschke},\ and\ \citenamefont
  {Keller}}]{Maier2000}%
  \BibitemOpen
  \bibfield  {author} {\bibinfo {author} {\bibfnamefont {T.}~\bibnamefont
  {Maier}}, \bibinfo {author} {\bibfnamefont {M.}~\bibnamefont {Jarrell}},
  \bibinfo {author} {\bibfnamefont {T.}~\bibnamefont {Pruschke}},\ and\
  \bibinfo {author} {\bibfnamefont {J.}~\bibnamefont {Keller}},\ }\bibfield
  {title} {\bibinfo {title} {A non-crossing approximation for the study of
  intersite correlations},\ }\href {https://doi.org/10.1007/s100510050077}
  {\bibfield  {journal} {\bibinfo  {journal} {The European Physical Journal B -
  Condensed Matter and Complex Systems}\ }\textbf {\bibinfo {volume} {13}},\
  \bibinfo {pages} {613} (\bibinfo {year} {2000})}\BibitemShut {NoStop}%
\bibitem [{\citenamefont {Huscroft}\ \emph {et~al.}(2001)\citenamefont
  {Huscroft}, \citenamefont {Jarrell}, \citenamefont {Maier}, \citenamefont
  {Moukouri},\ and\ \citenamefont {Tahvildarzadeh}}]{PhysRevLett.86.139}%
  \BibitemOpen
  \bibfield  {author} {\bibinfo {author} {\bibfnamefont {C.}~\bibnamefont
  {Huscroft}}, \bibinfo {author} {\bibfnamefont {M.}~\bibnamefont {Jarrell}},
  \bibinfo {author} {\bibfnamefont {T.}~\bibnamefont {Maier}}, \bibinfo
  {author} {\bibfnamefont {S.}~\bibnamefont {Moukouri}},\ and\ \bibinfo
  {author} {\bibfnamefont {A.~N.}\ \bibnamefont {Tahvildarzadeh}},\ }\bibfield
  {title} {\bibinfo {title} {Pseudogaps in the 2\uppercase{D}
  \uppercase{H}ubbard model},\ }\href
  {https://doi.org/10.1103/PhysRevLett.86.139} {\bibfield  {journal} {\bibinfo
  {journal} {Phys. Rev. Lett.}\ }\textbf {\bibinfo {volume} {86}},\ \bibinfo
  {pages} {139} (\bibinfo {year} {2001})}\BibitemShut {NoStop}%
\bibitem [{\citenamefont {Jarrell}\ \emph {et~al.}(2001)\citenamefont
  {Jarrell}, \citenamefont {Maier}, \citenamefont {Hettler},\ and\
  \citenamefont {Tahvildarzadeh}}]{M.Jarrell_2001}%
  \BibitemOpen
  \bibfield  {author} {\bibinfo {author} {\bibfnamefont {M.}~\bibnamefont
  {Jarrell}}, \bibinfo {author} {\bibfnamefont {T.}~\bibnamefont {Maier}},
  \bibinfo {author} {\bibfnamefont {M.~H.}\ \bibnamefont {Hettler}},\ and\
  \bibinfo {author} {\bibfnamefont {A.~N.}\ \bibnamefont {Tahvildarzadeh}},\
  }\bibfield  {title} {\bibinfo {title} {Phase diagram of the
  \uppercase{H}ubbard model: Beyond the dynamical mean field},\ }\href
  {https://doi.org/10.1209/epl/i2001-00557-x} {\bibfield  {journal} {\bibinfo
  {journal} {Europhysics Letters}\ }\textbf {\bibinfo {volume} {56}},\ \bibinfo
  {pages} {563} (\bibinfo {year} {2001})}\BibitemShut {NoStop}%
\bibitem [{\citenamefont {Stanescu}\ and\ \citenamefont
  {Phillips}(2003)}]{PhysRevLett.91.017002}%
  \BibitemOpen
  \bibfield  {author} {\bibinfo {author} {\bibfnamefont {T.~D.}\ \bibnamefont
  {Stanescu}}\ and\ \bibinfo {author} {\bibfnamefont {P.}~\bibnamefont
  {Phillips}},\ }\bibfield  {title} {\bibinfo {title} {Pseudogap in doped
  {\uppercase{m}ott} insulators is the near-neighbor analogue of the
  {\uppercase{m}ott} gap},\ }\href
  {https://doi.org/10.1103/PhysRevLett.91.017002} {\bibfield  {journal}
  {\bibinfo  {journal} {Phys. Rev. Lett.}\ }\textbf {\bibinfo {volume} {91}},\
  \bibinfo {pages} {017002} (\bibinfo {year} {2003})}\BibitemShut {NoStop}%
\bibitem [{\citenamefont {S\'en\'echal}\ and\ \citenamefont
  {Tremblay}(2004)}]{PhysRevLett.92.126401}%
  \BibitemOpen
  \bibfield  {author} {\bibinfo {author} {\bibfnamefont {D.}~\bibnamefont
  {S\'en\'echal}}\ and\ \bibinfo {author} {\bibfnamefont {A.-M.~S.}\
  \bibnamefont {Tremblay}},\ }\bibfield  {title} {\bibinfo {title} {Hot spots
  and pseudogaps for hole- and electron-doped high-temperature
  superconductors},\ }\href {https://doi.org/10.1103/PhysRevLett.92.126401}
  {\bibfield  {journal} {\bibinfo  {journal} {Phys. Rev. Lett.}\ }\textbf
  {\bibinfo {volume} {92}},\ \bibinfo {pages} {126401} (\bibinfo {year}
  {2004})}\BibitemShut {NoStop}%
\bibitem [{\citenamefont {Sadovskii}\ \emph {et~al.}(2005)\citenamefont
  {Sadovskii}, \citenamefont {Nekrasov}, \citenamefont {Kuchinskii},
  \citenamefont {Pruschke},\ and\ \citenamefont
  {Anisimov}}]{PhysRevB.72.155105}%
  \BibitemOpen
  \bibfield  {author} {\bibinfo {author} {\bibfnamefont {M.~V.}\ \bibnamefont
  {Sadovskii}}, \bibinfo {author} {\bibfnamefont {I.~A.}\ \bibnamefont
  {Nekrasov}}, \bibinfo {author} {\bibfnamefont {E.~Z.}\ \bibnamefont
  {Kuchinskii}}, \bibinfo {author} {\bibfnamefont {T.}~\bibnamefont
  {Pruschke}},\ and\ \bibinfo {author} {\bibfnamefont {V.~I.}\ \bibnamefont
  {Anisimov}},\ }\bibfield  {title} {\bibinfo {title} {{Pseudogaps in strongly
  correlated metals: A generalized dynamical mean-field theory approach}},\
  }\href {https://doi.org/10.1103/PhysRevB.72.155105} {\bibfield  {journal}
  {\bibinfo  {journal} {Phys. Rev. B}\ }\textbf {\bibinfo {volume} {72}},\
  \bibinfo {pages} {155105} (\bibinfo {year} {2005})}\BibitemShut {NoStop}%
\bibitem [{\citenamefont {Stanescu}\ and\ \citenamefont
  {Kotliar}(2006)}]{PhysRevB.74.125110}%
  \BibitemOpen
  \bibfield  {author} {\bibinfo {author} {\bibfnamefont {T.~D.}\ \bibnamefont
  {Stanescu}}\ and\ \bibinfo {author} {\bibfnamefont {G.}~\bibnamefont
  {Kotliar}},\ }\bibfield  {title} {\bibinfo {title} {{Fermi arcs and hidden
  zeros of the Green function in the pseudogap state}},\ }\href
  {https://doi.org/10.1103/PhysRevB.74.125110} {\bibfield  {journal} {\bibinfo
  {journal} {Phys. Rev. B}\ }\textbf {\bibinfo {volume} {74}},\ \bibinfo
  {pages} {125110} (\bibinfo {year} {2006})}\BibitemShut {NoStop}%
\bibitem [{\citenamefont {Tremblay}\ \emph {et~al.}(2006)\citenamefont
  {Tremblay}, \citenamefont {Kyung},\ and\ \citenamefont
  {Sénéchal}}]{10.1063/1.2199446}%
  \BibitemOpen
  \bibfield  {author} {\bibinfo {author} {\bibfnamefont {A.-M.~S.}\
  \bibnamefont {Tremblay}}, \bibinfo {author} {\bibfnamefont {B.}~\bibnamefont
  {Kyung}},\ and\ \bibinfo {author} {\bibfnamefont {D.}~\bibnamefont
  {Sénéchal}},\ }\bibfield  {title} {\bibinfo {title} {Pseudogap and
  high-temperature superconductivity from weak to strong coupling.
  \uppercase{T}owards a quantitative theory (review article)},\ }\href
  {https://doi.org/10.1063/1.2199446} {\bibfield  {journal} {\bibinfo
  {journal} {Low Temperature Physics}\ }\textbf {\bibinfo {volume} {32}},\
  \bibinfo {pages} {424} (\bibinfo {year} {2006})}\BibitemShut {NoStop}%
\bibitem [{\citenamefont {Kyung}\ \emph {et~al.}(2006)\citenamefont {Kyung},
  \citenamefont {Kancharla}, \citenamefont {S\'en\'echal}, \citenamefont
  {Tremblay}, \citenamefont {Civelli},\ and\ \citenamefont
  {Kotliar}}]{PhysRevB.73.165114}%
  \BibitemOpen
  \bibfield  {author} {\bibinfo {author} {\bibfnamefont {B.}~\bibnamefont
  {Kyung}}, \bibinfo {author} {\bibfnamefont {S.~S.}\ \bibnamefont
  {Kancharla}}, \bibinfo {author} {\bibfnamefont {D.}~\bibnamefont
  {S\'en\'echal}}, \bibinfo {author} {\bibfnamefont {A.-M.~S.}\ \bibnamefont
  {Tremblay}}, \bibinfo {author} {\bibfnamefont {M.}~\bibnamefont {Civelli}},\
  and\ \bibinfo {author} {\bibfnamefont {G.}~\bibnamefont {Kotliar}},\
  }\bibfield  {title} {\bibinfo {title} {{Pseudogap induced by short-range spin
  correlations in a doped \uppercase{M}ott insulator}},\ }\href
  {https://doi.org/10.1103/PhysRevB.73.165114} {\bibfield  {journal} {\bibinfo
  {journal} {Phys. Rev. B}\ }\textbf {\bibinfo {volume} {73}},\ \bibinfo
  {pages} {165114} (\bibinfo {year} {2006})}\BibitemShut {NoStop}%
\bibitem [{\citenamefont {Macridin}\ \emph {et~al.}(2006)\citenamefont
  {Macridin}, \citenamefont {Jarrell}, \citenamefont {Maier}, \citenamefont
  {Kent},\ and\ \citenamefont {D'Azevedo}}]{PhysRevLett.97.036401}%
  \BibitemOpen
  \bibfield  {author} {\bibinfo {author} {\bibfnamefont {A.}~\bibnamefont
  {Macridin}}, \bibinfo {author} {\bibfnamefont {M.}~\bibnamefont {Jarrell}},
  \bibinfo {author} {\bibfnamefont {T.}~\bibnamefont {Maier}}, \bibinfo
  {author} {\bibfnamefont {P.~R.~C.}\ \bibnamefont {Kent}},\ and\ \bibinfo
  {author} {\bibfnamefont {E.}~\bibnamefont {D'Azevedo}},\ }\bibfield  {title}
  {\bibinfo {title} {Pseudogap and antiferromagnetic correlations in the
  \uppercase{H}ubbard model},\ }\href
  {https://doi.org/10.1103/PhysRevLett.97.036401} {\bibfield  {journal}
  {\bibinfo  {journal} {Phys. Rev. Lett.}\ }\textbf {\bibinfo {volume} {97}},\
  \bibinfo {pages} {036401} (\bibinfo {year} {2006})}\BibitemShut {NoStop}%
\bibitem [{\citenamefont {Gull}\ \emph {et~al.}(2008)\citenamefont {Gull},
  \citenamefont {Werner}, \citenamefont {Wang}, \citenamefont {Troyer},\ and\
  \citenamefont {Millis}}]{Gull_2008}%
  \BibitemOpen
  \bibfield  {author} {\bibinfo {author} {\bibfnamefont {E.}~\bibnamefont
  {Gull}}, \bibinfo {author} {\bibfnamefont {P.}~\bibnamefont {Werner}},
  \bibinfo {author} {\bibfnamefont {X.}~\bibnamefont {Wang}}, \bibinfo {author}
  {\bibfnamefont {M.}~\bibnamefont {Troyer}},\ and\ \bibinfo {author}
  {\bibfnamefont {A.~J.}\ \bibnamefont {Millis}},\ }\bibfield  {title}
  {\bibinfo {title} {Local order and the gapped phase of the
  \uppercase{H}ubbard model: A plaquette dynamical mean-field investigation},\
  }\href {https://doi.org/10.1209/0295-5075/84/37009} {\bibfield  {journal}
  {\bibinfo  {journal} {Europhysics Letters}\ }\textbf {\bibinfo {volume}
  {84}},\ \bibinfo {pages} {37009} (\bibinfo {year} {2008})}\BibitemShut
  {NoStop}%
\bibitem [{\citenamefont {Liebsch}\ and\ \citenamefont
  {Tong}(2009)}]{PhysRevB.80.165126}%
  \BibitemOpen
  \bibfield  {author} {\bibinfo {author} {\bibfnamefont {A.}~\bibnamefont
  {Liebsch}}\ and\ \bibinfo {author} {\bibfnamefont {N.-H.}\ \bibnamefont
  {Tong}},\ }\bibfield  {title} {\bibinfo {title} {{Finite-temperature exact
  diagonalization cluster dynamical mean-field study of the two-dimensional
  \uppercase{H}ubbard model: Pseudogap, non-Fermi-liquid behavior, and
  particle-hole asymmetry}},\ }\href
  {https://doi.org/10.1103/PhysRevB.80.165126} {\bibfield  {journal} {\bibinfo
  {journal} {Phys. Rev. B}\ }\textbf {\bibinfo {volume} {80}},\ \bibinfo
  {pages} {165126} (\bibinfo {year} {2009})}\BibitemShut {NoStop}%
\bibitem [{\citenamefont {Vidhyadhiraja}\ \emph {et~al.}(2009)\citenamefont
  {Vidhyadhiraja}, \citenamefont {Macridin}, \citenamefont
  {\ifmmode~\mbox{\c{S}}\else \c{S}\fi{}en}, \citenamefont {Jarrell},\ and\
  \citenamefont {Ma}}]{PhysRevLett.102.206407}%
  \BibitemOpen
  \bibfield  {author} {\bibinfo {author} {\bibfnamefont {N.~S.}\ \bibnamefont
  {Vidhyadhiraja}}, \bibinfo {author} {\bibfnamefont {A.}~\bibnamefont
  {Macridin}}, \bibinfo {author} {\bibfnamefont {C.}~\bibnamefont
  {\ifmmode~\mbox{\c{S}}\else \c{S}\fi{}en}}, \bibinfo {author} {\bibfnamefont
  {M.}~\bibnamefont {Jarrell}},\ and\ \bibinfo {author} {\bibfnamefont
  {M.}~\bibnamefont {Ma}},\ }\bibfield  {title} {\bibinfo {title} {Quantum
  critical point at finite doping in the 2\uppercase{D} \uppercase{H}ubbard
  model: A dynamical cluster quantum {Monte Carlo} study},\ }\href
  {https://doi.org/10.1103/PhysRevLett.102.206407} {\bibfield  {journal}
  {\bibinfo  {journal} {Phys. Rev. Lett.}\ }\textbf {\bibinfo {volume} {102}},\
  \bibinfo {pages} {206407} (\bibinfo {year} {2009})}\BibitemShut {NoStop}%
\bibitem [{\citenamefont {Ferrero}\ \emph
  {et~al.}(2009{\natexlab{a}})\citenamefont {Ferrero}, \citenamefont
  {Cornaglia}, \citenamefont {De~Leo}, \citenamefont {Parcollet}, \citenamefont
  {Kotliar},\ and\ \citenamefont {Georges}}]{PhysRevB.80.064501}%
  \BibitemOpen
  \bibfield  {author} {\bibinfo {author} {\bibfnamefont {M.}~\bibnamefont
  {Ferrero}}, \bibinfo {author} {\bibfnamefont {P.~S.}\ \bibnamefont
  {Cornaglia}}, \bibinfo {author} {\bibfnamefont {L.}~\bibnamefont {De~Leo}},
  \bibinfo {author} {\bibfnamefont {O.}~\bibnamefont {Parcollet}}, \bibinfo
  {author} {\bibfnamefont {G.}~\bibnamefont {Kotliar}},\ and\ \bibinfo {author}
  {\bibfnamefont {A.}~\bibnamefont {Georges}},\ }\bibfield  {title} {\bibinfo
  {title} {{Pseudogap opening and formation of Fermi arcs as an
  orbital-selective \uppercase{M}ott transition in momentum space}},\ }\href
  {https://doi.org/10.1103/PhysRevB.80.064501} {\bibfield  {journal} {\bibinfo
  {journal} {Phys. Rev. B}\ }\textbf {\bibinfo {volume} {80}},\ \bibinfo
  {pages} {064501} (\bibinfo {year} {2009}{\natexlab{a}})}\BibitemShut
  {NoStop}%
\bibitem [{\citenamefont {Ferrero}\ \emph
  {et~al.}(2009{\natexlab{b}})\citenamefont {Ferrero}, \citenamefont
  {Cornaglia}, \citenamefont {{De Leo}}, \citenamefont {Parcollet},
  \citenamefont {Kotliar},\ and\ \citenamefont {Georges}}]{Ferrero_2009}%
  \BibitemOpen
  \bibfield  {author} {\bibinfo {author} {\bibfnamefont {M.}~\bibnamefont
  {Ferrero}}, \bibinfo {author} {\bibfnamefont {P.~S.}\ \bibnamefont
  {Cornaglia}}, \bibinfo {author} {\bibfnamefont {L.}~\bibnamefont {{De Leo}}},
  \bibinfo {author} {\bibfnamefont {O.}~\bibnamefont {Parcollet}}, \bibinfo
  {author} {\bibfnamefont {G.}~\bibnamefont {Kotliar}},\ and\ \bibinfo {author}
  {\bibfnamefont {A.}~\bibnamefont {Georges}},\ }\bibfield  {title} {\bibinfo
  {title} {{Valence bond dynamical mean-field theory of doped \uppercase{M}ott
  insulators with nodal/antinodal differentiation}},\ }\href
  {https://doi.org/10.1209/0295-5075/85/57009} {\bibfield  {journal} {\bibinfo
  {journal} {Europhysics Letters}\ }\textbf {\bibinfo {volume} {85}},\ \bibinfo
  {pages} {57009} (\bibinfo {year} {2009}{\natexlab{b}})}\BibitemShut {NoStop}%
\bibitem [{\citenamefont {Kohno}(2012)}]{PhysRevLett.108.076401}%
  \BibitemOpen
  \bibfield  {author} {\bibinfo {author} {\bibfnamefont {M.}~\bibnamefont
  {Kohno}},\ }\bibfield  {title} {\bibinfo {title} {\uppercase{M}ott transition
  in the two-dimensional \uppercase{H}ubbard model},\ }\href
  {https://doi.org/10.1103/PhysRevLett.108.076401} {\bibfield  {journal}
  {\bibinfo  {journal} {Phys. Rev. Lett.}\ }\textbf {\bibinfo {volume} {108}},\
  \bibinfo {pages} {076401} (\bibinfo {year} {2012})}\BibitemShut {NoStop}%
\bibitem [{\citenamefont {Sordi}\ \emph
  {et~al.}(2012{\natexlab{a}})\citenamefont {Sordi}, \citenamefont {S\'emon},
  \citenamefont {Haule},\ and\ \citenamefont
  {Tremblay}}]{PhysRevLett.108.216401}%
  \BibitemOpen
  \bibfield  {author} {\bibinfo {author} {\bibfnamefont {G.}~\bibnamefont
  {Sordi}}, \bibinfo {author} {\bibfnamefont {P.}~\bibnamefont {S\'emon}},
  \bibinfo {author} {\bibfnamefont {K.}~\bibnamefont {Haule}},\ and\ \bibinfo
  {author} {\bibfnamefont {A.-M.~S.}\ \bibnamefont {Tremblay}},\ }\bibfield
  {title} {\bibinfo {title} {Strong coupling superconductivity, pseudogap, and
  {\uppercase{m}ott} transition},\ }\href
  {https://doi.org/10.1103/PhysRevLett.108.216401} {\bibfield  {journal}
  {\bibinfo  {journal} {Phys. Rev. Lett.}\ }\textbf {\bibinfo {volume} {108}},\
  \bibinfo {pages} {216401} (\bibinfo {year} {2012}{\natexlab{a}})}\BibitemShut
  {NoStop}%
\bibitem [{\citenamefont {Sordi}\ \emph
  {et~al.}(2012{\natexlab{b}})\citenamefont {Sordi}, \citenamefont {S{\'e}mon},
  \citenamefont {Haule},\ and\ \citenamefont {Tremblay}}]{Sordi2012}%
  \BibitemOpen
  \bibfield  {author} {\bibinfo {author} {\bibfnamefont {G.}~\bibnamefont
  {Sordi}}, \bibinfo {author} {\bibfnamefont {P.}~\bibnamefont {S{\'e}mon}},
  \bibinfo {author} {\bibfnamefont {K.}~\bibnamefont {Haule}},\ and\ \bibinfo
  {author} {\bibfnamefont {A.-M.~S.}\ \bibnamefont {Tremblay}},\ }\bibfield
  {title} {\bibinfo {title} {{Pseudogap temperature as a Widom line in doped
  \uppercase{M}ott insulators}},\ }\href {https://doi.org/10.1038/srep00547}
  {\bibfield  {journal} {\bibinfo  {journal} {Scientific Reports}\ }\textbf
  {\bibinfo {volume} {2}},\ \bibinfo {pages} {547} (\bibinfo {year}
  {2012}{\natexlab{b}})}\BibitemShut {NoStop}%
\bibitem [{\citenamefont {Gull}\ \emph {et~al.}(2013)\citenamefont {Gull},
  \citenamefont {Parcollet},\ and\ \citenamefont
  {Millis}}]{PhysRevLett.110.216405}%
  \BibitemOpen
  \bibfield  {author} {\bibinfo {author} {\bibfnamefont {E.}~\bibnamefont
  {Gull}}, \bibinfo {author} {\bibfnamefont {O.}~\bibnamefont {Parcollet}},\
  and\ \bibinfo {author} {\bibfnamefont {A.~J.}\ \bibnamefont {Millis}},\
  }\bibfield  {title} {\bibinfo {title} {Superconductivity and the pseudogap in
  the two-dimensional \uppercase{H}ubbard model},\ }\href
  {https://doi.org/10.1103/PhysRevLett.110.216405} {\bibfield  {journal}
  {\bibinfo  {journal} {Phys. Rev. Lett.}\ }\textbf {\bibinfo {volume} {110}},\
  \bibinfo {pages} {216405} (\bibinfo {year} {2013})}\BibitemShut {NoStop}%
\bibitem [{\citenamefont {Sordi}\ \emph {et~al.}(2013)\citenamefont {Sordi},
  \citenamefont {S\'emon}, \citenamefont {Haule},\ and\ \citenamefont
  {Tremblay}}]{PhysRevB.87.041101}%
  \BibitemOpen
  \bibfield  {author} {\bibinfo {author} {\bibfnamefont {G.}~\bibnamefont
  {Sordi}}, \bibinfo {author} {\bibfnamefont {P.}~\bibnamefont {S\'emon}},
  \bibinfo {author} {\bibfnamefont {K.}~\bibnamefont {Haule}},\ and\ \bibinfo
  {author} {\bibfnamefont {A.-M.~S.}\ \bibnamefont {Tremblay}},\ }\bibfield
  {title} {\bibinfo {title} {$c$-axis resistivity, pseudogap,
  superconductivity, and {Widom} line in doped {\uppercase{m}ott} insulators},\
  }\href {https://doi.org/10.1103/PhysRevB.87.041101} {\bibfield  {journal}
  {\bibinfo  {journal} {Phys. Rev. B}\ }\textbf {\bibinfo {volume} {87}},\
  \bibinfo {pages} {041101} (\bibinfo {year} {2013})}\BibitemShut {NoStop}%
\bibitem [{\citenamefont {Kohno}(2014)}]{PhysRevB.90.035111}%
  \BibitemOpen
  \bibfield  {author} {\bibinfo {author} {\bibfnamefont {M.}~\bibnamefont
  {Kohno}},\ }\bibfield  {title} {\bibinfo {title} {Spectral properties near
  the {\uppercase{m}ott} transition in the two-dimensional \uppercase{H}ubbard
  model with next-nearest-neighbor hopping},\ }\href
  {https://doi.org/10.1103/PhysRevB.90.035111} {\bibfield  {journal} {\bibinfo
  {journal} {Phys. Rev. B}\ }\textbf {\bibinfo {volume} {90}},\ \bibinfo
  {pages} {035111} (\bibinfo {year} {2014})}\BibitemShut {NoStop}%
\bibitem [{\citenamefont {Kung}\ \emph {et~al.}(2014)\citenamefont {Kung},
  \citenamefont {Chen}, \citenamefont {Moritz}, \citenamefont {Johnston},
  \citenamefont {Thomale},\ and\ \citenamefont
  {Devereaux}}]{PhysRevB.90.224507}%
  \BibitemOpen
  \bibfield  {author} {\bibinfo {author} {\bibfnamefont {Y.~F.}\ \bibnamefont
  {Kung}}, \bibinfo {author} {\bibfnamefont {C.-C.}\ \bibnamefont {Chen}},
  \bibinfo {author} {\bibfnamefont {B.}~\bibnamefont {Moritz}}, \bibinfo
  {author} {\bibfnamefont {S.}~\bibnamefont {Johnston}}, \bibinfo {author}
  {\bibfnamefont {R.}~\bibnamefont {Thomale}},\ and\ \bibinfo {author}
  {\bibfnamefont {T.~P.}\ \bibnamefont {Devereaux}},\ }\bibfield  {title}
  {\bibinfo {title} {Numerical exploration of spontaneous broken symmetries in
  multiorbital \uppercase{H}ubbard models},\ }\href
  {https://doi.org/10.1103/PhysRevB.90.224507} {\bibfield  {journal} {\bibinfo
  {journal} {Phys. Rev. B}\ }\textbf {\bibinfo {volume} {90}},\ \bibinfo
  {pages} {224507} (\bibinfo {year} {2014})}\BibitemShut {NoStop}%
\bibitem [{\citenamefont {Gunnarsson}\ \emph {et~al.}(2015)\citenamefont
  {Gunnarsson}, \citenamefont {Sch\"afer}, \citenamefont {LeBlanc},
  \citenamefont {Gull}, \citenamefont {Merino}, \citenamefont {Sangiovanni},
  \citenamefont {Rohringer},\ and\ \citenamefont
  {Toschi}}]{PhysRevLett.114.236402}%
  \BibitemOpen
  \bibfield  {author} {\bibinfo {author} {\bibfnamefont {O.}~\bibnamefont
  {Gunnarsson}}, \bibinfo {author} {\bibfnamefont {T.}~\bibnamefont
  {Sch\"afer}}, \bibinfo {author} {\bibfnamefont {J.~P.~F.}\ \bibnamefont
  {LeBlanc}}, \bibinfo {author} {\bibfnamefont {E.}~\bibnamefont {Gull}},
  \bibinfo {author} {\bibfnamefont {J.}~\bibnamefont {Merino}}, \bibinfo
  {author} {\bibfnamefont {G.}~\bibnamefont {Sangiovanni}}, \bibinfo {author}
  {\bibfnamefont {G.}~\bibnamefont {Rohringer}},\ and\ \bibinfo {author}
  {\bibfnamefont {A.}~\bibnamefont {Toschi}},\ }\bibfield  {title} {\bibinfo
  {title} {Fluctuation diagnostics of the electron self-energy: Origin of the
  pseudogap physics},\ }\href {https://doi.org/10.1103/PhysRevLett.114.236402}
  {\bibfield  {journal} {\bibinfo  {journal} {Phys. Rev. Lett.}\ }\textbf
  {\bibinfo {volume} {114}},\ \bibinfo {pages} {236402} (\bibinfo {year}
  {2015})}\BibitemShut {NoStop}%
\bibitem [{\citenamefont {Yang}\ and\ \citenamefont
  {Feiguin}(2016)}]{PhysRevB.93.081107}%
  \BibitemOpen
  \bibfield  {author} {\bibinfo {author} {\bibfnamefont {C.}~\bibnamefont
  {Yang}}\ and\ \bibinfo {author} {\bibfnamefont {A.~E.}\ \bibnamefont
  {Feiguin}},\ }\bibfield  {title} {\bibinfo {title} {Spectral function of the
  two-dimensional \uppercase{H}ubbard model: A density matrix renormalization
  group plus cluster perturbation theory study},\ }\href
  {https://doi.org/10.1103/PhysRevB.93.081107} {\bibfield  {journal} {\bibinfo
  {journal} {Phys. Rev. B}\ }\textbf {\bibinfo {volume} {93}},\ \bibinfo
  {pages} {081107} (\bibinfo {year} {2016})}\BibitemShut {NoStop}%
\bibitem [{\citenamefont {Chen}\ \emph {et~al.}(2017)\citenamefont {Chen},
  \citenamefont {LeBlanc},\ and\ \citenamefont {Gull}}]{Chen2017}%
  \BibitemOpen
  \bibfield  {author} {\bibinfo {author} {\bibfnamefont {X.}~\bibnamefont
  {Chen}}, \bibinfo {author} {\bibfnamefont {J.~P.~F.}\ \bibnamefont
  {LeBlanc}},\ and\ \bibinfo {author} {\bibfnamefont {E.}~\bibnamefont
  {Gull}},\ }\bibfield  {title} {\bibinfo {title} {{Simulation of the NMR
  response in the pseudogap regime of the cuprates}},\ }\href
  {https://doi.org/10.1038/ncomms14986} {\bibfield  {journal} {\bibinfo
  {journal} {Nature Communications}\ }\textbf {\bibinfo {volume} {8}},\
  \bibinfo {pages} {14986} (\bibinfo {year} {2017})}\BibitemShut {NoStop}%
\bibitem [{\citenamefont {Wu}\ \emph {et~al.}(2017)\citenamefont {Wu},
  \citenamefont {Ferrero}, \citenamefont {Georges},\ and\ \citenamefont
  {Kozik}}]{PhysRevB.96.041105}%
  \BibitemOpen
  \bibfield  {author} {\bibinfo {author} {\bibfnamefont {W.}~\bibnamefont
  {Wu}}, \bibinfo {author} {\bibfnamefont {M.}~\bibnamefont {Ferrero}},
  \bibinfo {author} {\bibfnamefont {A.}~\bibnamefont {Georges}},\ and\ \bibinfo
  {author} {\bibfnamefont {E.}~\bibnamefont {Kozik}},\ }\bibfield  {title}
  {\bibinfo {title} {Controlling {Feynman} diagrammatic expansions: Physical
  nature of the pseudogap in the two-dimensional \uppercase{H}ubbard model},\
  }\href {https://doi.org/10.1103/PhysRevB.96.041105} {\bibfield  {journal}
  {\bibinfo  {journal} {Phys. Rev. B}\ }\textbf {\bibinfo {volume} {96}},\
  \bibinfo {pages} {041105} (\bibinfo {year} {2017})}\BibitemShut {NoStop}%
\bibitem [{\citenamefont {Kuz'min}\ \emph {et~al.}(2020)\citenamefont
  {Kuz'min}, \citenamefont {Visotin}, \citenamefont {Nikolaev},\ and\
  \citenamefont {Ovchinnikov}}]{PhysRevB.101.115141}%
  \BibitemOpen
  \bibfield  {author} {\bibinfo {author} {\bibfnamefont {V.~I.}\ \bibnamefont
  {Kuz'min}}, \bibinfo {author} {\bibfnamefont {M.~A.}\ \bibnamefont
  {Visotin}}, \bibinfo {author} {\bibfnamefont {S.~V.}\ \bibnamefont
  {Nikolaev}},\ and\ \bibinfo {author} {\bibfnamefont {S.~G.}\ \bibnamefont
  {Ovchinnikov}},\ }\bibfield  {title} {\bibinfo {title} {Doping and
  temperature evolution of pseudogap and spin-spin correlations in the
  two-dimensional \uppercase{H}ubbard model},\ }\href
  {https://doi.org/10.1103/PhysRevB.101.115141} {\bibfield  {journal} {\bibinfo
   {journal} {Phys. Rev. B}\ }\textbf {\bibinfo {volume} {101}},\ \bibinfo
  {pages} {115141} (\bibinfo {year} {2020})}\BibitemShut {NoStop}%
\bibitem [{\citenamefont {Huang}(2020)}]{huang2020strong}%
  \BibitemOpen
  \bibfield  {author} {\bibinfo {author} {\bibfnamefont {E.~W.}\ \bibnamefont
  {Huang}},\ }\href@noop {} {\bibinfo {title} {Strong-coupling mechanism of the
  pseudogap in small \uppercase{H}ubbard clusters}} (\bibinfo {year} {2020}),\
  \Eprint {https://arxiv.org/abs/2010.12601} {arXiv:2010.12601
  [cond-mat.str-el]} \BibitemShut {NoStop}%
\bibitem [{\citenamefont {Krien}\ \emph {et~al.}(2022)\citenamefont {Krien},
  \citenamefont {Worm}, \citenamefont {Chalupa-Gantner}, \citenamefont
  {Toschi},\ and\ \citenamefont {Held}}]{Krien2022}%
  \BibitemOpen
  \bibfield  {author} {\bibinfo {author} {\bibfnamefont {F.}~\bibnamefont
  {Krien}}, \bibinfo {author} {\bibfnamefont {P.}~\bibnamefont {Worm}},
  \bibinfo {author} {\bibfnamefont {P.}~\bibnamefont {Chalupa-Gantner}},
  \bibinfo {author} {\bibfnamefont {A.}~\bibnamefont {Toschi}},\ and\ \bibinfo
  {author} {\bibfnamefont {K.}~\bibnamefont {Held}},\ }\bibfield  {title}
  {\bibinfo {title} {Explaining the pseudogap through damping and antidamping
  on the \uppercase{F}ermi surface by imaginary spin scattering},\ }\href
  {https://doi.org/10.1038/s42005-022-01117-5} {\bibfield  {journal} {\bibinfo
  {journal} {Communications Physics}\ }\textbf {\bibinfo {volume} {5}},\
  \bibinfo {pages} {336} (\bibinfo {year} {2022})}\BibitemShut {NoStop}%
\bibitem [{\citenamefont {Rosenberg}\ \emph {et~al.}(2022)\citenamefont
  {Rosenberg}, \citenamefont {S\'en\'echal}, \citenamefont {Tremblay},\ and\
  \citenamefont {Charlebois}}]{PhysRevB.106.245132}%
  \BibitemOpen
  \bibfield  {author} {\bibinfo {author} {\bibfnamefont {P.}~\bibnamefont
  {Rosenberg}}, \bibinfo {author} {\bibfnamefont {D.}~\bibnamefont
  {S\'en\'echal}}, \bibinfo {author} {\bibfnamefont {A.-M.~S.}\ \bibnamefont
  {Tremblay}},\ and\ \bibinfo {author} {\bibfnamefont {M.}~\bibnamefont
  {Charlebois}},\ }\bibfield  {title} {\bibinfo {title} {Fermi arcs from
  dynamical variational \uppercase{M}onte \uppercase{C}arlo},\ }\href
  {https://doi.org/10.1103/PhysRevB.106.245132} {\bibfield  {journal} {\bibinfo
   {journal} {Phys. Rev. B}\ }\textbf {\bibinfo {volume} {106}},\ \bibinfo
  {pages} {245132} (\bibinfo {year} {2022})}\BibitemShut {NoStop}%
\bibitem [{\citenamefont {Huang}\ \emph {et~al.}(2023)\citenamefont {Huang},
  \citenamefont {Liu}, \citenamefont {Wang}, \citenamefont {Jiang},
  \citenamefont {Mai}, \citenamefont {Maier}, \citenamefont {Johnston},
  \citenamefont {Moritz},\ and\ \citenamefont
  {Devereaux}}]{PhysRevB.107.085126}%
  \BibitemOpen
  \bibfield  {author} {\bibinfo {author} {\bibfnamefont {E.~W.}\ \bibnamefont
  {Huang}}, \bibinfo {author} {\bibfnamefont {T.}~\bibnamefont {Liu}}, \bibinfo
  {author} {\bibfnamefont {W.~O.}\ \bibnamefont {Wang}}, \bibinfo {author}
  {\bibfnamefont {H.-C.}\ \bibnamefont {Jiang}}, \bibinfo {author}
  {\bibfnamefont {P.}~\bibnamefont {Mai}}, \bibinfo {author} {\bibfnamefont
  {T.~A.}\ \bibnamefont {Maier}}, \bibinfo {author} {\bibfnamefont
  {S.}~\bibnamefont {Johnston}}, \bibinfo {author} {\bibfnamefont
  {B.}~\bibnamefont {Moritz}},\ and\ \bibinfo {author} {\bibfnamefont {T.~P.}\
  \bibnamefont {Devereaux}},\ }\bibfield  {title} {\bibinfo {title}
  {Fluctuating intertwined stripes in the strange metal regime of the
  \uppercase{H}ubbard model},\ }\href
  {https://doi.org/10.1103/PhysRevB.107.085126} {\bibfield  {journal} {\bibinfo
   {journal} {Phys. Rev. B}\ }\textbf {\bibinfo {volume} {107}},\ \bibinfo
  {pages} {085126} (\bibinfo {year} {2023})}\BibitemShut {NoStop}%
\bibitem [{\citenamefont {Šimkovic}\ \emph {et~al.}(2024)\citenamefont
  {Šimkovic}, \citenamefont {Rossi}, \citenamefont {Georges},\ and\
  \citenamefont {Ferrero}}]{doi:10.1126/science.ade9194}%
  \BibitemOpen
  \bibfield  {author} {\bibinfo {author} {\bibfnamefont {F.}~\bibnamefont
  {Šimkovic}}, \bibinfo {author} {\bibfnamefont {R.}~\bibnamefont {Rossi}},
  \bibinfo {author} {\bibfnamefont {A.}~\bibnamefont {Georges}},\ and\ \bibinfo
  {author} {\bibfnamefont {M.}~\bibnamefont {Ferrero}},\ }\bibfield  {title}
  {\bibinfo {title} {Origin and fate of the pseudogap in the doped
  \uppercase{H}ubbard model},\ }\href {https://doi.org/10.1126/science.ade9194}
  {\bibfield  {journal} {\bibinfo  {journal} {Science}\ }\textbf {\bibinfo
  {volume} {385}},\ \bibinfo {pages} {eade9194} (\bibinfo {year}
  {2024})}\BibitemShut {NoStop}%
\bibitem [{\citenamefont {Worm}\ \emph {et~al.}(2024)\citenamefont {Worm},
  \citenamefont {Reitner}, \citenamefont {Held},\ and\ \citenamefont
  {Toschi}}]{PhysRevLett.133.166501}%
  \BibitemOpen
  \bibfield  {author} {\bibinfo {author} {\bibfnamefont {P.}~\bibnamefont
  {Worm}}, \bibinfo {author} {\bibfnamefont {M.}~\bibnamefont {Reitner}},
  \bibinfo {author} {\bibfnamefont {K.}~\bibnamefont {Held}},\ and\ \bibinfo
  {author} {\bibfnamefont {A.}~\bibnamefont {Toschi}},\ }\bibfield  {title}
  {\bibinfo {title} {Fermi and \uppercase{L}uttinger arcs: Two concepts,
  realized on one surface},\ }\href
  {https://doi.org/10.1103/PhysRevLett.133.166501} {\bibfield  {journal}
  {\bibinfo  {journal} {Phys. Rev. Lett.}\ }\textbf {\bibinfo {volume} {133}},\
  \bibinfo {pages} {166501} (\bibinfo {year} {2024})}\BibitemShut {NoStop}%
\bibitem [{\citenamefont {Meixner}\ \emph {et~al.}(2024)\citenamefont
  {Meixner}, \citenamefont {Menke}, \citenamefont {Klett}, \citenamefont
  {Heinzelmann}, \citenamefont {Andergassen}, \citenamefont {Hansmann},\ and\
  \citenamefont {Schäfer}}]{10.21468/SciPostPhys.16.2.059}%
  \BibitemOpen
  \bibfield  {author} {\bibinfo {author} {\bibfnamefont {M.}~\bibnamefont
  {Meixner}}, \bibinfo {author} {\bibfnamefont {H.}~\bibnamefont {Menke}},
  \bibinfo {author} {\bibfnamefont {M.}~\bibnamefont {Klett}}, \bibinfo
  {author} {\bibfnamefont {S.}~\bibnamefont {Heinzelmann}}, \bibinfo {author}
  {\bibfnamefont {S.}~\bibnamefont {Andergassen}}, \bibinfo {author}
  {\bibfnamefont {P.}~\bibnamefont {Hansmann}},\ and\ \bibinfo {author}
  {\bibfnamefont {T.}~\bibnamefont {Schäfer}},\ }\bibfield  {title} {\bibinfo
  {title} {{Mott transition and pseudogap of the square-lattice
  \uppercase{H}ubbard model: Results from center-focused cellular dynamical
  mean-field theory}},\ }\href {https://doi.org/10.21468/SciPostPhys.16.2.059}
  {\bibfield  {journal} {\bibinfo  {journal} {SciPost Phys.}\ }\textbf
  {\bibinfo {volume} {16}},\ \bibinfo {pages} {059} (\bibinfo {year}
  {2024})}\BibitemShut {NoStop}%
\bibitem [{\citenamefont {Blankenbecler}\ \emph {et~al.}(1981)\citenamefont
  {Blankenbecler}, \citenamefont {Scalapino},\ and\ \citenamefont
  {Sugar}}]{DQMC1}%
  \BibitemOpen
  \bibfield  {author} {\bibinfo {author} {\bibfnamefont {R.}~\bibnamefont
  {Blankenbecler}}, \bibinfo {author} {\bibfnamefont {D.~J.}\ \bibnamefont
  {Scalapino}},\ and\ \bibinfo {author} {\bibfnamefont {R.~L.}\ \bibnamefont
  {Sugar}},\ }\bibfield  {title} {\bibinfo {title} {\uppercase{M}onte
  \uppercase{C}arlo calculations of coupled boson-fermion systems. {I}},\
  }\href {https://doi.org/10.1103/PhysRevD.24.2278} {\bibfield  {journal}
  {\bibinfo  {journal} {Phys. Rev. D}\ }\textbf {\bibinfo {volume} {24}},\
  \bibinfo {pages} {2278} (\bibinfo {year} {1981})}\BibitemShut {NoStop}%
\bibitem [{\citenamefont {White}\ \emph {et~al.}(1989)\citenamefont {White},
  \citenamefont {Scalapino}, \citenamefont {Sugar}, \citenamefont {Loh},
  \citenamefont {Gubernatis},\ and\ \citenamefont {Scalettar}}]{DQMC2}%
  \BibitemOpen
  \bibfield  {author} {\bibinfo {author} {\bibfnamefont {S.~R.}\ \bibnamefont
  {White}}, \bibinfo {author} {\bibfnamefont {D.~J.}\ \bibnamefont
  {Scalapino}}, \bibinfo {author} {\bibfnamefont {R.~L.}\ \bibnamefont
  {Sugar}}, \bibinfo {author} {\bibfnamefont {E.~Y.}\ \bibnamefont {Loh}},
  \bibinfo {author} {\bibfnamefont {J.~E.}\ \bibnamefont {Gubernatis}},\ and\
  \bibinfo {author} {\bibfnamefont {R.~T.}\ \bibnamefont {Scalettar}},\
  }\bibfield  {title} {\bibinfo {title} {Numerical study of the two-dimensional
  \uppercase{H}ubbard model},\ }\href {https://doi.org/10.1103/PhysRevB.40.506}
  {\bibfield  {journal} {\bibinfo  {journal} {Phys. Rev. B}\ }\textbf {\bibinfo
  {volume} {40}},\ \bibinfo {pages} {506} (\bibinfo {year} {1989})}\BibitemShut
  {NoStop}%
\bibitem [{\citenamefont {Jarrell}\ and\ \citenamefont
  {Gubernatis}(1996)}]{maxent1}%
  \BibitemOpen
  \bibfield  {author} {\bibinfo {author} {\bibfnamefont {M.}~\bibnamefont
  {Jarrell}}\ and\ \bibinfo {author} {\bibfnamefont {J.~E.}\ \bibnamefont
  {Gubernatis}},\ }\bibfield  {title} {\bibinfo {title} {{Bayesian inference
  and the analytic continuation of imaginary-time quantum Monte Carlo data}},\
  }\href {https://doi.org/10.1016/0370-1573(95)00074-7} {\bibfield  {journal}
  {\bibinfo  {journal} {Physics Reports}\ }\textbf {\bibinfo {volume} {269}},\
  \bibinfo {pages} {133} (\bibinfo {year} {1996})}\BibitemShut {NoStop}%
\bibitem [{\citenamefont {Gunnarsson}\ \emph {et~al.}(2010)\citenamefont
  {Gunnarsson}, \citenamefont {Haverkort},\ and\ \citenamefont
  {Sangiovanni}}]{maxent2}%
  \BibitemOpen
  \bibfield  {author} {\bibinfo {author} {\bibfnamefont {O.}~\bibnamefont
  {Gunnarsson}}, \bibinfo {author} {\bibfnamefont {M.~W.}\ \bibnamefont
  {Haverkort}},\ and\ \bibinfo {author} {\bibfnamefont {G.}~\bibnamefont
  {Sangiovanni}},\ }\bibfield  {title} {\bibinfo {title} {Analytical
  continuation of imaginary axis data for optical conductivity},\ }\href
  {https://doi.org/10.1103/PhysRevB.82.165125} {\bibfield  {journal} {\bibinfo
  {journal} {Phys. Rev. B}\ }\textbf {\bibinfo {volume} {82}},\ \bibinfo
  {pages} {165125} (\bibinfo {year} {2010})}\BibitemShut {NoStop}%
\bibitem [{\citenamefont {Gros}(1996)}]{PhysRevB.53.6865}%
  \BibitemOpen
  \bibfield  {author} {\bibinfo {author} {\bibfnamefont {C.}~\bibnamefont
  {Gros}},\ }\bibfield  {title} {\bibinfo {title} {Control of the finite-size
  corrections in exact diagonalization studies},\ }\href
  {https://doi.org/10.1103/PhysRevB.53.6865} {\bibfield  {journal} {\bibinfo
  {journal} {Phys. Rev. B}\ }\textbf {\bibinfo {volume} {53}},\ \bibinfo
  {pages} {6865} (\bibinfo {year} {1996})}\BibitemShut {NoStop}%
\bibitem [{\citenamefont {Lin}\ \emph {et~al.}(2001)\citenamefont {Lin},
  \citenamefont {Zong},\ and\ \citenamefont {Ceperley}}]{PhysRevE.64.016702}%
  \BibitemOpen
  \bibfield  {author} {\bibinfo {author} {\bibfnamefont {C.}~\bibnamefont
  {Lin}}, \bibinfo {author} {\bibfnamefont {F.~H.}\ \bibnamefont {Zong}},\ and\
  \bibinfo {author} {\bibfnamefont {D.~M.}\ \bibnamefont {Ceperley}},\
  }\bibfield  {title} {\bibinfo {title} {Twist-averaged boundary conditions in
  continuum quantum monte carlo algorithms},\ }\href
  {https://doi.org/10.1103/PhysRevE.64.016702} {\bibfield  {journal} {\bibinfo
  {journal} {Phys. Rev. E}\ }\textbf {\bibinfo {volume} {64}},\ \bibinfo
  {pages} {016702} (\bibinfo {year} {2001})}\BibitemShut {NoStop}%
\bibitem [{\citenamefont {Koretsune}\ \emph {et~al.}(2007)\citenamefont
  {Koretsune}, \citenamefont {Motome},\ and\ \citenamefont
  {Furusaki}}]{doi:10.1143/JPSJ.76.074719}%
  \BibitemOpen
  \bibfield  {author} {\bibinfo {author} {\bibfnamefont {T.}~\bibnamefont
  {Koretsune}}, \bibinfo {author} {\bibfnamefont {Y.}~\bibnamefont {Motome}},\
  and\ \bibinfo {author} {\bibfnamefont {A.}~\bibnamefont {Furusaki}},\
  }\bibfield  {title} {\bibinfo {title} {Exact diagonalization study of
  \uppercase{M}ott transition in the \uppercase{H}ubbard model on an
  anisotropic triangular lattice},\ }\href
  {https://doi.org/10.1143/JPSJ.76.074719} {\bibfield  {journal} {\bibinfo
  {journal} {Journal of the Physical Society of Japan}\ }\textbf {\bibinfo
  {volume} {76}},\ \bibinfo {pages} {074719} (\bibinfo {year}
  {2007})}\BibitemShut {NoStop}%
\bibitem [{\citenamefont {Qin}\ \emph {et~al.}(2016)\citenamefont {Qin},
  \citenamefont {Shi},\ and\ \citenamefont {Zhang}}]{PhysRevB.94.085103}%
  \BibitemOpen
  \bibfield  {author} {\bibinfo {author} {\bibfnamefont {M.}~\bibnamefont
  {Qin}}, \bibinfo {author} {\bibfnamefont {H.}~\bibnamefont {Shi}},\ and\
  \bibinfo {author} {\bibfnamefont {S.}~\bibnamefont {Zhang}},\ }\bibfield
  {title} {\bibinfo {title} {Benchmark study of the two-dimensional
  \uppercase{H}ubbard model with auxiliary-field quantum monte carlo method},\
  }\href {https://doi.org/10.1103/PhysRevB.94.085103} {\bibfield  {journal}
  {\bibinfo  {journal} {Phys. Rev. B}\ }\textbf {\bibinfo {volume} {94}},\
  \bibinfo {pages} {085103} (\bibinfo {year} {2016})}\BibitemShut {NoStop}%
\bibitem [{\citenamefont {Karakuzu}\ \emph {et~al.}(2017)\citenamefont
  {Karakuzu}, \citenamefont {Tocchio}, \citenamefont {Sorella},\ and\
  \citenamefont {Becca}}]{PhysRevB.96.205145}%
  \BibitemOpen
  \bibfield  {author} {\bibinfo {author} {\bibfnamefont {S.}~\bibnamefont
  {Karakuzu}}, \bibinfo {author} {\bibfnamefont {L.~F.}\ \bibnamefont
  {Tocchio}}, \bibinfo {author} {\bibfnamefont {S.}~\bibnamefont {Sorella}},\
  and\ \bibinfo {author} {\bibfnamefont {F.}~\bibnamefont {Becca}},\ }\bibfield
   {title} {\bibinfo {title} {Superconductivity, charge-density waves,
  antiferromagnetism, and phase separation in the
  \uppercase{H}ubbard-\uppercase{H}olstein model},\ }\href
  {https://doi.org/10.1103/PhysRevB.96.205145} {\bibfield  {journal} {\bibinfo
  {journal} {Phys. Rev. B}\ }\textbf {\bibinfo {volume} {96}},\ \bibinfo
  {pages} {205145} (\bibinfo {year} {2017})}\BibitemShut {NoStop}%
\bibitem [{\citenamefont {Karakuzu}\ \emph {et~al.}(2018)\citenamefont
  {Karakuzu}, \citenamefont {Seki},\ and\ \citenamefont
  {Sorella}}]{PhysRevB.98.075156}%
  \BibitemOpen
  \bibfield  {author} {\bibinfo {author} {\bibfnamefont {S.}~\bibnamefont
  {Karakuzu}}, \bibinfo {author} {\bibfnamefont {K.}~\bibnamefont {Seki}},\
  and\ \bibinfo {author} {\bibfnamefont {S.}~\bibnamefont {Sorella}},\
  }\bibfield  {title} {\bibinfo {title} {Study of the superconducting order
  parameter in the two-dimensional negative-\uppercase{$U$} \uppercase{H}ubbard
  model by grand-canonical twist-averaged boundary conditions},\ }\href
  {https://doi.org/10.1103/PhysRevB.98.075156} {\bibfield  {journal} {\bibinfo
  {journal} {Phys. Rev. B}\ }\textbf {\bibinfo {volume} {98}},\ \bibinfo
  {pages} {075156} (\bibinfo {year} {2018})}\BibitemShut {NoStop}%
\bibitem [{\citenamefont {Mai}\ \emph {et~al.}(2026)\citenamefont {Mai},
  \citenamefont {Zhao}, \citenamefont {Tenkila}, \citenamefont {Hackner},
  \citenamefont {Kush}, \citenamefont {Pan},\ and\ \citenamefont
  {Phillips}}]{Mai2025}%
  \BibitemOpen
  \bibfield  {author} {\bibinfo {author} {\bibfnamefont {P.}~\bibnamefont
  {Mai}}, \bibinfo {author} {\bibfnamefont {J.}~\bibnamefont {Zhao}}, \bibinfo
  {author} {\bibfnamefont {G.}~\bibnamefont {Tenkila}}, \bibinfo {author}
  {\bibfnamefont {N.~A.}\ \bibnamefont {Hackner}}, \bibinfo {author}
  {\bibfnamefont {D.}~\bibnamefont {Kush}}, \bibinfo {author} {\bibfnamefont
  {D.}~\bibnamefont {Pan}},\ and\ \bibinfo {author} {\bibfnamefont {P.~W.}\
  \bibnamefont {Phillips}},\ }\bibfield  {title} {\bibinfo {title} {Twisting
  the \uppercase{H}ubbard model into the momentum-mixing
  \uppercase{H}atsugai--\uppercase{K}ohmoto model},\ }\href
  {https://doi.org/10.1038/s41567-025-03095-1} {\bibfield  {journal} {\bibinfo
  {journal} {Nature Physics}\ }\textbf {\bibinfo {volume} {22}},\ \bibinfo
  {pages} {81} (\bibinfo {year} {2026})}\BibitemShut {NoStop}%
\bibitem [{\citenamefont {Yang}\ \emph {et~al.}(2006)\citenamefont {Yang},
  \citenamefont {Rice},\ and\ \citenamefont {Zhang}}]{PhysRevB.73.174501}%
  \BibitemOpen
  \bibfield  {author} {\bibinfo {author} {\bibfnamefont {K.-Y.}\ \bibnamefont
  {Yang}}, \bibinfo {author} {\bibfnamefont {T.~M.}\ \bibnamefont {Rice}},\
  and\ \bibinfo {author} {\bibfnamefont {F.-C.}\ \bibnamefont {Zhang}},\
  }\bibfield  {title} {\bibinfo {title} {Phenomenological theory of the
  pseudogap state},\ }\href {https://doi.org/10.1103/PhysRevB.73.174501}
  {\bibfield  {journal} {\bibinfo  {journal} {Phys. Rev. B}\ }\textbf {\bibinfo
  {volume} {73}},\ \bibinfo {pages} {174501} (\bibinfo {year}
  {2006})}\BibitemShut {NoStop}%
\bibitem [{\citenamefont {Qi}\ and\ \citenamefont
  {Sachdev}(2010)}]{PhysRevB.81.115129}%
  \BibitemOpen
  \bibfield  {author} {\bibinfo {author} {\bibfnamefont {Y.}~\bibnamefont
  {Qi}}\ and\ \bibinfo {author} {\bibfnamefont {S.}~\bibnamefont {Sachdev}},\
  }\bibfield  {title} {\bibinfo {title} {{Effective theory of Fermi pockets in
  fluctuating antiferromagnets}},\ }\href
  {https://doi.org/10.1103/PhysRevB.81.115129} {\bibfield  {journal} {\bibinfo
  {journal} {Phys. Rev. B}\ }\textbf {\bibinfo {volume} {81}},\ \bibinfo
  {pages} {115129} (\bibinfo {year} {2010})}\BibitemShut {NoStop}%
\bibitem [{\citenamefont {Shen}\ \emph {et~al.}(2005)\citenamefont {Shen},
  \citenamefont {Ronning}, \citenamefont {Lu}, \citenamefont {Baumberger},
  \citenamefont {Ingle}, \citenamefont {Lee}, \citenamefont {Meevasana},
  \citenamefont {Kohsaka}, \citenamefont {Azuma}, \citenamefont {Takano},
  \citenamefont {Takagi},\ and\ \citenamefont
  {Shen}}]{doi:10.1126/science.1103627}%
  \BibitemOpen
  \bibfield  {author} {\bibinfo {author} {\bibfnamefont {K.~M.}\ \bibnamefont
  {Shen}}, \bibinfo {author} {\bibfnamefont {F.}~\bibnamefont {Ronning}},
  \bibinfo {author} {\bibfnamefont {D.~H.}\ \bibnamefont {Lu}}, \bibinfo
  {author} {\bibfnamefont {F.}~\bibnamefont {Baumberger}}, \bibinfo {author}
  {\bibfnamefont {N.~J.~C.}\ \bibnamefont {Ingle}}, \bibinfo {author}
  {\bibfnamefont {W.~S.}\ \bibnamefont {Lee}}, \bibinfo {author} {\bibfnamefont
  {W.}~\bibnamefont {Meevasana}}, \bibinfo {author} {\bibfnamefont
  {Y.}~\bibnamefont {Kohsaka}}, \bibinfo {author} {\bibfnamefont
  {M.}~\bibnamefont {Azuma}}, \bibinfo {author} {\bibfnamefont
  {M.}~\bibnamefont {Takano}}, \bibinfo {author} {\bibfnamefont
  {H.}~\bibnamefont {Takagi}},\ and\ \bibinfo {author} {\bibfnamefont {Z.-X.}\
  \bibnamefont {Shen}},\ }\bibfield  {title} {\bibinfo {title} {Nodal
  quasiparticles and antinodal charge ordering in
  {Ca$_{2-x}$Na$_x$CuO$_2$Cl$_2$}},\ }\href
  {https://doi.org/10.1126/science.1103627} {\bibfield  {journal} {\bibinfo
  {journal} {Science}\ }\textbf {\bibinfo {volume} {307}},\ \bibinfo {pages}
  {901} (\bibinfo {year} {2005})}\BibitemShut {NoStop}%
\bibitem [{\citenamefont {Lee}\ \emph {et~al.}(2007)\citenamefont {Lee},
  \citenamefont {Vishik}, \citenamefont {Tanaka}, \citenamefont {Lu},
  \citenamefont {Sasagawa}, \citenamefont {Nagaosa}, \citenamefont {Devereaux},
  \citenamefont {Hussain},\ and\ \citenamefont {Shen}}]{Lee2007}%
  \BibitemOpen
  \bibfield  {author} {\bibinfo {author} {\bibfnamefont {W.~S.}\ \bibnamefont
  {Lee}}, \bibinfo {author} {\bibfnamefont {I.~M.}\ \bibnamefont {Vishik}},
  \bibinfo {author} {\bibfnamefont {K.}~\bibnamefont {Tanaka}}, \bibinfo
  {author} {\bibfnamefont {D.~H.}\ \bibnamefont {Lu}}, \bibinfo {author}
  {\bibfnamefont {T.}~\bibnamefont {Sasagawa}}, \bibinfo {author}
  {\bibfnamefont {N.}~\bibnamefont {Nagaosa}}, \bibinfo {author} {\bibfnamefont
  {T.~P.}\ \bibnamefont {Devereaux}}, \bibinfo {author} {\bibfnamefont
  {Z.}~\bibnamefont {Hussain}},\ and\ \bibinfo {author} {\bibfnamefont {Z.-X.}\
  \bibnamefont {Shen}},\ }\bibfield  {title} {\bibinfo {title} {{Abrupt onset
  of a second energy gap at the superconducting transition of underdoped
  Bi2212}},\ }\href {https://doi.org/10.1038/nature06219} {\bibfield  {journal}
  {\bibinfo  {journal} {Nature}\ }\textbf {\bibinfo {volume} {450}},\ \bibinfo
  {pages} {81} (\bibinfo {year} {2007})}\BibitemShut {NoStop}%
\bibitem [{\citenamefont {Hashimoto}\ \emph {et~al.}(2010)\citenamefont
  {Hashimoto}, \citenamefont {He}, \citenamefont {Tanaka}, \citenamefont
  {Testaud}, \citenamefont {Meevasana}, \citenamefont {Moore}, \citenamefont
  {Lu}, \citenamefont {Yao}, \citenamefont {Yoshida}, \citenamefont {Eisaki},
  \citenamefont {Devereaux}, \citenamefont {Hussain},\ and\ \citenamefont
  {Shen}}]{Hashimoto2010}%
  \BibitemOpen
  \bibfield  {author} {\bibinfo {author} {\bibfnamefont {M.}~\bibnamefont
  {Hashimoto}}, \bibinfo {author} {\bibfnamefont {R.-H.}\ \bibnamefont {He}},
  \bibinfo {author} {\bibfnamefont {K.}~\bibnamefont {Tanaka}}, \bibinfo
  {author} {\bibfnamefont {J.-P.}\ \bibnamefont {Testaud}}, \bibinfo {author}
  {\bibfnamefont {W.}~\bibnamefont {Meevasana}}, \bibinfo {author}
  {\bibfnamefont {R.~G.}\ \bibnamefont {Moore}}, \bibinfo {author}
  {\bibfnamefont {D.}~\bibnamefont {Lu}}, \bibinfo {author} {\bibfnamefont
  {H.}~\bibnamefont {Yao}}, \bibinfo {author} {\bibfnamefont {Y.}~\bibnamefont
  {Yoshida}}, \bibinfo {author} {\bibfnamefont {H.}~\bibnamefont {Eisaki}},
  \bibinfo {author} {\bibfnamefont {T.~P.}\ \bibnamefont {Devereaux}}, \bibinfo
  {author} {\bibfnamefont {Z.}~\bibnamefont {Hussain}},\ and\ \bibinfo {author}
  {\bibfnamefont {Z.-X.}\ \bibnamefont {Shen}},\ }\bibfield  {title} {\bibinfo
  {title} {{Particle--hole symmetry breaking in the pseudogap state of
  Bi2201}},\ }\href {https://doi.org/10.1038/nphys1632} {\bibfield  {journal}
  {\bibinfo  {journal} {Nature Physics}\ }\textbf {\bibinfo {volume} {6}},\
  \bibinfo {pages} {414} (\bibinfo {year} {2010})}\BibitemShut {NoStop}%
\bibitem [{\citenamefont {Huang}\ \emph {et~al.}(2017)\citenamefont {Huang},
  \citenamefont {Mendl}, \citenamefont {Liu}, \citenamefont {Johnston},
  \citenamefont {Jiang}, \citenamefont {Moritz},\ and\ \citenamefont
  {Devereaux}}]{doi:10.1126/science.aak9546}%
  \BibitemOpen
  \bibfield  {author} {\bibinfo {author} {\bibfnamefont {E.~W.}\ \bibnamefont
  {Huang}}, \bibinfo {author} {\bibfnamefont {C.~B.}\ \bibnamefont {Mendl}},
  \bibinfo {author} {\bibfnamefont {S.}~\bibnamefont {Liu}}, \bibinfo {author}
  {\bibfnamefont {S.}~\bibnamefont {Johnston}}, \bibinfo {author}
  {\bibfnamefont {H.-C.}\ \bibnamefont {Jiang}}, \bibinfo {author}
  {\bibfnamefont {B.}~\bibnamefont {Moritz}},\ and\ \bibinfo {author}
  {\bibfnamefont {T.~P.}\ \bibnamefont {Devereaux}},\ }\bibfield  {title}
  {\bibinfo {title} {{Numerical evidence of fluctuating stripes in the normal
  state of high-$\uppercase{T}_c$ cuprate superconductors}},\ }\href
  {https://doi.org/10.1126/science.aak9546} {\bibfield  {journal} {\bibinfo
  {journal} {Science}\ }\textbf {\bibinfo {volume} {358}},\ \bibinfo {pages}
  {1161} (\bibinfo {year} {2017})}\BibitemShut {NoStop}%
\bibitem [{\citenamefont {Huang}\ \emph {et~al.}(2018)\citenamefont {Huang},
  \citenamefont {Mendl}, \citenamefont {Jiang}, \citenamefont {Moritz},\ and\
  \citenamefont {Devereaux}}]{Huang2018}%
  \BibitemOpen
  \bibfield  {author} {\bibinfo {author} {\bibfnamefont {E.~W.}\ \bibnamefont
  {Huang}}, \bibinfo {author} {\bibfnamefont {C.~B.}\ \bibnamefont {Mendl}},
  \bibinfo {author} {\bibfnamefont {H.-C.}\ \bibnamefont {Jiang}}, \bibinfo
  {author} {\bibfnamefont {B.}~\bibnamefont {Moritz}},\ and\ \bibinfo {author}
  {\bibfnamefont {T.~P.}\ \bibnamefont {Devereaux}},\ }\bibfield  {title}
  {\bibinfo {title} {{Stripe order from the perspective of the
  \uppercase{H}ubbard model}},\ }\href
  {https://doi.org/10.1038/s41535-018-0097-0} {\bibfield  {journal} {\bibinfo
  {journal} {npj Quantum Materials}\ }\textbf {\bibinfo {volume} {3}},\
  \bibinfo {pages} {22} (\bibinfo {year} {2018})}\BibitemShut {NoStop}%
\bibitem [{\citenamefont {Stock}\ \emph {et~al.}(2005)\citenamefont {Stock},
  \citenamefont {Buyers}, \citenamefont {Cowley}, \citenamefont {Clegg},
  \citenamefont {Coldea}, \citenamefont {Frost}, \citenamefont {Liang},
  \citenamefont {Peets}, \citenamefont {Bonn}, \citenamefont {Hardy},\ and\
  \citenamefont {Birgeneau}}]{PhysRevB.71.024522}%
  \BibitemOpen
  \bibfield  {author} {\bibinfo {author} {\bibfnamefont {C.}~\bibnamefont
  {Stock}}, \bibinfo {author} {\bibfnamefont {W.~J.~L.}\ \bibnamefont
  {Buyers}}, \bibinfo {author} {\bibfnamefont {R.~A.}\ \bibnamefont {Cowley}},
  \bibinfo {author} {\bibfnamefont {P.~S.}\ \bibnamefont {Clegg}}, \bibinfo
  {author} {\bibfnamefont {R.}~\bibnamefont {Coldea}}, \bibinfo {author}
  {\bibfnamefont {C.~D.}\ \bibnamefont {Frost}}, \bibinfo {author}
  {\bibfnamefont {R.}~\bibnamefont {Liang}}, \bibinfo {author} {\bibfnamefont
  {D.}~\bibnamefont {Peets}}, \bibinfo {author} {\bibfnamefont
  {D.}~\bibnamefont {Bonn}}, \bibinfo {author} {\bibfnamefont {W.~N.}\
  \bibnamefont {Hardy}},\ and\ \bibinfo {author} {\bibfnamefont {R.~J.}\
  \bibnamefont {Birgeneau}},\ }\bibfield  {title} {\bibinfo {title} {{From
  incommensurate to dispersive spin-fluctuations: The high-energy inelastic
  spectrum in superconducting
  $\mathrm{Y}{\mathrm{Ba}}_{2}{\mathrm{Cu}}_{3}{\mathrm{O}}_{6.5}$}},\ }\href
  {https://doi.org/10.1103/PhysRevB.71.024522} {\bibfield  {journal} {\bibinfo
  {journal} {Phys. Rev. B}\ }\textbf {\bibinfo {volume} {71}},\ \bibinfo
  {pages} {024522} (\bibinfo {year} {2005})}\BibitemShut {NoStop}%
\bibitem [{\citenamefont {Comin}\ and\ \citenamefont
  {Damascelli}(2016)}]{doi:10.1146/annurev-conmatphys-031115-011401}%
  \BibitemOpen
  \bibfield  {author} {\bibinfo {author} {\bibfnamefont {R.}~\bibnamefont
  {Comin}}\ and\ \bibinfo {author} {\bibfnamefont {A.}~\bibnamefont
  {Damascelli}},\ }\bibfield  {title} {\bibinfo {title} {Resonant {X}-ray
  scattering studies of charge order in cuprates},\ }\href
  {https://doi.org/10.1146/annurev-conmatphys-031115-011401} {\bibfield
  {journal} {\bibinfo  {journal} {Annual Review of Condensed Matter Physics}\
  }\textbf {\bibinfo {volume} {7}},\ \bibinfo {pages} {369} (\bibinfo {year}
  {2016})}\BibitemShut {NoStop}%
\bibitem [{\citenamefont {Fujita}\ \emph {et~al.}(2012)\citenamefont {Fujita},
  \citenamefont {Hiraka}, \citenamefont {Matsuda}, \citenamefont {Matsuura},
  \citenamefont {M.~Tranquada}, \citenamefont {Wakimoto}, \citenamefont {Xu},\
  and\ \citenamefont {Yamada}}]{doi:10.1143/JPSJ.81.011007}%
  \BibitemOpen
  \bibfield  {author} {\bibinfo {author} {\bibfnamefont {M.}~\bibnamefont
  {Fujita}}, \bibinfo {author} {\bibfnamefont {H.}~\bibnamefont {Hiraka}},
  \bibinfo {author} {\bibfnamefont {M.}~\bibnamefont {Matsuda}}, \bibinfo
  {author} {\bibfnamefont {M.}~\bibnamefont {Matsuura}}, \bibinfo {author}
  {\bibfnamefont {J.}~\bibnamefont {M.~Tranquada}}, \bibinfo {author}
  {\bibfnamefont {S.}~\bibnamefont {Wakimoto}}, \bibinfo {author}
  {\bibfnamefont {G.}~\bibnamefont {Xu}},\ and\ \bibinfo {author}
  {\bibfnamefont {K.}~\bibnamefont {Yamada}},\ }\bibfield  {title} {\bibinfo
  {title} {Progress in neutron scattering studies of spin excitations in
  high-{Tc} cuprates},\ }\href {https://doi.org/10.1143/JPSJ.81.011007}
  {\bibfield  {journal} {\bibinfo  {journal} {Journal of the Physical Society
  of Japan}\ }\textbf {\bibinfo {volume} {81}},\ \bibinfo {pages} {011007}
  (\bibinfo {year} {2012})}\BibitemShut {NoStop}%
\bibitem [{\citenamefont {Motoyama}\ \emph {et~al.}(2007)\citenamefont
  {Motoyama}, \citenamefont {Yu}, \citenamefont {Vishik}, \citenamefont {Vajk},
  \citenamefont {Mang},\ and\ \citenamefont {Greven}}]{Motoyama2007}%
  \BibitemOpen
  \bibfield  {author} {\bibinfo {author} {\bibfnamefont {E.~M.}\ \bibnamefont
  {Motoyama}}, \bibinfo {author} {\bibfnamefont {G.}~\bibnamefont {Yu}},
  \bibinfo {author} {\bibfnamefont {I.~M.}\ \bibnamefont {Vishik}}, \bibinfo
  {author} {\bibfnamefont {O.~P.}\ \bibnamefont {Vajk}}, \bibinfo {author}
  {\bibfnamefont {P.~K.}\ \bibnamefont {Mang}},\ and\ \bibinfo {author}
  {\bibfnamefont {M.}~\bibnamefont {Greven}},\ }\bibfield  {title} {\bibinfo
  {title} {{Spin correlations in the electron-doped high-transition-temperature
  superconductor Nd$_{2-x}$Ce$_x$CuO$_{4\pm\delta}$}},\ }\href
  {https://doi.org/10.1038/nature05437} {\bibfield  {journal} {\bibinfo
  {journal} {Nature}\ }\textbf {\bibinfo {volume} {445}},\ \bibinfo {pages}
  {186} (\bibinfo {year} {2007})}\BibitemShut {NoStop}%
\bibitem [{\citenamefont {Yamada}\ \emph {et~al.}(2003)\citenamefont {Yamada},
  \citenamefont {Kurahashi}, \citenamefont {Uefuji}, \citenamefont {Fujita},
  \citenamefont {Park}, \citenamefont {Lee},\ and\ \citenamefont
  {Endoh}}]{PhysRevLett.90.137004}%
  \BibitemOpen
  \bibfield  {author} {\bibinfo {author} {\bibfnamefont {K.}~\bibnamefont
  {Yamada}}, \bibinfo {author} {\bibfnamefont {K.}~\bibnamefont {Kurahashi}},
  \bibinfo {author} {\bibfnamefont {T.}~\bibnamefont {Uefuji}}, \bibinfo
  {author} {\bibfnamefont {M.}~\bibnamefont {Fujita}}, \bibinfo {author}
  {\bibfnamefont {S.}~\bibnamefont {Park}}, \bibinfo {author} {\bibfnamefont
  {S.-H.}\ \bibnamefont {Lee}},\ and\ \bibinfo {author} {\bibfnamefont
  {Y.}~\bibnamefont {Endoh}},\ }\bibfield  {title} {\bibinfo {title}
  {Commensurate spin dynamics in the superconducting state of an electron-doped
  cuprate superconductor},\ }\href
  {https://doi.org/10.1103/PhysRevLett.90.137004} {\bibfield  {journal}
  {\bibinfo  {journal} {Phys. Rev. Lett.}\ }\textbf {\bibinfo {volume} {90}},\
  \bibinfo {pages} {137004} (\bibinfo {year} {2003})}\BibitemShut {NoStop}%
\bibitem [{\citenamefont {Huang}\ \emph {et~al.}(2019)\citenamefont {Huang},
  \citenamefont {Sheppard}, \citenamefont {Moritz},\ and\ \citenamefont
  {Devereaux}}]{doi:10.1126/science.aau7063}%
  \BibitemOpen
  \bibfield  {author} {\bibinfo {author} {\bibfnamefont {E.~W.}\ \bibnamefont
  {Huang}}, \bibinfo {author} {\bibfnamefont {R.}~\bibnamefont {Sheppard}},
  \bibinfo {author} {\bibfnamefont {B.}~\bibnamefont {Moritz}},\ and\ \bibinfo
  {author} {\bibfnamefont {T.~P.}\ \bibnamefont {Devereaux}},\ }\bibfield
  {title} {\bibinfo {title} {Strange metallicity in the doped
  \uppercase{H}ubbard model},\ }\href {https://doi.org/10.1126/science.aau7063}
  {\bibfield  {journal} {\bibinfo  {journal} {Science}\ }\textbf {\bibinfo
  {volume} {366}},\ \bibinfo {pages} {987} (\bibinfo {year}
  {2019})}\BibitemShut {NoStop}%
\bibitem [{\citenamefont {Shastry}\ and\ \citenamefont
  {Mai}(2018)}]{Shastry_2018}%
  \BibitemOpen
  \bibfield  {author} {\bibinfo {author} {\bibfnamefont {B.~S.}\ \bibnamefont
  {Shastry}}\ and\ \bibinfo {author} {\bibfnamefont {P.}~\bibnamefont {Mai}},\
  }\bibfield  {title} {\bibinfo {title} {Extremely correlated \uppercase{F}ermi
  liquid theory of the t-\uppercase{J} model in 2 dimensions: low energy
  properties},\ }\href {https://doi.org/10.1088/1367-2630/aa9b74} {\bibfield
  {journal} {\bibinfo  {journal} {New Journal of Physics}\ }\textbf {\bibinfo
  {volume} {20}},\ \bibinfo {pages} {013027} (\bibinfo {year}
  {2018})}\BibitemShut {NoStop}%
\bibitem [{\citenamefont {Zemlji\ifmmode~\check{c}\else \v{c}\fi{}}\ and\
  \citenamefont {Prelov\ifmmode~\check{s}\else
  \v{s}\fi{}ek}(2005)}]{PhysRevB.72.075108}%
  \BibitemOpen
  \bibfield  {author} {\bibinfo {author} {\bibfnamefont {M.~M.}\ \bibnamefont
  {Zemlji\ifmmode~\check{c}\else \v{c}\fi{}}}\ and\ \bibinfo {author}
  {\bibfnamefont {P.}~\bibnamefont {Prelov\ifmmode~\check{s}\else
  \v{s}\fi{}ek}},\ }\bibfield  {title} {\bibinfo {title} {Resistivity and
  optical conductivity of cuprates within the
  $t\text{\ensuremath{-}}\uppercase{J}$ model},\ }\href
  {https://doi.org/10.1103/PhysRevB.72.075108} {\bibfield  {journal} {\bibinfo
  {journal} {Phys. Rev. B}\ }\textbf {\bibinfo {volume} {72}},\ \bibinfo
  {pages} {075108} (\bibinfo {year} {2005})}\BibitemShut {NoStop}%
\bibitem [{\citenamefont {Li}\ \emph {et~al.}(2016)\citenamefont {Li},
  \citenamefont {Tabis}, \citenamefont {Yu}, \citenamefont {Bari\ifmmode
  \check{s}\else \v{s}\fi{}i\ifmmode~\acute{c}\else \'{c}\fi{}},\ and\
  \citenamefont {Greven}}]{PhysRevLett.117.197001}%
  \BibitemOpen
  \bibfield  {author} {\bibinfo {author} {\bibfnamefont {Y.}~\bibnamefont
  {Li}}, \bibinfo {author} {\bibfnamefont {W.}~\bibnamefont {Tabis}}, \bibinfo
  {author} {\bibfnamefont {G.}~\bibnamefont {Yu}}, \bibinfo {author}
  {\bibfnamefont {N.}~\bibnamefont {Bari\ifmmode \check{s}\else
  \v{s}\fi{}i\ifmmode~\acute{c}\else \'{c}\fi{}}},\ and\ \bibinfo {author}
  {\bibfnamefont {M.}~\bibnamefont {Greven}},\ }\bibfield  {title} {\bibinfo
  {title} {Hidden \uppercase{F}ermi-liquid charge transport in the
  antiferromagnetic phase of the electron-doped cuprate superconductors},\
  }\href {https://doi.org/10.1103/PhysRevLett.117.197001} {\bibfield  {journal}
  {\bibinfo  {journal} {Phys. Rev. Lett.}\ }\textbf {\bibinfo {volume} {117}},\
  \bibinfo {pages} {197001} (\bibinfo {year} {2016})}\BibitemShut {NoStop}%
\bibitem [{\citenamefont {Onose}\ \emph {et~al.}(2004)\citenamefont {Onose},
  \citenamefont {Taguchi}, \citenamefont {Ishizaka},\ and\ \citenamefont
  {Tokura}}]{PhysRevB.69.024504}%
  \BibitemOpen
  \bibfield  {author} {\bibinfo {author} {\bibfnamefont {Y.}~\bibnamefont
  {Onose}}, \bibinfo {author} {\bibfnamefont {Y.}~\bibnamefont {Taguchi}},
  \bibinfo {author} {\bibfnamefont {K.}~\bibnamefont {Ishizaka}},\ and\
  \bibinfo {author} {\bibfnamefont {Y.}~\bibnamefont {Tokura}},\ }\bibfield
  {title} {\bibinfo {title} {Charge dynamics in underdoped
  ${\mathrm{\uppercase{n}d}}_{2\ensuremath{-}x}{\mathrm{\uppercase{c}e}}_{x}{\mathrm{\uppercase{c}u\uppercase{o}}}_{4}:$
  pseudogap and related phenomena},\ }\href
  {https://doi.org/10.1103/PhysRevB.69.024504} {\bibfield  {journal} {\bibinfo
  {journal} {Phys. Rev. B}\ }\textbf {\bibinfo {volume} {69}},\ \bibinfo
  {pages} {024504} (\bibinfo {year} {2004})}\BibitemShut {NoStop}%
\bibitem [{\citenamefont {Martin}\ \emph {et~al.}(1988)\citenamefont {Martin},
  \citenamefont {Fiory}, \citenamefont {Fleming}, \citenamefont {Schneemeyer},\
  and\ \citenamefont {Waszczak}}]{PhysRevLett.60.2194}%
  \BibitemOpen
  \bibfield  {author} {\bibinfo {author} {\bibfnamefont {S.}~\bibnamefont
  {Martin}}, \bibinfo {author} {\bibfnamefont {A.~T.}\ \bibnamefont {Fiory}},
  \bibinfo {author} {\bibfnamefont {R.~M.}\ \bibnamefont {Fleming}}, \bibinfo
  {author} {\bibfnamefont {L.~F.}\ \bibnamefont {Schneemeyer}},\ and\ \bibinfo
  {author} {\bibfnamefont {J.~V.}\ \bibnamefont {Waszczak}},\ }\bibfield
  {title} {\bibinfo {title} {Temperature dependence of the resistivity tensor
  in superconducting
  ${\mathrm{\uppercase{b}i}}_{2}$${\mathrm{\uppercase{s}r}}_{2.2}$${\mathrm{\uppercase{c}a}}_{0.8}$
  ${\mathrm{\uppercase{c}u}}_{2}$${\mathrm{\uppercase{o}}}_{8}$ crystals},\
  }\href {https://doi.org/10.1103/PhysRevLett.60.2194} {\bibfield  {journal}
  {\bibinfo  {journal} {Phys. Rev. Lett.}\ }\textbf {\bibinfo {volume} {60}},\
  \bibinfo {pages} {2194} (\bibinfo {year} {1988})}\BibitemShut {NoStop}%
\bibitem [{\citenamefont {Takagi}\ \emph {et~al.}(1989)\citenamefont {Takagi},
  \citenamefont {Ido}, \citenamefont {Ishibashi}, \citenamefont {Uota},
  \citenamefont {Uchida},\ and\ \citenamefont {Tokura}}]{PhysRevB.40.2254}%
  \BibitemOpen
  \bibfield  {author} {\bibinfo {author} {\bibfnamefont {H.}~\bibnamefont
  {Takagi}}, \bibinfo {author} {\bibfnamefont {T.}~\bibnamefont {Ido}},
  \bibinfo {author} {\bibfnamefont {S.}~\bibnamefont {Ishibashi}}, \bibinfo
  {author} {\bibfnamefont {M.}~\bibnamefont {Uota}}, \bibinfo {author}
  {\bibfnamefont {S.}~\bibnamefont {Uchida}},\ and\ \bibinfo {author}
  {\bibfnamefont {Y.}~\bibnamefont {Tokura}},\ }\bibfield  {title} {\bibinfo
  {title} {Superconductor-to-nonsuperconductor transition in
  (${\mathrm{\uppercase{l}a}}_{1\mathrm{\ensuremath{-}}\mathrm{x}}$${\mathrm{\uppercase{s}r}}_{\mathrm{x}}$${)}_{2}$${\mathrm{\uppercase{c}u\uppercase{o}}}_{4}$
  as investigated by transport and magnetic measurements},\ }\href
  {https://doi.org/10.1103/PhysRevB.40.2254} {\bibfield  {journal} {\bibinfo
  {journal} {Phys. Rev. B}\ }\textbf {\bibinfo {volume} {40}},\ \bibinfo
  {pages} {2254} (\bibinfo {year} {1989})}\BibitemShut {NoStop}%
\bibitem [{\citenamefont {Takagi}\ \emph {et~al.}(1992)\citenamefont {Takagi},
  \citenamefont {Batlogg}, \citenamefont {Kao}, \citenamefont {Kwo},
  \citenamefont {Cava}, \citenamefont {Krajewski},\ and\ \citenamefont
  {Peck}}]{PhysRevLett.69.2975}%
  \BibitemOpen
  \bibfield  {author} {\bibinfo {author} {\bibfnamefont {H.}~\bibnamefont
  {Takagi}}, \bibinfo {author} {\bibfnamefont {B.}~\bibnamefont {Batlogg}},
  \bibinfo {author} {\bibfnamefont {H.~L.}\ \bibnamefont {Kao}}, \bibinfo
  {author} {\bibfnamefont {J.}~\bibnamefont {Kwo}}, \bibinfo {author}
  {\bibfnamefont {R.~J.}\ \bibnamefont {Cava}}, \bibinfo {author}
  {\bibfnamefont {J.~J.}\ \bibnamefont {Krajewski}},\ and\ \bibinfo {author}
  {\bibfnamefont {W.~F.}\ \bibnamefont {Peck}},\ }\bibfield  {title} {\bibinfo
  {title} {Systematic evolution of temperature-dependent resistivity in
  ${\mathrm{\uppercase{l}a}}_{2\mathrm{\ensuremath{-}}\mathit{x}}$${\mathrm{\uppercase{s}r}}_{\mathit{x}}$${\mathrm{\uppercase{c}u\uppercase{o}}}_{4}$},\
  }\href {https://doi.org/10.1103/PhysRevLett.69.2975} {\bibfield  {journal}
  {\bibinfo  {journal} {Phys. Rev. Lett.}\ }\textbf {\bibinfo {volume} {69}},\
  \bibinfo {pages} {2975} (\bibinfo {year} {1992})}\BibitemShut {NoStop}%
\bibitem [{\citenamefont {Ando}\ \emph {et~al.}(2004)\citenamefont {Ando},
  \citenamefont {Komiya}, \citenamefont {Segawa}, \citenamefont {Ono},\ and\
  \citenamefont {Kurita}}]{PhysRevLett.93.267001}%
  \BibitemOpen
  \bibfield  {author} {\bibinfo {author} {\bibfnamefont {Y.}~\bibnamefont
  {Ando}}, \bibinfo {author} {\bibfnamefont {S.}~\bibnamefont {Komiya}},
  \bibinfo {author} {\bibfnamefont {K.}~\bibnamefont {Segawa}}, \bibinfo
  {author} {\bibfnamefont {S.}~\bibnamefont {Ono}},\ and\ \bibinfo {author}
  {\bibfnamefont {Y.}~\bibnamefont {Kurita}},\ }\bibfield  {title} {\bibinfo
  {title} {Electronic phase diagram of high-${\uppercase{t}}_{c}$ cuprate
  superconductors from a mapping of the in-plane resistivity curvature},\
  }\href {https://doi.org/10.1103/PhysRevLett.93.267001} {\bibfield  {journal}
  {\bibinfo  {journal} {Phys. Rev. Lett.}\ }\textbf {\bibinfo {volume} {93}},\
  \bibinfo {pages} {267001} (\bibinfo {year} {2004})}\BibitemShut {NoStop}%
\bibitem [{\citenamefont {Daou}\ \emph {et~al.}(2009)\citenamefont {Daou},
  \citenamefont {Doiron-Leyraud}, \citenamefont {LeBoeuf}, \citenamefont {Li},
  \citenamefont {Lalibert{\'e}}, \citenamefont {Cyr-Choini{\`e}re},
  \citenamefont {Jo}, \citenamefont {Balicas}, \citenamefont {Yan},
  \citenamefont {Zhou}, \citenamefont {Goodenough},\ and\ \citenamefont
  {Taillefer}}]{Daou2009}%
  \BibitemOpen
  \bibfield  {author} {\bibinfo {author} {\bibfnamefont {R.}~\bibnamefont
  {Daou}}, \bibinfo {author} {\bibfnamefont {N.}~\bibnamefont
  {Doiron-Leyraud}}, \bibinfo {author} {\bibfnamefont {D.}~\bibnamefont
  {LeBoeuf}}, \bibinfo {author} {\bibfnamefont {S.~Y.}\ \bibnamefont {Li}},
  \bibinfo {author} {\bibfnamefont {F.}~\bibnamefont {Lalibert{\'e}}}, \bibinfo
  {author} {\bibfnamefont {O.}~\bibnamefont {Cyr-Choini{\`e}re}}, \bibinfo
  {author} {\bibfnamefont {Y.~J.}\ \bibnamefont {Jo}}, \bibinfo {author}
  {\bibfnamefont {L.}~\bibnamefont {Balicas}}, \bibinfo {author} {\bibfnamefont
  {J.-Q.}\ \bibnamefont {Yan}}, \bibinfo {author} {\bibfnamefont {J.-S.}\
  \bibnamefont {Zhou}}, \bibinfo {author} {\bibfnamefont {J.~B.}\ \bibnamefont
  {Goodenough}},\ and\ \bibinfo {author} {\bibfnamefont {L.}~\bibnamefont
  {Taillefer}},\ }\bibfield  {title} {\bibinfo {title} {Linear temperature
  dependence of resistivity and change in the \uppercase{F}ermi surface at the
  pseudogap critical point of a high-\uppercase{T}c superconductor},\ }\href
  {https://doi.org/10.1038/nphys1109} {\bibfield  {journal} {\bibinfo
  {journal} {Nature Physics}\ }\textbf {\bibinfo {volume} {5}},\ \bibinfo
  {pages} {31} (\bibinfo {year} {2009})}\BibitemShut {NoStop}%
\bibitem [{\citenamefont {Cooper}\ \emph {et~al.}(2009)\citenamefont {Cooper},
  \citenamefont {Wang}, \citenamefont {Vignolle}, \citenamefont {Lipscombe},
  \citenamefont {Hayden}, \citenamefont {Tanabe}, \citenamefont {Adachi},
  \citenamefont {Koike}, \citenamefont {Nohara}, \citenamefont {Takagi},
  \citenamefont {Proust},\ and\ \citenamefont
  {Hussey}}]{doi:10.1126/science.1165015}%
  \BibitemOpen
  \bibfield  {author} {\bibinfo {author} {\bibfnamefont {R.~A.}\ \bibnamefont
  {Cooper}}, \bibinfo {author} {\bibfnamefont {Y.}~\bibnamefont {Wang}},
  \bibinfo {author} {\bibfnamefont {B.}~\bibnamefont {Vignolle}}, \bibinfo
  {author} {\bibfnamefont {O.~J.}\ \bibnamefont {Lipscombe}}, \bibinfo {author}
  {\bibfnamefont {S.~M.}\ \bibnamefont {Hayden}}, \bibinfo {author}
  {\bibfnamefont {Y.}~\bibnamefont {Tanabe}}, \bibinfo {author} {\bibfnamefont
  {T.}~\bibnamefont {Adachi}}, \bibinfo {author} {\bibfnamefont
  {Y.}~\bibnamefont {Koike}}, \bibinfo {author} {\bibfnamefont
  {M.}~\bibnamefont {Nohara}}, \bibinfo {author} {\bibfnamefont
  {H.}~\bibnamefont {Takagi}}, \bibinfo {author} {\bibfnamefont
  {C.}~\bibnamefont {Proust}},\ and\ \bibinfo {author} {\bibfnamefont {N.~E.}\
  \bibnamefont {Hussey}},\ }\bibfield  {title} {\bibinfo {title} {Anomalous
  criticality in the electrical resistivity of
  \uppercase{L}a$_{2-x}$\uppercase{S}r$_x$\uppercase{C}u\uppercase{O}$_4$},\
  }\href {https://doi.org/10.1126/science.1165015} {\bibfield  {journal}
  {\bibinfo  {journal} {Science}\ }\textbf {\bibinfo {volume} {323}},\ \bibinfo
  {pages} {603} (\bibinfo {year} {2009})}\BibitemShut {NoStop}%
\bibitem [{\citenamefont {Barišić}\ \emph {et~al.}(2013)\citenamefont
  {Barišić}, \citenamefont {Chan}, \citenamefont {Li}, \citenamefont {Yu},
  \citenamefont {Zhao}, \citenamefont {Dressel}, \citenamefont {Smontara},\
  and\ \citenamefont {Greven}}]{doi:10.1073/pnas.1301989110}%
  \BibitemOpen
  \bibfield  {author} {\bibinfo {author} {\bibfnamefont {N.}~\bibnamefont
  {Barišić}}, \bibinfo {author} {\bibfnamefont {M.~K.}\ \bibnamefont {Chan}},
  \bibinfo {author} {\bibfnamefont {Y.}~\bibnamefont {Li}}, \bibinfo {author}
  {\bibfnamefont {G.}~\bibnamefont {Yu}}, \bibinfo {author} {\bibfnamefont
  {X.}~\bibnamefont {Zhao}}, \bibinfo {author} {\bibfnamefont {M.}~\bibnamefont
  {Dressel}}, \bibinfo {author} {\bibfnamefont {A.}~\bibnamefont {Smontara}},\
  and\ \bibinfo {author} {\bibfnamefont {M.}~\bibnamefont {Greven}},\
  }\bibfield  {title} {\bibinfo {title} {Universal sheet resistance and revised
  phase diagram of the cuprate high-temperature superconductors},\ }\href
  {https://doi.org/10.1073/pnas.1301989110} {\bibfield  {journal} {\bibinfo
  {journal} {Proceedings of the National Academy of Sciences}\ }\textbf
  {\bibinfo {volume} {110}},\ \bibinfo {pages} {12235} (\bibinfo {year}
  {2013})}\BibitemShut {NoStop}%
\bibitem [{\citenamefont {Chan}\ \emph {et~al.}(2014)\citenamefont {Chan},
  \citenamefont {Veit}, \citenamefont {Dorow}, \citenamefont {Ge},
  \citenamefont {Li}, \citenamefont {Tabis}, \citenamefont {Tang},
  \citenamefont {Zhao}, \citenamefont {Bari\ifmmode \check{s}\else
  \v{s}\fi{}i\ifmmode~\acute{c}\else \'{c}\fi{}},\ and\ \citenamefont
  {Greven}}]{PhysRevLett.113.177005}%
  \BibitemOpen
  \bibfield  {author} {\bibinfo {author} {\bibfnamefont {M.~K.}\ \bibnamefont
  {Chan}}, \bibinfo {author} {\bibfnamefont {M.~J.}\ \bibnamefont {Veit}},
  \bibinfo {author} {\bibfnamefont {C.~J.}\ \bibnamefont {Dorow}}, \bibinfo
  {author} {\bibfnamefont {Y.}~\bibnamefont {Ge}}, \bibinfo {author}
  {\bibfnamefont {Y.}~\bibnamefont {Li}}, \bibinfo {author} {\bibfnamefont
  {W.}~\bibnamefont {Tabis}}, \bibinfo {author} {\bibfnamefont
  {Y.}~\bibnamefont {Tang}}, \bibinfo {author} {\bibfnamefont {X.}~\bibnamefont
  {Zhao}}, \bibinfo {author} {\bibfnamefont {N.}~\bibnamefont {Bari\ifmmode
  \check{s}\else \v{s}\fi{}i\ifmmode~\acute{c}\else \'{c}\fi{}}},\ and\
  \bibinfo {author} {\bibfnamefont {M.}~\bibnamefont {Greven}},\ }\bibfield
  {title} {\bibinfo {title} {In-plane magnetoresistance obeys
  \uppercase{K}ohler's rule in the pseudogap phase of cuprate
  superconductors},\ }\href {https://doi.org/10.1103/PhysRevLett.113.177005}
  {\bibfield  {journal} {\bibinfo  {journal} {Phys. Rev. Lett.}\ }\textbf
  {\bibinfo {volume} {113}},\ \bibinfo {pages} {177005} (\bibinfo {year}
  {2014})}\BibitemShut {NoStop}%
\bibitem [{\citenamefont {Collignon}\ \emph {et~al.}(2017)\citenamefont
  {Collignon}, \citenamefont {Badoux}, \citenamefont {Afshar}, \citenamefont
  {Michon}, \citenamefont {Lalibert\'e}, \citenamefont {Cyr-Choini\`ere},
  \citenamefont {Zhou}, \citenamefont {Licciardello}, \citenamefont {Wiedmann},
  \citenamefont {Doiron-Leyraud},\ and\ \citenamefont
  {Taillefer}}]{PhysRevB.95.224517}%
  \BibitemOpen
  \bibfield  {author} {\bibinfo {author} {\bibfnamefont {C.}~\bibnamefont
  {Collignon}}, \bibinfo {author} {\bibfnamefont {S.}~\bibnamefont {Badoux}},
  \bibinfo {author} {\bibfnamefont {S.~A.~A.}\ \bibnamefont {Afshar}}, \bibinfo
  {author} {\bibfnamefont {B.}~\bibnamefont {Michon}}, \bibinfo {author}
  {\bibfnamefont {F.}~\bibnamefont {Lalibert\'e}}, \bibinfo {author}
  {\bibfnamefont {O.}~\bibnamefont {Cyr-Choini\`ere}}, \bibinfo {author}
  {\bibfnamefont {J.-S.}\ \bibnamefont {Zhou}}, \bibinfo {author}
  {\bibfnamefont {S.}~\bibnamefont {Licciardello}}, \bibinfo {author}
  {\bibfnamefont {S.}~\bibnamefont {Wiedmann}}, \bibinfo {author}
  {\bibfnamefont {N.}~\bibnamefont {Doiron-Leyraud}},\ and\ \bibinfo {author}
  {\bibfnamefont {L.}~\bibnamefont {Taillefer}},\ }\bibfield  {title} {\bibinfo
  {title} {Fermi-surface transformation across the pseudogap critical point of
  the cuprate superconductor
  ${\mathrm{\uppercase{l}a}}_{1.6\ensuremath{-}x}{\mathrm{\uppercase{n}d}}_{0.4}{\mathrm{\uppercase{s}r}}_{x}{\mathrm{\uppercase{c}u\uppercase{o}}}_{4}$},\
  }\href {https://doi.org/10.1103/PhysRevB.95.224517} {\bibfield  {journal}
  {\bibinfo  {journal} {Phys. Rev. B}\ }\textbf {\bibinfo {volume} {95}},\
  \bibinfo {pages} {224517} (\bibinfo {year} {2017})}\BibitemShut {NoStop}%
\bibitem [{\citenamefont {Barišić}\ \emph {et~al.}(2019)\citenamefont
  {Barišić}, \citenamefont {Chan}, \citenamefont {Veit}, \citenamefont
  {Dorow}, \citenamefont {Ge}, \citenamefont {Li}, \citenamefont {Tabis},
  \citenamefont {Tang}, \citenamefont {Yu}, \citenamefont {Zhao},\ and\
  \citenamefont {Greven}}]{Barisic_2019}%
  \BibitemOpen
  \bibfield  {author} {\bibinfo {author} {\bibfnamefont {N.}~\bibnamefont
  {Barišić}}, \bibinfo {author} {\bibfnamefont {M.~K.}\ \bibnamefont {Chan}},
  \bibinfo {author} {\bibfnamefont {M.~J.}\ \bibnamefont {Veit}}, \bibinfo
  {author} {\bibfnamefont {C.~J.}\ \bibnamefont {Dorow}}, \bibinfo {author}
  {\bibfnamefont {Y.}~\bibnamefont {Ge}}, \bibinfo {author} {\bibfnamefont
  {Y.}~\bibnamefont {Li}}, \bibinfo {author} {\bibfnamefont {W.}~\bibnamefont
  {Tabis}}, \bibinfo {author} {\bibfnamefont {Y.}~\bibnamefont {Tang}},
  \bibinfo {author} {\bibfnamefont {G.}~\bibnamefont {Yu}}, \bibinfo {author}
  {\bibfnamefont {X.}~\bibnamefont {Zhao}},\ and\ \bibinfo {author}
  {\bibfnamefont {M.}~\bibnamefont {Greven}},\ }\bibfield  {title} {\bibinfo
  {title} {Evidence for a universal \uppercase{F}ermi-liquid scattering rate
  throughout the phase diagram of the copper-oxide superconductors},\ }\href
  {https://doi.org/10.1088/1367-2630/ab4d0f} {\bibfield  {journal} {\bibinfo
  {journal} {New Journal of Physics}\ }\textbf {\bibinfo {volume} {21}},\
  \bibinfo {pages} {113007} (\bibinfo {year} {2019})}\BibitemShut {NoStop}%
\bibitem [{\citenamefont {Coleman}(2015)}]{Coleman_2015}%
  \BibitemOpen
  \bibfield  {author} {\bibinfo {author} {\bibfnamefont {P.}~\bibnamefont
  {Coleman}},\ }\href@noop {} {\emph {\bibinfo {title} {Introduction to
  Many-Body Physics}}}\ (\bibinfo  {publisher} {Cambridge University Press},\
  \bibinfo {year} {2015})\BibitemShut {NoStop}%
\bibitem [{\citenamefont {Lee}(1989)}]{PhysRevLett.63.680}%
  \BibitemOpen
  \bibfield  {author} {\bibinfo {author} {\bibfnamefont {P.~A.}\ \bibnamefont
  {Lee}},\ }\bibfield  {title} {\bibinfo {title} {{Gauge field, Aharonov-Bohm
  flux, and high-${T}_{c}$ superconductivity}},\ }\href
  {https://doi.org/10.1103/PhysRevLett.63.680} {\bibfield  {journal} {\bibinfo
  {journal} {Phys. Rev. Lett.}\ }\textbf {\bibinfo {volume} {63}},\ \bibinfo
  {pages} {680} (\bibinfo {year} {1989})}\BibitemShut {NoStop}%
\bibitem [{\citenamefont {Martinez}\ and\ \citenamefont
  {Horsch}(1991)}]{PhysRevB.44.317}%
  \BibitemOpen
  \bibfield  {author} {\bibinfo {author} {\bibfnamefont {G.}~\bibnamefont
  {Martinez}}\ and\ \bibinfo {author} {\bibfnamefont {P.}~\bibnamefont
  {Horsch}},\ }\bibfield  {title} {\bibinfo {title} {Spin polarons in the
  $t$-{$J$} model},\ }\href {https://doi.org/10.1103/PhysRevB.44.317}
  {\bibfield  {journal} {\bibinfo  {journal} {Phys. Rev. B}\ }\textbf {\bibinfo
  {volume} {44}},\ \bibinfo {pages} {317} (\bibinfo {year} {1991})}\BibitemShut
  {NoStop}%
\bibitem [{\citenamefont {Manousakis}(2007)}]{PhysRevB.75.035106}%
  \BibitemOpen
  \bibfield  {author} {\bibinfo {author} {\bibfnamefont {E.}~\bibnamefont
  {Manousakis}},\ }\bibfield  {title} {\bibinfo {title} {String excitations of
  a hole in a quantum antiferromagnet and photoelectron spectroscopy},\ }\href
  {https://doi.org/10.1103/PhysRevB.75.035106} {\bibfield  {journal} {\bibinfo
  {journal} {Phys. Rev. B}\ }\textbf {\bibinfo {volume} {75}},\ \bibinfo
  {pages} {035106} (\bibinfo {year} {2007})}\BibitemShut {NoStop}%
\bibitem [{\citenamefont {Jia}\ \emph {et~al.}(2014)\citenamefont {Jia},
  \citenamefont {Nowadnick}, \citenamefont {Wohlfeld}, \citenamefont {Kung},
  \citenamefont {Chen}, \citenamefont {Johnston}, \citenamefont {Tohyama},
  \citenamefont {Moritz},\ and\ \citenamefont {Devereaux}}]{Jia2014}%
  \BibitemOpen
  \bibfield  {author} {\bibinfo {author} {\bibfnamefont {C.~J.}\ \bibnamefont
  {Jia}}, \bibinfo {author} {\bibfnamefont {E.~A.}\ \bibnamefont {Nowadnick}},
  \bibinfo {author} {\bibfnamefont {K.}~\bibnamefont {Wohlfeld}}, \bibinfo
  {author} {\bibfnamefont {Y.~F.}\ \bibnamefont {Kung}}, \bibinfo {author}
  {\bibfnamefont {C.-C.}\ \bibnamefont {Chen}}, \bibinfo {author}
  {\bibfnamefont {S.}~\bibnamefont {Johnston}}, \bibinfo {author}
  {\bibfnamefont {T.}~\bibnamefont {Tohyama}}, \bibinfo {author} {\bibfnamefont
  {B.}~\bibnamefont {Moritz}},\ and\ \bibinfo {author} {\bibfnamefont {T.~P.}\
  \bibnamefont {Devereaux}},\ }\bibfield  {title} {\bibinfo {title} {Persistent
  spin excitations in doped antiferromagnets revealed by resonant inelastic
  light scattering},\ }\href {https://doi.org/10.1038/ncomms4314} {\bibfield
  {journal} {\bibinfo  {journal} {Nature Communications}\ }\textbf {\bibinfo
  {volume} {5}},\ \bibinfo {pages} {3314} (\bibinfo {year} {2014})}\BibitemShut
  {NoStop}%
\bibitem [{\citenamefont {Verret}\ \emph {et~al.}(2022)\citenamefont {Verret},
  \citenamefont {Foley}, \citenamefont {S\'en\'echal}, \citenamefont
  {Tremblay},\ and\ \citenamefont {Charlebois}}]{PhysRevB.105.035117}%
  \BibitemOpen
  \bibfield  {author} {\bibinfo {author} {\bibfnamefont {S.}~\bibnamefont
  {Verret}}, \bibinfo {author} {\bibfnamefont {A.}~\bibnamefont {Foley}},
  \bibinfo {author} {\bibfnamefont {D.}~\bibnamefont {S\'en\'echal}}, \bibinfo
  {author} {\bibfnamefont {A.-M.~S.}\ \bibnamefont {Tremblay}},\ and\ \bibinfo
  {author} {\bibfnamefont {M.}~\bibnamefont {Charlebois}},\ }\bibfield  {title}
  {\bibinfo {title} {{Fermi arcs versus hole pockets: Periodization of a
  cellular two-band model}},\ }\href
  {https://doi.org/10.1103/PhysRevB.105.035117} {\bibfield  {journal} {\bibinfo
   {journal} {Phys. Rev. B}\ }\textbf {\bibinfo {volume} {105}},\ \bibinfo
  {pages} {035117} (\bibinfo {year} {2022})}\BibitemShut {NoStop}%
\bibitem [{\citenamefont {Senthil}\ and\ \citenamefont
  {Lee}(2009)}]{PhysRevB.79.245116}%
  \BibitemOpen
  \bibfield  {author} {\bibinfo {author} {\bibfnamefont {T.}~\bibnamefont
  {Senthil}}\ and\ \bibinfo {author} {\bibfnamefont {P.~A.}\ \bibnamefont
  {Lee}},\ }\bibfield  {title} {\bibinfo {title} {Synthesis of the
  phenomenology of the underdoped cuprates},\ }\href
  {https://doi.org/10.1103/PhysRevB.79.245116} {\bibfield  {journal} {\bibinfo
  {journal} {Phys. Rev. B}\ }\textbf {\bibinfo {volume} {79}},\ \bibinfo
  {pages} {245116} (\bibinfo {year} {2009})}\BibitemShut {NoStop}%
\bibitem [{\citenamefont {Micklitz}\ and\ \citenamefont
  {Norman}(2009)}]{PhysRevB.80.220513}%
  \BibitemOpen
  \bibfield  {author} {\bibinfo {author} {\bibfnamefont {T.}~\bibnamefont
  {Micklitz}}\ and\ \bibinfo {author} {\bibfnamefont {M.~R.}\ \bibnamefont
  {Norman}},\ }\bibfield  {title} {\bibinfo {title} {Nature of spectral gaps
  due to pair formation in superconductors},\ }\href
  {https://doi.org/10.1103/PhysRevB.80.220513} {\bibfield  {journal} {\bibinfo
  {journal} {Phys. Rev. B}\ }\textbf {\bibinfo {volume} {80}},\ \bibinfo
  {pages} {220513} (\bibinfo {year} {2009})}\BibitemShut {NoStop}%
\bibitem [{\citenamefont {Zhang}\ and\ \citenamefont
  {Weng}(2023)}]{PhysRevB.108.235156}%
  \BibitemOpen
  \bibfield  {author} {\bibinfo {author} {\bibfnamefont {J.-X.}\ \bibnamefont
  {Zhang}}\ and\ \bibinfo {author} {\bibfnamefont {Z.-Y.}\ \bibnamefont
  {Weng}},\ }\bibfield  {title} {\bibinfo {title} {{Crossover from Fermi arc to
  full \uppercase{F}ermi surface}},\ }\href
  {https://doi.org/10.1103/PhysRevB.108.235156} {\bibfield  {journal} {\bibinfo
   {journal} {Phys. Rev. B}\ }\textbf {\bibinfo {volume} {108}},\ \bibinfo
  {pages} {235156} (\bibinfo {year} {2023})}\BibitemShut {NoStop}%
\bibitem [{\citenamefont {Kunisada}\ \emph {et~al.}(2020)\citenamefont
  {Kunisada}, \citenamefont {Isono}, \citenamefont {Kohama}, \citenamefont
  {Sakai}, \citenamefont {Bareille}, \citenamefont {Sakuragi}, \citenamefont
  {Noguchi}, \citenamefont {Kurokawa}, \citenamefont {Kuroda}, \citenamefont
  {Ishida}, \citenamefont {Adachi}, \citenamefont {Sekine}, \citenamefont
  {Kim}, \citenamefont {Cacho}, \citenamefont {Shin}, \citenamefont {Tohyama},
  \citenamefont {Tokiwa},\ and\ \citenamefont
  {Kondo}}]{doi:10.1126/science.aay7311}%
  \BibitemOpen
  \bibfield  {author} {\bibinfo {author} {\bibfnamefont {S.}~\bibnamefont
  {Kunisada}}, \bibinfo {author} {\bibfnamefont {S.}~\bibnamefont {Isono}},
  \bibinfo {author} {\bibfnamefont {Y.}~\bibnamefont {Kohama}}, \bibinfo
  {author} {\bibfnamefont {S.}~\bibnamefont {Sakai}}, \bibinfo {author}
  {\bibfnamefont {C.}~\bibnamefont {Bareille}}, \bibinfo {author}
  {\bibfnamefont {S.}~\bibnamefont {Sakuragi}}, \bibinfo {author}
  {\bibfnamefont {R.}~\bibnamefont {Noguchi}}, \bibinfo {author} {\bibfnamefont
  {K.}~\bibnamefont {Kurokawa}}, \bibinfo {author} {\bibfnamefont
  {K.}~\bibnamefont {Kuroda}}, \bibinfo {author} {\bibfnamefont
  {Y.}~\bibnamefont {Ishida}}, \bibinfo {author} {\bibfnamefont
  {S.}~\bibnamefont {Adachi}}, \bibinfo {author} {\bibfnamefont
  {R.}~\bibnamefont {Sekine}}, \bibinfo {author} {\bibfnamefont {T.~K.}\
  \bibnamefont {Kim}}, \bibinfo {author} {\bibfnamefont {C.}~\bibnamefont
  {Cacho}}, \bibinfo {author} {\bibfnamefont {S.}~\bibnamefont {Shin}},
  \bibinfo {author} {\bibfnamefont {T.}~\bibnamefont {Tohyama}}, \bibinfo
  {author} {\bibfnamefont {K.}~\bibnamefont {Tokiwa}},\ and\ \bibinfo {author}
  {\bibfnamefont {T.}~\bibnamefont {Kondo}},\ }\bibfield  {title} {\bibinfo
  {title} {Observation of small \uppercase{F}ermi pockets protected by clean
  \uppercase{C}u\uppercase{O}$_2$ sheets of a high-$\uppercase{T}_c$
  superconductor},\ }\href {https://doi.org/10.1126/science.aay7311} {\bibfield
   {journal} {\bibinfo  {journal} {Science}\ }\textbf {\bibinfo {volume}
  {369}},\ \bibinfo {pages} {833} (\bibinfo {year} {2020})}\BibitemShut
  {NoStop}%
\bibitem [{\citenamefont {Bulut}\ \emph {et~al.}(1990)\citenamefont {Bulut},
  \citenamefont {Hone}, \citenamefont {Scalapino},\ and\ \citenamefont
  {Bickers}}]{PhysRevB.41.1797}%
  \BibitemOpen
  \bibfield  {author} {\bibinfo {author} {\bibfnamefont {N.}~\bibnamefont
  {Bulut}}, \bibinfo {author} {\bibfnamefont {D.~W.}\ \bibnamefont {Hone}},
  \bibinfo {author} {\bibfnamefont {D.~J.}\ \bibnamefont {Scalapino}},\ and\
  \bibinfo {author} {\bibfnamefont {N.~E.}\ \bibnamefont {Bickers}},\
  }\bibfield  {title} {\bibinfo {title} {{Knight shifts and
  nuclear-spin-relaxation rates for two-dimensional models of
  ${\mathrm{CuO}}_{2}$}},\ }\href {https://doi.org/10.1103/PhysRevB.41.1797}
  {\bibfield  {journal} {\bibinfo  {journal} {Phys. Rev. B}\ }\textbf {\bibinfo
  {volume} {41}},\ \bibinfo {pages} {1797} (\bibinfo {year}
  {1990})}\BibitemShut {NoStop}%
\bibitem [{\citenamefont {Millis}\ \emph {et~al.}(1990)\citenamefont {Millis},
  \citenamefont {Monien},\ and\ \citenamefont {Pines}}]{PhysRevB.42.167}%
  \BibitemOpen
  \bibfield  {author} {\bibinfo {author} {\bibfnamefont {A.~J.}\ \bibnamefont
  {Millis}}, \bibinfo {author} {\bibfnamefont {H.}~\bibnamefont {Monien}},\
  and\ \bibinfo {author} {\bibfnamefont {D.}~\bibnamefont {Pines}},\ }\bibfield
   {title} {\bibinfo {title} {{Phenomenological model of nuclear relaxation in
  the normal state of
  ${\mathrm{YBa}}_{2}$${\mathrm{Cu}}_{3}$${\mathrm{O}}_{7}$}},\ }\href
  {https://doi.org/10.1103/PhysRevB.42.167} {\bibfield  {journal} {\bibinfo
  {journal} {Phys. Rev. B}\ }\textbf {\bibinfo {volume} {42}},\ \bibinfo
  {pages} {167} (\bibinfo {year} {1990})}\BibitemShut {NoStop}%
\bibitem [{\citenamefont {Monien}\ \emph {et~al.}(1991)\citenamefont {Monien},
  \citenamefont {Pines},\ and\ \citenamefont {Takigawa}}]{PhysRevB.43.258}%
  \BibitemOpen
  \bibfield  {author} {\bibinfo {author} {\bibfnamefont {H.}~\bibnamefont
  {Monien}}, \bibinfo {author} {\bibfnamefont {D.}~\bibnamefont {Pines}},\ and\
  \bibinfo {author} {\bibfnamefont {M.}~\bibnamefont {Takigawa}},\ }\bibfield
  {title} {\bibinfo {title} {{Application of the antiferromagnetic-Fermi-liquid
  theory to NMR experiments on
  ${\mathrm{YBa}}_{2}$${\mathrm{Cu}}_{3}$${\mathrm{O}}_{6.63}$}},\ }\href
  {https://doi.org/10.1103/PhysRevB.43.258} {\bibfield  {journal} {\bibinfo
  {journal} {Phys. Rev. B}\ }\textbf {\bibinfo {volume} {43}},\ \bibinfo
  {pages} {258} (\bibinfo {year} {1991})}\BibitemShut {NoStop}%
\bibitem [{\citenamefont {Barzykin}\ and\ \citenamefont
  {Pines}(1995)}]{PhysRevB.52.13585}%
  \BibitemOpen
  \bibfield  {author} {\bibinfo {author} {\bibfnamefont {V.}~\bibnamefont
  {Barzykin}}\ and\ \bibinfo {author} {\bibfnamefont {D.}~\bibnamefont
  {Pines}},\ }\bibfield  {title} {\bibinfo {title} {Magnetic scaling in cuprate
  superconductors},\ }\href {https://doi.org/10.1103/PhysRevB.52.13585}
  {\bibfield  {journal} {\bibinfo  {journal} {Phys. Rev. B}\ }\textbf {\bibinfo
  {volume} {52}},\ \bibinfo {pages} {13585} (\bibinfo {year}
  {1995})}\BibitemShut {NoStop}%
\bibitem [{\citenamefont {Chuang}\ \emph {et~al.}(2004)\citenamefont {Chuang},
  \citenamefont {Gromko}, \citenamefont {Fedorov}, \citenamefont {Aiura},
  \citenamefont {Oka}, \citenamefont {Ando}, \citenamefont {Lindroos},
  \citenamefont {Markiewicz}, \citenamefont {Bansil},\ and\ \citenamefont
  {Dessau}}]{PhysRevB.69.094515}%
  \BibitemOpen
  \bibfield  {author} {\bibinfo {author} {\bibfnamefont {Y.-D.}\ \bibnamefont
  {Chuang}}, \bibinfo {author} {\bibfnamefont {A.~D.}\ \bibnamefont {Gromko}},
  \bibinfo {author} {\bibfnamefont {A.~V.}\ \bibnamefont {Fedorov}}, \bibinfo
  {author} {\bibfnamefont {Y.}~\bibnamefont {Aiura}}, \bibinfo {author}
  {\bibfnamefont {K.}~\bibnamefont {Oka}}, \bibinfo {author} {\bibfnamefont
  {Y.}~\bibnamefont {Ando}}, \bibinfo {author} {\bibfnamefont {M.}~\bibnamefont
  {Lindroos}}, \bibinfo {author} {\bibfnamefont {R.~S.}\ \bibnamefont
  {Markiewicz}}, \bibinfo {author} {\bibfnamefont {A.}~\bibnamefont {Bansil}},\
  and\ \bibinfo {author} {\bibfnamefont {D.~S.}\ \bibnamefont {Dessau}},\
  }\bibfield  {title} {\bibinfo {title} {{Bilayer splitting and coherence
  effects in optimal and underdoped
  ${\mathrm{Bi}}_{2}{\mathrm{Sr}}_{2}{\mathrm{CaCu}}_{2}{\mathrm{O}}_{8+\ensuremath{\delta}}$}},\
  }\href {https://doi.org/10.1103/PhysRevB.69.094515} {\bibfield  {journal}
  {\bibinfo  {journal} {Phys. Rev. B}\ }\textbf {\bibinfo {volume} {69}},\
  \bibinfo {pages} {094515} (\bibinfo {year} {2004})}\BibitemShut {NoStop}%
\bibitem [{\citenamefont {Kaminski}\ \emph {et~al.}(2006)\citenamefont
  {Kaminski}, \citenamefont {Rosenkranz}, \citenamefont {Fretwell},
  \citenamefont {Norman}, \citenamefont {Randeria}, \citenamefont {Campuzano},
  \citenamefont {Park}, \citenamefont {Li},\ and\ \citenamefont
  {Raffy}}]{PhysRevB.73.174511}%
  \BibitemOpen
  \bibfield  {author} {\bibinfo {author} {\bibfnamefont {A.}~\bibnamefont
  {Kaminski}}, \bibinfo {author} {\bibfnamefont {S.}~\bibnamefont
  {Rosenkranz}}, \bibinfo {author} {\bibfnamefont {H.~M.}\ \bibnamefont
  {Fretwell}}, \bibinfo {author} {\bibfnamefont {M.~R.}\ \bibnamefont
  {Norman}}, \bibinfo {author} {\bibfnamefont {M.}~\bibnamefont {Randeria}},
  \bibinfo {author} {\bibfnamefont {J.~C.}\ \bibnamefont {Campuzano}}, \bibinfo
  {author} {\bibfnamefont {J.-M.}\ \bibnamefont {Park}}, \bibinfo {author}
  {\bibfnamefont {Z.~Z.}\ \bibnamefont {Li}},\ and\ \bibinfo {author}
  {\bibfnamefont {H.}~\bibnamefont {Raffy}},\ }\bibfield  {title} {\bibinfo
  {title} {{Change of Fermi-surface topology in
  ${\mathrm{Bi}}_{2}{\mathrm{Sr}}_{2}{\mathrm{CaCu}}_{2}{\mathrm{O}}_{8+\ensuremath{\delta}}$
  with doping}},\ }\href {https://doi.org/10.1103/PhysRevB.73.174511}
  {\bibfield  {journal} {\bibinfo  {journal} {Phys. Rev. B}\ }\textbf {\bibinfo
  {volume} {73}},\ \bibinfo {pages} {174511} (\bibinfo {year}
  {2006})}\BibitemShut {NoStop}%
\bibitem [{\citenamefont {He}\ \emph {et~al.}(2018)\citenamefont {He},
  \citenamefont {Hashimoto}, \citenamefont {Song}, \citenamefont {Chen},
  \citenamefont {He}, \citenamefont {Vishik}, \citenamefont {Moritz},
  \citenamefont {Lee}, \citenamefont {Nagaosa}, \citenamefont {Zaanen},
  \citenamefont {Devereaux}, \citenamefont {Yoshida}, \citenamefont {Eisaki},
  \citenamefont {Lu},\ and\ \citenamefont
  {Shen}}]{doi:10.1126/science.aar3394}%
  \BibitemOpen
  \bibfield  {author} {\bibinfo {author} {\bibfnamefont {Y.}~\bibnamefont
  {He}}, \bibinfo {author} {\bibfnamefont {M.}~\bibnamefont {Hashimoto}},
  \bibinfo {author} {\bibfnamefont {D.}~\bibnamefont {Song}}, \bibinfo {author}
  {\bibfnamefont {S.-D.}\ \bibnamefont {Chen}}, \bibinfo {author}
  {\bibfnamefont {J.}~\bibnamefont {He}}, \bibinfo {author} {\bibfnamefont
  {I.~M.}\ \bibnamefont {Vishik}}, \bibinfo {author} {\bibfnamefont
  {B.}~\bibnamefont {Moritz}}, \bibinfo {author} {\bibfnamefont {D.-H.}\
  \bibnamefont {Lee}}, \bibinfo {author} {\bibfnamefont {N.}~\bibnamefont
  {Nagaosa}}, \bibinfo {author} {\bibfnamefont {J.}~\bibnamefont {Zaanen}},
  \bibinfo {author} {\bibfnamefont {T.~P.}\ \bibnamefont {Devereaux}}, \bibinfo
  {author} {\bibfnamefont {Y.}~\bibnamefont {Yoshida}}, \bibinfo {author}
  {\bibfnamefont {H.}~\bibnamefont {Eisaki}}, \bibinfo {author} {\bibfnamefont
  {D.~H.}\ \bibnamefont {Lu}},\ and\ \bibinfo {author} {\bibfnamefont {Z.-X.}\
  \bibnamefont {Shen}},\ }\bibfield  {title} {\bibinfo {title} {{Rapid change
  of superconductivity and electron-phonon coupling through critical doping in
  Bi-2212}},\ }\href {https://doi.org/10.1126/science.aar3394} {\bibfield
  {journal} {\bibinfo  {journal} {Science}\ }\textbf {\bibinfo {volume}
  {362}},\ \bibinfo {pages} {62} (\bibinfo {year} {2018})}\BibitemShut
  {NoStop}%
\bibitem [{\citenamefont {Drozdov}\ \emph {et~al.}(2018)\citenamefont
  {Drozdov}, \citenamefont {Pletikosi{\'{c}}}, \citenamefont {Kim},
  \citenamefont {Fujita}, \citenamefont {Gu}, \citenamefont {Davis},
  \citenamefont {Johnson}, \citenamefont {Bo{\v{z}}ovi{\'{c}}},\ and\
  \citenamefont {Valla}}]{Drozdov2018}%
  \BibitemOpen
  \bibfield  {author} {\bibinfo {author} {\bibfnamefont {I.~K.}\ \bibnamefont
  {Drozdov}}, \bibinfo {author} {\bibfnamefont {I.}~\bibnamefont
  {Pletikosi{\'{c}}}}, \bibinfo {author} {\bibfnamefont {C.-K.}\ \bibnamefont
  {Kim}}, \bibinfo {author} {\bibfnamefont {K.}~\bibnamefont {Fujita}},
  \bibinfo {author} {\bibfnamefont {G.~D.}\ \bibnamefont {Gu}}, \bibinfo
  {author} {\bibfnamefont {J.~C.~S.}\ \bibnamefont {Davis}}, \bibinfo {author}
  {\bibfnamefont {P.~D.}\ \bibnamefont {Johnson}}, \bibinfo {author}
  {\bibfnamefont {I.}~\bibnamefont {Bo{\v{z}}ovi{\'{c}}}},\ and\ \bibinfo
  {author} {\bibfnamefont {T.}~\bibnamefont {Valla}},\ }\bibfield  {title}
  {\bibinfo {title} {{Phase diagram of Bi$_2$Sr$_2$CaCu$_2$O$_{8+\delta}$
  revisited}},\ }\href {https://doi.org/10.1038/s41467-018-07686-w} {\bibfield
  {journal} {\bibinfo  {journal} {Nature Communications}\ }\textbf {\bibinfo
  {volume} {9}},\ \bibinfo {pages} {5210} (\bibinfo {year} {2018})}\BibitemShut
  {NoStop}%
\bibitem [{\citenamefont {Horio}\ \emph {et~al.}(2018)\citenamefont {Horio},
  \citenamefont {Hauser}, \citenamefont {Sassa}, \citenamefont {Mingazheva},
  \citenamefont {Sutter}, \citenamefont {Kramer}, \citenamefont {Cook},
  \citenamefont {Nocerino}, \citenamefont {Forslund}, \citenamefont
  {Tjernberg}, \citenamefont {Kobayashi}, \citenamefont {Chikina},
  \citenamefont {Schr\"oter}, \citenamefont {Krieger}, \citenamefont {Schmitt},
  \citenamefont {Strocov}, \citenamefont {Pyon}, \citenamefont {Takayama},
  \citenamefont {Takagi}, \citenamefont {Lipscombe}, \citenamefont {Hayden},
  \citenamefont {Ishikado}, \citenamefont {Eisaki}, \citenamefont {{Neupert,
  T.}}, \citenamefont {{M$\mathring{\mathrm{a}}$nsson, M.}}, \citenamefont
  {Matt},\ and\ \citenamefont {Chang}}]{PhysRevLett.121.077004}%
  \BibitemOpen
  \bibfield  {author} {\bibinfo {author} {\bibfnamefont {M.}~\bibnamefont
  {Horio}}, \bibinfo {author} {\bibfnamefont {K.}~\bibnamefont {Hauser}},
  \bibinfo {author} {\bibfnamefont {Y.}~\bibnamefont {Sassa}}, \bibinfo
  {author} {\bibfnamefont {Z.}~\bibnamefont {Mingazheva}}, \bibinfo {author}
  {\bibfnamefont {D.}~\bibnamefont {Sutter}}, \bibinfo {author} {\bibfnamefont
  {K.}~\bibnamefont {Kramer}}, \bibinfo {author} {\bibfnamefont
  {A.}~\bibnamefont {Cook}}, \bibinfo {author} {\bibfnamefont {E.}~\bibnamefont
  {Nocerino}}, \bibinfo {author} {\bibfnamefont {O.~K.}\ \bibnamefont
  {Forslund}}, \bibinfo {author} {\bibfnamefont {O.}~\bibnamefont {Tjernberg}},
  \bibinfo {author} {\bibfnamefont {M.}~\bibnamefont {Kobayashi}}, \bibinfo
  {author} {\bibfnamefont {A.}~\bibnamefont {Chikina}}, \bibinfo {author}
  {\bibfnamefont {N.~B.~M.}\ \bibnamefont {Schr\"oter}}, \bibinfo {author}
  {\bibfnamefont {J.~A.}\ \bibnamefont {Krieger}}, \bibinfo {author}
  {\bibfnamefont {T.}~\bibnamefont {Schmitt}}, \bibinfo {author} {\bibfnamefont
  {V.~N.}\ \bibnamefont {Strocov}}, \bibinfo {author} {\bibfnamefont
  {S.}~\bibnamefont {Pyon}}, \bibinfo {author} {\bibfnamefont {T.}~\bibnamefont
  {Takayama}}, \bibinfo {author} {\bibfnamefont {H.}~\bibnamefont {Takagi}},
  \bibinfo {author} {\bibfnamefont {O.~J.}\ \bibnamefont {Lipscombe}}, \bibinfo
  {author} {\bibfnamefont {S.~M.}\ \bibnamefont {Hayden}}, \bibinfo {author}
  {\bibfnamefont {M.}~\bibnamefont {Ishikado}}, \bibinfo {author}
  {\bibfnamefont {H.}~\bibnamefont {Eisaki}}, \bibinfo {author} {\bibnamefont
  {{Neupert, T.}}}, \bibinfo {author} {\bibnamefont
  {{M$\mathring{\mathrm{a}}$nsson, M.}}}, \bibinfo {author} {\bibfnamefont
  {C.~E.}\ \bibnamefont {Matt}},\ and\ \bibinfo {author} {\bibfnamefont
  {J.}~\bibnamefont {Chang}},\ }\bibfield  {title} {\bibinfo {title}
  {Three-dimensional \uppercase{F}ermi surface of overdoped {La}-based
  cuprates},\ }\href {https://doi.org/10.1103/PhysRevLett.121.077004}
  {\bibfield  {journal} {\bibinfo  {journal} {Phys. Rev. Lett.}\ }\textbf
  {\bibinfo {volume} {121}},\ \bibinfo {pages} {077004} (\bibinfo {year}
  {2018})}\BibitemShut {NoStop}%
\bibitem [{\citenamefont {Wang}\ \emph
  {et~al.}(2023{\natexlab{a}})\citenamefont {Wang}, \citenamefont {Ding},
  \citenamefont {Huang}, \citenamefont {Moritz},\ and\ \citenamefont
  {Devereaux}}]{thermopower}%
  \BibitemOpen
  \bibfield  {author} {\bibinfo {author} {\bibfnamefont {W.~O.}\ \bibnamefont
  {Wang}}, \bibinfo {author} {\bibfnamefont {J.~K.}\ \bibnamefont {Ding}},
  \bibinfo {author} {\bibfnamefont {E.~W.}\ \bibnamefont {Huang}}, \bibinfo
  {author} {\bibfnamefont {B.}~\bibnamefont {Moritz}},\ and\ \bibinfo {author}
  {\bibfnamefont {T.~P.}\ \bibnamefont {Devereaux}},\ }\bibfield  {title}
  {\bibinfo {title} {Quantitative assessment of the universal thermopower in
  the \uppercase{H}ubbard model},\ }\href
  {https://doi.org/10.1038/s41467-023-42772-8} {\bibfield  {journal} {\bibinfo
  {journal} {Nature Communications}\ }\textbf {\bibinfo {volume} {14}},\
  \bibinfo {pages} {7064} (\bibinfo {year} {2023}{\natexlab{a}})}\BibitemShut
  {NoStop}%
\bibitem [{\citenamefont {Chien}\ \emph {et~al.}(1991)\citenamefont {Chien},
  \citenamefont {Wang},\ and\ \citenamefont {Ong}}]{PhysRevLett.67.2088}%
  \BibitemOpen
  \bibfield  {author} {\bibinfo {author} {\bibfnamefont {T.~R.}\ \bibnamefont
  {Chien}}, \bibinfo {author} {\bibfnamefont {Z.~Z.}\ \bibnamefont {Wang}},\
  and\ \bibinfo {author} {\bibfnamefont {N.~P.}\ \bibnamefont {Ong}},\
  }\bibfield  {title} {\bibinfo {title} {Effect of \uppercase{Z}n impurities on
  the normal-state \uppercase{H}all angle in single-crystal
  ${\mathrm{\uppercase{yb}a}}_{2}$${\mathrm{\uppercase{c}u}}_{3\mathrm{\ensuremath{-}}\mathit{x}}$${\mathrm{\uppercase{z}n}}_{\mathit{x}}$${\mathrm{\uppercase{o}}}_{7\mathrm{\ensuremath{-}}\mathrm{\ensuremath{\delta}}}$},\
  }\href {https://doi.org/10.1103/PhysRevLett.67.2088} {\bibfield  {journal}
  {\bibinfo  {journal} {Phys. Rev. Lett.}\ }\textbf {\bibinfo {volume} {67}},\
  \bibinfo {pages} {2088} (\bibinfo {year} {1991})}\BibitemShut {NoStop}%
\bibitem [{\citenamefont {Carrington}\ \emph {et~al.}(1992)\citenamefont
  {Carrington}, \citenamefont {Mackenzie}, \citenamefont {Lin},\ and\
  \citenamefont {Cooper}}]{PhysRevLett.69.2855}%
  \BibitemOpen
  \bibfield  {author} {\bibinfo {author} {\bibfnamefont {A.}~\bibnamefont
  {Carrington}}, \bibinfo {author} {\bibfnamefont {A.~P.}\ \bibnamefont
  {Mackenzie}}, \bibinfo {author} {\bibfnamefont {C.~T.}\ \bibnamefont {Lin}},\
  and\ \bibinfo {author} {\bibfnamefont {J.~R.}\ \bibnamefont {Cooper}},\
  }\bibfield  {title} {\bibinfo {title} {Temperature dependence of the
  \uppercase{H}all angle in single-crystal
  ${\mathrm{\uppercase{yb}a}}_{2}$(${\mathrm{\uppercase{c}u}}_{1\mathrm{\ensuremath{-}}\mathit{x}}$${\mathrm{\uppercase{c}o}}_{\mathit{x}}$${)}_{3}$${\mathrm{\uppercase{o}}}_{7\mathrm{\ensuremath{-}}\mathrm{\ensuremath{\delta}}}$},\
  }\href {https://doi.org/10.1103/PhysRevLett.69.2855} {\bibfield  {journal}
  {\bibinfo  {journal} {Phys. Rev. Lett.}\ }\textbf {\bibinfo {volume} {69}},\
  \bibinfo {pages} {2855} (\bibinfo {year} {1992})}\BibitemShut {NoStop}%
\bibitem [{\citenamefont {Hwang}\ \emph {et~al.}(1994)\citenamefont {Hwang},
  \citenamefont {Batlogg}, \citenamefont {Takagi}, \citenamefont {Kao},
  \citenamefont {Kwo}, \citenamefont {Cava}, \citenamefont {Krajewski},\ and\
  \citenamefont {Peck}}]{PhysRevLett.72.2636}%
  \BibitemOpen
  \bibfield  {author} {\bibinfo {author} {\bibfnamefont {H.~Y.}\ \bibnamefont
  {Hwang}}, \bibinfo {author} {\bibfnamefont {B.}~\bibnamefont {Batlogg}},
  \bibinfo {author} {\bibfnamefont {H.}~\bibnamefont {Takagi}}, \bibinfo
  {author} {\bibfnamefont {H.~L.}\ \bibnamefont {Kao}}, \bibinfo {author}
  {\bibfnamefont {J.}~\bibnamefont {Kwo}}, \bibinfo {author} {\bibfnamefont
  {R.~J.}\ \bibnamefont {Cava}}, \bibinfo {author} {\bibfnamefont {J.~J.}\
  \bibnamefont {Krajewski}},\ and\ \bibinfo {author} {\bibfnamefont {W.~F.}\
  \bibnamefont {Peck}},\ }\bibfield  {title} {\bibinfo {title} {Scaling of the
  temperature dependent \uppercase{H}all effect in
  ${\mathrm{\uppercase{l}a}}_{2\mathrm{\ensuremath{-}}\mathit{x}}$${\mathrm{\uppercase{s}r}}_{\mathit{x}}$${\mathrm{\uppercase{c}u\uppercase{o}}}_{4}$},\
  }\href {https://doi.org/10.1103/PhysRevLett.72.2636} {\bibfield  {journal}
  {\bibinfo  {journal} {Phys. Rev. Lett.}\ }\textbf {\bibinfo {volume} {72}},\
  \bibinfo {pages} {2636} (\bibinfo {year} {1994})}\BibitemShut {NoStop}%
\bibitem [{\citenamefont {Kubo}\ and\ \citenamefont
  {Manako}(1992)}]{KUBO1992378}%
  \BibitemOpen
  \bibfield  {author} {\bibinfo {author} {\bibfnamefont {Y.}~\bibnamefont
  {Kubo}}\ and\ \bibinfo {author} {\bibfnamefont {T.}~\bibnamefont {Manako}},\
  }\bibfield  {title} {\bibinfo {title} {T2-dependence of inverse
  \uppercase{H}all mobility observed in overdoped \uppercase{T}l-cuprates},\
  }\href {https://doi.org/10.1016/0921-4534(92)90020-D} {\bibfield  {journal}
  {\bibinfo  {journal} {Physica C: Superconductivity}\ }\textbf {\bibinfo
  {volume} {197}},\ \bibinfo {pages} {378} (\bibinfo {year}
  {1992})}\BibitemShut {NoStop}%
\bibitem [{\citenamefont {Badoux}\ \emph {et~al.}(2016)\citenamefont {Badoux},
  \citenamefont {Tabis}, \citenamefont {Lalibert{\'e}}, \citenamefont
  {Grissonnanche}, \citenamefont {Vignolle}, \citenamefont {Vignolles},
  \citenamefont {B{\'e}ard}, \citenamefont {Bonn}, \citenamefont {Hardy},
  \citenamefont {Liang}, \citenamefont {Doiron-Leyraud}, \citenamefont
  {Taillefer},\ and\ \citenamefont {Proust}}]{Badoux2016}%
  \BibitemOpen
  \bibfield  {author} {\bibinfo {author} {\bibfnamefont {S.}~\bibnamefont
  {Badoux}}, \bibinfo {author} {\bibfnamefont {W.}~\bibnamefont {Tabis}},
  \bibinfo {author} {\bibfnamefont {F.}~\bibnamefont {Lalibert{\'e}}}, \bibinfo
  {author} {\bibfnamefont {G.}~\bibnamefont {Grissonnanche}}, \bibinfo {author}
  {\bibfnamefont {B.}~\bibnamefont {Vignolle}}, \bibinfo {author}
  {\bibfnamefont {D.}~\bibnamefont {Vignolles}}, \bibinfo {author}
  {\bibfnamefont {J.}~\bibnamefont {B{\'e}ard}}, \bibinfo {author}
  {\bibfnamefont {D.~A.}\ \bibnamefont {Bonn}}, \bibinfo {author}
  {\bibfnamefont {W.~N.}\ \bibnamefont {Hardy}}, \bibinfo {author}
  {\bibfnamefont {R.}~\bibnamefont {Liang}}, \bibinfo {author} {\bibfnamefont
  {N.}~\bibnamefont {Doiron-Leyraud}}, \bibinfo {author} {\bibfnamefont
  {L.}~\bibnamefont {Taillefer}},\ and\ \bibinfo {author} {\bibfnamefont
  {C.}~\bibnamefont {Proust}},\ }\bibfield  {title} {\bibinfo {title} {Change
  of carrier density at the pseudogap critical point of a cuprate
  superconductor},\ }\href {https://doi.org/10.1038/nature16983} {\bibfield
  {journal} {\bibinfo  {journal} {Nature}\ }\textbf {\bibinfo {volume} {531}},\
  \bibinfo {pages} {210} (\bibinfo {year} {2016})}\BibitemShut {NoStop}%
\bibitem [{\citenamefont {Pelc}\ \emph {et~al.}(2020)\citenamefont {Pelc},
  \citenamefont {Veit}, \citenamefont {Dorow}, \citenamefont {Ge},
  \citenamefont {Bari\ifmmode \check{s}\else \v{s}\fi{}i\ifmmode~\acute{c}\else
  \'{c}\fi{}},\ and\ \citenamefont {Greven}}]{PhysRevB.102.075114}%
  \BibitemOpen
  \bibfield  {author} {\bibinfo {author} {\bibfnamefont {D.}~\bibnamefont
  {Pelc}}, \bibinfo {author} {\bibfnamefont {M.~J.}\ \bibnamefont {Veit}},
  \bibinfo {author} {\bibfnamefont {C.~J.}\ \bibnamefont {Dorow}}, \bibinfo
  {author} {\bibfnamefont {Y.}~\bibnamefont {Ge}}, \bibinfo {author}
  {\bibfnamefont {N.}~\bibnamefont {Bari\ifmmode \check{s}\else
  \v{s}\fi{}i\ifmmode~\acute{c}\else \'{c}\fi{}}},\ and\ \bibinfo {author}
  {\bibfnamefont {M.}~\bibnamefont {Greven}},\ }\bibfield  {title} {\bibinfo
  {title} {Resistivity phase diagram of cuprates revisited},\ }\href
  {https://doi.org/10.1103/PhysRevB.102.075114} {\bibfield  {journal} {\bibinfo
   {journal} {Phys. Rev. B}\ }\textbf {\bibinfo {volume} {102}},\ \bibinfo
  {pages} {075114} (\bibinfo {year} {2020})}\BibitemShut {NoStop}%
\bibitem [{\citenamefont {Wang}\ \emph {et~al.}(2022)\citenamefont {Wang},
  \citenamefont {Ding}, \citenamefont {Moritz}, \citenamefont {Huang},\ and\
  \citenamefont {Devereaux}}]{PhysRevB.105.L161103}%
  \BibitemOpen
  \bibfield  {author} {\bibinfo {author} {\bibfnamefont {W.~O.}\ \bibnamefont
  {Wang}}, \bibinfo {author} {\bibfnamefont {J.~K.}\ \bibnamefont {Ding}},
  \bibinfo {author} {\bibfnamefont {B.}~\bibnamefont {Moritz}}, \bibinfo
  {author} {\bibfnamefont {E.~W.}\ \bibnamefont {Huang}},\ and\ \bibinfo
  {author} {\bibfnamefont {T.~P.}\ \bibnamefont {Devereaux}},\ }\bibfield
  {title} {\bibinfo {title} {{Magnon heat transport in a two-dimensional
  \uppercase{M}ott insulator}},\ }\href
  {https://doi.org/10.1103/PhysRevB.105.L161103} {\bibfield  {journal}
  {\bibinfo  {journal} {Phys. Rev. B}\ }\textbf {\bibinfo {volume} {105}},\
  \bibinfo {pages} {L161103} (\bibinfo {year} {2022})}\BibitemShut {NoStop}%
\bibitem [{\citenamefont {Loram}\ \emph {et~al.}(1993)\citenamefont {Loram},
  \citenamefont {Mirza}, \citenamefont {Cooper},\ and\ \citenamefont
  {Liang}}]{PhysRevLett.71.1740}%
  \BibitemOpen
  \bibfield  {author} {\bibinfo {author} {\bibfnamefont {J.~W.}\ \bibnamefont
  {Loram}}, \bibinfo {author} {\bibfnamefont {K.~A.}\ \bibnamefont {Mirza}},
  \bibinfo {author} {\bibfnamefont {J.~R.}\ \bibnamefont {Cooper}},\ and\
  \bibinfo {author} {\bibfnamefont {W.~Y.}\ \bibnamefont {Liang}},\ }\bibfield
  {title} {\bibinfo {title} {Electronic specific heat of
  ${\mathrm{\uppercase{yb}a}}_{2}$${\mathrm{\uppercase{c}u}}_{3}$${\mathrm{\uppercase{o}}}_{6+\mathit{x}}$
  from 1.8 to 300 \uppercase{K}},\ }\href
  {https://doi.org/10.1103/PhysRevLett.71.1740} {\bibfield  {journal} {\bibinfo
   {journal} {Phys. Rev. Lett.}\ }\textbf {\bibinfo {volume} {71}},\ \bibinfo
  {pages} {1740} (\bibinfo {year} {1993})}\BibitemShut {NoStop}%
\bibitem [{\citenamefont {Loram}\ \emph {et~al.}(1998)\citenamefont {Loram},
  \citenamefont {Mirza}, \citenamefont {Cooper},\ and\ \citenamefont
  {Tallon}}]{LORAMA19982091}%
  \BibitemOpen
  \bibfield  {author} {\bibinfo {author} {\bibfnamefont {J.~W.}\ \bibnamefont
  {Loram}}, \bibinfo {author} {\bibfnamefont {K.~A.}\ \bibnamefont {Mirza}},
  \bibinfo {author} {\bibfnamefont {J.~R.}\ \bibnamefont {Cooper}},\ and\
  \bibinfo {author} {\bibfnamefont {J.~L.}\ \bibnamefont {Tallon}},\ }\bibfield
   {title} {\bibinfo {title} {Specific heat evidence on the normal state
  pseudogap},\ }\href {https://doi.org/10.1016/S0022-3697(98)00180-2}
  {\bibfield  {journal} {\bibinfo  {journal} {Journal of Physics and Chemistry
  of Solids}\ }\textbf {\bibinfo {volume} {59}},\ \bibinfo {pages} {2091}
  (\bibinfo {year} {1998})}\BibitemShut {NoStop}%
\bibitem [{\citenamefont {Loram}\ \emph {et~al.}(2001)\citenamefont {Loram},
  \citenamefont {Luo}, \citenamefont {Cooper}, \citenamefont {Liang},\ and\
  \citenamefont {Tallon}}]{LORAM200159}%
  \BibitemOpen
  \bibfield  {author} {\bibinfo {author} {\bibfnamefont {J.~W.}\ \bibnamefont
  {Loram}}, \bibinfo {author} {\bibfnamefont {J.}~\bibnamefont {Luo}}, \bibinfo
  {author} {\bibfnamefont {J.~R.}\ \bibnamefont {Cooper}}, \bibinfo {author}
  {\bibfnamefont {W.~Y.}\ \bibnamefont {Liang}},\ and\ \bibinfo {author}
  {\bibfnamefont {J.~L.}\ \bibnamefont {Tallon}},\ }\bibfield  {title}
  {\bibinfo {title} {Evidence on the pseudogap and condensate from the
  electronic specific heat},\ }\href
  {https://doi.org/10.1016/S0022-3697(00)00101-3} {\bibfield  {journal}
  {\bibinfo  {journal} {Journal of Physics and Chemistry of Solids}\ }\textbf
  {\bibinfo {volume} {62}},\ \bibinfo {pages} {59} (\bibinfo {year}
  {2001})}\BibitemShut {NoStop}%
\bibitem [{\citenamefont {Matsuzaki}\ \emph {et~al.}(2004)\citenamefont
  {Matsuzaki}, \citenamefont {Momono}, \citenamefont {Oda},\ and\ \citenamefont
  {Ido}}]{doi:10.1143/JPSJ.73.2232}%
  \BibitemOpen
  \bibfield  {author} {\bibinfo {author} {\bibfnamefont {T.}~\bibnamefont
  {Matsuzaki}}, \bibinfo {author} {\bibfnamefont {N.}~\bibnamefont {Momono}},
  \bibinfo {author} {\bibfnamefont {M.}~\bibnamefont {Oda}},\ and\ \bibinfo
  {author} {\bibfnamefont {M.}~\bibnamefont {Ido}},\ }\bibfield  {title}
  {\bibinfo {title} {Electronic specific heat of
  $\mathrm{\uppercase{l}a_{2-x}\uppercase{s}r_x\uppercase{c}u\uppercase{o}_4}$:
  Pseudogap formation and reduction of the superconducting condensation
  energy},\ }\href {https://doi.org/10.1143/JPSJ.73.2232} {\bibfield  {journal}
  {\bibinfo  {journal} {Journal of the Physical Society of Japan}\ }\textbf
  {\bibinfo {volume} {73}},\ \bibinfo {pages} {2232} (\bibinfo {year}
  {2004})}\BibitemShut {NoStop}%
\bibitem [{\citenamefont {Girod}\ \emph {et~al.}(2021)\citenamefont {Girod},
  \citenamefont {LeBoeuf}, \citenamefont {Demuer}, \citenamefont {Seyfarth},
  \citenamefont {Imajo}, \citenamefont {Kindo}, \citenamefont {Kohama},
  \citenamefont {Lizaire}, \citenamefont {Legros}, \citenamefont {Gourgout},
  \citenamefont {Takagi}, \citenamefont {Kurosawa}, \citenamefont {Oda},
  \citenamefont {Momono}, \citenamefont {Chang}, \citenamefont {Ono},
  \citenamefont {Zheng}, \citenamefont {Marcenat}, \citenamefont {Taillefer},\
  and\ \citenamefont {Klein}}]{PhysRevB.103.214506}%
  \BibitemOpen
  \bibfield  {author} {\bibinfo {author} {\bibfnamefont {C.}~\bibnamefont
  {Girod}}, \bibinfo {author} {\bibfnamefont {D.}~\bibnamefont {LeBoeuf}},
  \bibinfo {author} {\bibfnamefont {A.}~\bibnamefont {Demuer}}, \bibinfo
  {author} {\bibfnamefont {G.}~\bibnamefont {Seyfarth}}, \bibinfo {author}
  {\bibfnamefont {S.}~\bibnamefont {Imajo}}, \bibinfo {author} {\bibfnamefont
  {K.}~\bibnamefont {Kindo}}, \bibinfo {author} {\bibfnamefont
  {Y.}~\bibnamefont {Kohama}}, \bibinfo {author} {\bibfnamefont
  {M.}~\bibnamefont {Lizaire}}, \bibinfo {author} {\bibfnamefont
  {A.}~\bibnamefont {Legros}}, \bibinfo {author} {\bibfnamefont
  {A.}~\bibnamefont {Gourgout}}, \bibinfo {author} {\bibfnamefont
  {H.}~\bibnamefont {Takagi}}, \bibinfo {author} {\bibfnamefont
  {T.}~\bibnamefont {Kurosawa}}, \bibinfo {author} {\bibfnamefont
  {M.}~\bibnamefont {Oda}}, \bibinfo {author} {\bibfnamefont {N.}~\bibnamefont
  {Momono}}, \bibinfo {author} {\bibfnamefont {J.}~\bibnamefont {Chang}},
  \bibinfo {author} {\bibfnamefont {S.}~\bibnamefont {Ono}}, \bibinfo {author}
  {\bibfnamefont {G.-q.}\ \bibnamefont {Zheng}}, \bibinfo {author}
  {\bibfnamefont {C.}~\bibnamefont {Marcenat}}, \bibinfo {author}
  {\bibfnamefont {L.}~\bibnamefont {Taillefer}},\ and\ \bibinfo {author}
  {\bibfnamefont {T.}~\bibnamefont {Klein}},\ }\bibfield  {title} {\bibinfo
  {title} {Normal state specific heat in the cuprate superconductors
  ${\mathrm{\uppercase{l}a}}_{2\ensuremath{-}x}{\mathrm{\uppercase{s}r}}_{x}{\mathrm{\uppercase{c}u\uppercase{o}}}_{4}$
  and
  ${\mathrm{\uppercase{b}i}}_{2+y}{\mathrm{\uppercase{s}r}}_{2\ensuremath{-}x\ensuremath{-}y}{\mathrm{\uppercase{l}a}}_{x}{\mathrm{\uppercase{c}u\uppercase{o}}}_{6+\ensuremath{\delta}}$
  near the critical point of the pseudogap phase},\ }\href
  {https://doi.org/10.1103/PhysRevB.103.214506} {\bibfield  {journal} {\bibinfo
   {journal} {Phys. Rev. B}\ }\textbf {\bibinfo {volume} {103}},\ \bibinfo
  {pages} {214506} (\bibinfo {year} {2021})}\BibitemShut {NoStop}%
\bibitem [{Pho(2026)}]{PhotoemissionDataZenodo}%
  \BibitemOpen
  \href@noop {} {\bibinfo {title} {{The data and code are available at
  \url{https://doi.org/10.5281/zenodo.15537628}}}} (\bibinfo {year}
  {2026})\BibitemShut {NoStop}%
\bibitem [{\citenamefont {Mahan}(2000)}]{mahan2013many}%
  \BibitemOpen
  \bibfield  {author} {\bibinfo {author} {\bibfnamefont {G.~D.}\ \bibnamefont
  {Mahan}},\ }\href {https://doi.org/10.1007/978-1-4757-5714-9} {\emph
  {\bibinfo {title} {Many-particle physics}}}\ (\bibinfo  {publisher}
  {Springer},\ \bibinfo {year} {2000})\BibitemShut {NoStop}%
\bibitem [{Note1()}]{Note1}%
  \BibitemOpen
  \bibinfo {note} {At high temperatures ($T/t > 1/3$), we apply $15$ twisted
  boundary conditions for $t''/t=0.15$, but only the standard periodic boundary
  condition ($\theta =0$) for $t''/t=0$. This choice does not affect the
  interpolated results because the spectral weight varies smoothly at these
  high temperatures.}\BibitemShut {Stop}%
\bibitem [{\citenamefont {Negele}\ and\ \citenamefont
  {Orland}(2018)}]{negele2018quantum}%
  \BibitemOpen
  \bibfield  {author} {\bibinfo {author} {\bibfnamefont {J.~W.}\ \bibnamefont
  {Negele}}\ and\ \bibinfo {author} {\bibfnamefont {H.}~\bibnamefont
  {Orland}},\ }\href {https://doi.org/10.1201/9780429497926} {\emph {\bibinfo
  {title} {Quantum Many-Particle Systems}}}\ (\bibinfo  {publisher} {CRC
  Press},\ \bibinfo {year} {2018})\BibitemShut {NoStop}%
\bibitem [{\citenamefont {Abrikosov}\ \emph {et~al.}(1975)\citenamefont
  {Abrikosov}, \citenamefont {Gorkov},\ and\ \citenamefont
  {Dzyaloshinski}}]{abrikosov2012methods}%
  \BibitemOpen
  \bibfield  {author} {\bibinfo {author} {\bibfnamefont {A.~A.}\ \bibnamefont
  {Abrikosov}}, \bibinfo {author} {\bibfnamefont {L.~P.}\ \bibnamefont
  {Gorkov}},\ and\ \bibinfo {author} {\bibfnamefont {I.~E.}\ \bibnamefont
  {Dzyaloshinski}},\ }\href@noop {} {\emph {\bibinfo {title} {Methods of
  Quantum Field Theory in Statistical Physics}}}\ (\bibinfo  {publisher} {Dover
  Publications},\ \bibinfo {address} {New York},\ \bibinfo {year}
  {1975})\BibitemShut {NoStop}%
\bibitem [{\citenamefont {{\v{Z}}itko}\ \emph {et~al.}(2015)\citenamefont
  {{\v{Z}}itko}, \citenamefont {Osolin},\ and\ \citenamefont
  {Jegli{\v{c}}}}]{PhysRevB.91.155111}%
  \BibitemOpen
  \bibfield  {author} {\bibinfo {author} {\bibfnamefont {R.}~\bibnamefont
  {{\v{Z}}itko}}, \bibinfo {author} {\bibfnamefont {{\v{Z}}.}~\bibnamefont
  {Osolin}},\ and\ \bibinfo {author} {\bibfnamefont {P.}~\bibnamefont
  {Jegli{\v{c}}}},\ }\bibfield  {title} {\bibinfo {title} {Repulsive versus
  attractive \uppercase{H}ubbard model: Transport properties and spin-lattice
  relaxation rate},\ }\href {https://doi.org/10.1103/PhysRevB.91.155111}
  {\bibfield  {journal} {\bibinfo  {journal} {Phys. Rev. B}\ }\textbf {\bibinfo
  {volume} {91}},\ \bibinfo {pages} {155111} (\bibinfo {year}
  {2015})}\BibitemShut {NoStop}%
\bibitem [{\citenamefont {Mu\ss{}hoff}\ \emph {et~al.}(2021)\citenamefont
  {Mu\ss{}hoff}, \citenamefont {Kiani},\ and\ \citenamefont
  {Pavarini}}]{PhysRevB.103.075136}%
  \BibitemOpen
  \bibfield  {author} {\bibinfo {author} {\bibfnamefont {J.}~\bibnamefont
  {Mu\ss{}hoff}}, \bibinfo {author} {\bibfnamefont {A.}~\bibnamefont {Kiani}},\
  and\ \bibinfo {author} {\bibfnamefont {E.}~\bibnamefont {Pavarini}},\
  }\bibfield  {title} {\bibinfo {title} {{Magnetic response trends in cuprates
  and the $t$-$t'$ \uppercase{H}ubbard model}},\ }\href
  {https://doi.org/10.1103/PhysRevB.103.075136} {\bibfield  {journal} {\bibinfo
   {journal} {Phys. Rev. B}\ }\textbf {\bibinfo {volume} {103}},\ \bibinfo
  {pages} {075136} (\bibinfo {year} {2021})}\BibitemShut {NoStop}%
\bibitem [{\citenamefont {Mila}\ and\ \citenamefont
  {Rice}(1989)}]{MILA1989561}%
  \BibitemOpen
  \bibfield  {author} {\bibinfo {author} {\bibfnamefont {F.}~\bibnamefont
  {Mila}}\ and\ \bibinfo {author} {\bibfnamefont {T.~M.}\ \bibnamefont
  {Rice}},\ }\bibfield  {title} {\bibinfo {title} {{Analysis of magnetic
  resonance experiments in YBa$_2$Cu$_3$O$_7$}},\ }\href
  {https://doi.org/10.1016/0921-4534(89)90286-4} {\bibfield  {journal}
  {\bibinfo  {journal} {Physica C: Superconductivity}\ }\textbf {\bibinfo
  {volume} {157}},\ \bibinfo {pages} {561} (\bibinfo {year}
  {1989})}\BibitemShut {NoStop}%
\bibitem [{\citenamefont {Shastry}(1989)}]{PhysRevLett.63.1288}%
  \BibitemOpen
  \bibfield  {author} {\bibinfo {author} {\bibfnamefont {B.~S.}\ \bibnamefont
  {Shastry}},\ }\bibfield  {title} {\bibinfo {title} {{$t$-{$J$} model and
  nuclear magnetic relaxation in high-${\mathrm{T}}_{\mathrm{c}}$ materials}},\
  }\href {https://doi.org/10.1103/PhysRevLett.63.1288} {\bibfield  {journal}
  {\bibinfo  {journal} {Phys. Rev. Lett.}\ }\textbf {\bibinfo {volume} {63}},\
  \bibinfo {pages} {1288} (\bibinfo {year} {1989})}\BibitemShut {NoStop}%
\bibitem [{\citenamefont {Wang}\ \emph
  {et~al.}(2023{\natexlab{b}})\citenamefont {Wang}, \citenamefont {Ding},
  \citenamefont {Schattner}, \citenamefont {Huang}, \citenamefont {Moritz},\
  and\ \citenamefont {Devereaux}}]{doi:10.1126/science.ade3232}%
  \BibitemOpen
  \bibfield  {author} {\bibinfo {author} {\bibfnamefont {W.~O.}\ \bibnamefont
  {Wang}}, \bibinfo {author} {\bibfnamefont {J.~K.}\ \bibnamefont {Ding}},
  \bibinfo {author} {\bibfnamefont {Y.}~\bibnamefont {Schattner}}, \bibinfo
  {author} {\bibfnamefont {E.~W.}\ \bibnamefont {Huang}}, \bibinfo {author}
  {\bibfnamefont {B.}~\bibnamefont {Moritz}},\ and\ \bibinfo {author}
  {\bibfnamefont {T.~P.}\ \bibnamefont {Devereaux}},\ }\bibfield  {title}
  {\bibinfo {title} {{The Wiedemann-Franz law in doped \uppercase{M}ott
  insulators without quasiparticles}},\ }\href
  {https://doi.org/10.1126/science.ade3232} {\bibfield  {journal} {\bibinfo
  {journal} {Science}\ }\textbf {\bibinfo {volume} {382}},\ \bibinfo {pages}
  {1070} (\bibinfo {year} {2023}{\natexlab{b}})}\BibitemShut {NoStop}%
\bibitem [{\citenamefont {Bergeron}\ and\ \citenamefont
  {Tremblay}(2016)}]{PhysRevE.94.023303}%
  \BibitemOpen
  \bibfield  {author} {\bibinfo {author} {\bibfnamefont {D.}~\bibnamefont
  {Bergeron}}\ and\ \bibinfo {author} {\bibfnamefont {A.-M.~S.}\ \bibnamefont
  {Tremblay}},\ }\bibfield  {title} {\bibinfo {title} {Algorithms for optimized
  maximum entropy and diagnostic tools for analytic continuation},\ }\href
  {https://doi.org/10.1103/PhysRevE.94.023303} {\bibfield  {journal} {\bibinfo
  {journal} {Phys. Rev. E}\ }\textbf {\bibinfo {volume} {94}},\ \bibinfo
  {pages} {023303} (\bibinfo {year} {2016})}\BibitemShut {NoStop}%
\end{thebibliography}%

\end{document}